\documentclass[11pt]{article}% добавляется twoside для двусторонней печати

\usepackage[T1]{fontenc}
\usepackage[cp1251]{inputenc}
\usepackage{textcomp}
\usepackage[centertags]{amsmath}
\usepackage[mediummath]{nccmath}
\usepackage{amsfonts}
\usepackage{amssymb}
\usepackage[pdftex]{hyperref}
\usepackage{graphicx}
\usepackage{graphbox}
\usepackage[numbers,sort&compress]{natbib}% упорядочивает ссылки

\usepackage{paperinitial}% Установка параметров страницы

\paperinitialization{15mm}{15mm}{15mm}{15mm}{2pt}{10pt}

\DeclareMathOperator{\re}{Re}
\DeclareMathOperator{\im}{Im}
\DeclareMathOperator{\sgn}{sgn}

\DeclareMathOperator{\Texp}{Texp}
\DeclareMathOperator{\Sp}{Sp}
\DeclareMathOperator{\rot}{rot}
\DeclareMathOperator{\diag}{diag}

\newcommand{\lan}{\langle}
\newcommand{\ran}{\rangle}
\newcommand{\bs}{\boldsymbol}

\newcommand{\e}{\varepsilon}
\newcommand{\vf}{\varphi}

\newcommand{\s}{\sigma}

\newcommand{\al}{\alpha}
\newcommand{\be}{\beta}
\newcommand{\ga}{\gamma}

\newcommand{\de}{\delta}
\newcommand{\De}{\Delta}

\newcommand{\la}{\lambda}

\newcommand{\spx}{\mathbf{x}}
\newcommand{\spy}{\mathbf{y}}

\newcommand{\spk}{\mathbf{k}}
\newcommand{\spe}{\mathbf{e}}
\newcommand{\xio}{\xi^{\text{out}}}
\newcommand{\xii}{\xi^{\text{in}}}
\newcommand{\R}{\mathbb{R}}

\begin{document}
%\sle
\allowdisplaybreaks[4]% позволяет переносить многострочные формулы
\frenchspacing% уменьшение пробелов после запятых
\setlength{\unitlength}{1pt}% устанавливает единицу длины в окружении picture

%Scattering of photons by helical media
%Scattering of plane-wave and twisted photons by helical media
\title{{\Large\textbf{Scattering of plane-wave and twisted photons by helical media}}}

\date{}

\author{P.O. Kazinski\thanks{E-mail: \texttt{kpo@phys.tsu.ru}},\;
P.S. Korolev\thanks{E-mail: \texttt{kizorph.d@gmail.com}}\\[0.5em]
{\normalsize  Physics Faculty, Tomsk State University, Tomsk 634050, Russia}}

\maketitle

\begin{abstract}

By using quantum electrodynamics in a dispersive medium, we describe scattering of plane-wave and twisted photons by a slab made of a helical medium, the helix axis being normal to the slab plane and the medium being not translation invariant in this plane, in general. In the particular cases, the permittivity tensor of a helical medium corresponds to cholesteric liquid crystals, $C^*$-smectics, biaxial chiral nematics and smectics, $q$-plates, chiral sculptured thin films, and helical dislocations. Both perturbative and nonperturbative approaches are considered. The explicit expressions for scattering amplitudes, probabilities, and Stokes parameters of photons are found taking into account the form of the photon wave packet. The selection rules are established showing that the helical medium transfers the momentum and the angular momentum to scattered photons. This property can be employed for production of twisted photons with large projection of the total angular momentum. We describe the device for shifting the projection of the total angular momentum of a photon and the principal scheme for signal coding in terms of twisted photons.

\end{abstract}

\section{Introduction}

Twisted photons, which are the excitations of a quantum electromagnetic field with given projections of the total angular momentum different from their helicity, provide a promising tool for parallel signal coding in (quantum) telecommunication \cite{Roadmap16,New18,OAMPM}. They can also be used in optical tweezers and in manipulating of rotational degrees of freedom of nano-particles, molecules, atoms, and nuclei \cite{PadgOAM25,Roadmap16,SerboNew,New19}. As for detection of twisted photons in the optical range, at the present moment there are elaborated techniques allowing one to decompose an almost arbitrary electromagnetic wave in terms of twisted photons even at a single photon level \cite{BLCBP,RGMMSCFR,Walsh,RGMMSCFR2}. In this spectral range, the twisted photons can routinely be produced by numerous means (see, for review, \cite{Barboza2,Fang21,LiZX21,SerboNew,PadgOAM25,OAMPM,Roadmap16}) but on-chip miniaturization of their conversion or generation process resulting in large angular momenta remains an open issue. Apart from the general theory of scattering of photons by helical media, one of the aims of the present paper is to propose the mechanism for generation of twisted photons with large projections of the total angular momentum that can be miniaturized to chip scales, at least, in principle.

The helical media, i.e., the media with permittivity tensor invariant under rotation and simultaneous shift along the rotation axis, are ubiquitous in nature and can be artificially fabricated down to nanometer scale and below. Some of the examples of such media are cholesteric liquid crystals, chiral nematics and smectics \cite{Barboza2,deGennProst,YangWu06,BelyakovBook,VetTimShab20}, $q$-plates \cite{LiZX21,Barboza2,MarManPap06,Barboza1,Naidoo16,Brasselet18}, certain types of chiral metamaterials and chiral sculptured thin films \cite{LakhMess05,MacLakh,AskZhaAl14,FarLakh14,SemKhakh,HTTFB05,RobBroeBret,HodgWu01,SitBroeBret00,HPSB05,SchmSchuSchu13,OhHess15}, and helical defects in ordered media \cite{Weert57,GrilheThes,Friedel64,KRS21}. Various aspects of the electromagnetic properties of these structures were thoroughly investigated \cite{deGennProst,YangWu06,BelyakovBook,LakhMess05,MacLakh,Barboza2,LiZX21,MarManPap06,KPMS09} but the complete quantum theory of scattering of photons by such structures of a general form has not been constructed yet, especially for scattering of twisted photons. In \cite{BKL5}, the general arguments based on Feynman diagrams and conservation laws were given showing that helical media can be employed for increasing the projection of the total angular momentum of a scattered photon. In the present paper, we develop this idea and describe scattering of twisted photons by such media. The results of our study confirm the general observation made in \cite{BKL5} and pave the way for construction of devices producing twisted photons with large projection of the total angular momentum. As for scattering of plane-wave photons by helical media, we find several properties of this process that have been unknown before, to our knowledge. The numerical simulations of scattering of plane-wave and twisted photons that we also present in the paper corroborate the analytical results.

In particular, developing the perturbation theory with respect to dielectric susceptibility, we find the explicit expressions for the scattering amplitudes and probabilities of plane-wave and twisted photons in the first Born approximation. The generalization to higher orders of perturbation theory is straightforward. As regards the plane-wave photons, we restrict our consideration to the case of the helical medium invariant under translations in the plane normal to the axis of the helical symmetry. The shape of the wave packet of the initial photon is assumed to be arbitrary whereas the detected photon is plane. The explicit expressions for the scattering amplitude of twisted photons reveal the selection rules reading that the medium transfers the momentum and the angular momentum to the photon. We also investigate this scattering process nonperturbatively by solving the Maxwell equations for the mode functions of the quantum electromagnetic field. As is known \cite{LakhWeigh95}, in the paraxial limit, the Maxwell equations are exactly solvable for a helical medium invariant with respect to translations perpendicular to the helical symmetry axis. We analyze the band gaps and the polarization properties of the modes in this case. In particular, we obtain that, in addition to the well-known real band gaps of the photon dispersion law, there exist peculiarities of the photon spectrum that we refer to as the imaginary band gaps. These imaginary band gaps are related to the branch points of the Riemann surface of the dispersion law that lie out of the real axis of the photon energy complex plane. The imaginary band gaps appear for a general helical medium. For example, they do not exist in the dispersion law of photons propagating in $C^*$-smectics. The properties of scattered twisted photons change rapidly at energies near these branch points when the branch points are close to the real axis.

The paper is organized as follows. In Sec. \ref{Gener_Form}, we derive the general form of the permittivity tensor possessing the helical symmetry. In Sec. \ref{Kin_Approach}, we evolve the perturbative approach to scattering of plane-wave and twisted photons by the slab made of the helical medium and obtain the explicit expressions for the scattering amplitudes. Section \ref{Exact_Solut} is devoted to a nonperturbative approach to the scattering problem for the case of the helical medium invariant under translations perpendicular to the helical symmetry axis. In particular, in Sec. \ref{Num_Sim}, the numerical procedure for simulation of the scattering process is described. The results of the numerical simulation are given in Appendix \ref{Scat_Data_Ap}. In Conclusion we summarize the results.

Throughout the text we the system of units such that $\hbar=c=1$ and $e^2=4\pi\al$, where $\al$ is the fine structure constant. We also use interchangeably the notation for the axes $1$, $2$, $3$ and $x$, $y$, $z$.

\section{Helical medium}\label{Gener_Form}

Let us find a general form of the permittivity tensor possessing the helical symmetry. To this end, we introduce the operator of the total angular momentum
\begin{equation}\label{ang_mom}
\begin{split}
    J_{l ij}&:=L_{l ij}+S_{l ij},\\
    L_{l ij}&:=\e_{lmn}x_m k_n\de_{ij},\qquad S_{l ij}:=-i\e_{l ij},\qquad k_n:=-i\partial_n,
\end{split}
\end{equation}
where the index $l$ numerates the components of the angular momentum operators, $L_{lij}$ is the operator of an orbital angular momentum, and $S_{l ij}$ is the operator of a photon spin. Denote the operator of rotation around the $z$ axis by an angle of $\psi$ as
\begin{equation}
    R_\psi=R^L_\psi R^S_\psi=e^{i\psi J_3}=e^{i\psi L_3}e^{i\psi S_3}.
\end{equation}
Let $\{\spe_1,\spe_2,\spe_3\}$ be a right-handed orthonormal basis. Then
\begin{equation}\label{epm_e3}
    S_3 \spe_\pm=\pm\spe_\pm,\qquad S_3\spe_3=0,
\end{equation}
where
\begin{equation}
    \spe_\pm:=\spe_1\pm i\spe_2.
\end{equation}
Any vector is decomposed in terms of the eigenvectors of the operator $S_3$ as
\begin{equation}\label{pm_basis}
    \spx=\frac12(x_-\spe_++x_+\spe_-)+x_3\spe_3.
\end{equation}

Let $T_{a}$ be the operator of translations along the $z$ axis: $z\rightarrow z+a$. Then the permittivity tensor, $\e_{ij}(k_0;\spx)$, possesses the helical symmetry provided that
\begin{equation}\label{hel_permit}
    R_\psi T_{\psi/q} \e (R_\psi T_{\psi/q})^{-1}=\e,\quad\forall\psi\in \R.
\end{equation}
Henceforth, the tensor indices of $\e$ are not explicitly shown. Since $R_{2\pi n}=1$, $n\in \mathbb{Z}$, such a permittivity tensor is periodic with the period $2\pi/q$. Representing $\e$ as a Fourier series with respect to the variable $z$ and imposing the condition \eqref{hel_permit}, we arrive at
\begin{equation}
    \e=\sum_{l,s=-\infty}^\infty e^{il\vf}e^{-iq(l+s)z}\e_{ls},
\end{equation}
where $\vf:=\arg x_+$ and $\e_{ls}$ depends only on $r:=|x_+|$. Note that the indices $l$ and $s$ in this formula are not the spatial ones but specify the orbital and spin angular momenta, respectively. Furthermore, the following relation must hold
\begin{equation}\label{irred_cond}
    R_\psi\e_{ls}R^{-1}_\psi= R^S_\psi\e_{ls}(R^{S}_\psi)^{-1}=e^{is\psi}\e_{ls},
\end{equation}
i.e., the tensor $\e_{ls}$ is an irreducible tensor of the spin $s$ of the group of rotations around the $z$ axis (see also \cite{BelyakovBook,PontReyOld02}).

It is not difficult to find the explicit expressions for $\e_{ls}$. These tensors are constructed as the tensor products of the corresponding spin of the vectors $\spe_\pm$ and $\spe_3$. The spins of the vectors $\spe_\pm$ and $\spe_3$ are given in formula \eqref{epm_e3}. It follows from \eqref{irred_cond} for $|s|>2$ that
\begin{equation}
    \e_{ls}=0.
\end{equation}
For $s=\pm2$, we have
\begin{equation}\label{eps_hel_s2}
    \e_{l,\pm2}=A^\pm_{l}\spe_\pm\otimes\spe_\pm,
\end{equation}
respectively. Notice that the tensors \eqref{eps_hel_s2} have the form of the polarization tensors of a gravitational wave, as expected. For $s=\pm1$, we obtain
\begin{equation}\label{eps_hel_s1}
    \e_{l,\pm1}=\al^\pm_l\spe_\pm\otimes\spe_3+\be^\pm_l\spe_3\otimes\spe_\pm.
\end{equation}
For $s=0$, we come to
\begin{equation}\label{eps_hel_s0}
    \e_{l0}=\e_l\spe_+\spe_- +y_l\spe_+\wedge\spe_- +\e_{\perp l}\spe_3\otimes\spe_3,
\end{equation}
where
\begin{equation}
    \mathbf{a}\mathbf{b}:=\frac12(\mathbf{a}\otimes\mathbf{b}+\mathbf{b}\otimes\mathbf{a}),\qquad
    \mathbf{a}\wedge\mathbf{b}:=\frac12(\mathbf{a}\otimes\mathbf{b}-\mathbf{b}\otimes\mathbf{a}).
\end{equation}
Demanding that the medium is transparent and so the permittivity tensor $\e$ is Hermitian, we are left with
\begin{equation}\label{eps_hel_herm}
\begin{aligned}
    \e_{l2}&=A_{l}\spe_+\otimes\spe_+,&\qquad \e_{l,-2}&=A^*_{-l}\spe_-\otimes\spe_-,\\
    \e_{l1}&=\al_l\spe_+\otimes\spe_3+\be_l\spe_3\otimes\spe_+,&\qquad \e_{l,-1}&=\be^*_{-l}\spe_-\otimes\spe_3+\al^*_{-l}\spe_3\otimes\spe_-.
\end{aligned}
\end{equation}
As far as $\e_{l0}$ is concerned, the Hermiticity condition implies $\e_l=\e^*_{-l}$, $y_l=y^*_{-l}$, and $\e_{\perp l}=\e^*_{\perp,- l}$.

If the components of the permittivity tensor are infinitely smooth functions in the vicinity of the $z$ axis, then all the coefficients with the index $l$ in formulas \eqref{eps_hel_s2}, \eqref{eps_hel_s1}, \eqref{eps_hel_s0}, and \eqref{eps_hel_herm} should have the form
\begin{equation}
    c_l(r)=r^{|l|} f_l(r^2),
\end{equation}
where $c_l(r)$ is a shorthand notation for all these coefficients and $f_l(x)$ are some infinitely smooth functions in the neighborhood of the point $x=0$. Of course, the functions $f_l(x)$ are, in general, different for the different coefficients in the abovementioned formulas. In particular, if the helical medium is also invariant with respect to translations in the $(x,y)$ plane, then all the components of the permittivity tensor with $l\neq0$ vanish.

As a result, the general expression for the permittivity tensor of a transparent medium possessing a helical symmetry is written as
\begin{equation}\label{eps_hel_gen}
    \e=\sum_{s=-2}^2\sum_{l=-\infty}^\infty e^{il\vf}e^{-iq(l+s)z}\e_{ls}.
\end{equation}
The irreducible tensors of the spin $s$ are given in formulas \eqref{eps_hel_s0} and \eqref{eps_hel_herm}. Notice that the permittivity tensor of $q$-plates \cite{MarManPap06} is a particular case of \eqref{eps_hel_gen} with $q=0$. In this case, the variable part of the permittivity  tensor corresponds to $|s|=2$ and $A_l=0$ except $l=2q_p$, where $q_p$ is the parameter of the $q$-plate\footnote{Hereinafter, we denote the parameter of a $q$-plate as $q_p$ since the letter $q$ has been already reserved for the periodicity parameter of a helical medium.}.

Let us dwell on the case of a helical medium with the permittivity tensor invariant with respect to translations perpendicular to the $z$ axis. As it was mentioned above, in that case the terms with $l=0$ should only be kept in formulas \eqref{eps_hel_s0}, \eqref{eps_hel_herm}, and \eqref{eps_hel_gen}. Besides, $\e_0$, $y_0$, and $\e_{\perp 0}$ are to be real numbers. Then
\begin{equation}\label{permittivity_comps}
\begin{gathered}
    \e_{02}=A
    \left[
      \begin{array}{ccc}
        1 & i & 0 \\
        i & -1 & 0 \\
        0 & 0 & 0 \\
      \end{array}
    \right],\qquad
    \e_{0,-2}=A^*
    \left[
      \begin{array}{ccc}
        1 & -i & 0 \\
        -i & -1 & 0 \\
        0 & 0 & 0 \\
      \end{array}
    \right],\\
    \e_{01}=
    \left[
      \begin{array}{ccc}
        0 & 0 & \al \\
        0 & 0 & i\al \\
        \be & i\be & 0 \\
      \end{array}
    \right],\qquad
    \e_{0,-1}=
    \left[
      \begin{array}{ccc}
        0 & 0 & \be^* \\
        0 & 0 & -i\be^* \\
        \al^* & -i\al^* & 0 \\
      \end{array}
    \right],\qquad
        \e_{00}=
    \left[
      \begin{array}{ccc}
        \e & -iy & 0 \\
        iy & \e & 0 \\
        0 & 0 & \e_\perp \\
      \end{array}
    \right],
\end{gathered}
\end{equation}
where for brevity the index $0$ at the components of the permittivity tensor is omitted. Such a permittivity tensor is positive-definite if and only if
\begin{equation}
    \e>0, \qquad\e_\perp>0,\qquad\e^2-y^2-4|A|^2>0, \qquad\det(\e_{ij})=\frac{\e_\perp}{4}(ac-|b|^2)>0,
\end{equation}
where
\begin{equation}\label{abc}
    a:=\e-y-2\frac{|\be|^2}{\e_\perp},\qquad b^*:=2\big(A-\frac{\al\be}{\e_\perp}\big),\qquad  c:=\e+y-2\frac{|\al|^2}{\e_\perp}.
\end{equation}
Many optical properties of the media with permittivity tensor of such a form were considered in \cite{LakhWeigh95,FarLakh14,MacLakh,LakhMess05,LaVeMcC00,BitThom05,FurAlex08,Lakht10,TChJhHu17}. In the particular case $\al=\be$ and $y=0$, this permittivity tensor describes the electromagnetic properties of $C^*$-smectics, whose optical characteristics were studied in \cite{BelyakovBook}. For $\al=\be=y=0$, this permittivity tensor corresponds to cholesterics \cite{Barboza2,deGennProst,YangWu06,BelyakovBook,VetTimShab20}.

In the general case, one can provide the following physical interpretation to the permittivity tensor with components \eqref{permittivity_comps}. Let us introduce the vector $\spe_3$, the vector $\bs\tau(z)$ invariant under the action of the helical symmetry
\begin{equation}
    \bs\tau(z)=(\cos(qz), \sin(qz), \tau_3),\qquad \tau_3=const,
\end{equation}
and the vector $\bs\xi=[\spe_3,\bs\tau]$, where the square brackets denote the cross product of vectors. Then the tensor \eqref{eps_hel_gen} with $l=0$ is constructed as a linear combination of every possible tensor product of these vectors
\begin{equation}\label{eps_biax}
    \e=b_1\bs\tau\bs\tau+b_2\bs\xi\bs\xi+b_3\spe_3\spe_3+b_4\bs\tau\bs\xi+b_5\bs\tau\bs\spe_3+b_6\bs\xi\bs\spe_3+ib_7\bs\tau\wedge\bs\xi +ib_8\bs\tau\wedge\spe_3+ib_9\bs\xi\wedge\bs\spe_3,
\end{equation}
where $b_k\in \R$, $k=\overline{1,9}$, are some arbitrary parameters. As long as $\tau_3=(\bs\tau\spe_3)$, the parameter $\tau_3$ in \eqref{eps_biax} can be made arbitrary and be set, for example, to zero. Evidently, expression \eqref{eps_biax} is invariant with respect to the helical symmetry and contains the same number of independent real parameters as the tensor \eqref{eps_hel_gen} for $l=0$. Therefore, these tensors coincide. Such a permittivity tensor is realized for the medium with the structure similar to a $C^*$-smectic where the effective permittivity ellipsoids of molecules or nano-objects constituting the medium have three different axes and these axes are the same for all the points of the layer at fixed $z$. For example, this permittivity tensor describes the electromagnetic properties of biaxial chiral nematics and smectics \cite{deGennProst} or of chiral sculptured thin films \cite{LakhMess05}.

\section{Scattering by a helical medium}
\subsection{Kinematic approach}\label{Kin_Approach}

Consider scattering of the electromagnetic wave by the helical medium constituting a thin plate orthogonal to the $z$ axis. For definiteness, we suppose that the plate is placed at $z\in[-L/2,L/2]$, where $L$ is the plate thickness, in the homogeneous isotropic transparent medium with permittivity $\e_v(k_0)$. The fact that the plate is thin allows one to describe such scattering perturbatively with respect to the dielectric susceptibility
\begin{equation}
    \chi_{ij}(k_0):=\e_{ij}(k_0)-\e_v(k_0)\de_{ij}.
\end{equation}
To describe such scattering, it is convenient to construct quantum electrodynamics in the Coulomb gauge (see for details, e.g., \cite{BKL5,AbrGorDzyal,GinzbThPhAstr,BKL2,parax,wkb_chol,KazLaz20}). The interaction Hamiltonian in the interaction representation is given by
\begin{equation}
    \hat{H}_{int}=-\frac12\int d\spx \hat{E}_i\chi_{ij}(\hat k_0) \hat{E}_j,\qquad \hat{k}_0:=i\partial_t,
\end{equation}
where $\hat{E}_i=\dot{\hat{A}}_i$ is the electric field strength operator and $\hat{A}_i$ is the electromagnetic potential operator
\begin{equation}\label{field_oper}
    \hat{A}_i(t,\spx)=\sum_\al \hat{c}_\al f_\al^{1/2} \psi_{\al i}(\spx) e^{-ik_{0_\al}t}+ \sum_\al \hat{c}^\dag_\al f_\al^{1/2} \psi^*_{\al i}(\spx) e^{ik_{0_\al}t}.
\end{equation}
The normalization coefficients take the form
\begin{equation}
    f^{-1}_\al=[k_0^2\e_v(k_0)]'|_{k_0=k_{0\al}}=2 [k_0 n_v(k_0) c_v^{-1}(k_0)]_{k_0=k_{0\al}},\qquad c_v^{-1}(k_0)=[k_0 n_v(k_0)]',
\end{equation}
where $n_v(k_0):=\e_v^{1/2}(k_0)$ is the refraction index of the medium where the plate is placed and $c_v$ is the group velocity of the electromagnetic waves in this medium. The creation-annihilation operators obey the standard commutation relations,
\begin{equation}
    [\hat{c}_\al,\hat{c}_\be]=0,\qquad [\hat{c}_\al,\hat{c}^\dag_\be]=\de_{\al\be},
\end{equation}
and the mode functions are the stationary solutions of the Maxwell equations. The mode functions constitute a complete orthonormal set in the Hilbert space of complex divergence-free vector fields
\begin{equation}
    \lan \psi_\al,\psi_\be\ran=\de_{\al\be},\qquad\sum_\al \psi_{\al i}(\spx) \psi^*_{\al j}(\spy)=\de^\perp_{ij}(\spx-\spy):=(\de_{ij}-\partial_i\De^{-1}\partial_j)\de(\spx-\spy).
\end{equation}
Hereinafter we imply the ``box'' normalization of the mode functions.

Let the photons be prepared at the instant of time $t=t_1$ in the state
\begin{equation}\label{in_state}
    |\vf_{\text{in}}\ran:=\sum_\ga e^{-ik_{0\ga}t_1}\vf^\text{in}_\ga\hat{c}^\dag_\ga|0\ran,\qquad \sum_\ga (\vf^\text{in}_\ga)^*\vf^\text{in}_\ga=1,
\end{equation}
and be recorded at the instant of time $t=t_2$ in the state
\begin{equation}\label{out_state}
    |\vf_{\text{out}}\ran:=\sum_{\ga'} e^{-ik_{0\ga'}t_2}\vf^\text{out}_{\ga'}\hat{c}^\dag_{\ga'}|0\ran,\qquad\sum_\ga (\vf^\text{out}_\ga)^* \vf^\text{out}_\ga=1.
\end{equation}
The functions $\vf^\text{in}_\ga$, $\vf^\text{out}_\ga$ specify the forms of the corresponding wave packets at the instant of time $t=0$. Then taking into account the commutation relation
\begin{equation}
    \hat{U}^0_{0,t}\hat{c}_\al \hat{U}^0_{t,0}=e^{-ik_{0\al}t} \hat{c}_\al,
\end{equation}
and the relation between the evolution operator and the $S$-operator
\begin{equation}
    \hat{U}_{t_2,t_1}=\hat{U}^0_{t_2,0}\hat{S}_{t_2,t_1}\hat{U}^0_{0,t_1},\qquad \hat{S}_{t_2,t_1}=\Texp\Big\{-i\int_{t_1}^{t_2}dt \hat{H}_{int}(t)\Big\},
\end{equation}
where $\hat{U}^0_{t_2,t_1}$ is the free evolution operator, we obtain in the first Born approximation
\begin{equation}\label{amplitude_gen}
    \lan\vf_{\text{out}}|\hat{U}_{t_2,t_1}|\vf_{\text{in}}\ran=e^{-i E_0(t_2-t_1)}\sum_{\ga',\ga}(\vf^\text{out}_{\ga'})^* \Big[\de_{\ga'\ga} +ik_{0\ga'} f_{\ga'}^{1/2} k_{0\ga}f_{\ga}^{1/2}\int_{t_1}^{t_2} dt d\spx e^{it (k_{0\ga'}-k_{0\ga})} \psi^\dag_{\ga'}\chi(k_{0\ga})\psi_\ga +\cdots\Big]\vf^\text{in}_\ga,
\end{equation}
where $E_0$ is the energy of the vacuum state and the ellipses denote the higher terms of the perturbation theory that are proportional to $\chi$ raised to a higher power. Henceforth, we suppose that $E_0=0$ and do not write the omission points.

The choice of the initial and final states in the form \eqref{in_state}, \eqref{out_state} allows one to pass readily to the limit $t_2\rightarrow+\infty$, $t_1\rightarrow-\infty$ in expression \eqref{amplitude_gen}. As a result, we have
\begin{equation}\label{trans_ampl_1born}
    \lan\vf_{\text{out}}|\hat{U}_{\infty,-\infty}|\vf_{\text{in}}\ran_{\text{1st Born}}= \sum_\ga(\vf^\text{out}_{\ga})^* \vf^\text{in}_\ga +2\pi i\sum_{\ga',\ga}k_{0\ga}^2f_{\ga}\de(k_{0\ga'}-k_{0\ga})\int d\spx (\vf^\text{out}_{\ga'})^*  \psi^\dag_{\ga'}\chi(k_{0\ga})\psi_\ga \vf^\text{in}_\ga.
\end{equation}
As is seen, the leading nontrivial contribution to scattering is determined by the amplitude
\begin{equation}\label{scat_ampl}
    F_{\ga'\ga}:=2\pi i k_{0\ga}^2f_{\ga}\de(k_{0\ga'}-k_{0\ga})\int d\spx\psi^\dag_{\ga'}\chi(k_{0\ga})\psi_\ga,
\end{equation}
where the mode functions are taken at the same photon energy $k_{0\ga}$. It is not difficult to write out the higher terms of the perturbation series with respect to the dielectric susceptibility $\chi$.

\subsubsection{Scattering of plane-wave photons}

Let us start an investigation of the transition amplitude \eqref{trans_ampl_1born} with scattering of plane-wave photons. We restrict ourselves to the case of scattering by a helical medium with permittivity tensor invariant under translations in the $(x,y)$ plane. In the particular case of $C^*$-smectics, such an analysis was carried out in \cite{BelyakovBook}, see also \cite{FurAlex08}. The mode functions in the expansion \eqref{field_oper} take the form
\begin{equation}\label{mode_func_med}
    \psi_\ga(\spx)=\frac{\mathbf{f}_\la(\spk)}{\sqrt{V}}e^{i\spk\spx},
\end{equation}
where $\mathbf{f}_\la(\spk)$ are the photon polarization vectors
\begin{equation}\label{polariz_vect}
    \mathbf{f}_1(\spk)=(\cos\phi\cos\theta,\sin\phi\cos\theta,-\sin\theta),\qquad \mathbf{f}_2(\spk)=(-\sin\phi,\cos\phi,0),
\end{equation}
and
\begin{equation}\label{n_defn}
    \spk=|\spk|\mathbf{n}=|\spk|(\sin\theta\cos\phi,\sin\theta\sin\phi,\cos\theta).
\end{equation}
The polarization vectors $\mathbf{f}_1$ and $\mathbf{f}_2$ specify the $\pi$- and $\s$-polarization, respectively. Furthermore,
\begin{equation}
    \sum_\ga\equiv\sum_\la\int\frac{Vd\spk}{(2\pi)^3},\qquad \de_{\ga'\ga}=\frac{(2\pi)^3}{V}\de_{\la'\la}\de(\spk'-\spk).
\end{equation}
The mass-shell condition becomes
\begin{equation}
    k_0^2\e_v(k_0)=\spk^2.
\end{equation}
Then the scattering amplitude \eqref{scat_ampl} is written as
\begin{equation}\label{amplitude_1}
    F_{\ga'\ga}=i\frac{(2\pi)^3k^2_0}{2V|k_3|}\sum_{r=\pm1}\de(k_3'-rk_3)\de(\spk_\perp'-\spk_\perp)\mathbf{f}^\dag_{\la'}(\spk') \tilde{\chi}\big((1-r)k_3\big) \mathbf{f}_{\la}(\spk),
\end{equation}
where
\begin{equation}\label{chi_fourier}
    \tilde{\chi}(k_3):=\int dz e^{ik_3z}\chi(z)=2\pi\sum_{s_h=-2}^2\de_L(k_3-qs_h)\chi_{0s_h},\qquad \de_L(k_3):=\frac{\sin(k_3L/2)}{\pi k_3},
\end{equation}
and
\begin{equation}
    (\chi_{l_hs_h})_{ij}:=(\e_{l_hs_h})_{ij}-\de_{l_h0}\de_{s_h0}\e_v\de_{ij}.
\end{equation}
The term with $r=1$ describes the wave transmitted through the plate, whereas the term with $r=-1$ is responsible for the reflected wave. Hereinafter, we suppose that $L$ is much larger than the photon wavelength such that $\de_L(k_3)$ is close to a delta function.

We also assume that the final photon state is an eigenstate of the momentum operator. This state is supposed to be mixed with respect to spin with the density matrix,
\begin{equation}\label{rho_out}
    \rho^{\text{spin}}_{\text{out}}=\frac12(1+\bs\s\bs\xi_{\text{out}}),
\end{equation}
defined in the basis of the polarization vectors \eqref{polariz_vect}, where $\bs\xi_{\text{out}}$ is the Stokes vector and $\s_i$ are the Pauli matrices. The corresponding density operator is given by
\begin{equation}
    \hat{R}_{out}=\sum_{\ga_1',\ga_2'} (\rho^{\text{spin}}_{\text{out}})_{\la'_1\la'_2} e^{-ik_{0\ga'_1}t_2} \frac{(2\pi)^3}{V}\de(\spk_{\ga'_1}-\spk')  \hat{c}^\dag_{\ga'_1}|0\ran\lan0|\hat{c}_{\ga'_2} \frac{(2\pi)^3}{V}\de(\spk_{\ga'_2}-\spk') e^{ik_{0\ga'_2}t_2}.
\end{equation}
The initial photon state is taken in the general form \eqref{in_state} with the assumption that $\vf^{\text{in}}_\la(\spk)$ vanish for $k_3<0$. In other words, before scattering, the photon wave packet moves from left to right. The presence of the three delta functions in the amplitude \eqref{amplitude_1} entails that the probability to record a photon is determined by the momentum space diagonal of the density matrix of the initial photon state
\begin{equation}\label{rho_in}
    \rho^\text{in}_{\la'\la}(\spk,\spk)=\vf^{\text{in}}_{\la'}(\spk) [\vf^{\text{in}}_\la(\spk)]^*.
\end{equation}
It is clear that \eqref{rho_in} does not depend on a common phase of the photon wave function, which can be nonconstant and defines, for example, the orbital angular momentum of the state. Let
\begin{equation}\label{c_defn}
    c_{\text{in}}(\spk):=\sum_\la \rho^\text{in}_{\la\la}(\spk,\spk),\qquad \int\frac{V d\spk}{(2\pi)^3} c_{\text{in}}(\spk)=1.
\end{equation}
Introduce the spin density matrix
\begin{equation}\label{rho_spin}
    (\rho^\text{spin}_\text{in})_{\la'\la}(\spk):=\rho^\text{in}_{\la'\la}(\spk,\spk)/c_{\text{in}}(\spk).
\end{equation}
This matrix is Hermitian, nonnegative definite, and possesses a unit trace. Consequently, it can be cast into the form
\begin{equation}
    \rho^\text{spin}_\text{in}=\frac12(1+\bs\s\bs\xi^{\text{in}}),
\end{equation}
where $\bs\xi_{\text{in}}$ is the effective Stokes vector, $|\bs\xi_{\text{in}}|\leqslant1$ and $|\bs\xi_{\text{in}}|=1$ for a pure state. In the case of a pure initial state, the explicit expression for $\bs\xi_{\text{in}}$ is given in \cite{KazLaz20}, Sec. 5.B. Later, for brevity, we will refer to $\bs\xii$ as the Stokes vector. Notice that the initial photon state can be mixed. In that case, one should suppose that $\rho^\text{in}_{\la'\la}(\spk,\spk)$ is the momentum space diagonal of the density matrix of the initial photon state, and $c_{\text{in}}$ and $\rho^\text{spin}_\text{in}$ are defined as in \eqref{c_defn} and \eqref{rho_spin}.

Consider separately the cases $k_3'>0$ and $k_3'<0$. For $k_3'>0$, the nonzero contribution to the amplitude \eqref{amplitude_1} comes from the term with $r=1$. Substituting \eqref{amplitude_1} into \eqref{trans_ampl_1born}, summing over $\ga$, and squaring the modulus of the result, we obtain the probability to record a photon in the state characterized by the spin density matrix \eqref{rho_out}:
\begin{equation}\label{dP_forward}
    dP(\spk)=\Sp\Big[\rho^\text{spin}_{\text{out}} \Big(1+\frac{i k^2_0}{2k_3} \mathbf{f}^\dag(\spk) \tilde{\chi}(0) \mathbf{f}(\spk) \Big) \rho^{\text{spin}}_{\text{in}}(\spk) \Big(1 -\frac{i k^2_0}{2k_3} \mathbf{f}^\dag(\spk) \tilde{\chi}^\dag(0) \mathbf{f}(\spk)\Big)\Big] c_{\text{in}}(\spk) \frac{Vd\spk}{(2\pi)^3},
\end{equation}
where the matrix multiplication is implied and the trace is carried out with respect to the indices enumerating the photon polarization vectors. Expression \eqref{dP_forward} does not depend on $V$ because $c_{\text{in}}(\spk)$ contains the factor $V^{-1}$. Supposing that the second term in the parenthesis in \eqref{dP_forward} is small in comparison with unity and saving only the term with $s_h=0$ in the expansion \eqref{chi_fourier}, we arrive at
\begin{equation}
    dP(\spk)\approx\frac12\Big\{1+(\bs\xi^{\text{out}}\bs\xi^\text{in}) -\frac{k_0^2L}{2k_3} \big[(\e-\e_\perp) [\bs\xi^\text{out},\bs\xi^\text{in}]_3 n_\perp^2 -2y[\bs\xi^\text{out},\bs\xi^\text{in}]_2 n_3\big]  \Big\} c_{\text{in}}(\spk) \frac{Vd\spk}{(2\pi)^3},
\end{equation}
where the square brackets denote the cross product of vectors.

Analogously, for $k_3'<0$, the nonzero contribution to the amplitude \eqref{amplitude_1} stems from the term with $r=-1$. So we have
\begin{equation}\label{dP_plane_refl}
    dP(\spk')= \Sp\Big[\rho^\text{spin}_{\text{out}} \mathbf{f}^\dag(\spk') \tilde{\chi}(2k_3) \mathbf{f}(\spk) \rho^{\text{spin}}_\text{in}(\spk) \mathbf{f}^\dag(\spk) \tilde{\chi}^\dag(2k_3) \mathbf{f}(\spk')\Big] c_{\text{in}}(\spk) \frac{k_0^4}{4k_3^2} \frac{Vd\spk'}{(2\pi)^3},
\end{equation}
where $\spk_\perp=\spk_\perp'$ and $k_3=-k_3'$. Substituting the expansion \eqref{chi_fourier} into \eqref{dP_plane_refl} and neglecting the cross terms, we obtain
\begin{equation}\label{dP_plane_refl1}%dP_forward,dP_plane_refl1
    dP(\spk')= \sum_{s_h=-2}^2 \de^2_L(2k_3-qs_h) H_{s_h}(\bs\xi^\text{out},\bs\xi^\text{in}) c_{\text{in}}(\spk) \frac{k_0^4}{4k_3^2} \frac{Vd\spk'}{2\pi}.
\end{equation}
where the following notation has been introduced
\begin{equation}
    H_{s_h}(\bs\xi_\text{out},\bs\xi_\text{in}):=\Sp\Big[\rho^\text{spin}_{\text{out}} \mathbf{f}^\dag(\spk') \chi_{0s_h} \mathbf{f}(\spk) \rho^{\text{spin}}_\text{in}(\spk) \mathbf{f}^\dag(\spk) \chi^\dag_{0s_h} \mathbf{f}(\spk')\Big].
\end{equation}
The main contribution to expression \eqref{dP_plane_refl1} is made by the terms with $qs_h>0$. Then
\begin{equation}\label{H21}
\begin{split}
    H_{\pm2}(\bs\xi^\text{out},\bs\xi^\text{in})=&\,\frac{|A|^2}4\big[1-\xi_3^{\text{out}} \pm2n_3'\xi_2^{\text{out}} +(n_3')^2(1+\xi_3^{\text{out}})\big] \big[1-\xi_3^{\text{in}} \pm2n_3'\xi_2^{\text{in}} +(n_3')^2(1+\xi_3^{\text{in}})\big],\\
    H_{1}(\bs\xi^\text{out},\bs\xi^\text{in})=&\,\frac{n^2_\perp}4\Big\{|\al|^2(1-\xio_3)(1+\xii_3) +|\be|^2(1+\xio_3)(1-\xii_3) +2\re(\al\be^*\xio_-\xii_-) +\\
    &+2n_3'\Big[|\al|^2\xio_2(1+\xii_3) +|\be|^2\xii_2(1+\xio_3)
    +\im\{\al\be^*[(1+\xii_3)\xio_-+(1+\xio_3)\xii_-]\}\Big]+\\
    & +(n_3')^2|\al-\be|^2 (1+\xio_3)(1+\xii_3) \Big\}.
\end{split}
\end{equation}
The expression for $H_{-1}$ is obtained from the expression for $H_{1}$ by the replacement $\al\leftrightarrow\be$ and $\xi^\text{out}_2\rightarrow-\xi^\text{out}_2$, $\xi^\text{in}_2\rightarrow-\xi^\text{in}_2$.

Let us consider some properties of expressions \eqref{H21}. In the particular case of $C^*$-smectics, the features of scattering of electromagnetic waves were studied in \cite{BelyakovBook}. As regards the contribution of the components of the susceptibility tensor with spin $s_h=\pm2$, we see that the maximum scattering is achieved at the Stokes vector of the initial pure state
\begin{equation}\label{Stokes_max_2}
    s_h=\pm2:\qquad\xii_1=0,\qquad\xii_2=\pm\frac{2n'_3}{1+(n'_3)^2},\qquad\xii_3=-\frac{1-(n'_3)^2}{1+(n'_3)^2}.
\end{equation}
The Stokes vector $\bs\xii$ with opposite sign leads to vanishing of the corresponding contribution to scattering probability. Inasmuch as the expression for $H_{\pm2}$ is symmetric under the replacement $\bs\xii\leftrightarrow\bs\xio$, the Stokes vector of the final state, $\bs\xio$, resulting in maximum scattering probability has the form \eqref{Stokes_max_2}. Notice that expression \eqref{Stokes_max_2} admits a geometric interpretation. Formula \eqref{Stokes_max_2} describes the stereographic projection of the Poincar\'{e} sphere from the north pole $(0,0,1)$ to the plane $(x,y)$ where the $y$ axis is identified with $n_3'$ or $-n_3'$ depending on the sign of $s_h$.

As far as the terms with $s_h=\pm1$ are concerned, we only mention some particular cases:
\begin{enumerate}
  \item If the electromagnetic wave with Stokes vector $\bs\xii=(0,0,-1)$ falls onto the plate made of a helical medium, then the maximum of scattering probability is realized at $\bs\xio=(0,0,1)$, i.e., the $\s$-polarization turns into the $\pi$-polarization. This is valid for both signs of $s_h=\pm1$;
  \item In the paraxial limit, $n_3=-n_3'\approx1$, the maximum of scattering probability for the photon with Stokes vector $\bs\xii=(0,-s_h,0)$ is achieved for the final photon state with Stokes vector $\bs\xio=(0,s_h,0)$;
  \item If $|\be/\al|\gg1$ or $|\be/\al|\ll1$, then the unpolarized electromagnetic wave, $\bs\xii=0$, grazing the plate of a helical medium, $n_3\approx0$, is reflected into the electromagnetic wave, which is to a high degree linearly polarized with Stokes vector $\bs\xio=(0,0,\pm s_h)$, where the upper sign is taken for $|\be/\al|\gg1$ and the lower sign is chosen for $|\be/\al|\ll1$. In the case $\al=\be$, i.e., for a $C^*$-smectic, the unpolarized light remains unpolarized in scattering by the susceptibility tensor component $\chi_{0,\pm1}$;
  \item For $\al=0$, $s_h=1$ or $\be=0$, $s_h=-1$, the maximum of photon reflection probability is realized at the Stokes vector \eqref{Stokes_max_2} for any $\bs\xio$, the sign in \eqref{Stokes_max_2} agreeing with the sign of $s_h=\pm1$. The photon reflection probability is zero for the incident photons with Stokes vector $\bs\xii$ opposite to \eqref{Stokes_max_2};
  \item For $\be=0$, $s_h=1$ or $\al=0$, $s_h=-1$, the maximum of scattering probability is achieved at the Stokes vector
    \begin{equation}\label{Stokes_max_3}
        \bs\xii=(0,0,1),
    \end{equation}
    for any $\bs\xio$. The photons with Stokes vector $\bs\xii$ opposite to \eqref{Stokes_max_3} are not scattered by the given component of the dielectric susceptibility within the approximations made in deriving \eqref{dP_plane_refl};
    \item If the Stokes vectors of the initial and final pure states of the photon are varied, then the maximum of scattering probability is realized at
    \begin{equation}\label{xii_xio_s1}
    \begin{split}
        \bs\xii&=(2n'_3\im(\al\be^*),2n'_3(|\be|^2-\re(\al\be^*)),|\al|^2-|\be|^2+(n'_3)^2|\al-\be|^2 )/d,\\
        \bs\xio&=(2n'_3\im(\al\be^*),2n'_3(|\al|^2-\re(\al\be^*)),|\be|^2-|\al|^2+(n'_3)^2|\al-\be|^2 )/d,\\
    \end{split}
    \end{equation}
    for $s_h=1$, and at
    \begin{equation}\label{xii_xio_sm1}
    \begin{split}
        \bs\xii&=-(2n'_3\im(\al\be^*),2n'_3(|\al|^2-\re(\al\be^*)),|\al|^2-|\be|^2-(n'_3)^2|\al-\be|^2 )/d,\\
        \bs\xio&=-(2n'_3\im(\al\be^*),2n'_3(|\be|^2-\re(\al\be^*)),|\be|^2-|\al|^2-(n'_3)^2|\al-\be|^2 )/d,\\
    \end{split}
    \end{equation}
    for $s_h=-1$, where
    \begin{equation}
        d:=\sqrt{\big(|\al|^2+|\be|^2+(n'_3)^2|\al-\be|^2\big)^2-4|\al\be|^2}.
    \end{equation}
    The contribution to the scattering probability vanishes in the case when $\bs\xii$, $\bs\xio$ take the form \eqref{xii_xio_s1}, \eqref{xii_xio_sm1}, where the sign of $\bs\xii$ or of $\bs\xio$ (but not of both) is flipped.
\end{enumerate}

\subsubsection{Scattering of twisted photons}

It follows from the general symmetry considerations and conservation laws given in \cite{BKL5} that scattering of twisted photons by a helical medium results in a change of the projection of the photon total angular momentum, $m$, onto the $z$ axis. Let us find the explicit expression for the probability to record a twisted photon in such scattering.

The mode functions of twisted photons in a homogeneous isotropic dispersive medium are written as \cite{BKL5,BKL6}
\begin{equation}\label{psi_decomp}
    \psi(s,m,k_3,k_\perp;\spx)=\frac12(\psi_-\spe_++\psi_+\spe_-)+\psi_3\spe_3,
\end{equation}
where
\begin{equation}\label{mode_func_an}
\begin{split}
    \psi_3(m,k_3,k_\perp;\spx)&=\sqrt{\frac{2k_0n_v}{RL_z}}\big(\frac{n_\perp}{2}\big)^{3/2}J_m(k_\perp|x_+|)e^{im\arg x_+ +ik_3x_3},\\ \psi_\pm(s,m,k_3,k_\perp;\spx)&=\frac{i n_\perp}{s\pm n_3}\psi_3(m\pm1,k_3,k_\perp;\spx),
\end{split}
\end{equation}
and the vector $\mathbf{n}$ is defined in \eqref{n_defn}. Furthermore,
\begin{equation}
    \sum_\ga\equiv\sum_{s=\pm1}\sum_{m=-\infty}^\infty\int_{-\infty}^\infty\frac{L_z dk_3}{2\pi}\int_0^\infty\frac{Rdk_\perp}{\pi},\qquad\de_{\ga'\ga}=\de_{s's}\de_{m'm} \frac{2\pi^2}{RL_z}\de(k_\perp'-k_\perp)\de(k_3'-k_3).
\end{equation}
The quantities $R$ and $L_z$ characterize the normalization volume and $s$ is the photon helicity.

\begin{figure}[tp]
\centering
\includegraphics*[width=0.49\linewidth]{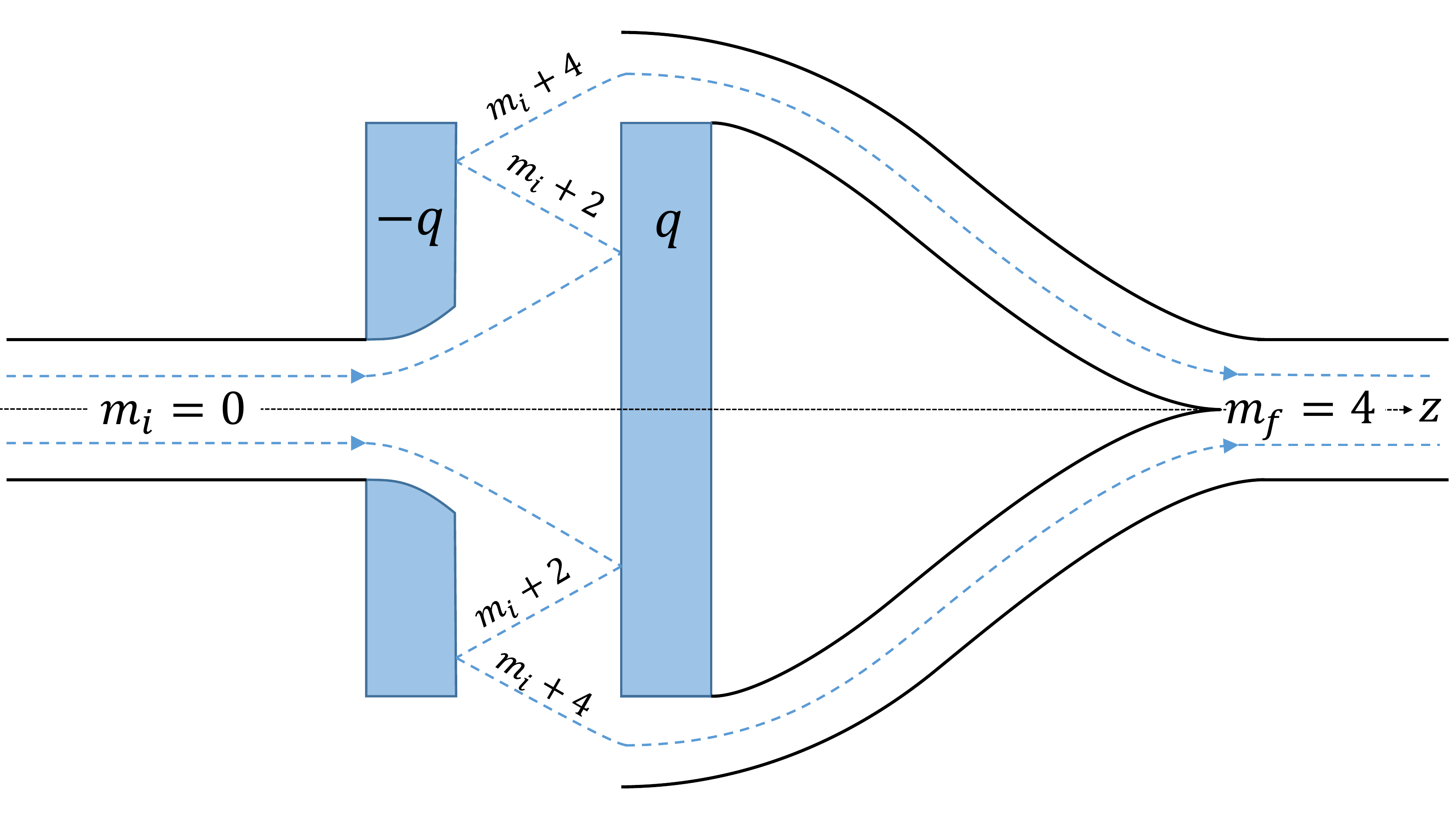}%\,
\caption{{\footnotesize A schematic representation of the axial section of the angular momentum shifting device (AMSD). This device consists of the two parallel plates made of helical media with opposite $q$, whereas the other parameters are the same. The plates are immersed into the homogeneous isotropic transparent medium with permittivity $\e_v(k_0)$. If one neglects the chirality of these plates, then the whole picture is invariant under rotations around the $z$ axis.}}
\label{Scheme_AMSD_plots}
\end{figure}

Then the amplitude \eqref{scat_ampl} becomes
\begin{equation}\label{amplitude_tw_0}
\begin{split}
    F_{\ga'\ga}=&\,i\frac{(2\pi)^3}{RL_z}\big(\frac{n'_\perp n_\perp}{4}\big)^{3/2} c_v k_0^2\de(k_0'-k_0)\times\\ &\times\sum_{l_h=-\infty}^\infty\sum_{s_h=-2}^2\de_L\big(k_3'-k_3+(l+s)q\big)\de_{m',m+l_h+s_h}\int_0^\infty drr a^\dag_{\ga'}(r)\chi_{l_hs_h}(r)a_{\ga}(r),
\end{split}
\end{equation}
where
\begin{equation}
    \psi_{\ga i}(\spx)=:\sqrt{\frac{2k_0n_v}{RL_z}}\big(\frac{n_\perp}{2}\big)^{3/2}e^{ik_3z}a_{\ga i}(x_+).
\end{equation}
Notice that the argument of the functions $a_{\ga i}(r)$ entering into expression \eqref{amplitude_tw_0} is real. The Kroneker delta in \eqref{amplitude_tw_0} comes from evaluation of the integral over $\arg x_+$. Taking together all the terms at the delta functions with the same argument and substituting the explicit expressions for $\chi_{l_hs_h}$, we obtain
\begin{equation}\label{amplitude_tw}
\begin{split}
    F_{\ga'\ga}=&\,i\frac{(2\pi)^3}{RL_z}\big(\frac{n'_\perp n_\perp}{4}\big)^{3/2} c_vk_0^2\de(k_0'-k_0)\sum_{m_h=-\infty}^\infty \de_L\big(k_3'-k_3+m_hq\big)\de_{m',m+m_h}\int_0^\infty drr\times\\
    &\times\Big\{\frac12\Big[\frac{n'_\perp n_\perp (\chi_{m_h}+y_{m_h})}{(s'-n'_3)(s-n_3)} J_{m'-1}J_{m-1} + \frac{n'_\perp n_\perp (\chi_{m_h}-y_{m_h})}{(s'+n'_3)(s+n_3)} J_{m'+1}J_{m+1}\Big] +\chi_{\perp m_h}J_{m'}J_m-\\
    &-\frac{in'_\perp\al_{m_h-1}}{s'-n_3'}J_{m'-1}J_m +\frac{in_\perp\be_{m_h-1}}{s+n_3}J_{m'}J_{m+1} -\frac{in'_\perp\be^*_{-m_h-1}}{s'+n_3'}J_{m'+1}J_m +\frac{in_\perp\al^*_{-m_h-1}}{s-n_3}J_{m'}J_{m-1}+\\
    &+\frac{n'_\perp n_\perp A_{m_h-2}}{(s'-n'_3)(s+n_3)} J_{m'-1}J_{m+1} +\frac{n'_\perp n_\perp A^*_{-m_h-2}}{(s'+n'_3)(s-n_3)} J_{m'+1}J_{m-1}  \Big\},
\end{split}
\end{equation}
where the Bessel functions whose indices contain $m$ depend on $k_\perp r$, whereas the Bessel functions whose indices contain $m'$ depend on $k'_\perp r$. The notation has also been introduced: $\chi_{l}:=\e_{l}-\e_v\de_{0l}$ and $\chi_{\perp l}:=\e_{\perp l}-\e_v\de_{0l}$. The terms on the second line of \eqref{amplitude_tw} describe the contribution of the component of the dielectric susceptibility tensor with spin $0$ to the scattering amplitude, the terms on the third line of \eqref{amplitude_tw} are responsible for the contribution of spin  $\pm1$ components, and the terms on the fourth line of \eqref{amplitude_tw} stem from the spin $\pm2$ components of the dielectric susceptibility tensor.

We see that scattering of photons by a helical medium leads to a transfer of the momentum component along the $z$ axis and of the projection of the total angular momentum onto this axis to a photon. By the same reasons, the photons passing through a $q$-plate acquire the additional angular momentum -- the medium transfers this angular momentum to the photon. However, in the case of $q$-plates, the parameter $q=0$ and so there is no transfer of the momentum component $k_3$ from the medium to the photon.

The selection rule \cite{BKL5},
\begin{equation}\label{sel_rule}
    k_3'=k_3-m_hq,\qquad m'=m+m_h,\quad m_h\in \mathbb{Z},
\end{equation}
is satisfied at the maximum of the scattering probability. Moreover, the energy conservation law, $k_0'=k_0$, relates the perpendicular momentum components of a photon before and after scattering. At the maximum of the scattering probability, we have
\begin{equation}\label{k_perp'}
    k_\perp'^2=k_\perp^2+m_hq(2k_3-m_hq).
\end{equation}
It is not difficult to verify that the selection rule \eqref{sel_rule} and the relation \eqref{k_perp'} hold on accounting for the higher orders of perturbation theory.

Consider some particular cases of the general formula \eqref{amplitude_tw}. It is clear that if $\chi_{l_hs_h}(r)$ are different from zero only in a small vicinity of the point $r_0$ and the extent of this vicinity, $\De r$, is such that
\begin{equation}
    k_\perp\De r\ll1,\qquad k'_\perp\De r\ll1,
\end{equation}
then the integral over $r$ in expression \eqref{amplitude_tw} is removed and all the coefficients whose index contains $m_h$ are replaced in accordance with the rule
\begin{equation}
    c_{l}(r)\rightarrow\lan c_{l}\ran:=\int_0^\infty drrc_{l}(r).
\end{equation}
The resulting formula describes scattering of twisted photons by a thin cylinder or a helix with the dielectric susceptibility $\chi$ obeying the helical symmetry. In particular, this formula is applicable for description of scattering of twisted photons by helical dislocations in crystals \cite{Weert57,GrilheThes,Friedel64,KRS21}.

\begin{figure}[tp]
\centering
\raisebox{-0.5\height}{\includegraphics*[width=0.49\linewidth]{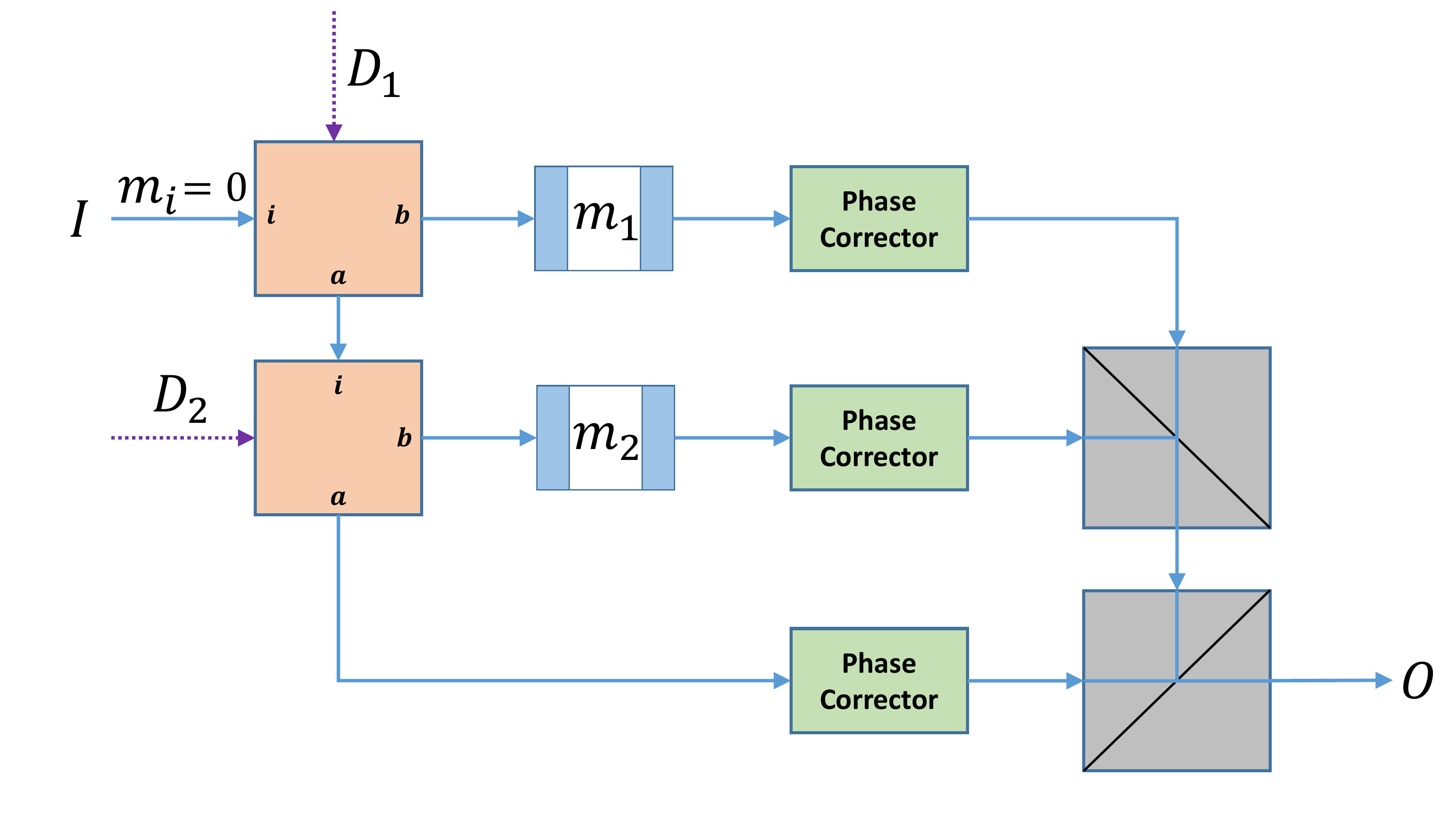}}\,
\raisebox{-0.5\height}{\includegraphics*[width=0.49\linewidth]{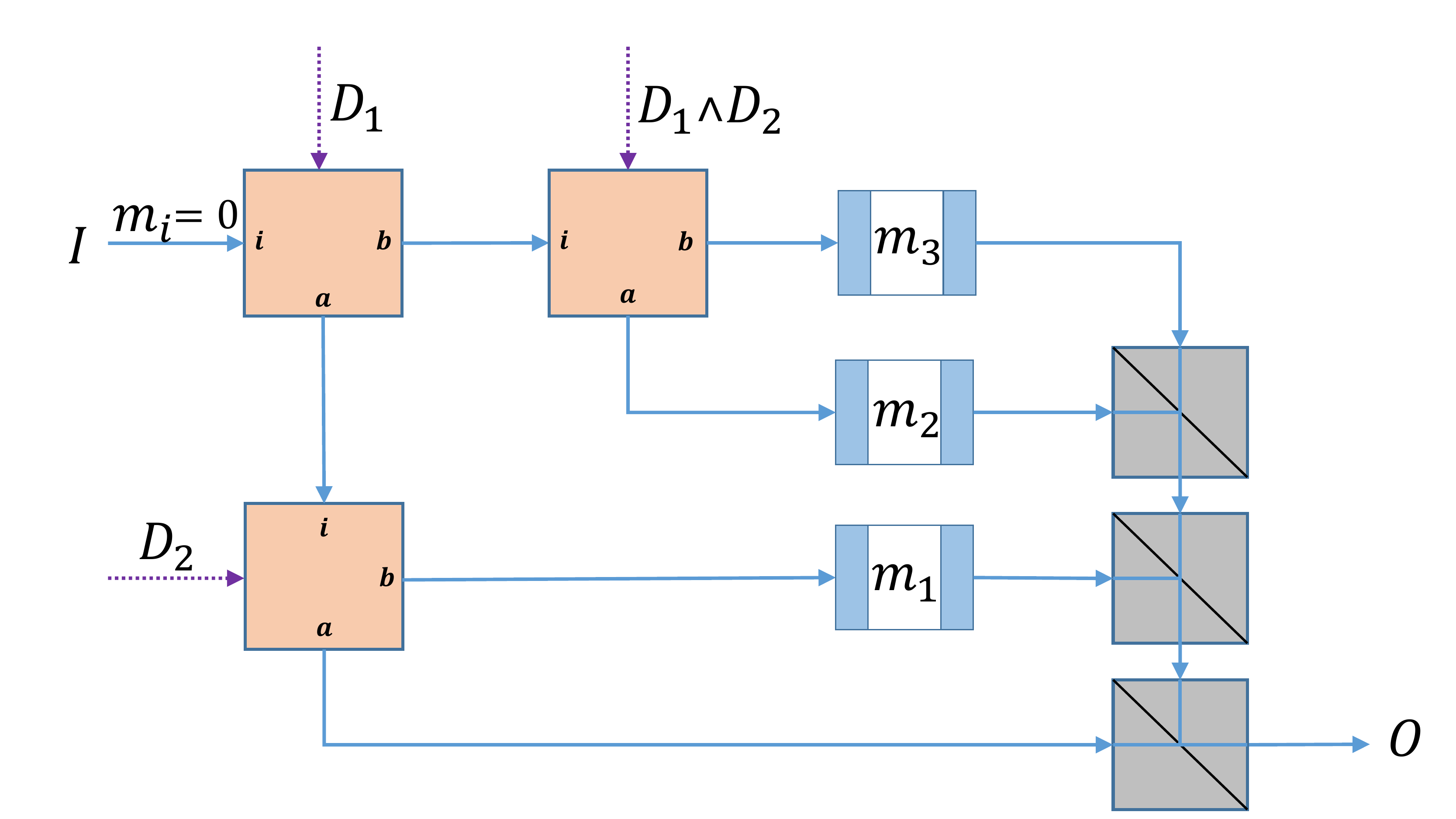}}
\caption{{\footnotesize The principal schemes for parallel coding of a given binary signal $D_1D_2$ by a single photon with a certain spectrum of the projections of the total angular momentum.}}
\label{Scheme_plots}
\end{figure}

The other particular case is scattering of a twisted photon by a helical medium invariant with respect to translations in the $(x,y)$ plane. In this case, only the terms standing at the coefficients $c_{l}$ with $l=0$ are left in expression \eqref{amplitude_tw} and these coefficients are to be independent of $r$. For brevity, we will not mark these coefficients by the index $0$. We also suppose that $q\neq0$, otherwise the problem at issue is trivial. Then the integral over $r$ boils down to
\begin{equation}
    \int_0^\infty drrJ_\nu(k_\perp'r)J_\nu(k_\perp r)=\frac{\de(k_\perp'-k_\perp)}{k_\perp},\qquad \re\nu>-1.
\end{equation}
As a result, the amplitude \eqref{amplitude_tw} is reduced to
\begin{equation}\label{amplitude_tw_plane}
    F_{\ga'\ga}=\frac{i\pi^3}{RL_z}\frac{c_v}{n_v} n^2_\perp  k_0\de(k_0'-k_0)\de(k_\perp'-k_\perp)\sum_{s_h=-2}^2 \de_L\big(k_3'-k_3+s_hq\big)\de_{m',m+s_h} F_{s_h},
\end{equation}
where
\begin{equation}\label{F_s_h}
\begin{split}
    F_0=&\,\chi_{\perp}+\frac{n_\perp^2}{2}\Big[\frac{\chi+y}{(s'-n_3')(s-n_3)} +\frac{\chi -y}{(s'+n_3')(s+n_3)} \Big]=\e_\perp-\e +2\de_{s's}\frac{\chi+sn_3y}{n_\perp^2},\\
    F_1=&\,in_\perp\Big(\frac{\be}{s+n_3} -\frac{\al}{s'-n_3'}\Big)=in_\perp\Big(\frac{\be}{s+n_3} -\frac{\al}{s'+n_3}\Big),\\
    F_{-1}=&\,in_\perp\Big(\frac{\al^*}{s-n_3} -\frac{\be^*}{s'+n_3'}\Big)=in_\perp\Big(\frac{\al^*}{s-n_3} -\frac{\be^*}{s'-n_3}\Big),\\
    F_2=&\frac{n^2_\perp A}{(s'-n_3')(s+n_3)} =\Big(2\de_{s's}\frac{1-sn_3}{n_\perp^2}-1\Big)A,\\
    F_{-2}=&\,\frac{n^2_\perp A^*}{(s'+n_3')(s-n_3)} =\Big(2\de_{s's}\frac{1+sn_3}{n_\perp^2}-1\Big)A^*,
\end{split}
\end{equation}
where, in the second equalities, the expressions are taken at the maximum of the scattering probability. In this case, $n_3'=n_3$ for the spin $0$ contribution, i.e., this contribution describes the transmission of a photon through a plate, and $n_3'=-n_3$ for the contributions of other spins, i.e., these contributions correspond to a reflected wave. In the latter case,
\begin{equation}
    k_3=s_hq/2.
\end{equation}
We see that the reflected twisted photon acquires an additional amount of the projection of the total angular momentum due to scattering by a helical medium.

\begin{figure}[tp]
\centering
\raisebox{-0.5\height}{\includegraphics*[width=0.49\linewidth]{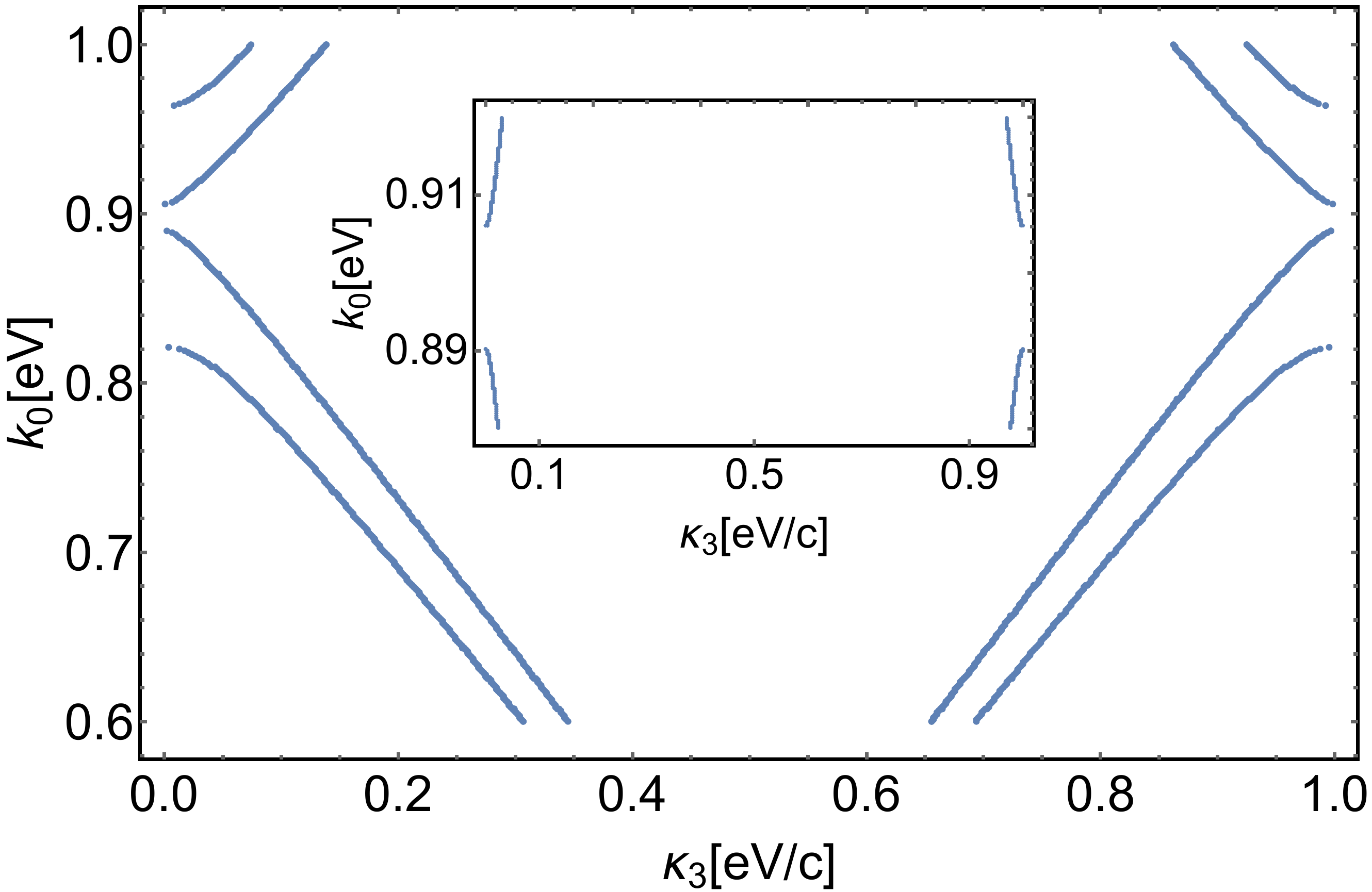}}
\caption{{\footnotesize The dispersion law brought to the first Brillouin zone for photons in the cholesteric plate immersed into the medium with permittivity $\e_v=2.65$. The parameters of the cholesteric plate are as follows: $q=1$ eV, the number of helix turns $N_h=40$ and so the plate width $L=49.6$ $\mu$m, $A\equiv(\e_\parallel-\e_\perp)/4=0.125$, $\e\equiv(\e_\parallel+\e_\perp)/2=\e_v=2.65$, and $\e_\perp=2.4$. The photon falls onto the plate at an angle of $\arcsin(k_\perp/|\spk|)\approx \pi/4$. The total band gap is clearly seen. Inset: The same as on the main plot but near the total band gap.}}
\label{Disp_Chol_plots}
\end{figure}

For example, on reflecting by the cholesteric plate, $\al=\be=y=0$, the twisted photon gains an additional projection of the total angular momentum $\pm2$, where the sign is determined by chirality of the cholesteric, viz., by the sign of $q$. This property can be used for construction of a compact device shifting the projection of the total angular momentum of an incident photon by a given integer number $m_1$, which is even in the case of cholesterics (see Fig. \ref{Scheme_plots}). Indeed, if one launches the twisted photon with given projection of the total angular momentum $m_i$ between two parallel plates made of cholesterics with opposite values of $q$, then after $N$-fold reflection from these plates the twisted photon acquires the additional projection of the total angular momentum $\pm2N$. Furthermore, the intensity loss in scattering can be reduced to a minimum provided $n_\perp$ and the energy of the twisted photon are such that its state belongs to the forbidden band gap of the photon dispersion law for both photon helicities in the helical medium. Such band gaps in the photon dispersion law do exist in cholesterics in the nonparaxial regime (\cite{BerrSchef,RisSchm19}, see also Fig. \ref{Disp_Chol_plots}) when the cholesteric plates are placed in a homogeneous isotropic medium with permittivity
\begin{equation}
    \e_v=\e.
\end{equation}
It is clear that one can shift the projection of total angular momentum of the photon by scattering it on any other helical media and not only on cholesterics. Notice also that one of the two plates can be replaced by the mirror. On reflecting from a mirror, the projections of the total and orbital angular momenta of a twisted photon are conserved whereas the helicity sign is flipped.

The scheme of a possible angular momentum shifting device is presented in Fig. \ref{Scheme_AMSD_plots}. The twisted photon with $m=m_i$ ($m_i=0$, on the scheme) is sent to the input over the optical fiber. The maximum of the probability density of the twisted photon wave packet follows the thin dashed blue lines -- the axial section of a tube. When this wave packet passes through the hole in the first plate, it is refracted by a proper refracting element so that the maximum of the probability density should propagate along the dashed lines, as depicted. This refracting element can be some kind of a lens or a mere axial deformation of the optical fiber. This element changes the parameter $k_\perp/k_0$ of the twisted photon. Then this twisted photon is reflected by the parallel plates several times and gets the additional projection ($+4$, on the scheme) of the total angular momentum as it is explained above. Eventually, the twisted photon with $m=m_f$ ($+4$, on the scheme) is captured by the refracting element that returns it to the optical fiber.

Combining these AMSDs with the beam splitters and the electro-optical switchers for each angular momentum channel, one can construct a scheme for parallel coding of a given signal in terms of twisted photons and obtain thereby a coherent parallel transmission of the signal over a single channel. Indeed, the initial twisted photon with the projection of the total angular momentum $m_i=0$ is sent to the controlled splitter with the input $I$ (see the left panel in Fig. \ref{Scheme_plots}). This controlled splitter sends the optical signal from the port $i$ to the port $a$ when $D_1=0$ and works as $50/50$ splitter of $i$ into $a$ and $b$ when $D_1=1$. Such a splitter can be realized, for example, by using the Mach-Zehnder interferometer where the electro-optical total phase shifter is inserted into one of its arms. Then the part of the wave packet moving along the upper arm on the scheme is transformed by the AMSD. The internal structure of this device is irrelevant. It can be such a device as depicted in Fig. \ref{Scheme_AMSD_plots}, or a $q$-plate, or any other device shifting the projection of total angular momentum of a single photon by a given quantity $m_1\neq0$. After that, the total phase of the part of the wave packet this arm of the scheme is corrected in order to compensate the different traveling times. The similar operations are carried out on the part of the wave packet sent to the other channels of the scheme. Finally, the parts of the initial wave packet are joined by the splitter on the right and are sent to the output $O$. As a result, for example, the electric signal $D_1D_2=11$ leads to the photon at the output $O$ with the spectrum of the projections of the total angular momentum: $\{0, m_1\}$, whereas the electric signal $D_1D_2=01$ gives rise to the spectrum $\{m_1\}$. It is clear that one can add other arms containing the AMSDs with different $m_k$ into this scheme and realize coding of an arbitrary binary number $D_1D_2\cdots D_n$.

Such a coding results in the photon states that are not orthogonal to each other. The scheme on the right panel in Fig. \ref{Scheme_plots} leads to orthogonal states. In this scheme, the controlled splitters send the optical signal from the port $i$ to the port $a$ when $D_1=0$ and send the optical signal from the port $i$ to the port $b$ when $D_1=1$, Such splitters can also be realized with the aid of the Mach-Zehnder interferometers. Then it is not difficult to see that $D_1D_2=00$ results in the twisted photon with $m=0$, $D_1D_2=01$ leads to the twisted photon with $m=m_1$, $D_1D_2=10$ gives the twisted photon with $m=m_2$, and $D_1D_2=11$ provides the twisted photon with $m=m_3$. Of course, the schemes described above can be ameliorated in various directions but we will not dwell on it here.

Consider in more detail the amplitudes \eqref{F_s_h} in the paraxial regime $n_\perp\rightarrow0$ ($n_3>0$). In this case, keeping only singular at $n_\perp\rightarrow0$ terms in \eqref{F_s_h}, we are left with
\begin{equation}\label{F_s_h_parax}
\begin{split}
    F_0\approx&\,2\de_{s's}\frac{\chi+sy}{n_\perp^2},\\
    F_1\approx&\,\frac{2i}{n_\perp}\big[\al\de_{s,1}\de_{s',-1} -\be\de_{s,-1}\de_{s',1} +
    (\al-\be)\de_{s,-1}\de_{s',-1}\big],\\
    F_{-1}\approx&\,\frac{2i}{n_\perp}\big[\al^*\de_{s,1}\de_{s',-1} -\be^*\de_{s,-1}\de_{s',1} +
    (\al^*-\be^*)\de_{s,1}\de_{s',1}\big],\\
    F_2\approx\,&\frac{4A}{n_\perp^2}\de_{s,-1}\de_{s',-1},\\
    F_{-2}\approx&\,\frac{4A^*}{n_\perp^2}\de_{s,1}\de_{s',1}.
\end{split}
\end{equation}
In the paraxial approximation, one can introduce the projection of the orbital angular momentum onto the $z$ axis as $l=m-\sgn(n_3)s$. Then, as follows from \eqref{F_s_h_parax}, the projection of the orbital angular momentum of a transmitted photon is conserved
\begin{equation}
    l'=m'-s'=m-s=l.
\end{equation}
As regards the reflected twisted photon,
\begin{equation}
    l'=m'+s',\qquad l=m-s.
\end{equation}
Expressions \eqref{F_s_h_parax} imply that, in scattering by the component of the susceptibility tensor with spin $s_h=\pm2$, the projection of the orbital angular momentum is conserved,
\begin{equation}
    l'=l,
\end{equation}
in the paraxial approximation. This, in particular, entails that the aforementioned mechanism for enlarging the projection of the total angular momentum with the aid of reflection of a twisted photon from parallel cholesteric plates does not work in the paraxial regime. The twisted photon reflected from the first plate with $q>0$ is not reflected by the second plate with $q<0$.

As for the contributions with $s_h=\pm1$, we have
\begin{equation}
    l'-l=\sgn(q)(\de_{s',-s}-\de_{s's}),
\end{equation}
where $s'$ takes only such values that the case $s'=s=\sgn(q)$ is not realized, for a given $s$. In particular, in such scattering, the twisted photon with helicity $s=\sgn(q)$ turns into the twisted photon with helicity $s'=-\sgn(q)$ and additional projection of the orbital angular momentum $l'-l=\sgn(q)$. Hence, launching the twisted photon with helicity $s=\sgn(q)$ between two parallel plates made of helical media with opposite $q$, one obtains a gain of the projections of the total and orbital angular momenta at the output after a multiple reflection from the component of the dielectric susceptibility tensor with spin $s_h=\pm1$.

In the case of a general helical medium noninvariant under translations in the $(x,y)$ plane, it is convenient to cast expression \eqref{amplitude_tw} into a different form -- to pass into the momentum representation for the coefficient functions. Namely, we write the coefficients $c_{l}$ entering into \eqref{amplitude_tw} in the form of the Hankel transform
\begin{equation}\label{Hankel_transform}
    c_{l}(r)=\int_0^\infty dppJ_{l}(pr)\tilde{c}_{l}(p),\qquad \tilde{c}_{l}(p)=\int_0^\infty drrJ_{l}(pr) c_{l}(r).
\end{equation}
Then the integrals over $r$ in \eqref{amplitude_tw} are reduced to \cite{Vilenkin3J,JackMax71}
\begin{equation}\label{I_ml}
    I_{ml}(k_\perp',k_\perp,p)=\int_0^\infty drr J_{m+l}(k_\perp'r)J_{m}(k_\perp r)J_{l}(pr)=
    \left\{
      \begin{array}{ll}
        \frac{2}{\pi\De}\cos(m'\vf-m\vf_2), & \hbox{$|k_\perp'-k_\perp|<p<k_\perp'+k_\perp$;} \\[0.5em]
        0, & \hbox{\text{otherwise},}
      \end{array}
    \right.
\end{equation}
where $m'=m+l$, the indices $m, l\in \mathbb{Z}$, and
\begin{equation}
    \vf:=\arccos\frac{k'^2_\perp+p^2-k_\perp^2}{2pk'_\perp},\qquad \vf_2:=\arccos\frac{k'^2_\perp-p^2-k_\perp^2}{2pk_\perp}, \qquad\De:=\sqrt{4p^2k_\perp^2 -(k'^2_\perp-p^2-k_\perp^2)^2}.
\end{equation}
In fact, $\De$ is the area of the triangle with sides $p$, $k_\perp$, and $k_\perp'$; $\vf$ is the angle opposite to the side $k_\perp$; and $(\pi-\vf_2)$ is the angle opposite to the side $k_\perp'$. Expression \eqref{I_ml}, when it is not zero, can be rewritten in terms of the Chebyshev polynomials of the first and second kinds
\begin{equation}\label{I_ml1}
    I_{ml}(k_\perp',k_\perp,p)=\frac{2}{\pi\De}\Big[T_{|m'|}(x)T_{|m|}(y)+\frac{\sgn(m'm)\De^2}{4p^2k_\perp k_\perp'} U_{|m'|-1}(x)U_{|m|-1}(y) \Big],
\end{equation}
where $x=\cos\vf$, $y=\cos\vf_2$, and $U_{-1}(x):=0$. Then
\begin{equation}\label{amplitude_tw_mom}
\begin{split}
    F_{\ga'\ga}=&\,i\frac{(2\pi)^3}{RL_z}\big(\frac{n'_\perp n_\perp}{4}\big)^{3/2} c_vk_0^2\de(k_0'-k_0)\sum_{m_h=-\infty}^\infty \de_L\big(k_3'-k_3+m_hq\big)\de_{m',m+m_h}\int_{|k_\perp'-k_\perp|}^{k_\perp'+k_\perp} dpp\times\\
    &\times\Big\{\frac12\Big[\frac{n'_\perp n_\perp (\tilde{\chi}_{m_h}+\tilde{y}_{m_h})}{(s'-n'_3)(s-n_3)} I_{m-1,m_h} + \frac{n'_\perp n_\perp (\tilde{\chi}_{m_h}-\tilde{y}_{m_h})}{(s'+n'_3)(s+n_3)} I_{m+1,m_h}\Big] +\tilde{\chi}_{\perp m_h} I_{mm_h}-\\
    &-\frac{in'_\perp\tilde{\al}_{m_h-1}}{s'-n_3'} I_{m,m_h-1} +\frac{in_\perp\tilde{\be}_{m_h-1}}{s+n_3} I_{m+1,m_h-1}+\\
    &+(-1)^{m_h} \frac{in'_\perp\tilde{\be}^*_{-m_h-1}}{s'+n_3'} I_{m,m_h+1} -(-1)^{m_h}\frac{in_\perp\tilde{\al}^*_{-m_h-1}}{s-n_3} I_{m-1,m_h+1}+\\
    &+\frac{n'_\perp n_\perp \tilde{A}_{m_h-2}}{(s'-n'_3)(s+n_3)} I_{m+1,m_h-2} +(-1)^{m_h}\frac{n'_\perp n_\perp \tilde{A}^*_{-m_h-2}}{(s'+n'_3)(s-n_3)} I_{m-1,m_h+2}  \Big\}.
\end{split}
\end{equation}
Expression \eqref{I_ml}, \eqref{I_ml1} diverges when the triangle area, $\De$, tends to zero. Therefore, in the case when the coefficients $\tilde{c}_{l}$ are different from zero in a small neighborhood of the point $p_0$, the scattering amplitude \eqref{amplitude_tw_mom} possesses sharp maxima at
\begin{equation}
    k_\perp'=k_\perp\pm p_0>0.
\end{equation}
This property can be employed for amplification of twisted photon scattering by helical media, in particular, by $q$-plates. Such Bessel profiles of the coefficients of the dielectric susceptibility tensor \eqref{Hankel_transform} can be created by exciting cylindrical sound waves in the helical medium.

\subsection{Exact solution of the Maxwell equations}\label{Exact_Solut}
\subsubsection{General formulas}

A nonperturbative analysis of scattering of photons by a helical medium usually requires a knowledge of a complete set of stationary solutions of the Maxwell equations in this medium (see for details, e.g., \cite{BKL5,parax,wkb_chol,KazLaz20,BKL6})
\begin{equation}\label{Max_eqns}
    (\rot^2_{ij}-\e_{ij}(k_0)k_0^2)A_j=0,\qquad \partial_i(\e_{ij}A_j)=0.
\end{equation}
The second equation in this system is the Coulomb gauge, which follows from the first one under the assumption that $k_0\neq0$. We will consider only the case of a helical medium with the permittivity tensor invariant under translations in the $(x,y)$ plane.

In this case, it is convenient to seek for a solution of the first equation in \eqref{Max_eqns} in the form
\begin{equation}\label{vect_pot}
    \mathbf{A}(\spx)=\Big[\frac12(\spe_+ a_-(z) e^{-i\phi} +\spe_- a_+(z) e^{i\phi})+\spe_3 a_3(z)\Big]e^{i\spk_\perp\spx_\perp},
\end{equation}
where the basis \eqref{pm_basis} has been used and $\phi:=\arg(k_1+ik_2)$. Having substituted this expression into \eqref{Max_eqns}, the third component of the first equation becomes
\begin{equation}
    a_3=\bar{k}_3^{-2}\Big[\frac{ik_\perp}{2}\partial_z(a_++a_-)-k_0^2(\al^*e^{i\theta} a_- +\be e^{-i\theta} a_+)\Big],
\end{equation}
where $\bar{k}_3^2:=\e_\perp k_0^2-k_\perp^2$ and $\theta:=qz-\phi$. Substituting the above expression into the remaining two equations, we come to the system of equations
\begin{equation}\label{Max_eqns_1}
    \Big[M_2\partial_z^2+\frac{k_\perp k_0^2}{\bar{k}_3^2}M_1 i\partial_z+V_0+\frac{qk_\perp k_0^2}{\bar{k}_3^2} V_1 -2\frac{k_0^4}{\bar{k}_3^2}V_2\Big]a(z)=0,
\end{equation}
where
\begin{equation}
\begin{gathered}
    M_2=\left[
          \begin{array}{cc}
            1+\frac{k_\perp^2}{2\bar{k}_3^2} & \frac{k_\perp^2}{2\bar{k}_3^2} \\
            \frac{k_\perp^2}{2\bar{k}_3^2} & 1+\frac{k_\perp^2}{2\bar{k}_3^2} \\
          \end{array}
        \right],\qquad
    M_1=\left[
          \begin{array}{cc}
            \be^* e^{i\theta} +\be e^{-i\theta} & (\al^*+\be^*)e^{i\theta} \\
            (\al+\be)e^{-i\theta} & \al^* e^{i\theta} +\al e^{-i\theta} \\
          \end{array}
        \right],\\
    V_0=\left[
          \begin{array}{cc}
            k_0^2(\e-y)-\frac{k_\perp^2}{2} & \frac{k_\perp^2}{2}+2A^*k_0^2e^{2i\theta} \\
            \frac{k_\perp^2}{2}+2Ak_0^2e^{-2i\theta} & k_0^2(\e+y)-\frac{k_\perp^2}{2} \\
          \end{array}
        \right],\qquad
    V_1=\left[
          \begin{array}{cc}
            \be e^{-i\theta} & -\al^* e^{i\theta} \\
            \be e^{-i\theta} & -\al^* e^{i\theta} \\
          \end{array}
        \right],\\
    V_2=\left[
          \begin{array}{cc}
            |\be|^2 & (\al\be)^* \\
            \al\be & |\al|^2 \\
          \end{array}
        \right],
    \end{gathered}
\end{equation}
and $a(z):=(a_+(z),a_-(z))$. Notice that, in a general case, the matrix operator in the square brackets in \eqref{Max_eqns_1} is not Hermitian. The system of equations \eqref{Max_eqns_1} is comprised of two linear ordinary differential equations of the second order and so there are four linearly independent solutions to it.

Inasmuch as the coefficients of Eq. \eqref{Max_eqns_1} are periodic functions of $\theta$, we seek for a solution of \eqref{Max_eqns_1} in the form of the Fourier series
\begin{equation}\label{Fourier_series}
    a_\pm^{(i)}(z)=\sum_{l=-\infty}^\infty a_{l\pm}^{(i)} e^{i(p_3^{(i)}+ql)\theta/q},
\end{equation}
where $a_{l \pm}^{(i)}:=a_\pm^{(i)}(p_3^{(i)}+ql)$ and $i=\overline{1,4}$ numerates the solutions to Eq. \eqref{Max_eqns_1}. On substituting the series \eqref{Fourier_series} into \eqref{Max_eqns_1}, one arrives at the infinite system of entangled linear equations. It has a nontrivial solution only at certain values of momenta $p_3^{(i)}(k_0)$, the momenta differing by a multiple of $q$ corresponding to the same solution. The general solution of the Maxwell equations  \eqref{Max_eqns_1} is a linear combination of solutions \eqref{Fourier_series} with some constant coefficients $b_i$. In order to impose the boundary conditions, we also need
\begin{equation}
    \rot a^{(i)}_\pm=\mp\sum_{l=-\infty}^\infty\Big\{(p_3^{(i)}+ql) \big[a^{(i)}_{l\pm} +\frac{k_\perp^2}{2\bar{k}_3^2}(a^{(i)}_{l+}+a^{(i)}_{l-}) \big] +\frac{k_0^2k_\perp}{\bar{k}_3^2}(\al^* a^{(i)}_{l-1,-} +\be a^{(i)}_{l+1,+}) \Big\} e^{i(p_3^{(i)}+ql)\theta/q},
\end{equation}
where we have used the same notation as in formula \eqref{vect_pot}.

Suppose that the helical medium constitutes the plate parallel to the $(x,y)$ plane and $z\in[-L,0]$ in the plate. As in the previous sections, we assume that the plate is placed in the homogeneous isotropic medium with permittivity $\e_v(k_0)$. We also suppose that the incident electromagnetic wave falls onto the plate from left to right. The standard boundary conditions are satisfied on the facets of the plate:
\begin{equation}\label{bound_conds}
    [a_\pm]_{z=-L}=[a_\pm]_{z=0}=0,\qquad [\rot a_\pm]_{z=-L}=[\rot a_\pm]_{z=0}=0,
\end{equation}
where the square brackets denote a jump of the quantity in the brackets at the respective boundary. Introduce the matrix $U_{ij}$,
\begin{equation}\label{U_matrix}
    U_{1i}(z)=a_+^{(i)}(z),\qquad U_{2i}(z)=a_-^{(i)}(z),\qquad U_{3i}(z)=-\rot a_+^{(i)}(z),\qquad U_{4i}(z)=\rot a_-^{(i)}(z),
\end{equation}
and the column $g(s)$,
\begin{equation}
    g^T(s)=\left[
           \begin{array}{cccc}
             n'_3-s\e_v^{1/2} & n'_3+s\e_v^{1/2} & k_0(\e_v-s\e_v^{1/2}n'_3) & k_0(\e_v+s\e_v^{1/2}n'_3) \\
           \end{array}
         \right]/\sqrt{2},
\end{equation}
where $n'_3:=\sqrt{\e_v-n_\perp^2}$ and $n_\perp:=k_\perp/k_0$. The column $g(s)$ is a collection of the values of the components \eqref{U_matrix} at $z=0$ for the mode function \eqref{mode_func_med} in the homogeneous isotropic medium with permittivity $\e_v(k_0)$ corresponding to the helicity $s$ and $n_3'>0$. The four linear independent mode functions in the medium surrounding the plate comprise  the matrix
\begin{equation}
    U_0=\frac{1}{\sqrt{2}}\left[
          \begin{array}{cccc}
            n'_3-\e_v^{1/2} & n'_3+\e_v^{1/2} & -n'_3-\e_v^{1/2} & -n'_3+\e_v^{1/2} \\
            n'_3+\e_v^{1/2} & n'_3-\e_v^{1/2} & -n'_3+\e_v^{1/2} & -n'_3-\e_v^{1/2} \\
            k_0(\e_v-\e_v^{1/2}n'_3) & k_0(\e_v+\e_v^{1/2}n'_3) & k_0(\e_v+\e_v^{1/2}n'_3) & k_0(\e_v-\e_v^{1/2}n'_3) \\
            k_0(\e_v+\e_v^{1/2}n'_3) & k_0(\e_v-\e_v^{1/2}n'_3) & k_0(\e_v-\e_v^{1/2}n'_3) & k_0(\e_v+\e_v^{1/2}n'_3) \\
          \end{array}
        \right],
\end{equation}
at $z=0$, which is made up of the columns $g(s)$ with $s=\pm1$ and the different signs of $n_3'$. Let
\begin{equation}
    T_{-L}=\diag\big(e^{-ik'_3L},e^{-ik'_3L},e^{ik'_3L},e^{ik'_3L}),
\end{equation}
where $k'_3=k_0n'_3$. Then the conditions \eqref{bound_conds} for the scattering problem at issue are written as
\begin{equation}\label{joining_conds}
    U_0 T_{-L} b_l=U(-L)b,\qquad U(0)b=t_+ g(1) +t_-g(-1),
\end{equation}
where
\begin{equation}
    b^T_l:=\left[
             \begin{array}{cccc}
               i_+ & i_- & r_+ & r_- \\
             \end{array}
           \right].
\end{equation}
Here $i_s$ is the amplitude of an incident wave with helicity $s$, $t_s$ is the amplitude of a transmitted wave with helicity $s$, and $r_s$ is the amplitude of a reflected wave with helicity $s$.

The unitarity relation for the scattering matrix is fulfilled:
\begin{equation}
    |i_+|^2+|i_-|^2=|t_+|^2+|t_-|^2+|r_+|^2+|r_-|^2.
\end{equation}
The transmission and reflection coefficients read
\begin{equation}\label{trans_refl_coeff}
    T_c=\frac{|t_+|^2+|t_-|^2}{|i_+|^2+|i_-|^2},\qquad R_c=\frac{|r_+|^2+|r_-|^2}{|i_+|^2+|i_-|^2}.
\end{equation}
The Stokes parameters for the transmitted and reflected waves take the form
\begin{equation}\label{Stokes_defn}
\begin{aligned}
    \xi_{1,tr}&=i\frac{t_+t^*_--t_+^*t_-}{|t_+|^2+|t_-|^2},&\qquad \xi_{2,tr}&=\frac{|t_+|^2-|t_-|^2}{|t_+|^2+|t_-|^2},&\qquad \xi_{3,tr}&=\frac{t_+t^*_-+t_+^*t_-}{|t_+|^2+|t_-|^2},\\
    \xi_{1,r}&=i\frac{r_+r^*_--r_+^*r_-}{|r_+|^2+|r_-|^2},&\qquad \xi_{2,r}&=\frac{|r_+|^2-|r_-|^2}{|r_+|^2+|r_-|^2},&\qquad \xi_{3,r}&=\frac{r_+r^*_-+r_+^*r_-}{|r_+|^2+|r_-|^2}.
\end{aligned}
\end{equation}

\begin{figure}[tp]
\centering
\raisebox{-0.5\height}{\includegraphics*[width=0.49\linewidth]{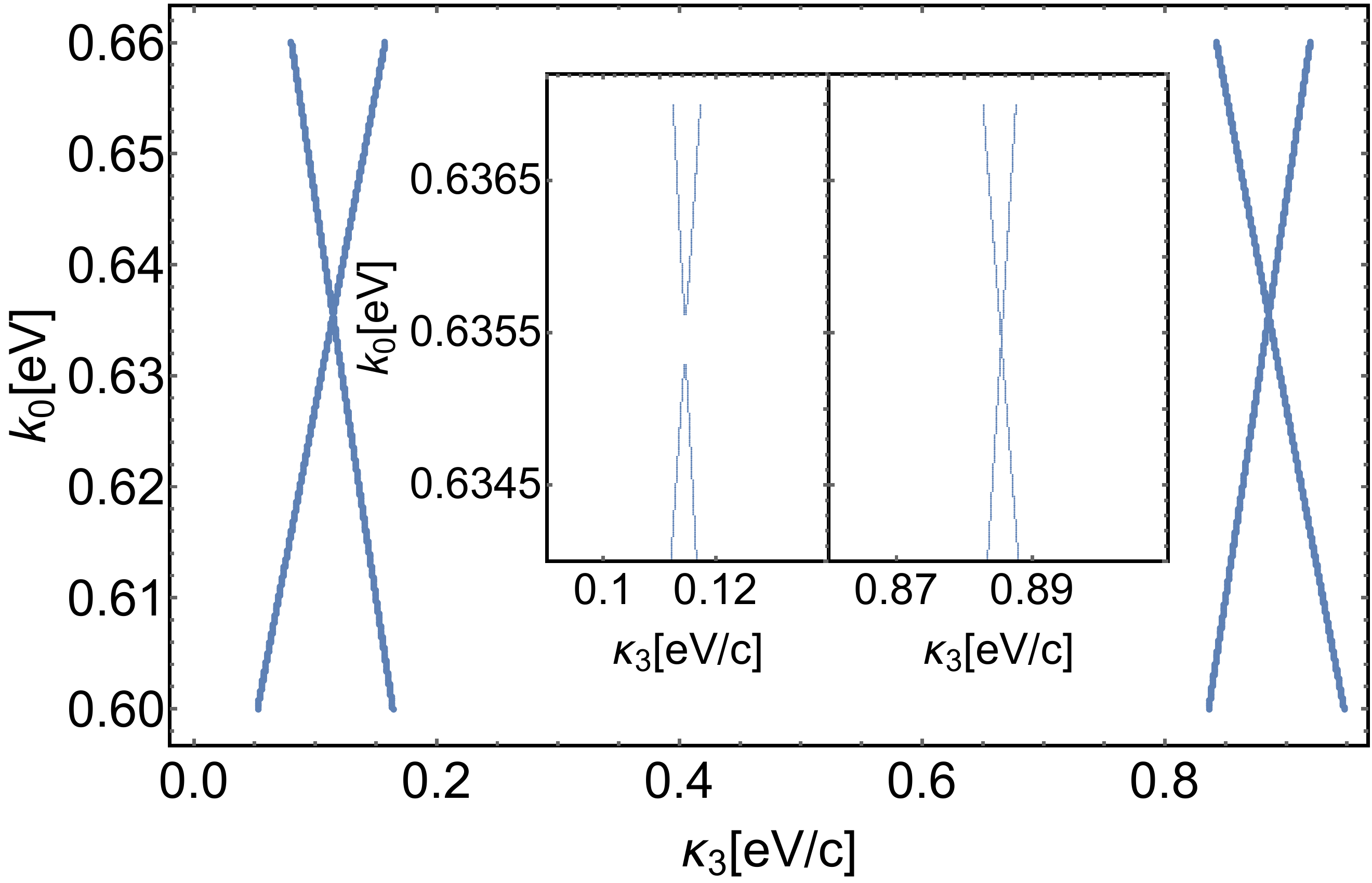}}\;
\raisebox{-0.5\height}{\includegraphics*[width=0.49\linewidth]{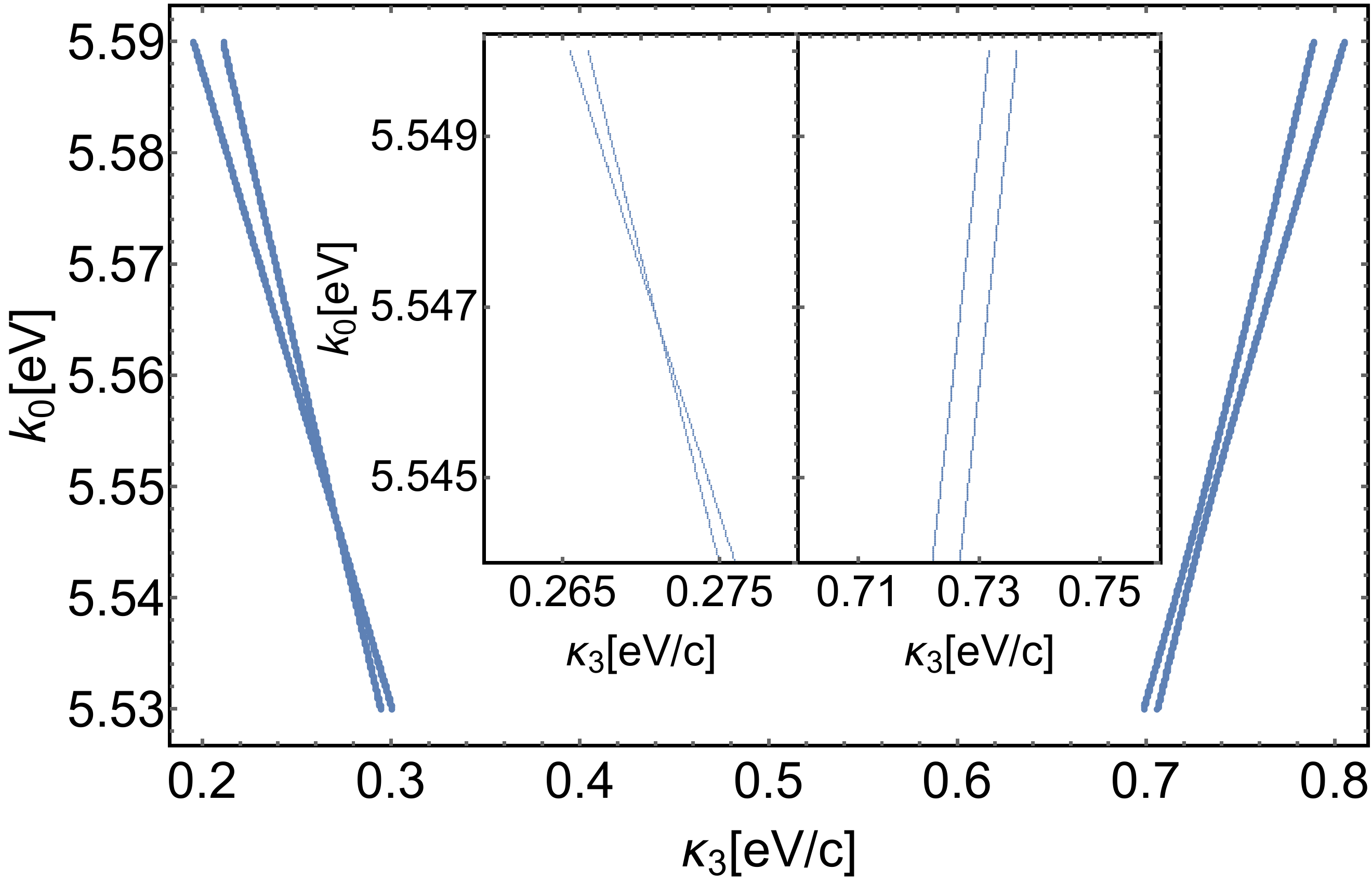}}\;
\caption{{\footnotesize The dispersion law brought to the first Brillouin zone for photons in the plate made of the helical medium near the band gaps corresponding to $s_h=\pm2$. The plate is immersed into the medium with permittivity $\e_v=2.65$. The parameters of the helical medium are as follows: $q=1$ eV, the number of helix turns $N_h=80$ and so the plate width $L=99.2$ $\mu$m, $A=6.25\times 10^{-2}$, $\al=0.5$, $\be=0.297$, $\e=\e_v=2.65$, $\e_\perp=2.4$, and $y=-0.5$. For simplicity, we do not take into account the dependence of the permittivity tensor coefficients on the photon energy $k_0$. The photon falls onto the plate at an angle of $6.14\times10^{-3}$. Hence, the paraxial regime is realized with the parameters $a=3.076$, $b=1.25\times 10^{-3}$, and $c=1.942$. Formula \eqref{band_gaps_s2} give $k_0=0.6354$ eV for the real gap and $k_0=5.547$ eV for the imaginary gap. Insets: The same as on the main plots but near the band gaps.}}
\label{Disp_Hels2_plots}
\end{figure}

In order to describe scattering of twisted photons by the plate made of a helical medium, we employ the relation between the plane-wave and twisted photons. Let
\begin{equation}
    t_s=T_{ss'}(\phi)i_{s'},\qquad r_s=R_{ss'}(\phi)i_{s'},
\end{equation}
where the summation over $s'$ is understood. The explicit expressions for the matrices $T$ and $R$ readily follow from the system of equations \eqref{joining_conds}. It is clear that
\begin{equation}
    T_c=i^\dag T^\dag Ti/(i^\dag i),\qquad R_c=i^\dag R^\dag Ri/(i^\dag i).
\end{equation}
The matrices $T$ and $R$ obey the unitarity relation
\begin{equation}\label{unit_rels}
    T^\dag T+R^\dag R=1.
\end{equation}
Let us take
\begin{equation}
    i_s\equiv i_s(\phi,m)=i_{m,s}i^{-m}e^{im\phi},
\end{equation}
where $i_{s,m}$ is the amplitude of a twisted photon impinging on the plate, $s$ is its helicity and $m$ is the projection of its total angular momentum. Define the amplitudes of the transmitted and reflected twisted photons as
\begin{equation}
\begin{split}
    t_{m,s}&=i^{m}\int_{-\pi}^\pi\frac{d\phi}{2\pi}e^{-im\phi} T_{ss'}(\phi)i_{s'}(\phi,m') =i^{m-m'}\int_{-\pi}^\pi\frac{d\phi}{2\pi} e^{-i(m-m')\phi}T_{ss'}(\phi)i_{m',s'},\\
    r_{m,s}&=i^{m}\int_{-\pi}^\pi\frac{d\phi}{2\pi}e^{-im\phi} R_{ss'}(\phi)i_{s'}(\phi,m') =i^{m-m'}\int_{-\pi}^\pi\frac{d\phi}{2\pi} e^{-i(m-m')\phi}R_{ss'}(\phi)i_{m',s'}.
\end{split}
\end{equation}
Introducing the matrices
\begin{equation}
    (T_m)_{ss'}:=i^{m}\int_{-\pi}^\pi\frac{d\phi}{2\pi} e^{-im\phi}T_{ss'}(\phi),\qquad (R_m)_{ss'}:=i^{m}\int_{-\pi}^\pi\frac{d\phi}{2\pi} e^{-im\phi}R_{ss'}(\phi),
\end{equation}
we can write
\begin{equation}\label{t_r_twisted}
    t_{m,s}=(T_{m-m'})_{ss'}i_{m',s'},\qquad r_{m,s}=(R_{m-m'})_{ss'}i_{m',s'}.
\end{equation}
The unitarity relation \eqref{unit_rels} implies
\begin{equation}
\begin{gathered}
    \sum_{m=-\infty}^\infty(T^\dag_m T_m +R^\dag_m R_m)=1,\\
    \sum_{s=\pm1}\sum_{m=-\infty}^\infty(|t_{m,s}|^2+|r_{m,s}|^2)=\sum_{s'=\pm1}|i_{m',s'}|^2.
\end{gathered}
\end{equation}
The transmission and reflection coefficients and the Stokes vectors for scattered twisted photons are defined by formulas  \eqref{trans_refl_coeff}, \eqref{Stokes_defn}, where $t_\pm$ and $r_\pm$ are given in \eqref{t_r_twisted}.

\subsubsection{Paraxial limit}

In the general case, it seems impossible to construct an explicit solution in a closed form to the system of equations \eqref{Max_eqns_1}. However, in the case $k_\perp=0$, the system \eqref{Max_eqns_1} is exactly solvable. Such solutions were investigated in \cite{LakhWeigh95,FarLakh14,MacLakh,LakhMess05,LaVeMcC00,BitThom05,Lakht10,TChJhHu17,McCLakh04}. As for $q$-plates, the paraxial propagation of electromagnetic waves in them was studied in \cite{KPMS09}. In this section, we will study some additional properties of this solution and analyze the features of scattering and propagation of photons in helical media ensuing from this solution.

\begin{figure}[tp]
\centering
\raisebox{-0.5\height}{\includegraphics*[width=0.49\linewidth]{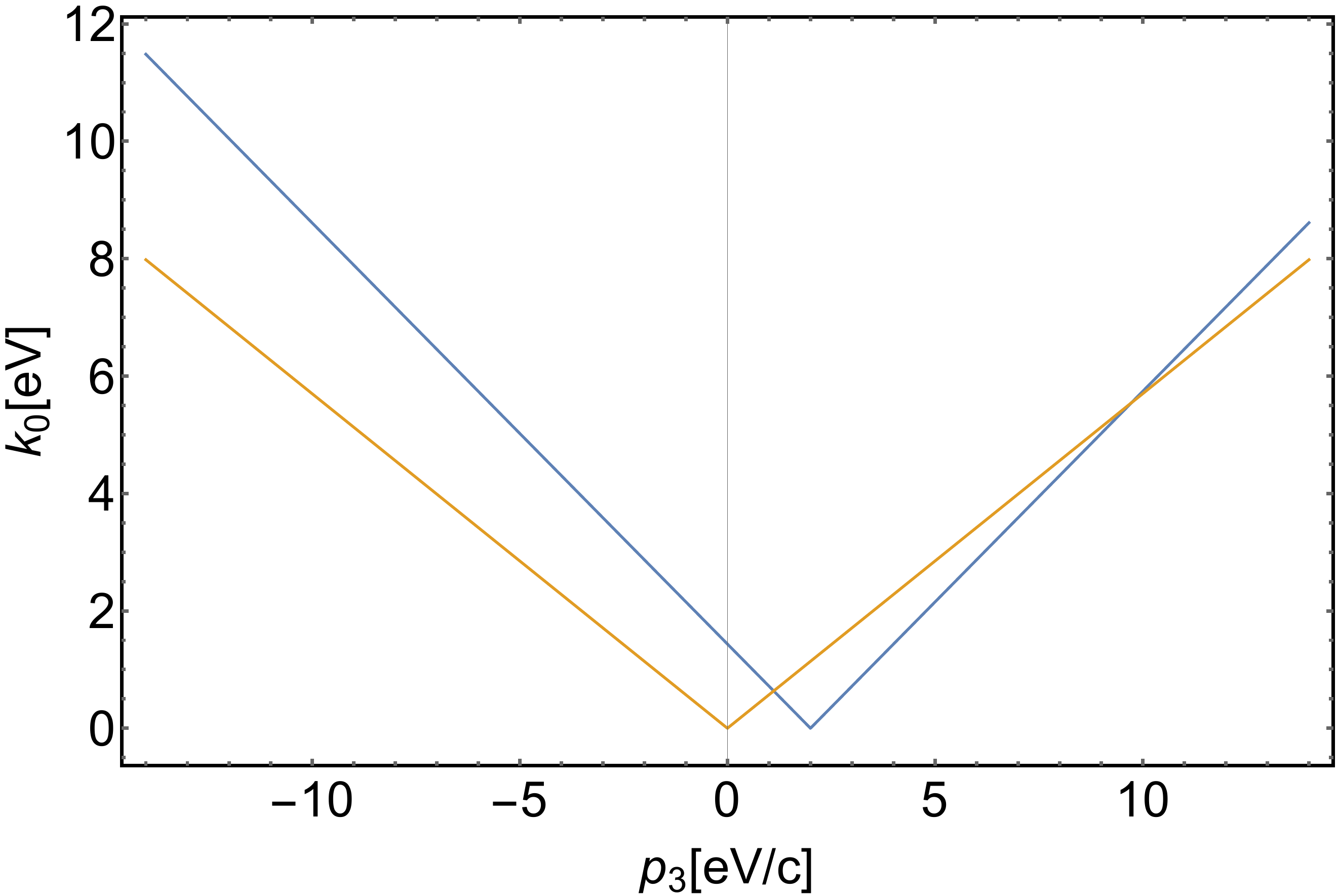}}\,
\raisebox{-0.5\height}{\includegraphics*[width=0.49\linewidth]{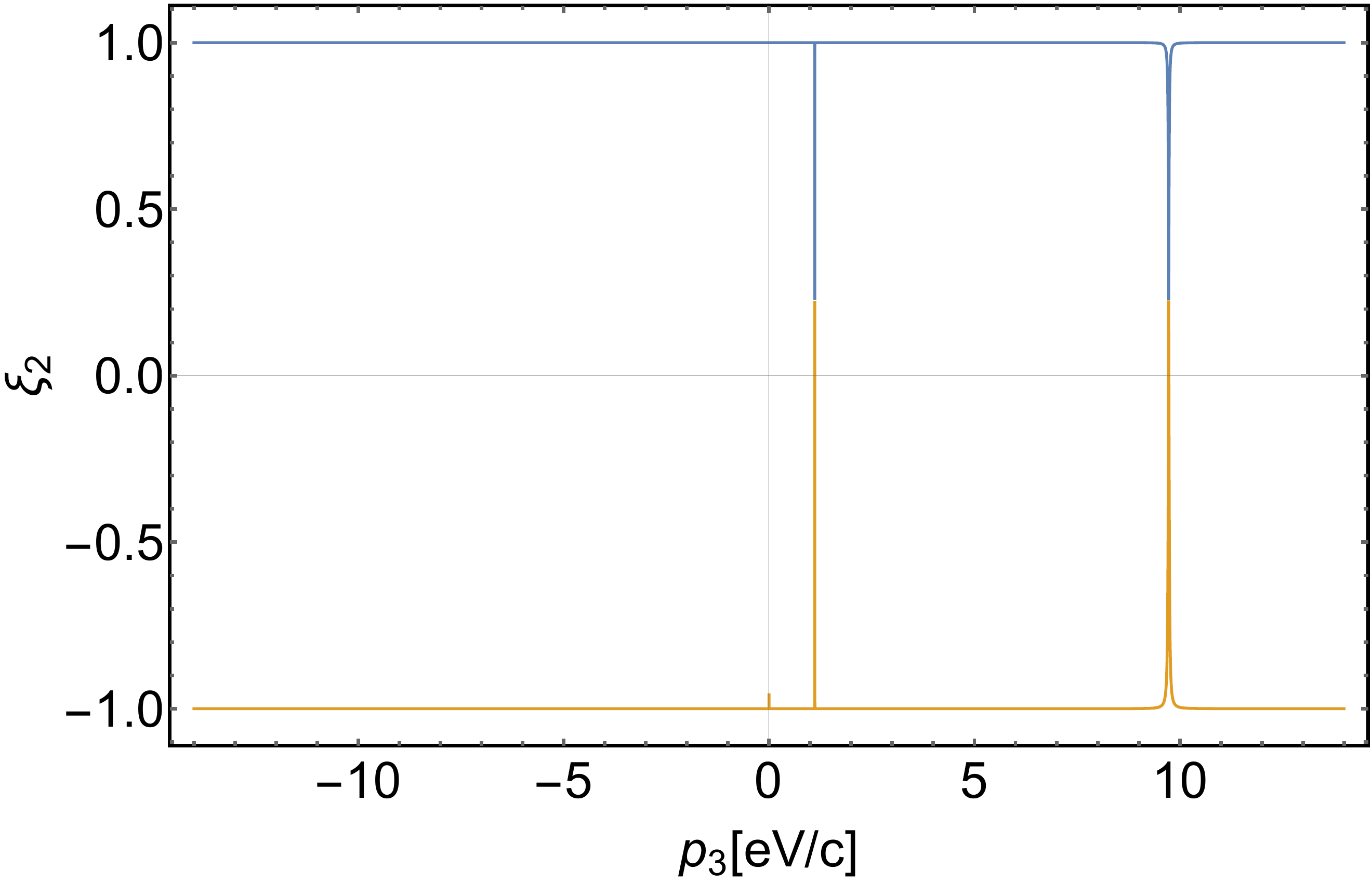}}
\caption{{\footnotesize The dispersion law and the Stokes parameter $\xi_2$ in the paraxial limit $k_\perp=0$ for the natural choice of the branches. Different colors on the plots correspond to different branches. The parameters of the helical medium are the same as in Fig. \ref{Disp_Hels2_plots}. Formulas \eqref{band_gaps_s2}, \eqref{band_gaps_s1} give the positions of the band gaps (i) $k_0=0.635$ eV, (ii) $k_0=5.547$ eV for $|s_h|=2$ and (i) $k_0=0.318$ eV, (ii) $k_0=2.773$ eV for $|s_h|=1$. The asymmetry is positive in this case. If the asymmetry is negative, the branches will approach each other for negative $p_3$.}}
\label{Disp_Law_Parax_plots}
\end{figure}

Substituting the expansion \eqref{Fourier_series} into Eq. \eqref{Max_eqns_1}, we obtain
\begin{equation}\label{parax_eqs}
    \left[
      \begin{array}{cc}
        a-\frac{(p_3+ql)^2}{k_0^2} & b \\
        b^* & c-\frac{(p_3+q(l-2))^2}{k_0^2} \\
      \end{array}
    \right]
    \left[
      \begin{array}{c}
        a_{l,+} \\
        a_{l-2,-} \\
      \end{array}
    \right]
    =0,
\end{equation}
where, for brevity, we do not show the index $i$ enumerating different solutions and employ the notation introduced in \eqref{abc}. The solutions of \eqref{parax_eqs} with different $l$ correspond to the same mode function \eqref{Fourier_series}. Consequently, for definiteness, we put $l=0$ in \eqref{parax_eqs}. The system of equations \eqref{parax_eqs} possess a nontrivial solution provided that
\begin{equation}\label{disp_law}
    k_0^2=\frac{a(p_3-2q)^2+cp_3^2\pm\s \sqrt{\big(a(p_3-2q)^2-cp_3^2\big)^2+4|b|^2p_3^2(p_3-2q)^2}}{2(ac-|b|^2)},
\end{equation}
where $\s:=\sgn(a(p_3-2q)^2-cp_3^2)$. The expression on the right-hand side is nonnegative for any $p_3\in \mathbb{R}$ if and only if
\begin{equation}
    a>0,\qquad c>0,\qquad ac-|b|^2>0.
\end{equation}
Moreover, expression \eqref{disp_law} is unchanged under the replacement $p_3\rightarrow 2q-p_3$, $a\leftrightarrow c$. In the case when the dispersion relation \eqref{disp_law} holds, the system \eqref{parax_eqs} admits the solution
\begin{equation}\label{parax_sol}
    a_{-2,-}=\frac{1}{b}\Big(\frac{p_3^2}{k_0^2} -a\Big)a_{0,+}=\frac{b^*}{(p_3-2q)^2/k_0^2-c}a_{0,+}.
\end{equation}

The function $p_3(k_0)$, which is the inverse to the function $k_0(p_3)$ given in \eqref{disp_law}, possesses, in general, four branch points of the square root type in the complex $k_0$ plane. Therefore, there are four different values of $p_3$ at fixed $k_0$, as expected. At the branch points, the group velocity $\partial k_0/\partial p_3=0$.  In virtue of real analyticity of $p_3(k_0)$, the branch points are located symmetrically with respect to the real axis of the complex $k_0$ plane. Simple expressions for the positions of the branch points can be obtained in the case $|b|\ll1$. Then
\begin{equation}\label{band_gaps_s2}
\begin{aligned}
    (i)&&\frac{k_0}{q}&=\frac{2}{a^{1/2}+c^{1/2}}\pm\frac{2|b|}{(ac)^{1/4}(a^{1/2}+c^{1/2})^2}+\cdots,&\qquad \frac{p_3}{q}&=\frac{2a^{1/2}}{a^{1/2}+c^{1/2}} \pm\frac{|b|(a^{1/2}-c^{1/2})}{(ac)^{1/4}(a^{1/2}+c^{1/2})^2}+\cdots,\\
    (ii)&&\frac{k_0}{q}&=\frac{2}{|a^{1/2}-c^{1/2}|}\pm\frac{2i|b|}{(ac)^{1/4}(a^{1/2}-c^{1/2})^2}+\cdots,&\qquad \frac{p_3}{q}&=\frac{2a^{1/2}}{a^{1/2}-c^{1/2}} \pm \frac{i|b|(a^{1/2}+c^{1/2})}{(ac)^{1/4}(a^{1/2}-c^{1/2})^2}+\cdots.\\
\end{aligned}
\end{equation}
The branch points (i) on the $k_0$ plane specify the boundaries of a (real) band gap in the dispersion law, whereas the branch points (ii) on the $k_0$ plane define the position of the imaginary band gap (see Fig. \ref{Disp_Hels2_plots}). We will refer to $|b|$ as the gap parameter and to $a^{1/2}-c^{1/2}$ as the asymmetry of the dispersion law.

It is natural to specify the branches of $p_3(k_0)$ by two cuts symmetric with respect to the real axis: the first cut lies on the real axis and connects the branch points (i), the second cut connects the branch points (ii) and intersects the real axis. Such a choice of the branches ensures that the branch points ``annihilate'' at $b=0$ and there appear the branches without singularities on the complex $k_0$ plane. The branches of the plot of $k_0(p_3)$ intersect when the gap parameter vanishes (see Fig. \ref{Disp_Law_Parax_plots}). The solution \eqref{parax_sol} corresponding to the given branch goes continuously to the solution of Eq. \eqref{parax_eqs} taken at $b=0$, and vice versa, the solution corresponding to the given branch at a finite $b$ is a continuous deformation of the solution with $b=0$. Notice also that the imaginary gap disappears for $\al=\be$ and $y=0$, i.e., in the case of $C^*$-smectics and cholesterics, since $a=c$ and the asymmetry vanishes in the case. Other non-Bragg band gaps in three dimensional dielectric helix structures were discussed in \cite{TChJhHu17}. The real gap is a well-known forbidden band gap of the dispersion law of photons propagating in $C^*$-smectics or cholesterics \cite{deGennProst,YangWu06,BelyakovBook}. The study of scattering of twisted photons in the vicinity of the gaps \eqref{band_gaps_s2} reveals that these peculiarities of the dispersion law are formed due to the contribution to the permittivity tensor with spin $s_h=\pm2$ (see Fig. \ref{Scatt_Hels2c_plots} and the selection rule in \eqref{amplitude_tw_plane}).

\begin{figure}[tp]
\centering
\raisebox{-0.5\height}{\includegraphics*[width=0.49\linewidth]{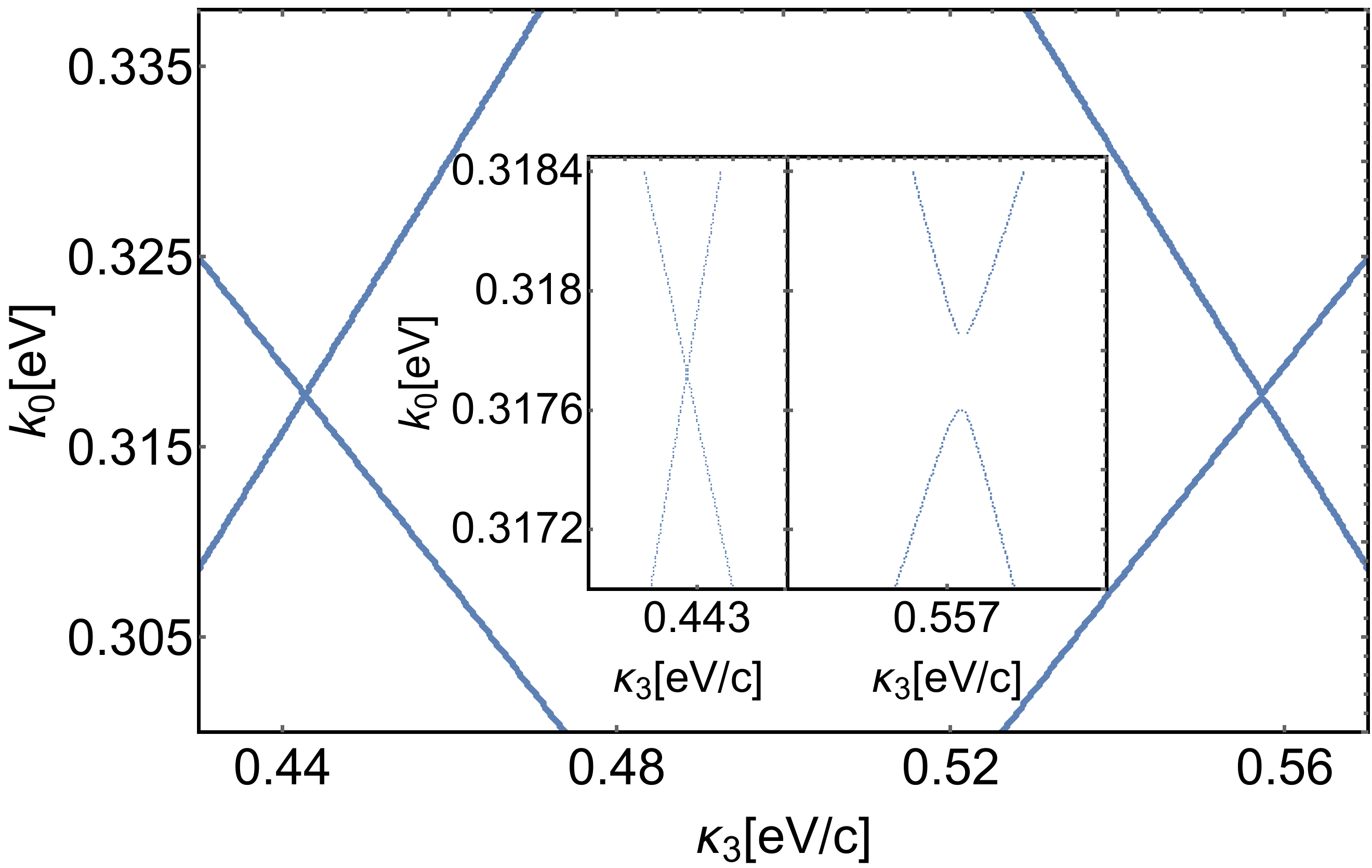}}\;
\raisebox{-0.5\height}{\includegraphics*[width=0.49\linewidth]{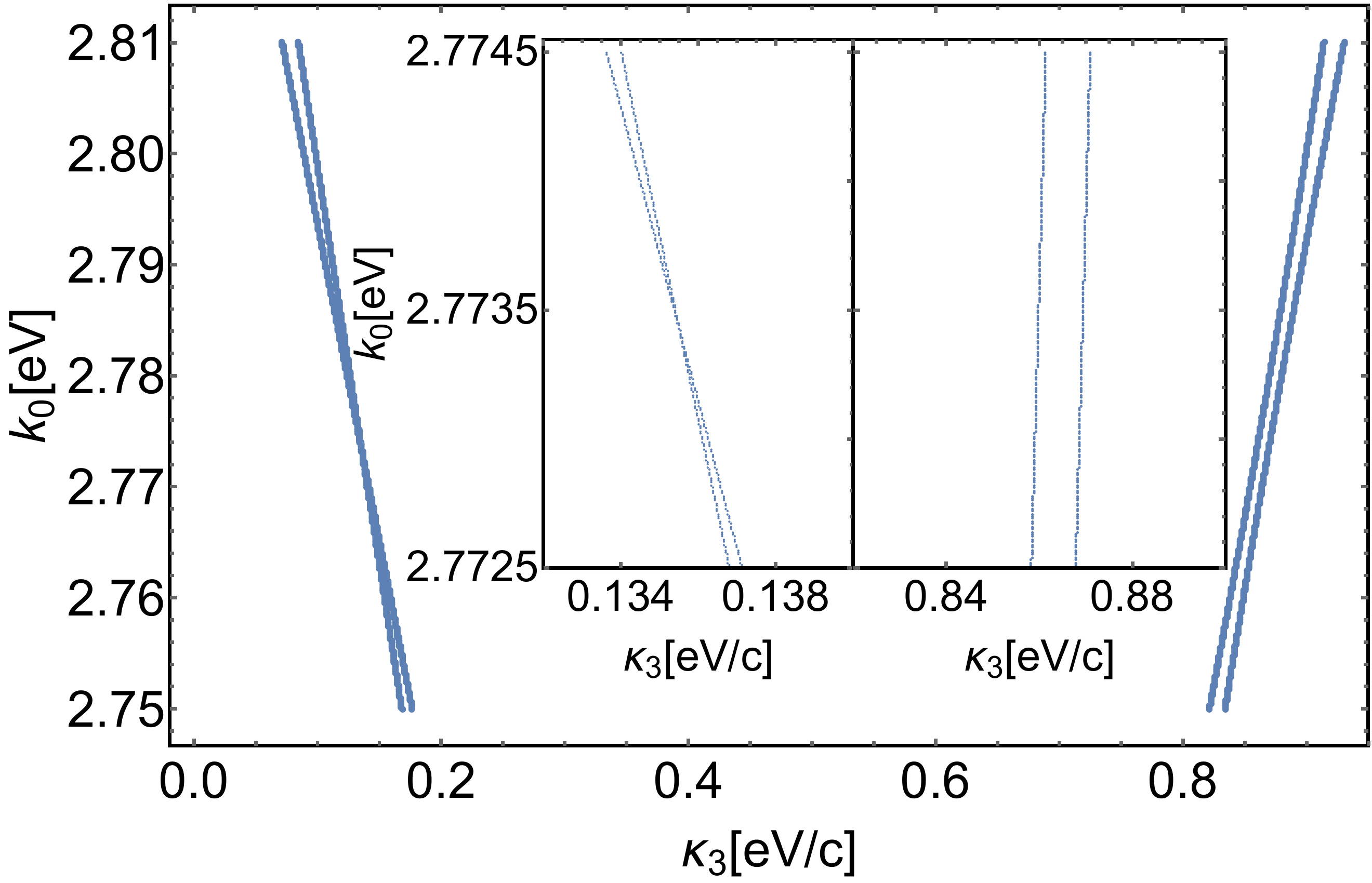}}\;
\caption{{\footnotesize  The dispersion law brought to the first Brillouin zone for photons in the plate made of the helical medium near the band gaps corresponding to $s_h=\pm1$. The plate is immersed into the medium with permittivity $\e_v=2.65$. The parameters of the helical medium and the incident photon are the same as in Fig. \ref{Disp_Hels2_plots}. Formula \eqref{band_gaps_s1} give $k_0=0.3177$ eV for the real gap and $k_0=2.7335$ eV for the imaginary gap. Insets: The same as on the main plots but near the band gaps.}}
\label{Disp_Hels1_plots}
\end{figure}

Since the natural choice of the branches assures that they are continuous deformations of the branches at zero gap parameter, the properties of the mode functions at $b=0$ are inherited by the respective branches, at sufficiently small $b$. Consider, for example, the Stokes parameters
\begin{equation}\label{xi2_medium}
    \xi_1:=\frac{2\im(A_+A^*_-)}{|A_+|^2+|A_-|^2},\qquad\xi_2:=\frac{|A_-|^2-|A_+|^2}{|A_+|^2+|A_-|^2},\qquad \xi_3:=\frac{2\re(A_+A^*_-)}{|A_+|^2+|A_-|^2},
\end{equation}
where
\begin{equation}
    A_\pm=a_\pm(z)e^{\pm i\phi}e^{i\spk_\perp\spx_\perp}.
\end{equation}
It is evident from Eq. \eqref{parax_eqs} that there are two solutions when the gap parameter is zero: ($+$), where $a_{0,+}=0$; and ($-$), where $a_{-2,-}=0$. The first solution has $\xi_2=1$, whereas the second solution possesses $\xi_2=-1$. Therefore, for the natural choice of the branches and for $b$ small, $\xi_2\approx1$ on the deformation of the branch ($+$) and $\xi_2\approx-1$ on the deformation of the branch ($-$) (see Fig. \ref{Disp_Law_Parax_plots}). It is easy to check that the ($+$) branch corresponds to ``$+$'' in \eqref{disp_law}, whereas the ($-$) branch is described by ``$-$'' in \eqref{disp_law}.

However, the conclusions on polarization of the mode functions become invalid for  $|p_3/q|\gg1$ when the asymmetry is zero, because in this case the eigenvalues of the matrix \eqref{parax_eqs} are degenerate at $b=0$. It is not difficult to see that for $|p_3/q|\gg1$ the following asymptotics take place
\begin{equation}\label{Stokes_parax}
    \xi_3+i \xi_1\approx\frac{\pm2\s b}{\sqrt{(a-c)^2+4|b|^2}} e^{2iqz},\qquad\xi_2\approx1-\frac{8|b|^2}{4|b|^2+\big(\sqrt{(a-c)^2+4|b|^2}\pm|a-c|\big)^2}.
\end{equation}
If $a=c$, then $\xi_2=0$, i.e., in this case the mode functions are linearly polarized for $|p_3/q|\gg1$. Furthermore, the conclusions on polarization of mode functions are not valid for the energies close to the band gaps since, for $|b|\ll1$, the branches are not well separated and strongly interact (are mixed) in this region of energies (see Fig. \ref{Disp_Law_Parax_plots}).

Notice that the paraxial limit we are considering does not reproduce the resonance Bragg reflection resulting from the component of the permittivity tensor with spin $s_h=\pm1$ \cite{FarLakh14}. Such scattering must lead to the appearance of the additional band gaps in the dispersion law at energies lower than the energy (i) in \eqref{band_gaps_s2}. The absence of these additional gaps in the paraxial limit can be expected already from the perturbative approach. In the leading order of perturbation theory, the scattering probability \eqref{H21} vanishes for $s_h=\pm1$ contribution to the permittivity tensor when $n_\perp=0$ \cite{BelyakovBook}. Nevertheless, one can find the approximate positions of the additional gaps in the dispersion law for small $n_\perp$. When $n_\perp$ is small, the Maxwell equations take approximately the form \eqref{parax_eqs}. The gaps in the dispersion law appear when the branches of the dispersion laws following from the nulling of the determinants \eqref{parax_eqs} with different $l$ intersect (see for details \cite{LLPhysKin}). The sought forbidden gap lies at the intersection of the branches of the dispersion laws corresponding to $l=0$ and $l=1$. Solving the respective system of equations for $p_3$ and $k_0$, we come to the analog of \eqref{band_gaps_s2} but for scattering by the component of the permittivity tensor with $s_h=\pm1$:
\begin{equation}\label{band_gaps_s1}
\begin{aligned}
    (i)&&\frac{k_0}{q}&=\frac{1}{a^{1/2}+c^{1/2}}+\cdots,&\qquad \frac{p_3}{q}&=\frac{a^{1/2}}{a^{1/2}+c^{1/2}}+\cdots,\\
    (ii)&&\frac{k_0}{q}&=\frac{1}{|a^{1/2}-c^{1/2}|}+\cdots,&\qquad \frac{p_3}{q}&=\frac{a^{1/2}}{a^{1/2}-c^{1/2}}+\cdots.\\
\end{aligned}
\end{equation}
As in the case $s_h=\pm2$, the branch points (i) describe the real band gap, whereas the branch points (ii) are responsible for the imaginary gap. Formulas \eqref{band_gaps_s1} are in a good agreement with numerical simulations (see Fig. \ref{Disp_Hels1_plots}). Notice that the imaginary gaps of the dispersion law become apparent in the scattering data only for sufficiently small gap parameter, when the corresponding branch points are close to the real axis of the complex $k_0$ plane. In this case, the branches of the plot of the dispersion law come close to each other near the values of momentum given in item (ii) of formulas \eqref{band_gaps_s2}, \eqref{band_gaps_s1}. If the asymmetry is positive, $a>c$, then $\re p_3>0$ and the peculiarities of scattering at the given energy are observed in the transmitted wave. If the asymmetry is negative, $a<c$, then $\re p_3<0$ and, as a result, the peaks appear in the characteristics of the reflected wave. It follows from the analysis of scattering of twisted photons near the band gaps \eqref{band_gaps_s1} that these gaps of the dispersion law are provided by the component of the permittivity tensor with spin $s_h=\pm1$ (see Fig. \ref{Scatt_Hels1c_plots}).

\subsubsection{Numerical simulations}\label{Num_Sim}

As it has been already mentioned above, the system of equations \eqref{Max_eqns_1} does not seem to have an exact solution in a closed form for $k_\perp>0$. Nevertheless, the mode functions can be found numerically by employing, for example, the $N$-wave approximation.

In this approximate procedure, one substitutes the Fourier series \eqref{Fourier_series} into Eq. \eqref{Max_eqns_1}. As a result, Eq. \eqref{Max_eqns_1} is reduced to the infinite system of entangled linear equations for the coefficients $a_{l,+}$ and $a_{l-2,-}$. Then one singles out the block from the system of equations corresponding to these coefficients with $|l|\leqslant N$. The remaining coefficients are set to zero. The resulting finite dimensional matrix possesses nontrivial null vectors for the momenta $p^{(a)}_3(k_0)$, $a=\overline{1,4N}$, which are found by solving the respective polynomial equation. After that, one picks out the four physical vectors from these null vectors requiring that the contribution of the coefficients with $l$ close to zero dominates in these vectors. Notice that if we had the exact solutions of the infinite system of equations, then the choice of the four physical null vectors would be irrelevant -- the only requirement is that they correspond to the different quasimomenta $\kappa_3(k_0)=p_3(k_0)\mod q$. However, as long as the infinite chain of equations is broken in numerical simulations, one needs to use a certain criterion for the choice of the physical null vectors assuming that the maximum contribution to the norm of these vectors comes from the components near $l=0$. The components with large $|l|\sim N$ are poorly described by the $N$-wave approximation as they are close to the boundary of the chosen matrix block out of which the coefficients  $a_{l,+}$ and $a_{l-2,-}$ are set to zero. Having found the mode functions, their linear combination with the coefficients $b_i$ is constructed and Eqs. \eqref{joining_conds} resulting from the joining conditions for the mode functions \eqref{bound_conds} are solved. The results of application of this numerical procedure are presented in Appendix \ref{Scat_Data_Ap}.

\section{Conclusion}

Let us sum up the results. We investigated the properties of scattering of plane-wave and twisted photons by helical media. At first, we found the general form \eqref{eps_hel_gen} of the permittivity tensor obeying the helical symmetry. In the particular cases, it describes the electromagnetic properties of cholesteric liquid crystals, $C^*$-smectics, chiral sculptured thin films, and $q$-plates. In the first Born approximation, we derived the explicit expressions for the scattering probability of plane-wave photons \eqref{dP_forward}, \eqref{dP_plane_refl1} by the helical medium invariant under translations perpendicular to the $z$ axis (the axis of the helical symmetry) and for the scattering amplitude of twisted photons \eqref{amplitude_tw}, \eqref{amplitude_tw_mom} by the helical medium of a general form. The scattering amplitude \eqref{F_s_h} of twisted photons for the particular case of a helical medium invariant with respect to translations in the $(x,y)$ plane was also obtained. The helical medium is supposed to form a plate normal to the $z$ axis and immersed into a homogeneous isotropic medium with permittivity $\e_v(k_0)$.

The general formulas imply, in particular, that the selection rules expressing the momentum and angular momentum conservation laws are satisfied, the projection of the momentum onto the $z$ axis and the projection of the total angular momentum onto this axis being transferred from the medium to the photon. We proposed to use such a process for production of photons with given projection of total angular momentum. In fact, such a mechanism works in $q$-plates \cite{LiZX21,Barboza2,MarManPap06,Barboza1,Naidoo16,Brasselet18} and we proposed an alternative scheme for shifting the total angular momentum of a twisted photon. This procedure is based on multiple reflection of a twisted photon propagating between two parallel plates made of helical media with opposite periodicity parameters $q$, for example, between two cholesteric plates with the same pitches but opposite chiralities (for other schemes see, e.g., \cite{FarLakh14,RafBrass18,LinT19}). The proposed AMSD can be made rather compact, can produce twisted photons with large projection of the total angular momentum, and, at least in principle, allows one to construct the mechanism for parallel coding of a given signal in terms of twisted photons (see Fig. \ref{Scheme_plots}).

As far as a nonperturbative analysis is concerned, we investigated scattering of photons by the plate made of helical medium invariant under translations in the $(x,y)$ plane. Many properties of such scattering were already investigated in the literature \cite{LakhWeigh95,FarLakh14,MacLakh,LakhMess05,LaVeMcC00,BitThom05,Lakht10,TChJhHu17,McCLakh04}. Nevertheless, we established several new, to our knowledge, results. First, we developed a general formalism for description of scattering of twisted photons by introducing the respective transmission and reflection matrices. Second, we considered the paraxial limit, when the corresponding Maxwell equations are exactly solvable \cite{LakhWeigh95}, and found simple expressions \eqref{band_gaps_s2}, \eqref{band_gaps_s1} for the positions of the real and imaginary band gaps for both second and first orders of Bragg scattering. As for imaginary gaps, they appear only for helical media where the dispersion law of photons has a certain asymmetry such that the branches of the plot $p_3(k_0)$ come close to each other near this gap. Third, we described the polarization properties of the modes corresponding to different branches of the dispersion law and obtained explicit expressions \eqref{Stokes_parax} for the Stokes vectors of these modes in the short wavelength approximation.

With the aid of numerical simulations, we corroborated the conclusions drawn from analytical results. Furthermore, we investigated scattering of twisted photons by the plate of helical medium invariant with respect to translations in the $(x,y)$ plane in the nonperturbative regime. As expected, the scattered twisted photons with energies lying near the band gaps acquire an additional amount of projection of the total angular momentum as it is predicted by the selection rules. Scattering near the band gaps related to the spin $\pm1$ component of the permittivity tensor gives rise to the shift of the total angular momentum by $\pm1$ (see Figs. \ref{Scatt_Hels1r_plots}, \ref{Scatt_Hels1c_plots}), where the sign is determined by the medium chirality. Scattering near the band gaps related to the spin $\pm2$ component of the permittivity tensor results in the shift of the total angular momentum by $\pm2$ (see Figs. \ref{Scatt_Hels2r_plots}, \ref{Scatt_Hels2c_plots}). Notice that, in the nonparaxial regime, the latter band gap reflects the photons of both helicities, i.e., it is the total band gap (as for cholesterics, see \cite{BerrSchef,RisSchm19} and Fig. \ref{Disp_Chol_plots}). This allows one to construct the AMSD by using the two cholesteric plates with opposite chiralities. On reflecting, the first order Bragg band gap, i.e., the band gap of the spin $\pm1$, reverts the helicity of the incident photon in the paraxial limit and transfers the projection of the orbital angular momentum to it. Therefore, this band gap can be employed in the AMSD even in the paraxial regime.

\paragraph{Acknowledgments.}

We are indebted to O.V. Bogdanov for the help with numerical simulations and to G.Yu. Lazarenko for useful comments. This study was supported by the Tomsk State University Development Programme (Priority-2030).

%newpage
\appendix
\section{Scattering data}\label{Scat_Data_Ap}

See the next pages.
%\newpage

\begin{figure}[tp]
\centering
\raisebox{-0.5\height}{\includegraphics*[width=0.24\linewidth]{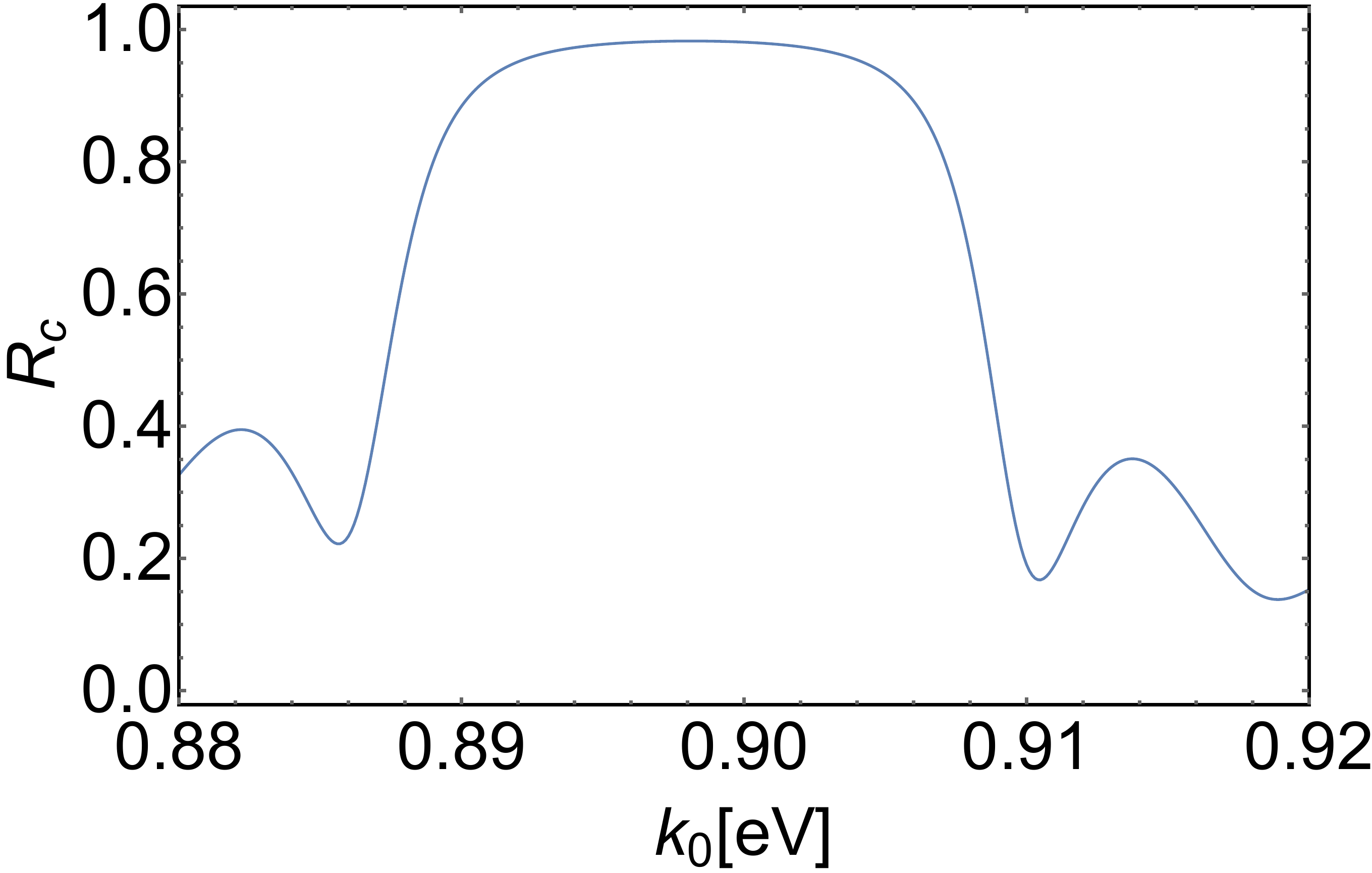}}\,
\raisebox{-0.5\height}{\includegraphics*[width=0.24\linewidth]{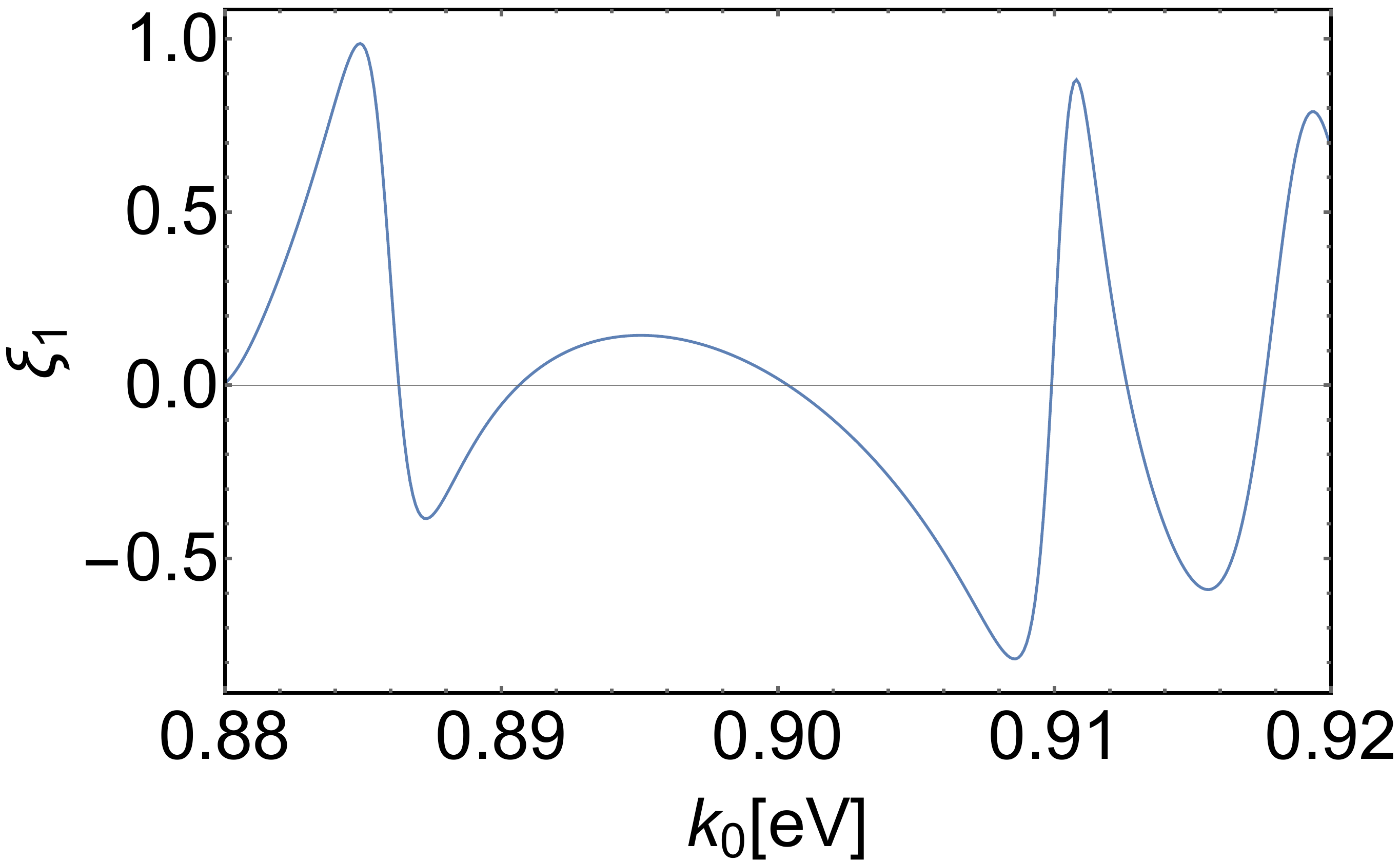}}\,
\raisebox{-0.5\height}{\includegraphics*[width=0.24\linewidth]{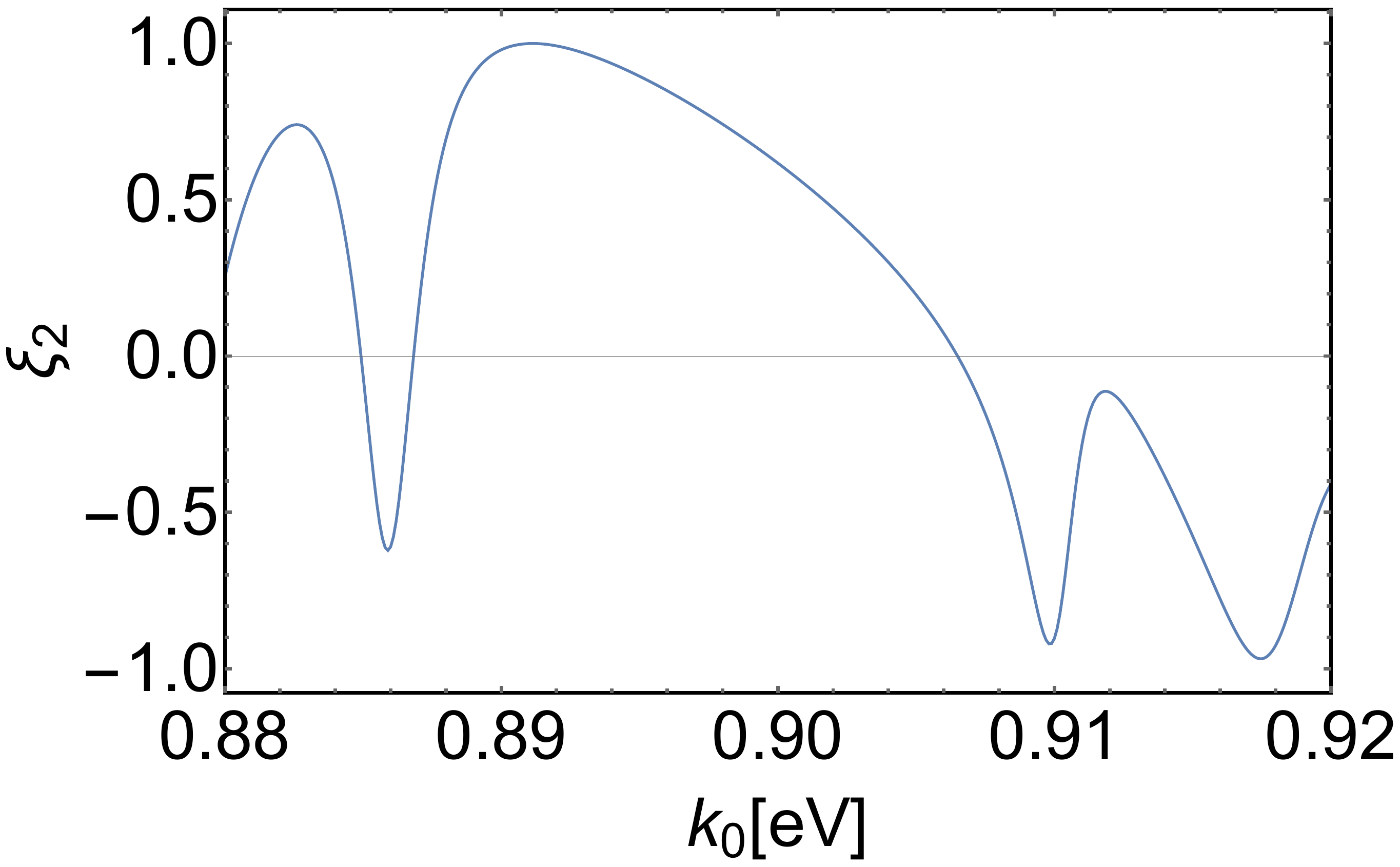}}\,
\raisebox{-0.5\height}{\includegraphics*[width=0.24\linewidth]{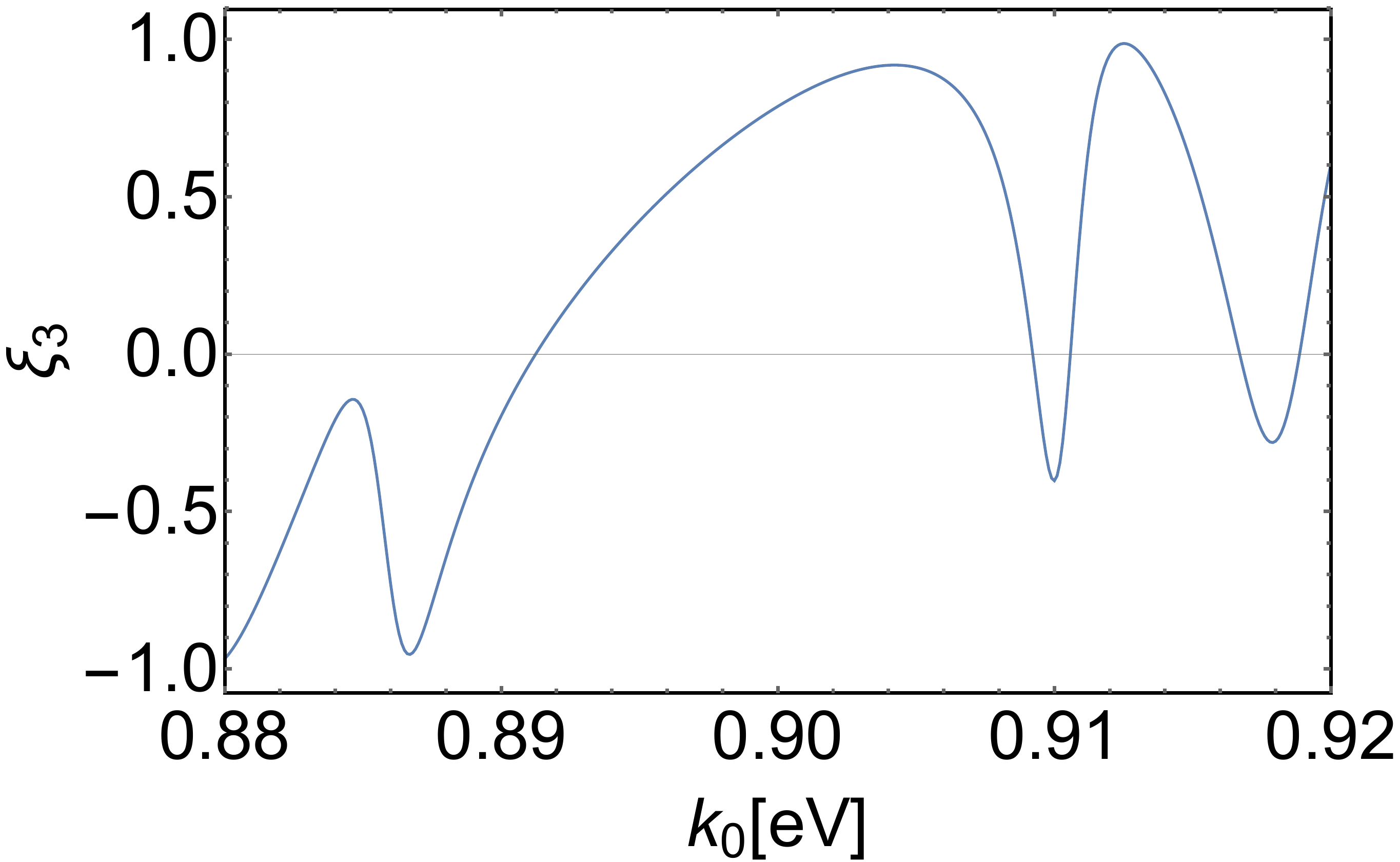}}\\
\raisebox{-0.5\height}{\includegraphics*[width=0.24\linewidth]{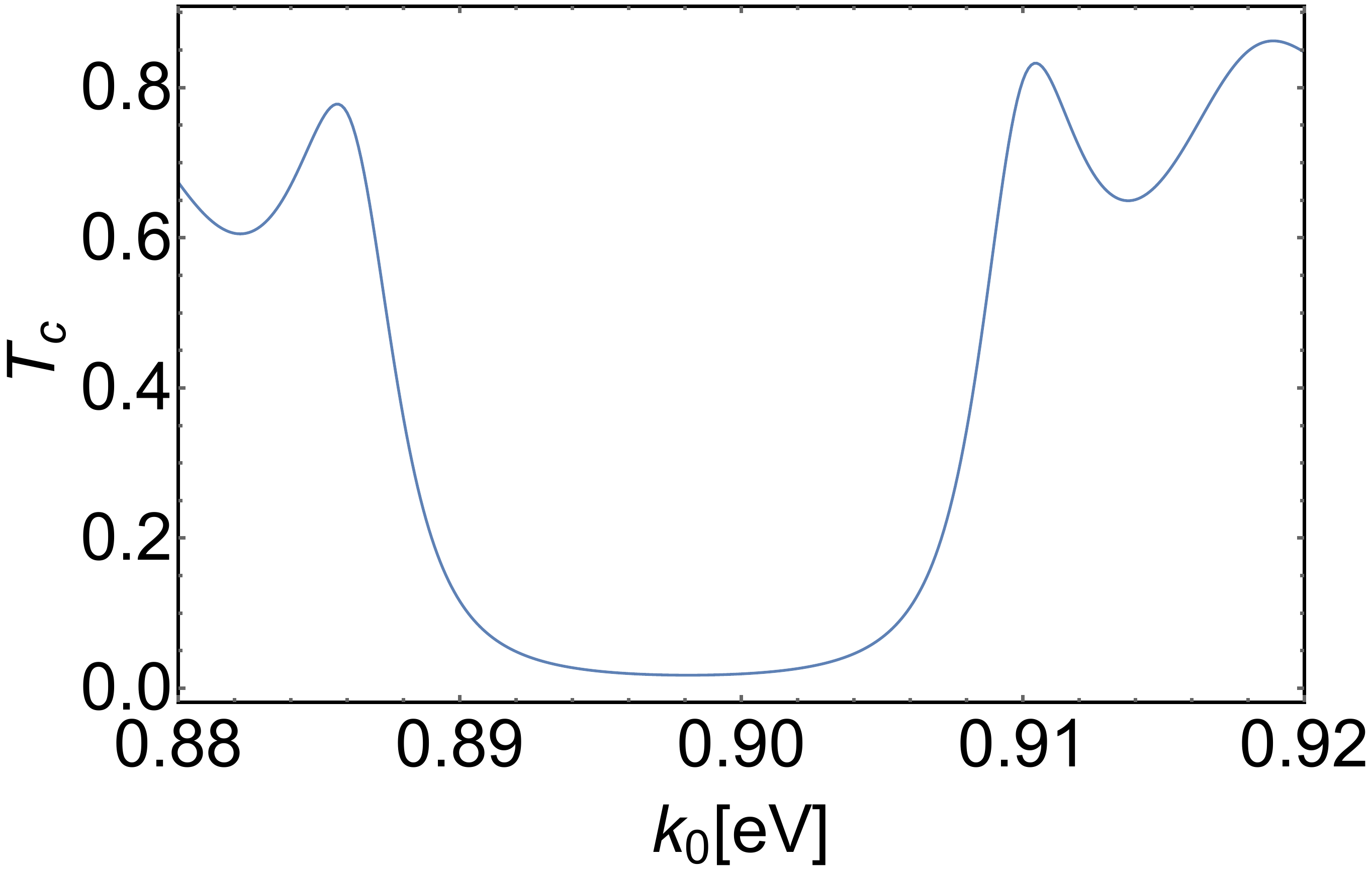}}\,
\raisebox{-0.5\height}{\includegraphics*[width=0.24\linewidth]{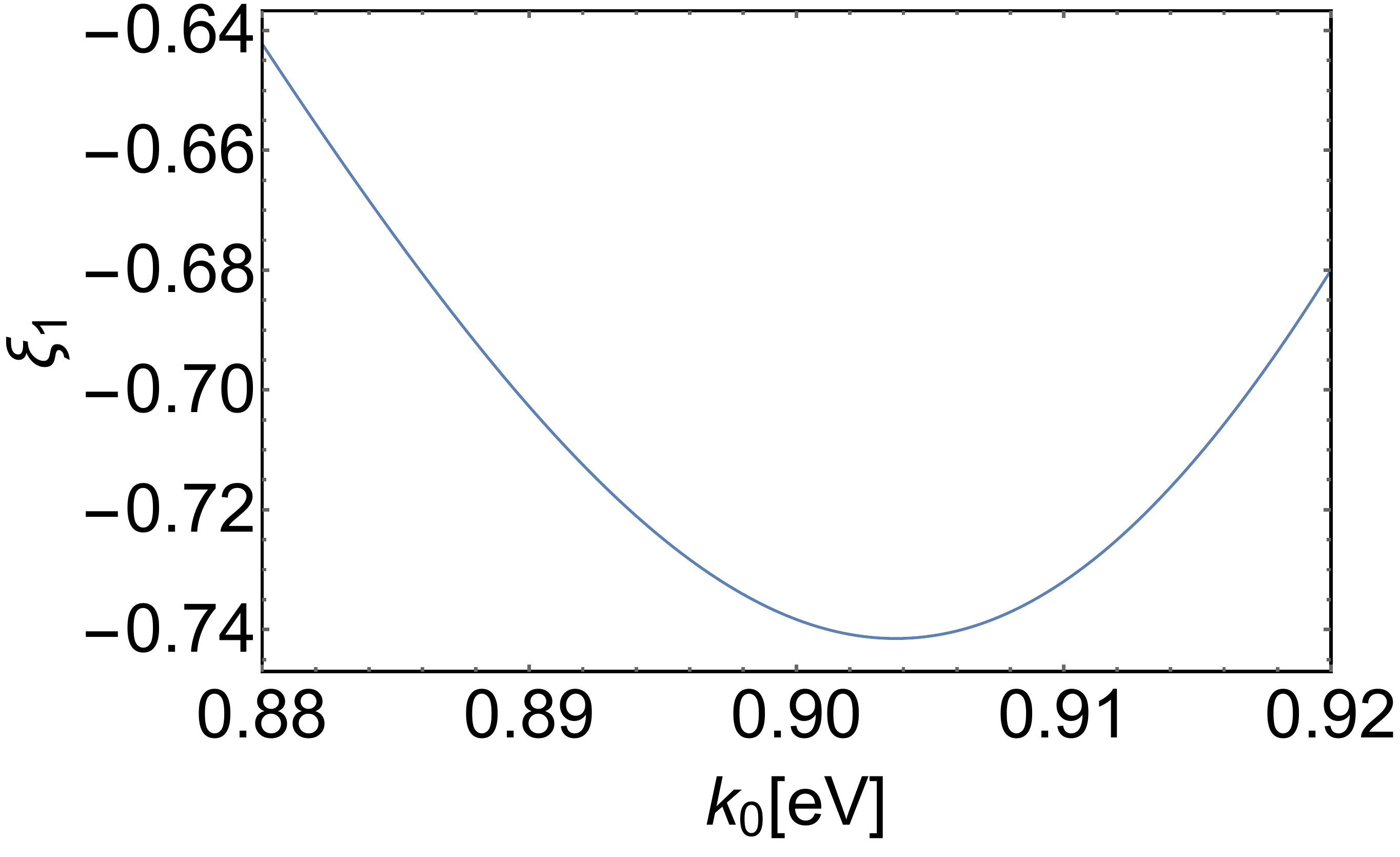}}\,
\raisebox{-0.5\height}{\includegraphics*[width=0.24\linewidth]{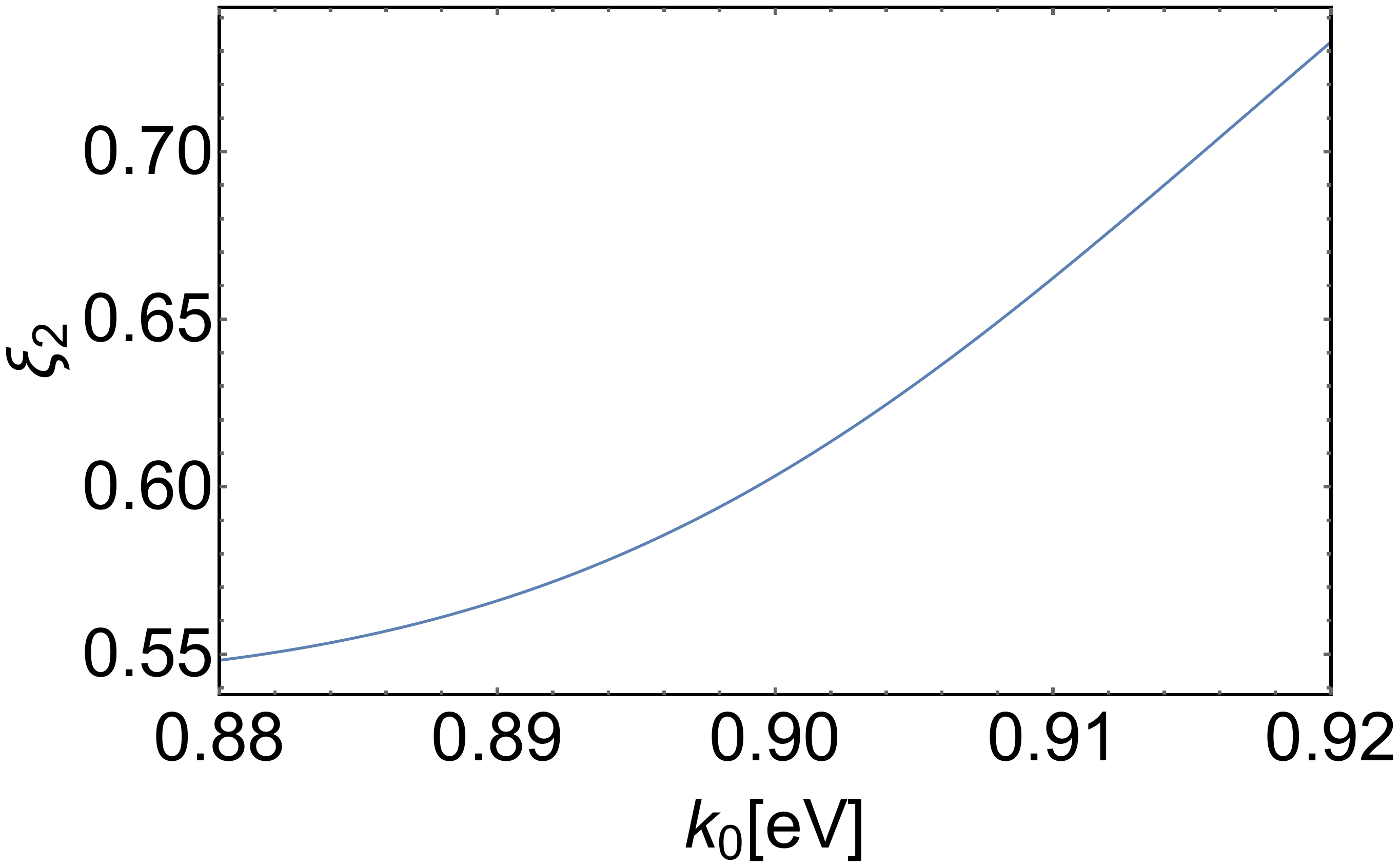}}\,
\raisebox{-0.5\height}{\includegraphics*[width=0.24\linewidth]{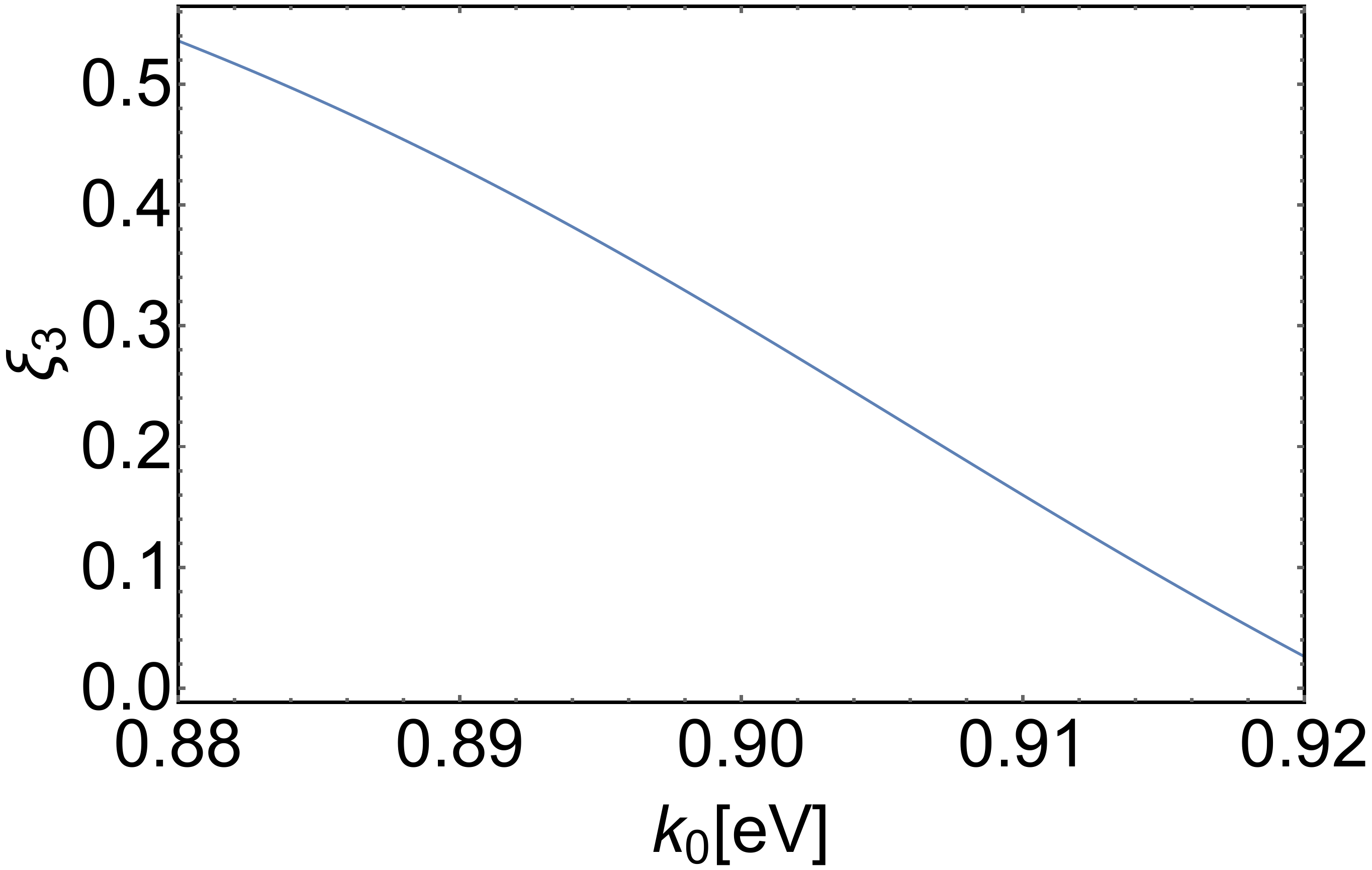}}\\
\raisebox{-0.5\height}{\includegraphics*[width=0.24\linewidth]{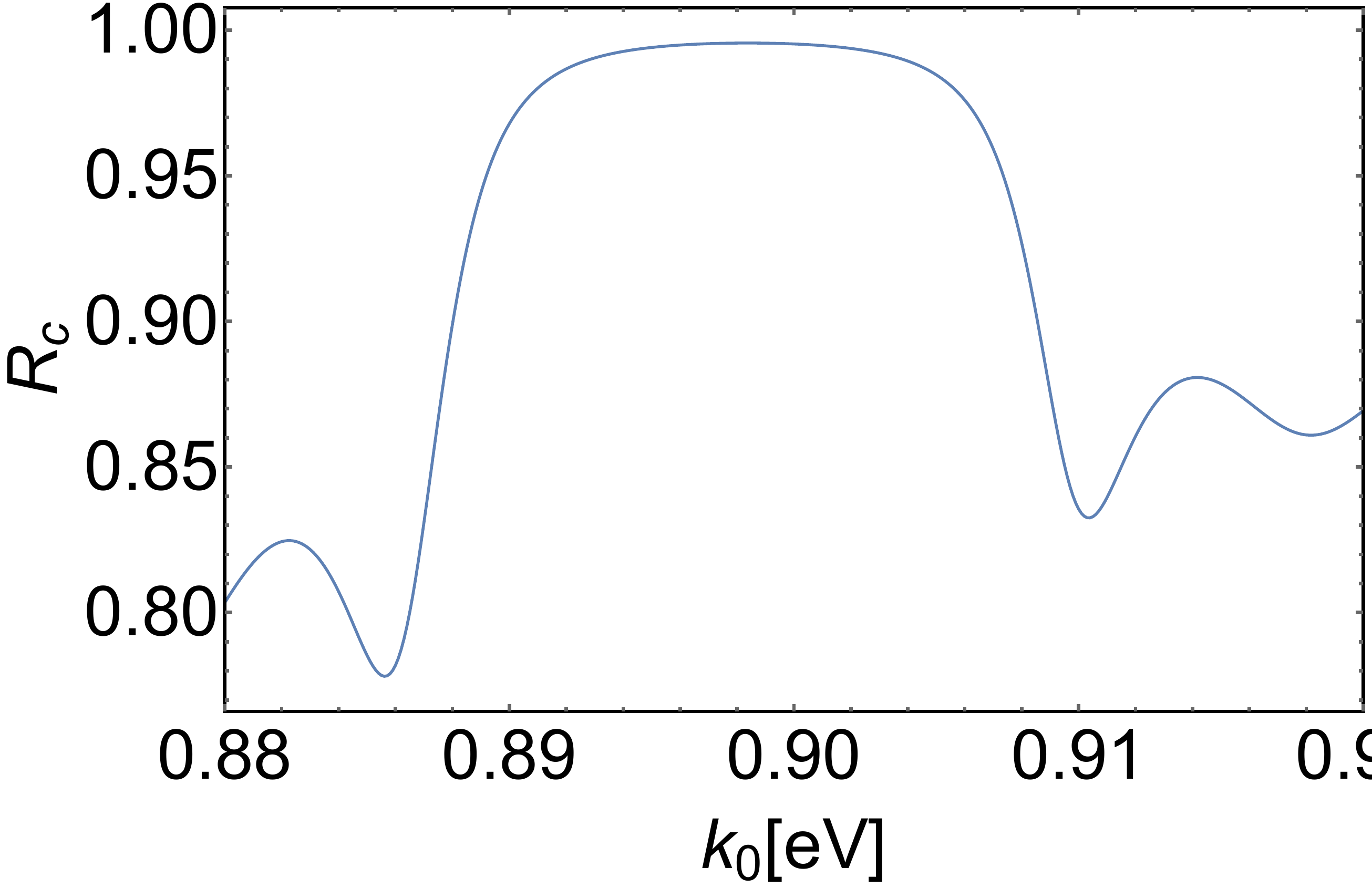}}\,
\raisebox{-0.5\height}{\includegraphics*[width=0.24\linewidth]{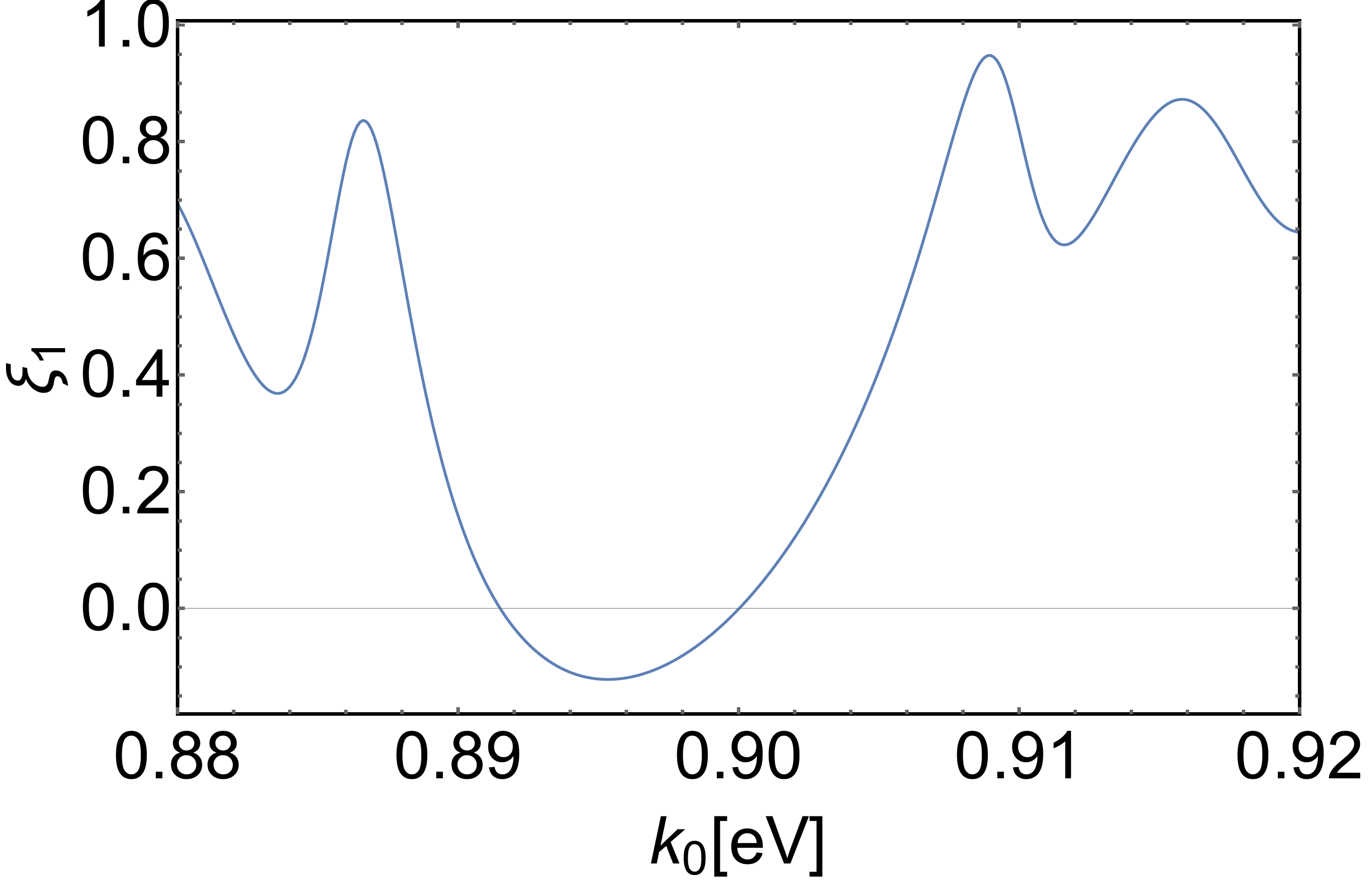}}\,
\raisebox{-0.5\height}{\includegraphics*[width=0.24\linewidth]{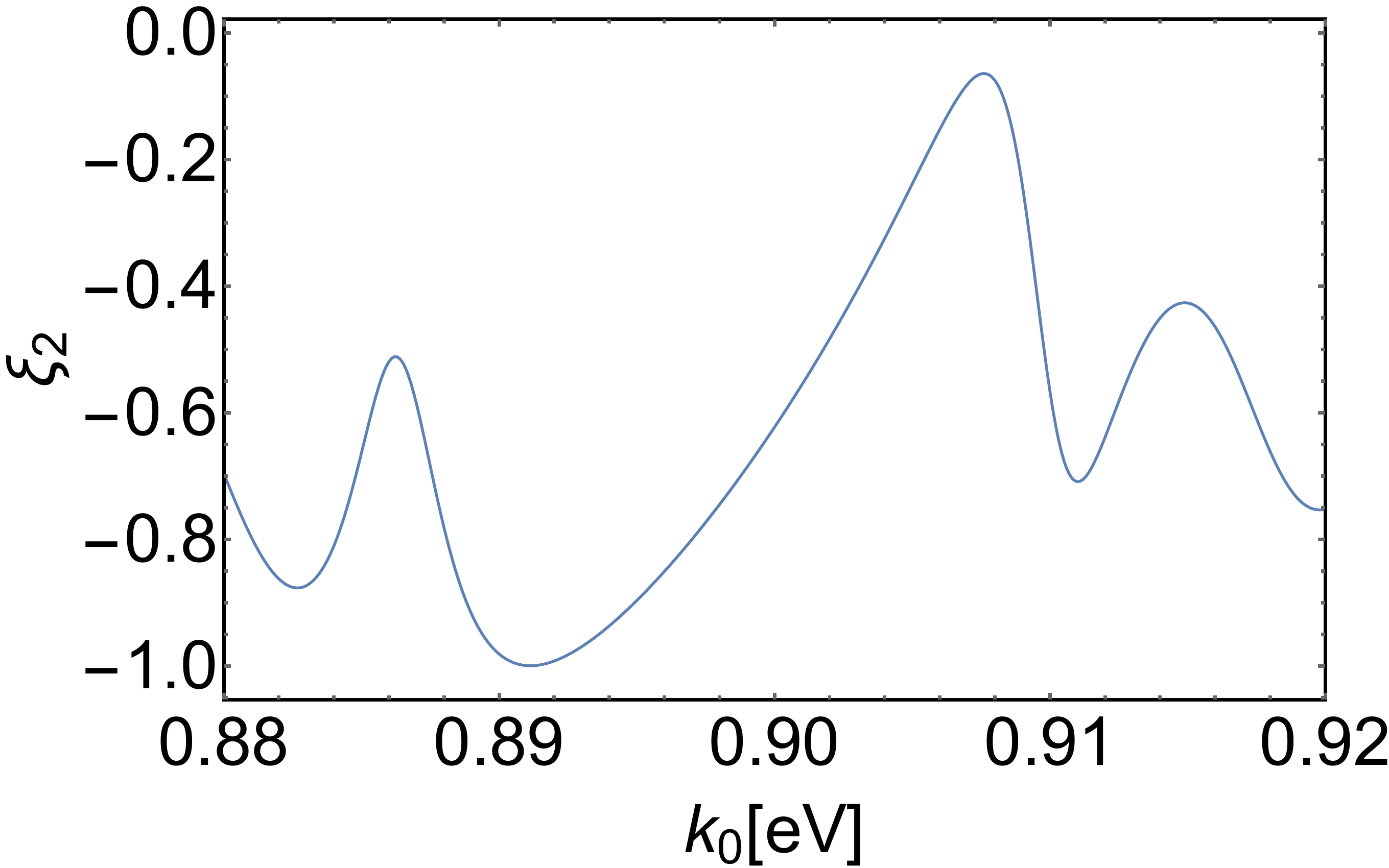}}\,
\raisebox{-0.5\height}{\includegraphics*[width=0.24\linewidth]{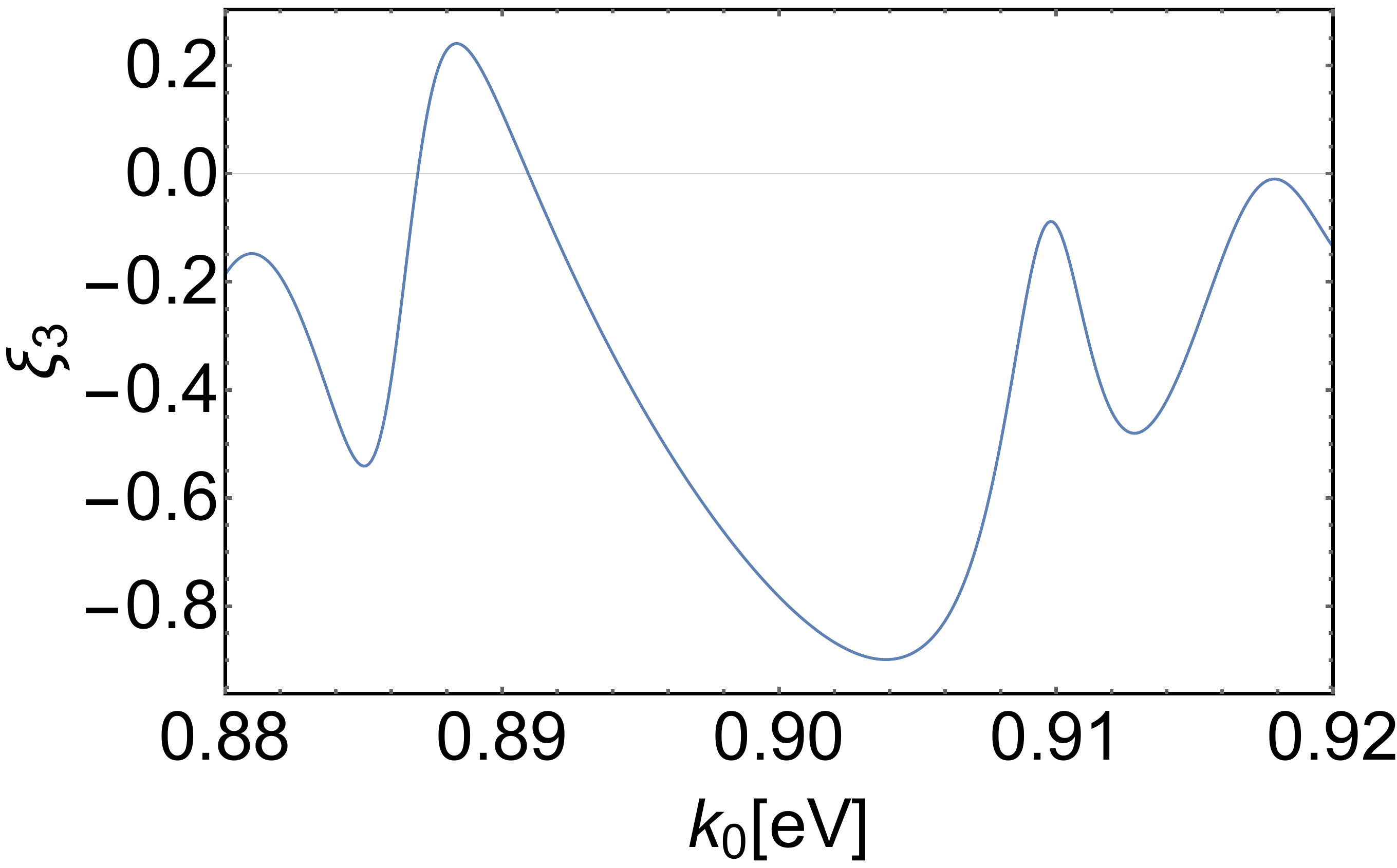}}\\
\raisebox{-0.5\height}{\includegraphics*[width=0.24\linewidth]{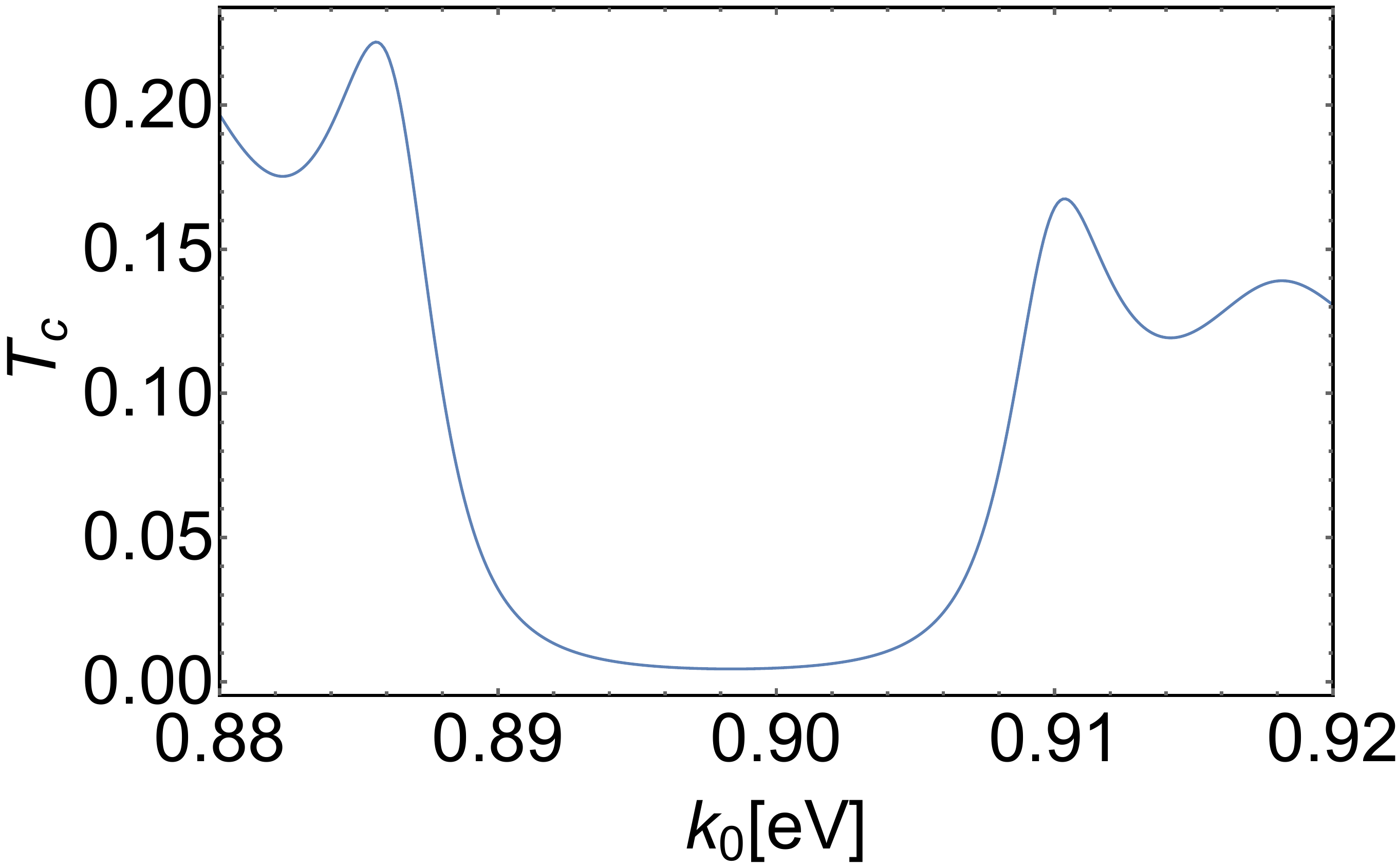}}\,
\raisebox{-0.5\height}{\includegraphics*[width=0.24\linewidth]{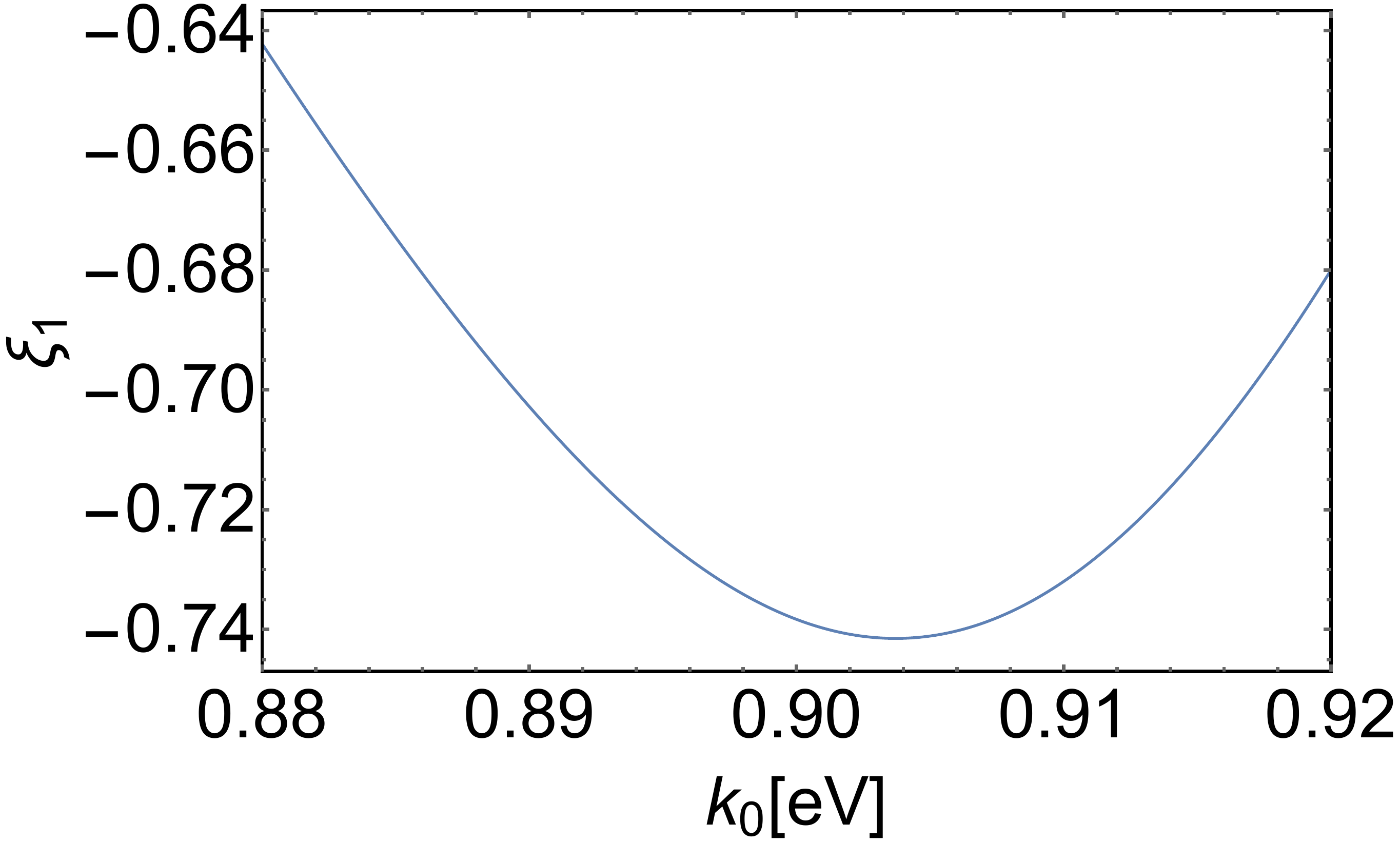}}\,
\raisebox{-0.5\height}{\includegraphics*[width=0.24\linewidth]{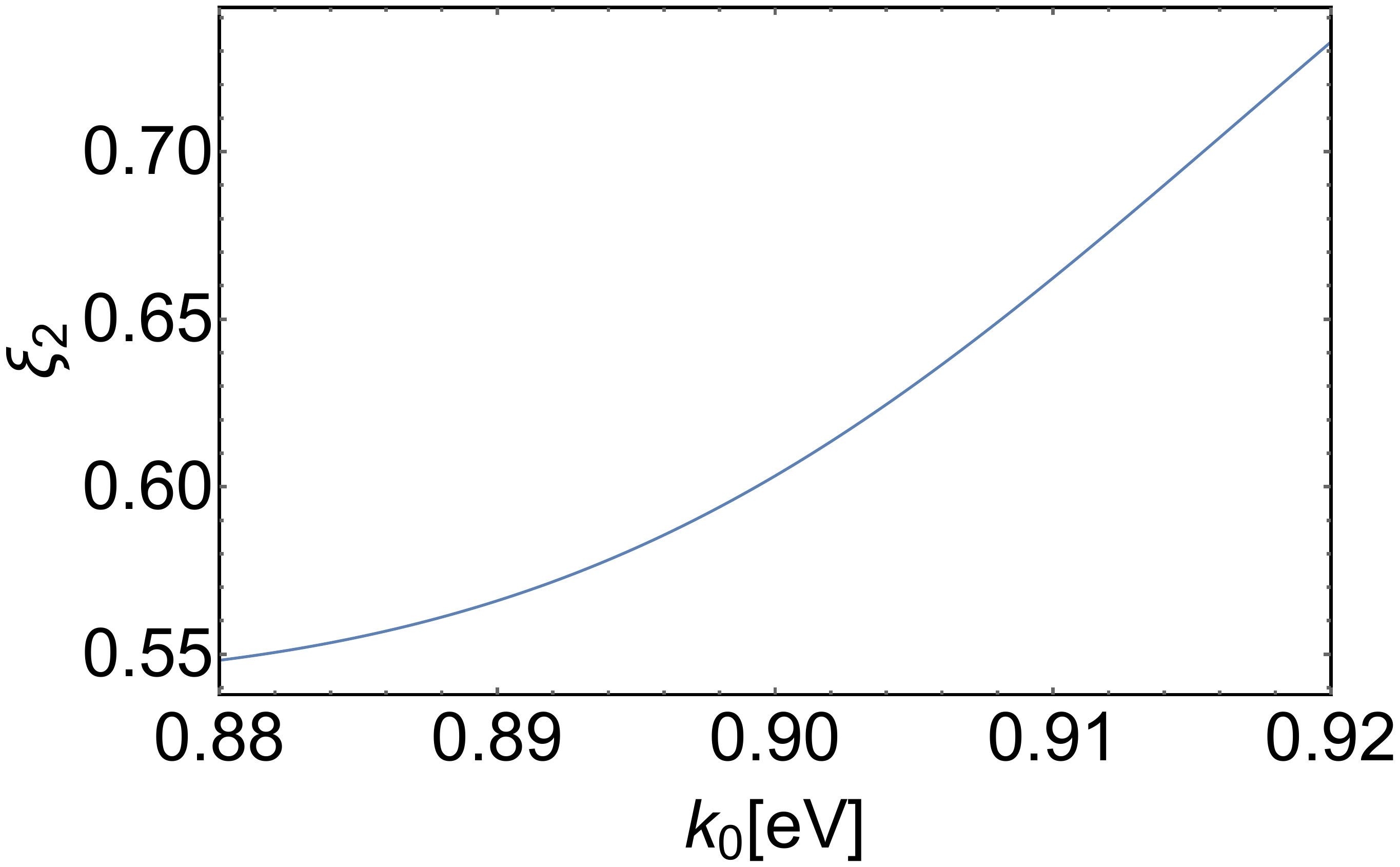}}\,
\raisebox{-0.5\height}{\includegraphics*[width=0.24\linewidth]{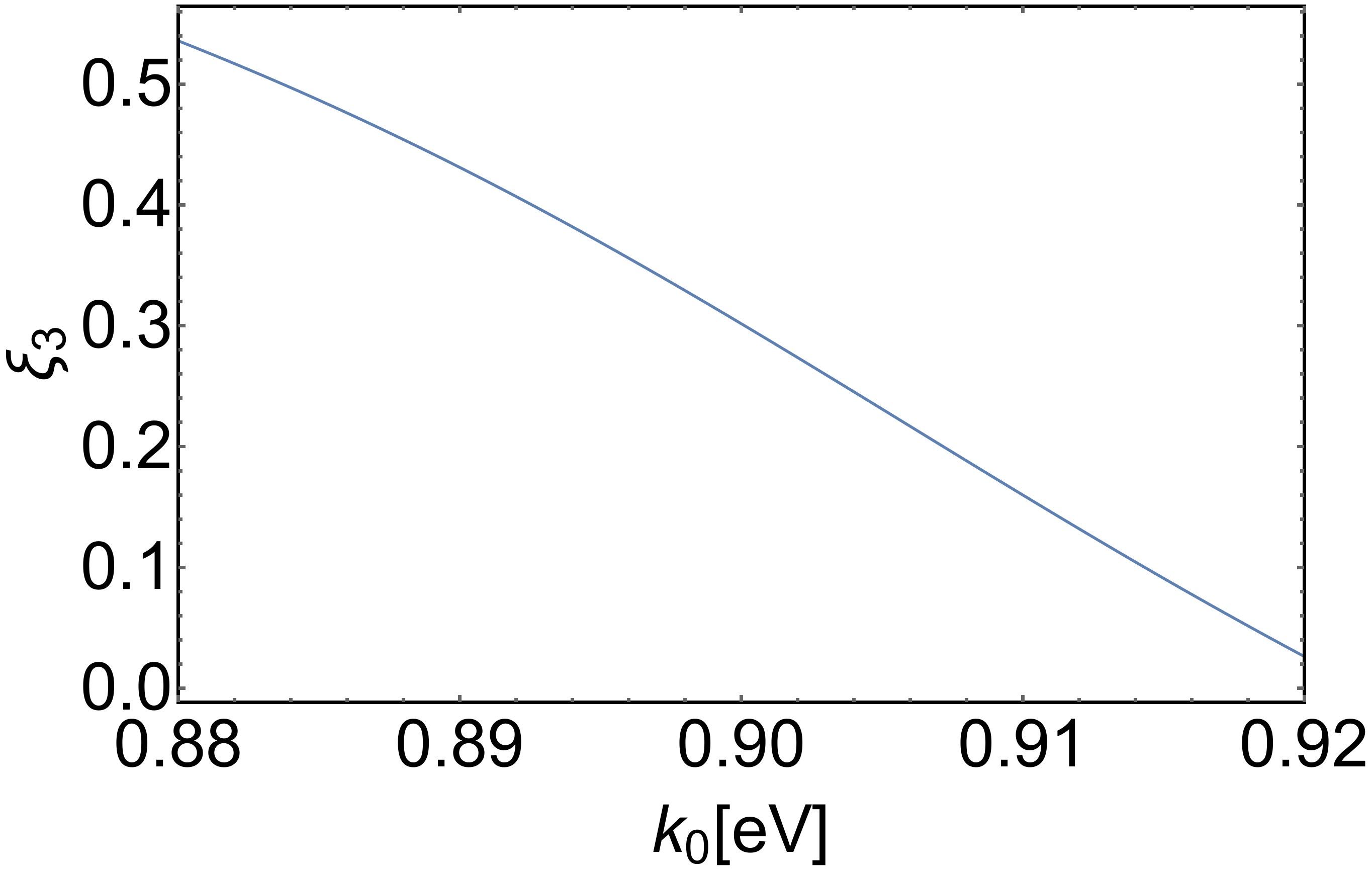}}\\
\raisebox{-0.5\height}{\includegraphics*[width=0.24\linewidth]{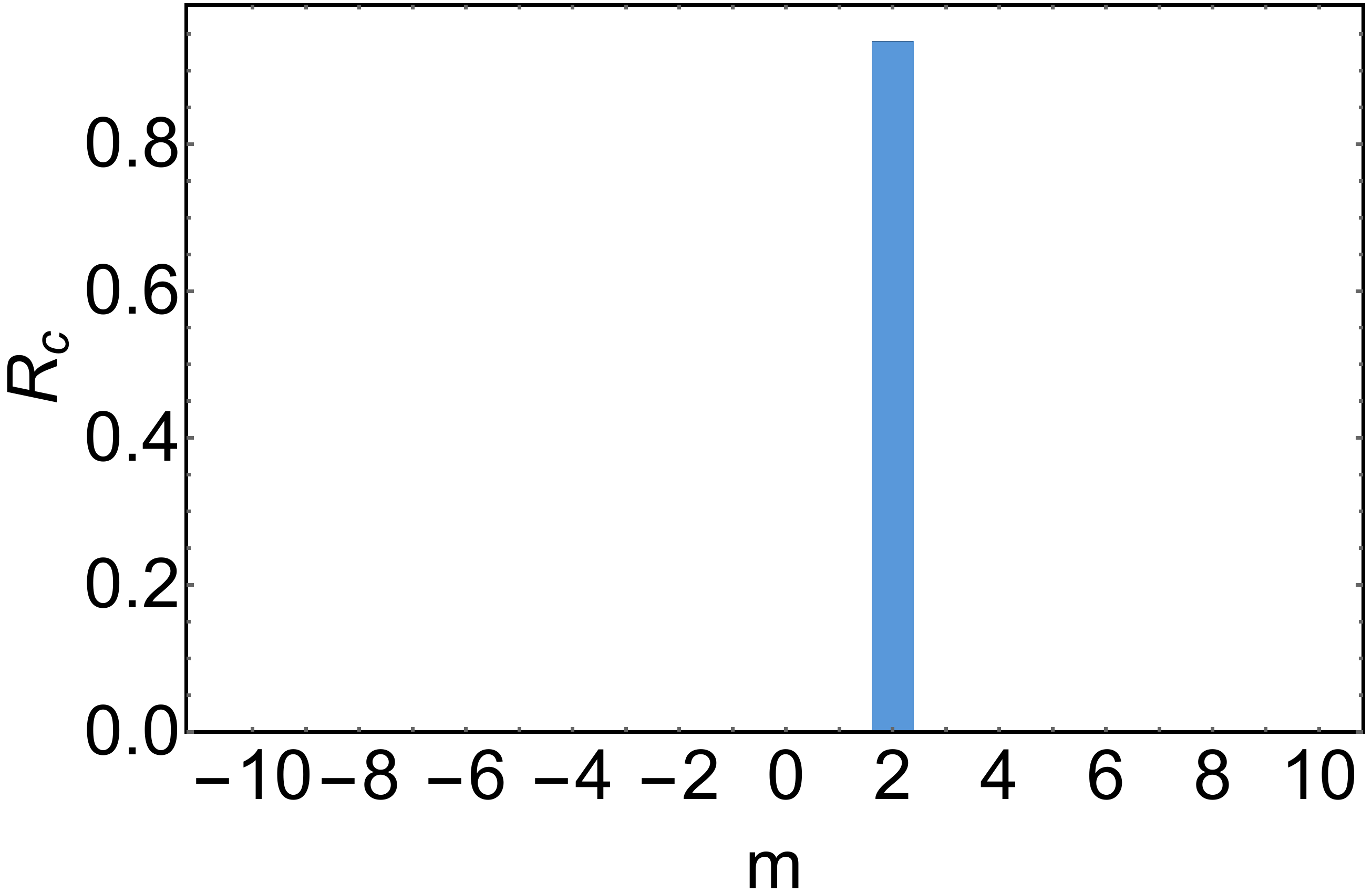}}\,
\raisebox{-0.5\height}{\includegraphics*[width=0.24\linewidth]{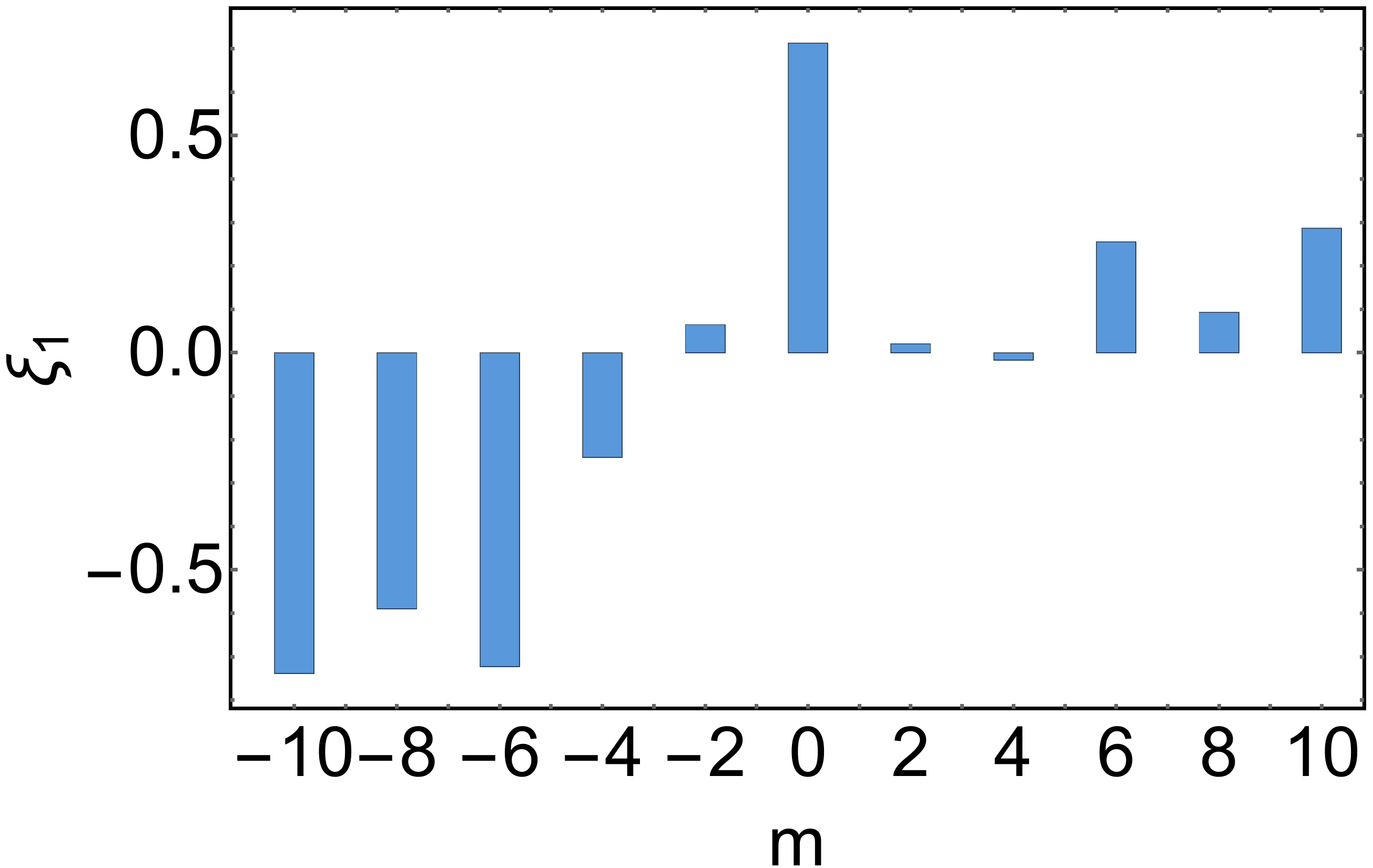}}\,
\raisebox{-0.5\height}{\includegraphics*[width=0.24\linewidth]{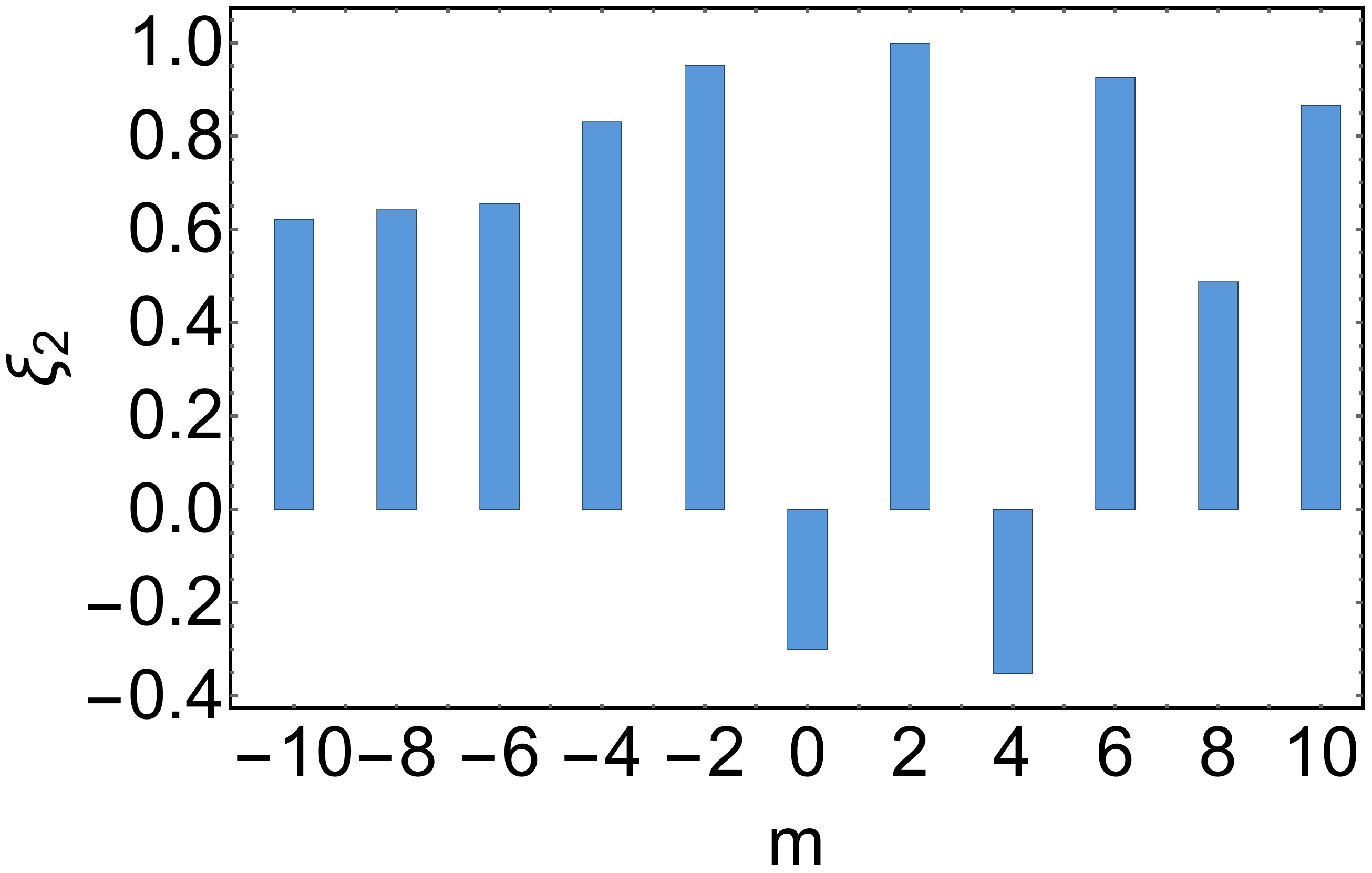}}\,
\raisebox{-0.5\height}{\includegraphics*[width=0.24\linewidth]{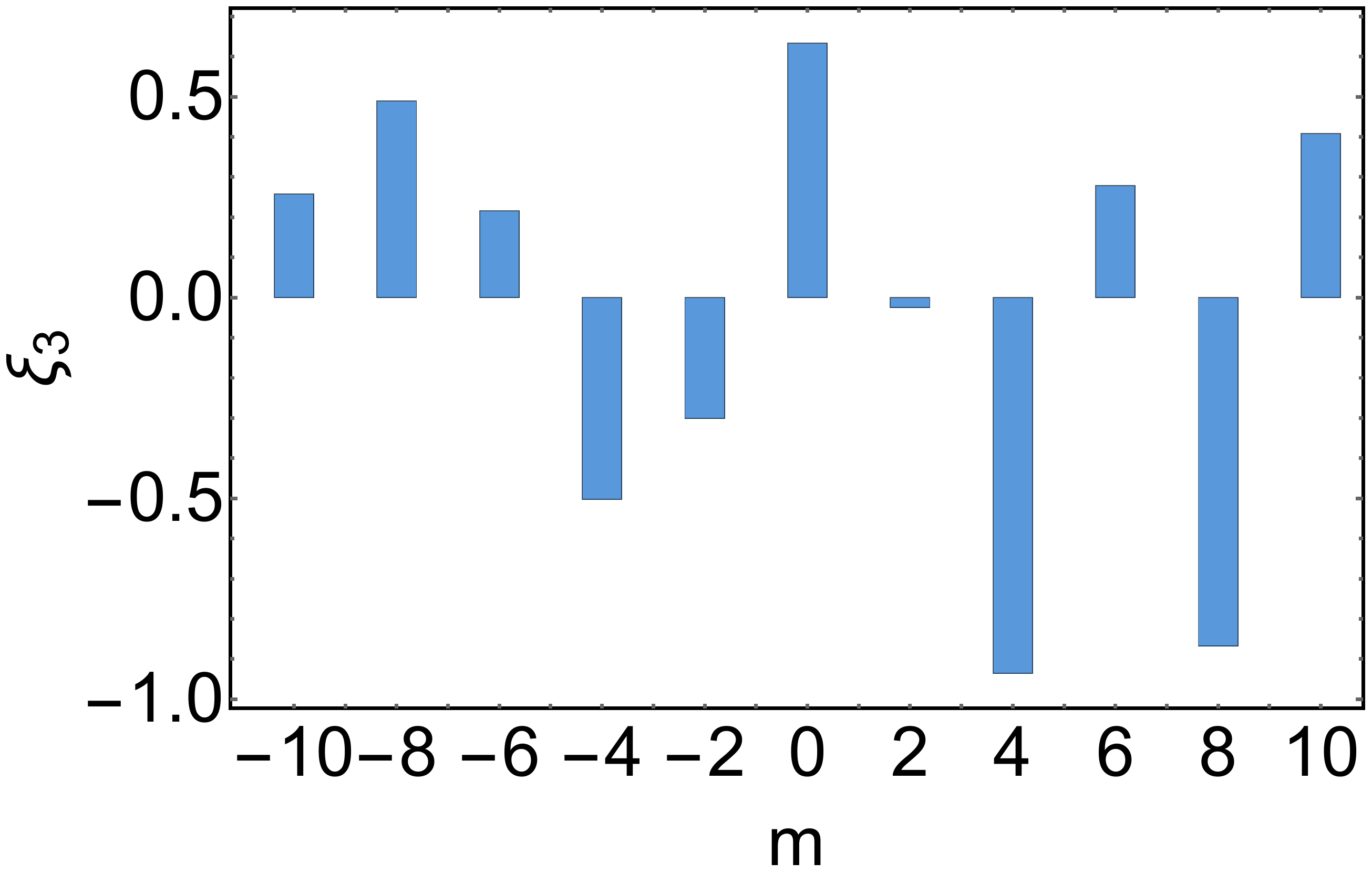}}\\
\raisebox{-0.5\height}{\includegraphics*[width=0.24\linewidth]{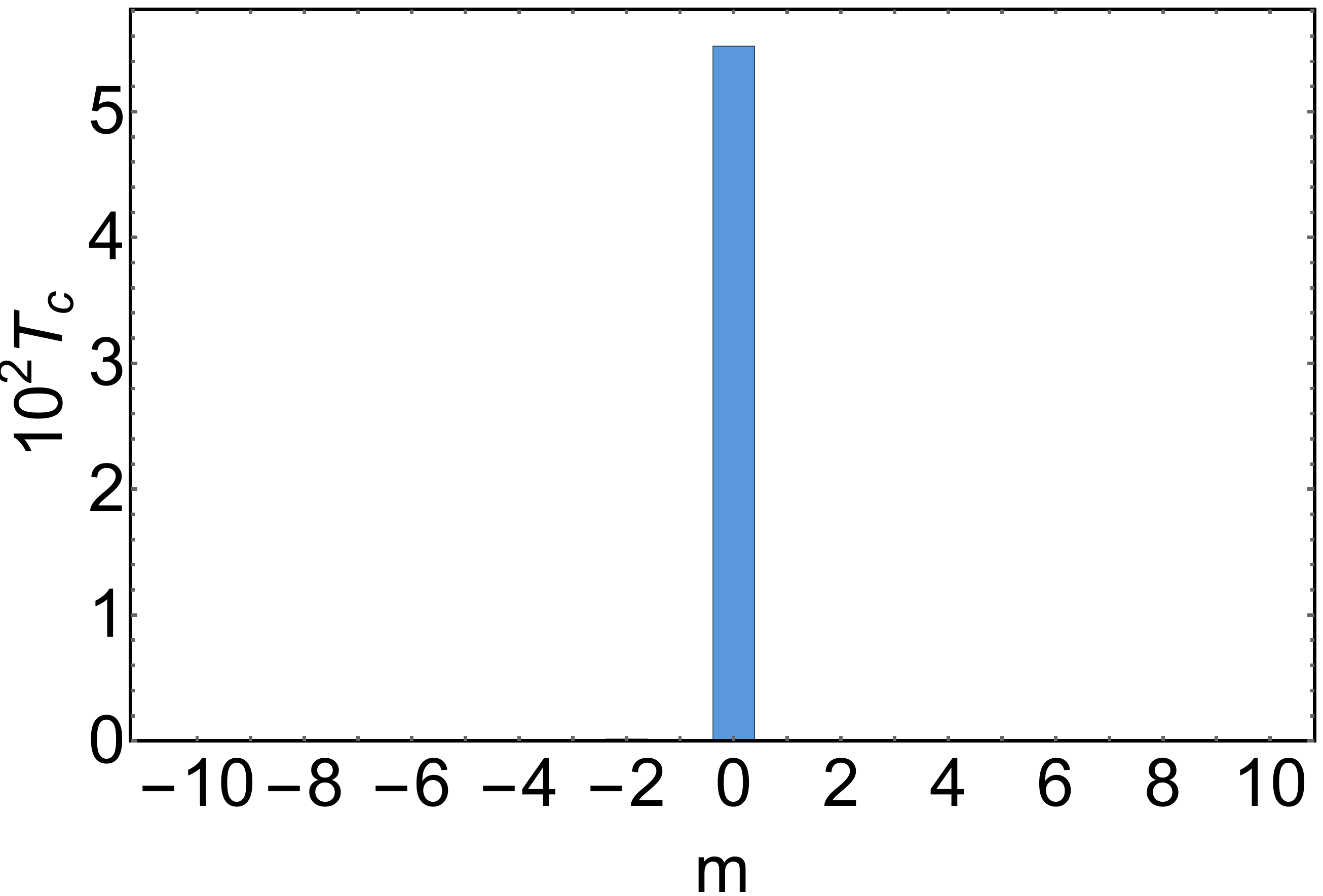}}\,
\raisebox{-0.5\height}{\includegraphics*[width=0.24\linewidth]{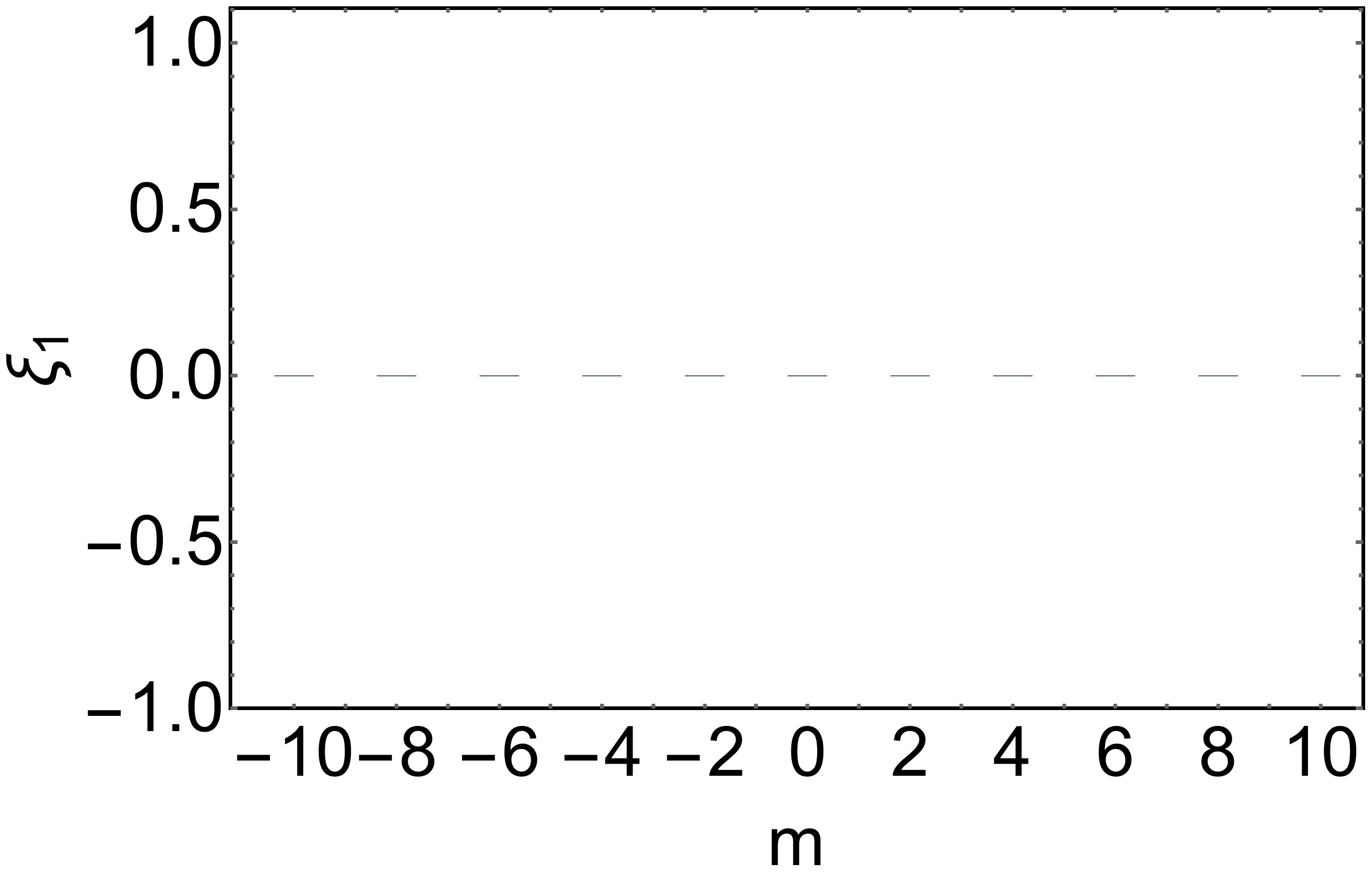}}\,
\raisebox{-0.5\height}{\includegraphics*[width=0.24\linewidth]{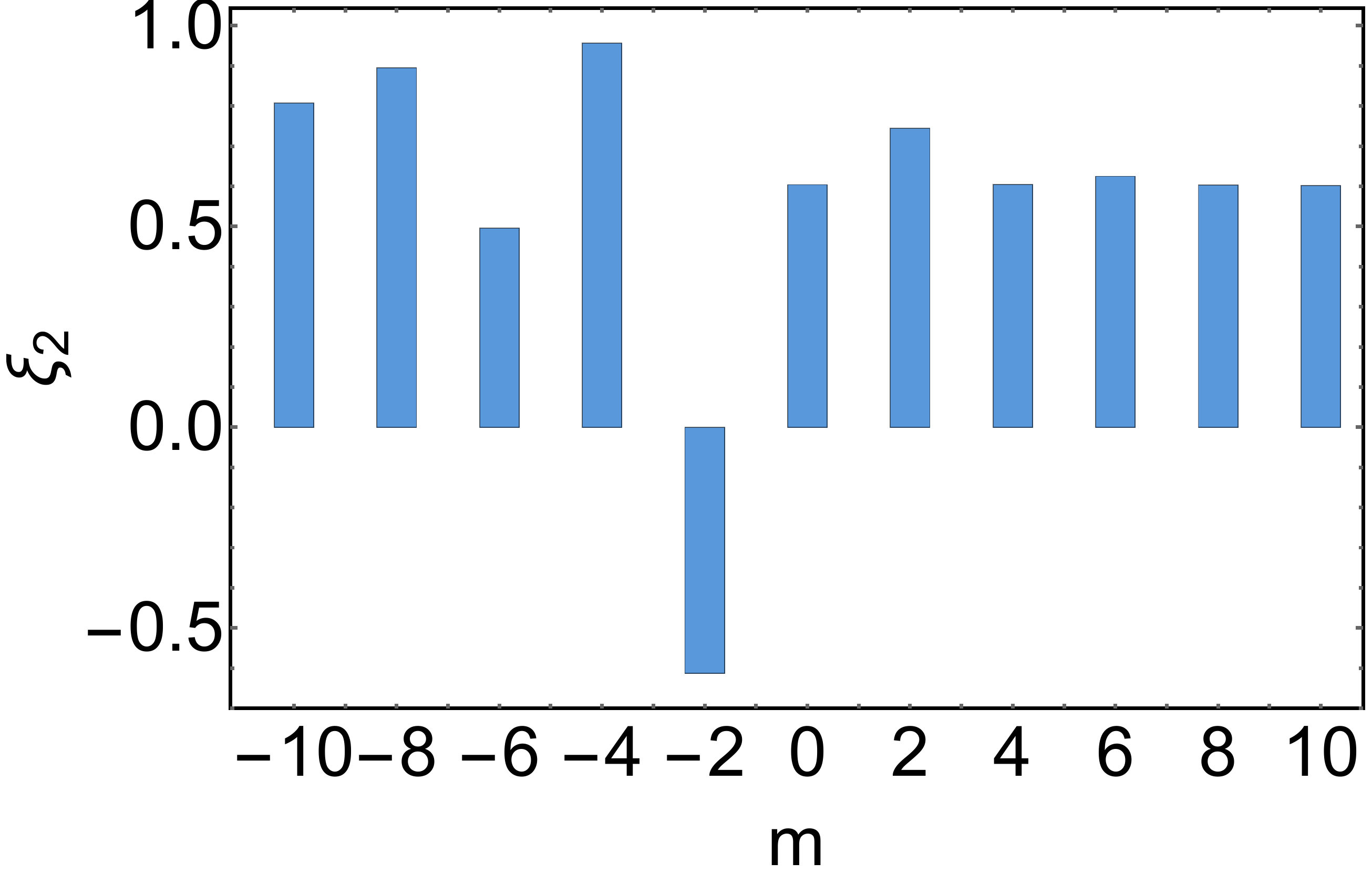}}\,
\raisebox{-0.5\height}{\includegraphics*[width=0.24\linewidth]{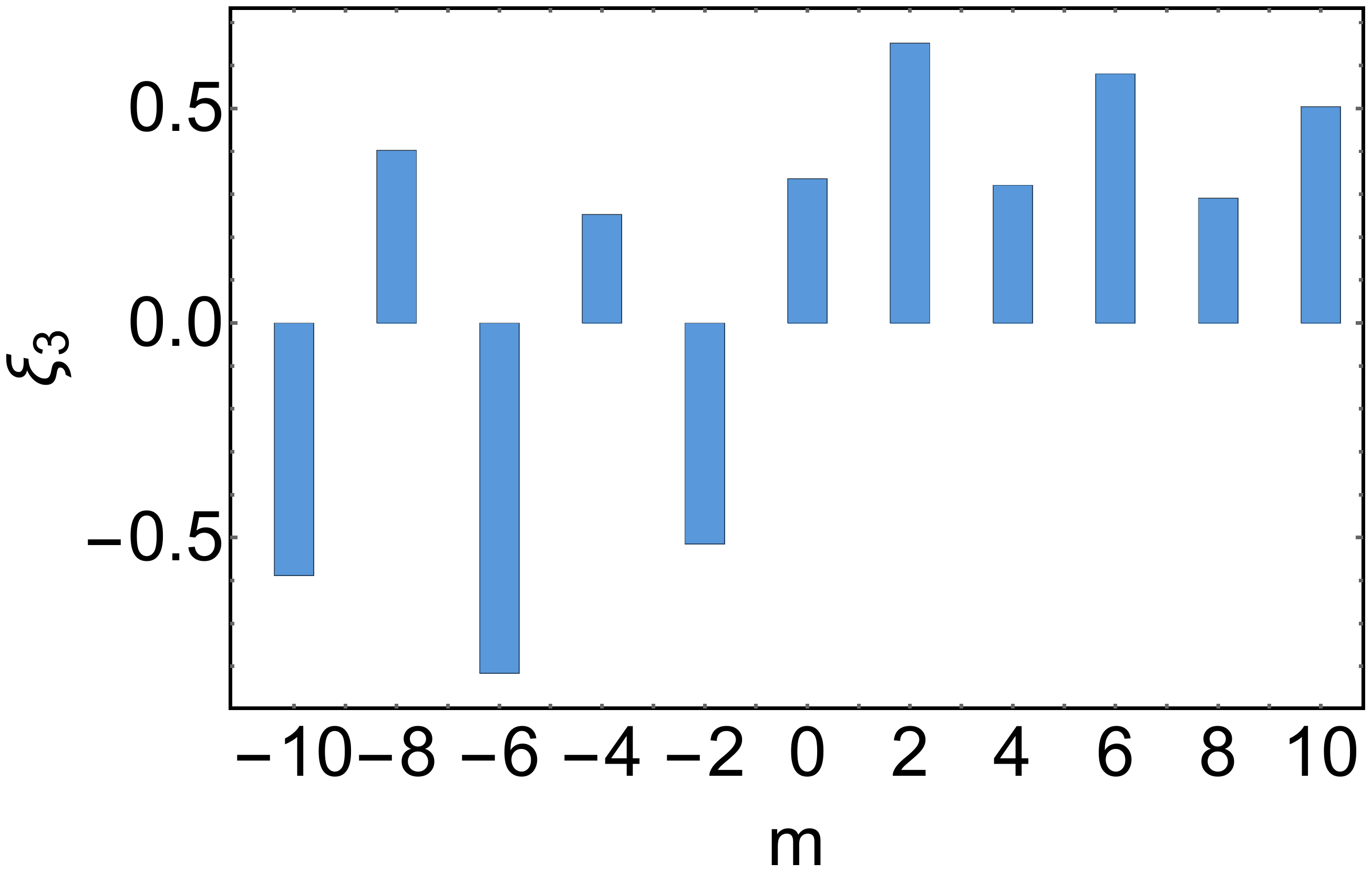}}\\
\raisebox{-0.5\height}{\includegraphics*[width=0.24\linewidth]{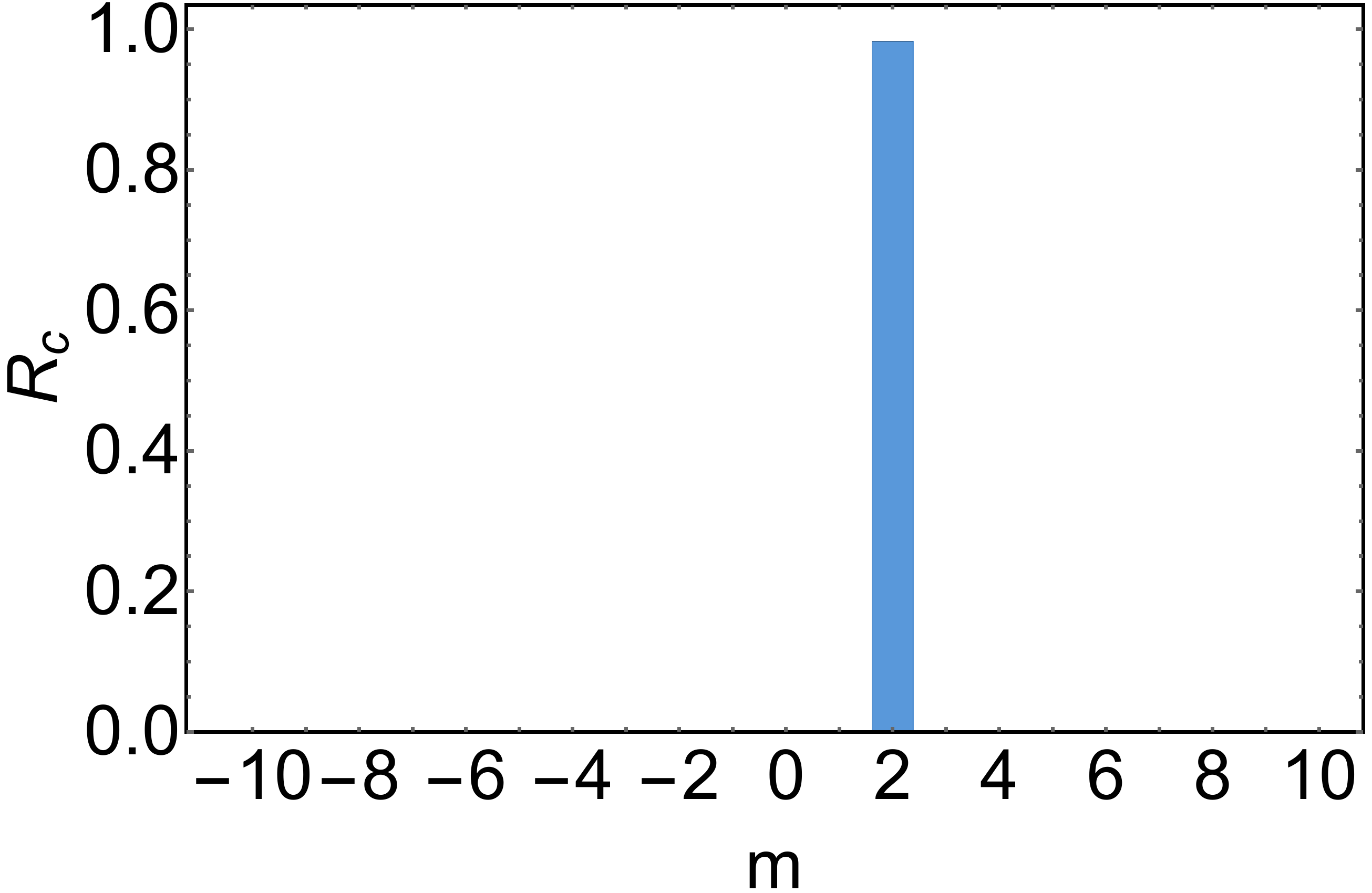}}\,
\raisebox{-0.5\height}{\includegraphics*[width=0.24\linewidth]{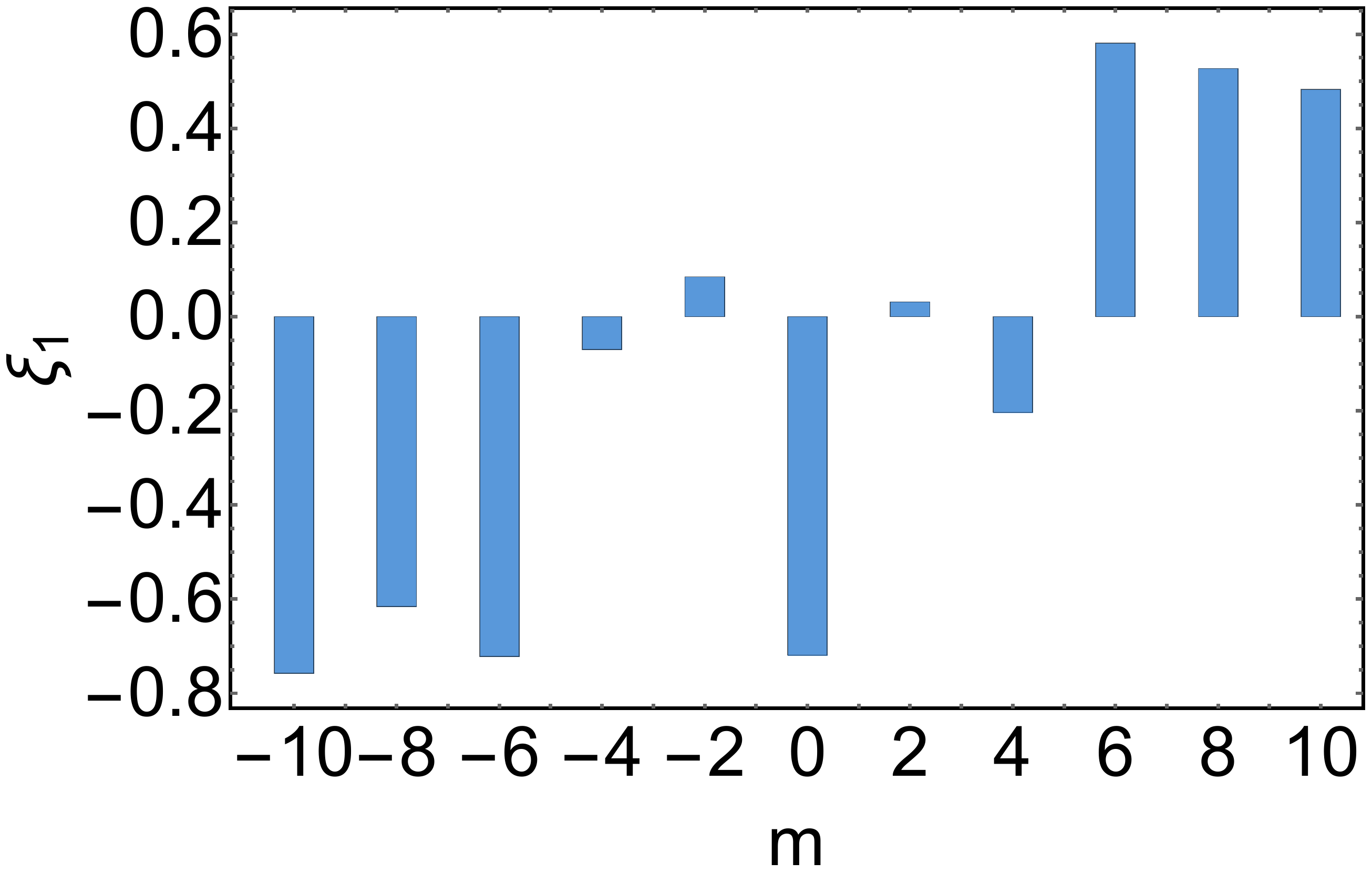}}\,
\raisebox{-0.5\height}{\includegraphics*[width=0.24\linewidth]{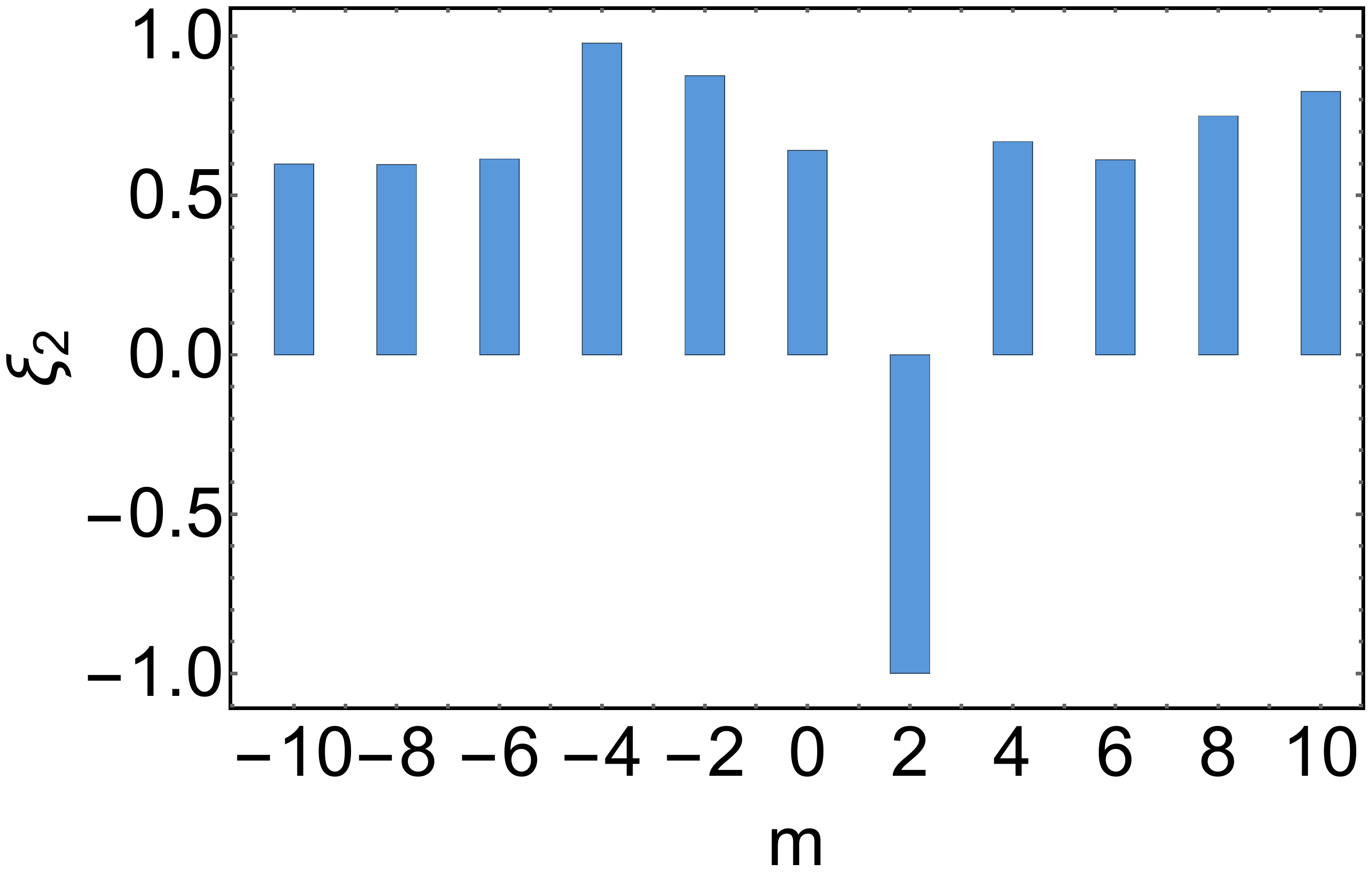}}\,
\raisebox{-0.5\height}{\includegraphics*[width=0.24\linewidth]{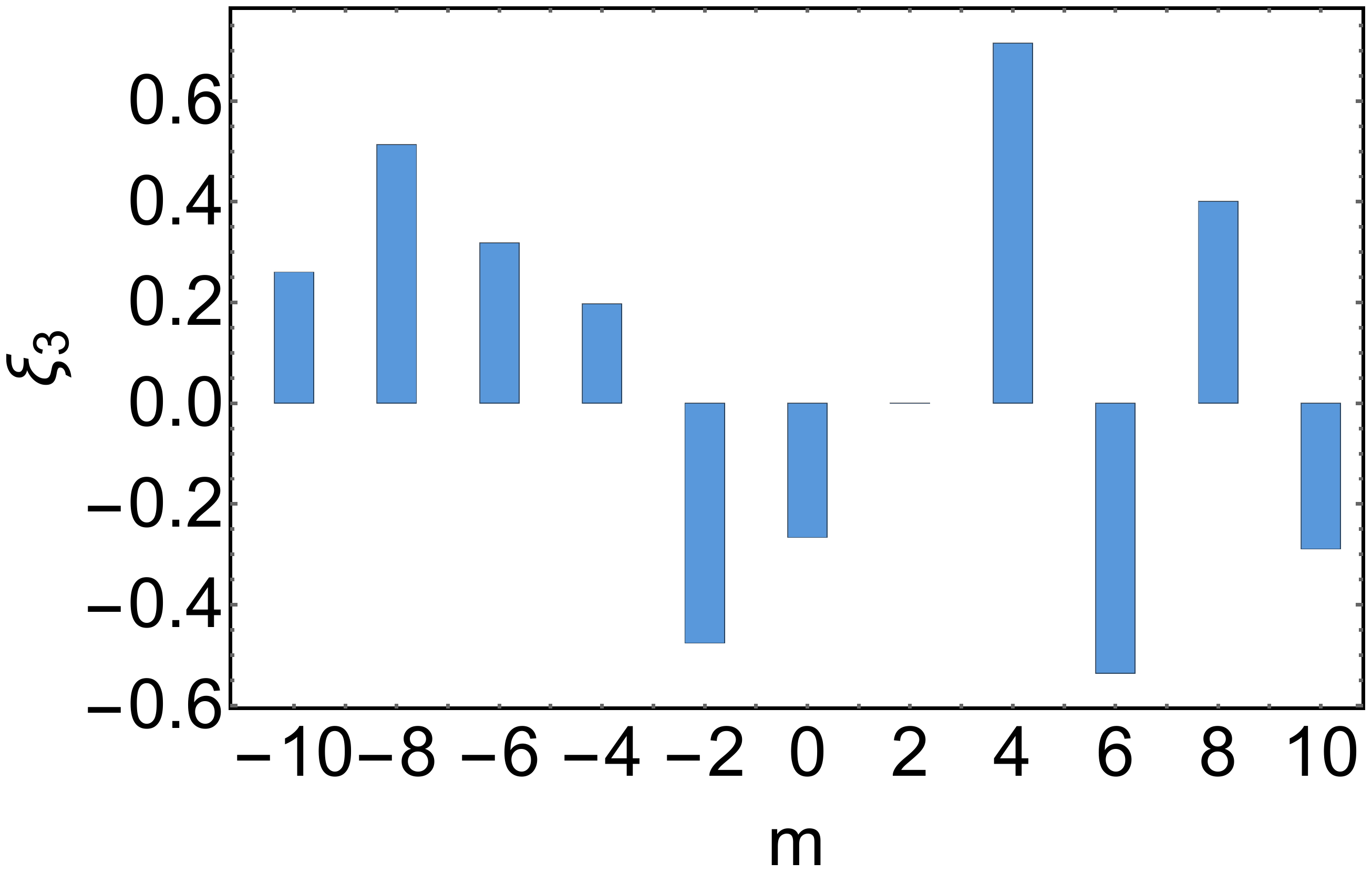}}\\
\raisebox{-0.5\height}{\includegraphics*[width=0.24\linewidth]{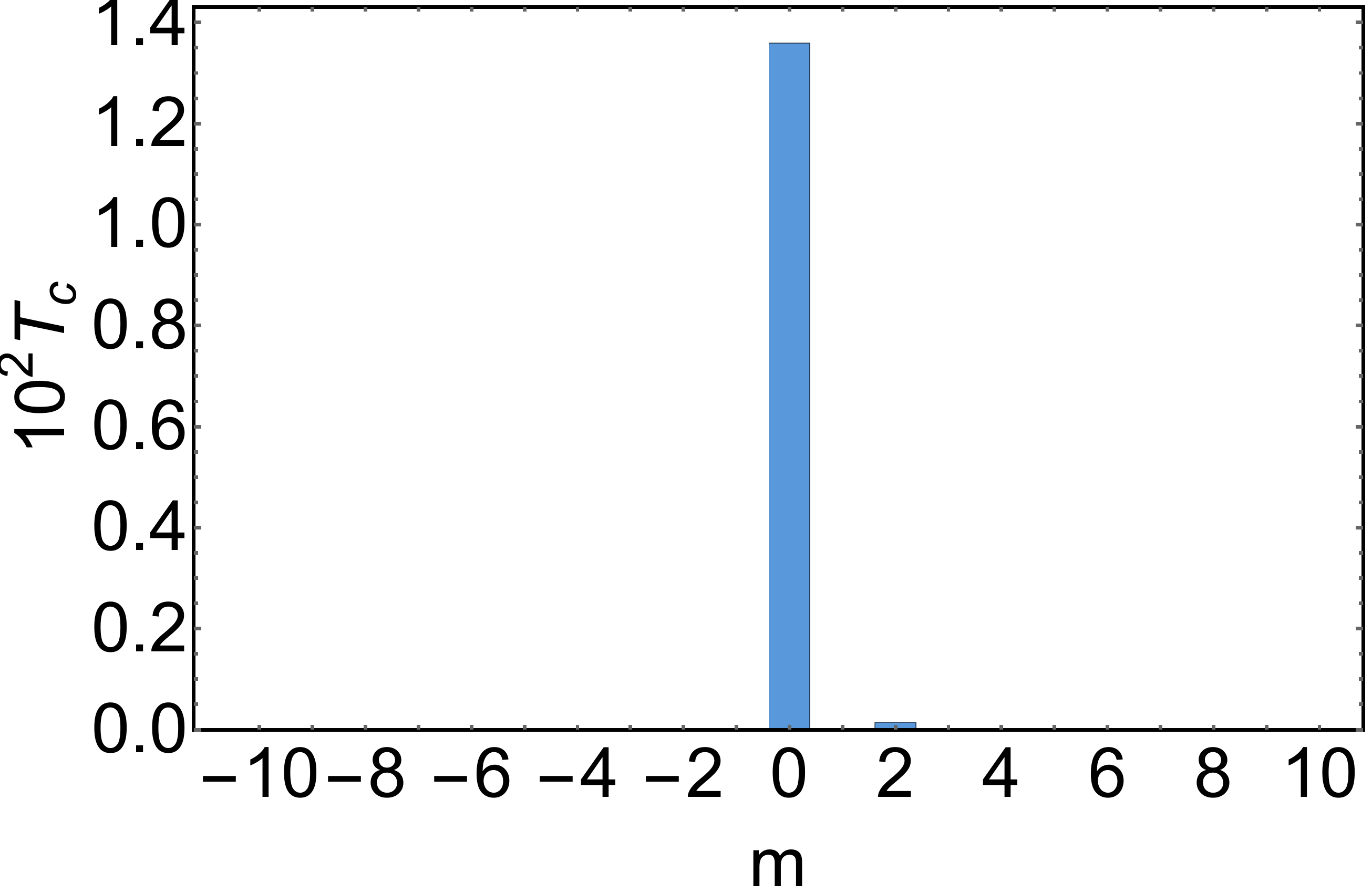}}\,
\raisebox{-0.5\height}{\includegraphics*[width=0.24\linewidth]{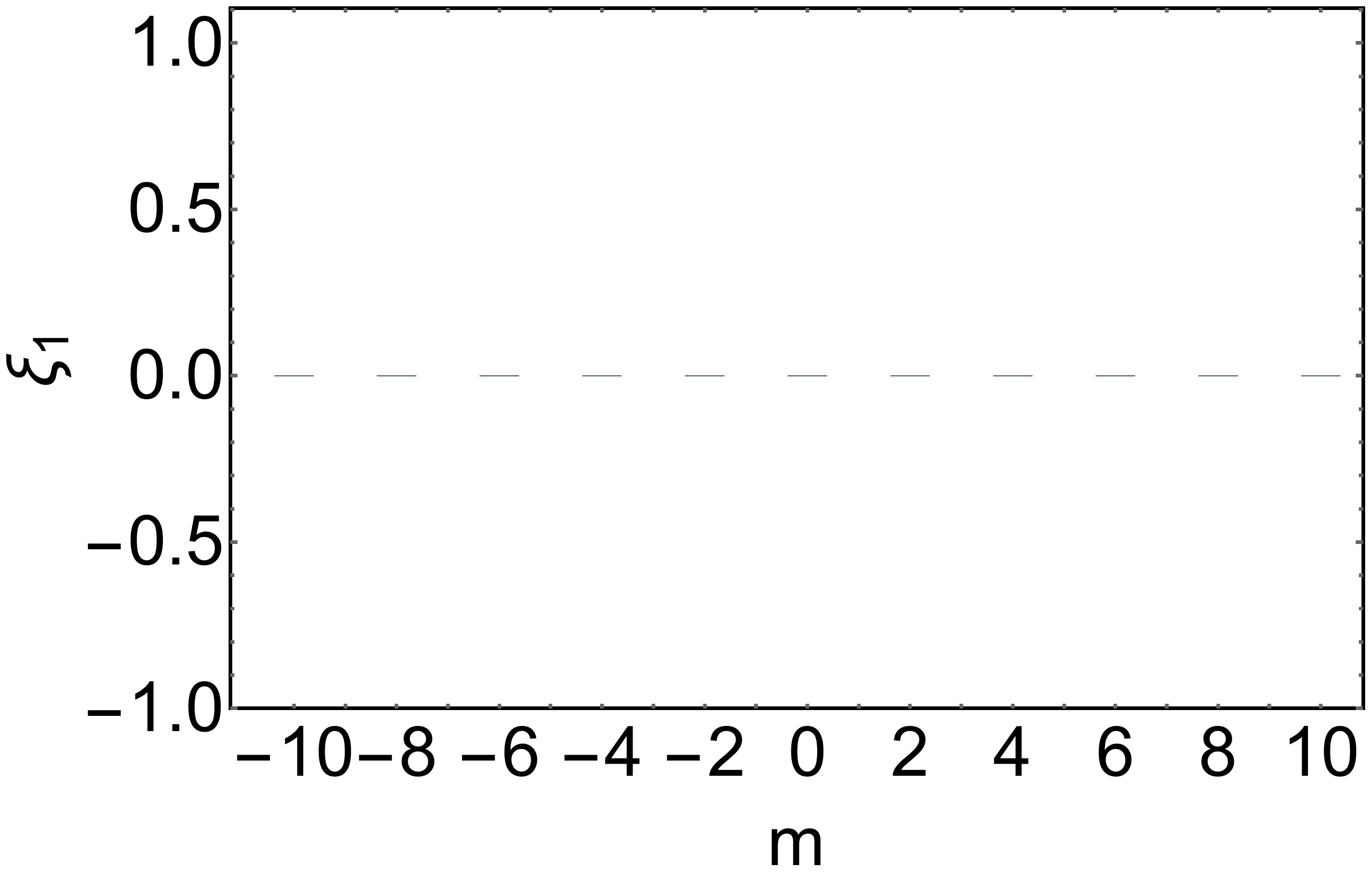}}\,
\raisebox{-0.5\height}{\includegraphics*[width=0.24\linewidth]{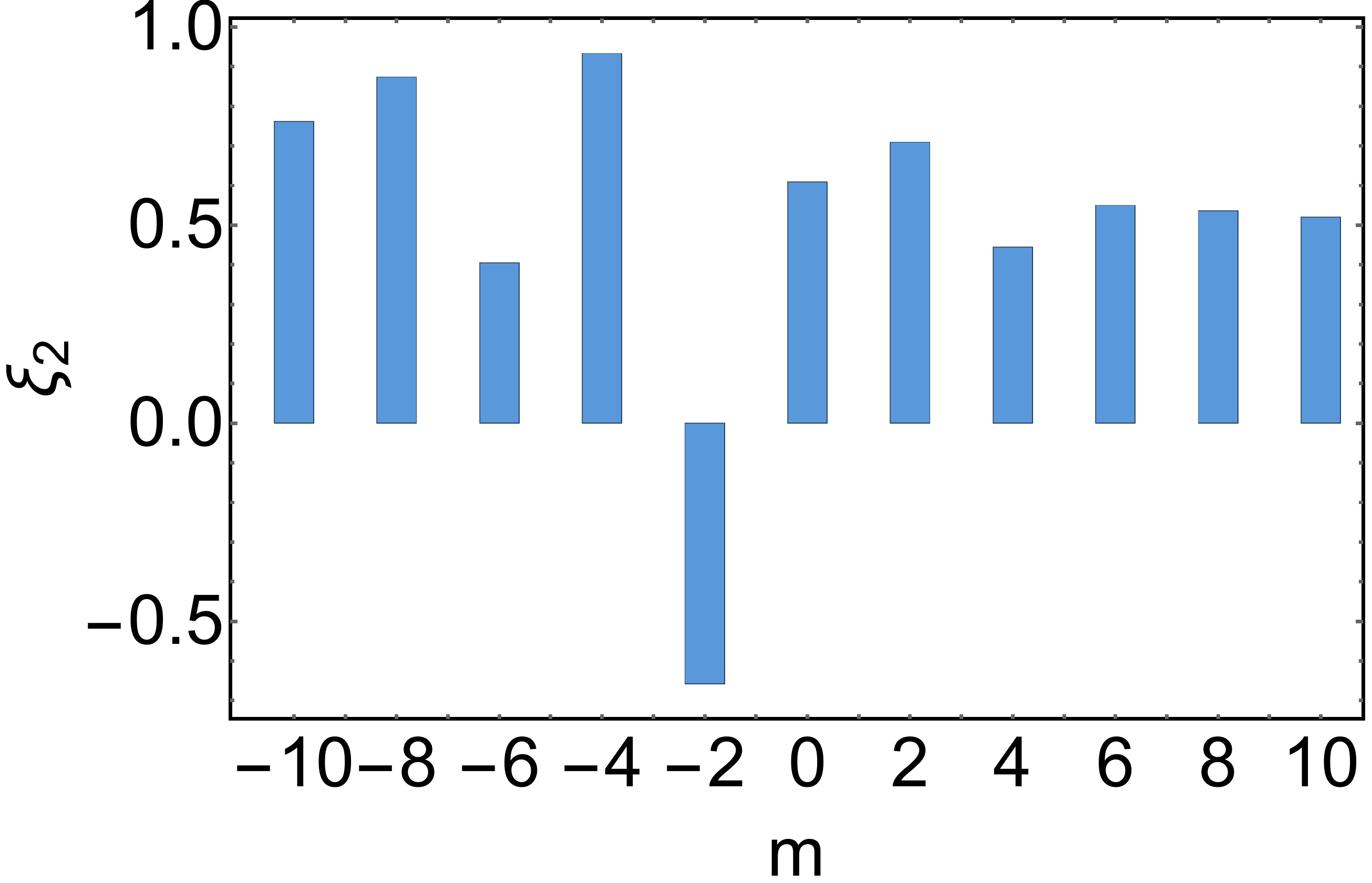}}\,
\raisebox{-0.5\height}{\includegraphics*[width=0.24\linewidth]{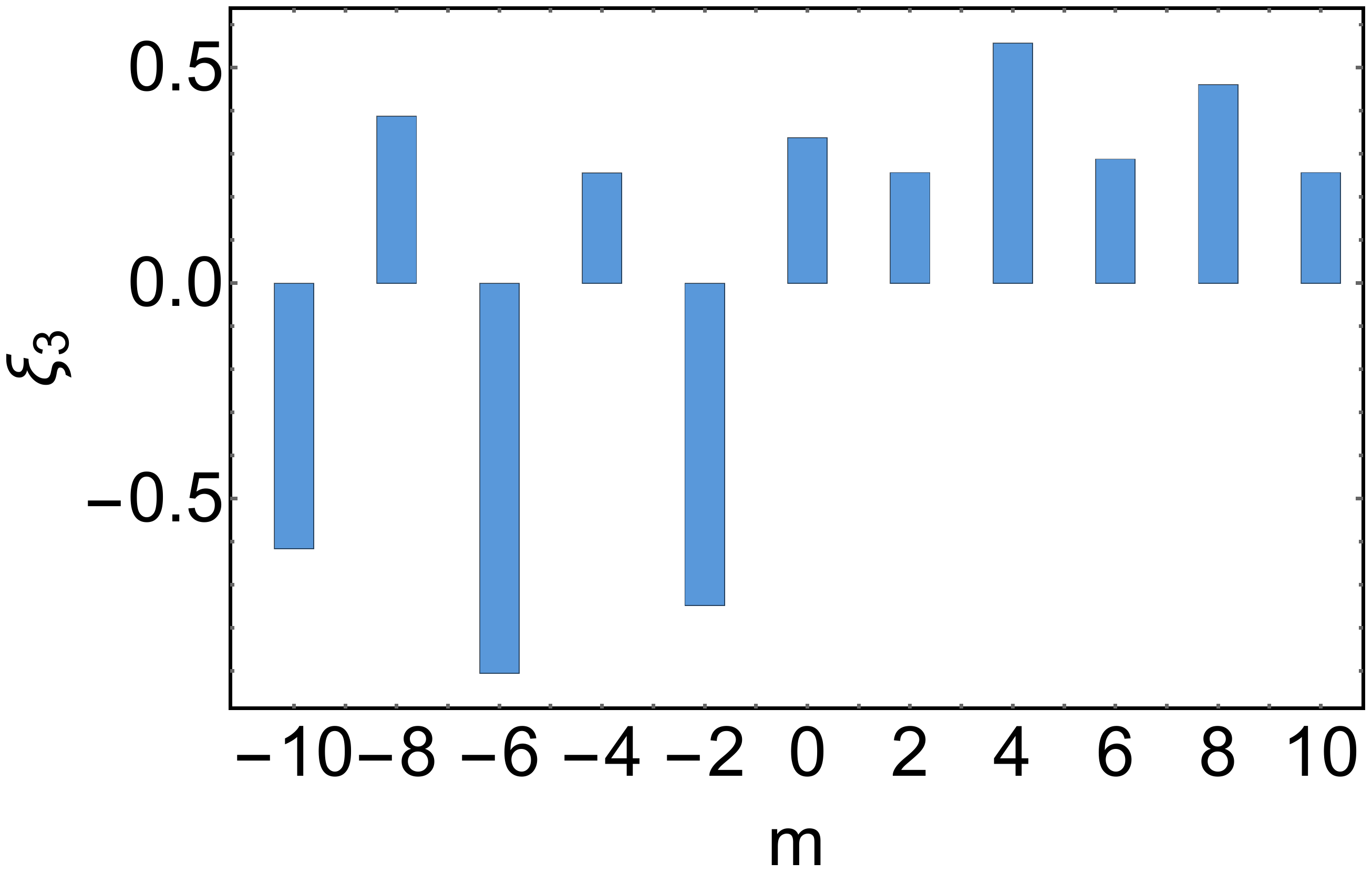}}
\caption{{\footnotesize The transmission and reflection coefficients and the Stokes parameters for the photons transmitted or reflected by the cholesteric plate immersed into the dielectric medium. The parameters are the same as in Fig. \ref{Disp_Chol_plots}. The lines $1$-$4$: The case of plane-wave photons scattered in the $(x,z)$ plane is considered. The lines $1$-$2$: The initial photon possesses the helicity $s=1$. The lines $3$-$4$: It has $s=-1$. The lines $5$-$8$: Scattering of the twisted photon with $m=0$ is considered at the energy $k_0=0.892$ eV belonging to the total band gap. The lines $5$-$6$ corresponds to $s=1$, whereas the lines $7$-$8$ are for $s=-1$. The reason for that only even $m$ are realized for scattered twisted photons is that a cholesteric can transfer only an even multiple of $\hbar$ of the projection of the total angular momentum to a photon. As is seen from the plots, the total band gap reflects twisted photons of both helicities with the shifted projection of the total angular momentum in accordance with the selection rule \eqref{sel_rule}.}}
\label{Scatt_Chol_plots}
\end{figure}

%\newpage
\begin{figure}[tp]
\centering
\raisebox{-0.5\height}{\includegraphics*[width=0.24\linewidth]{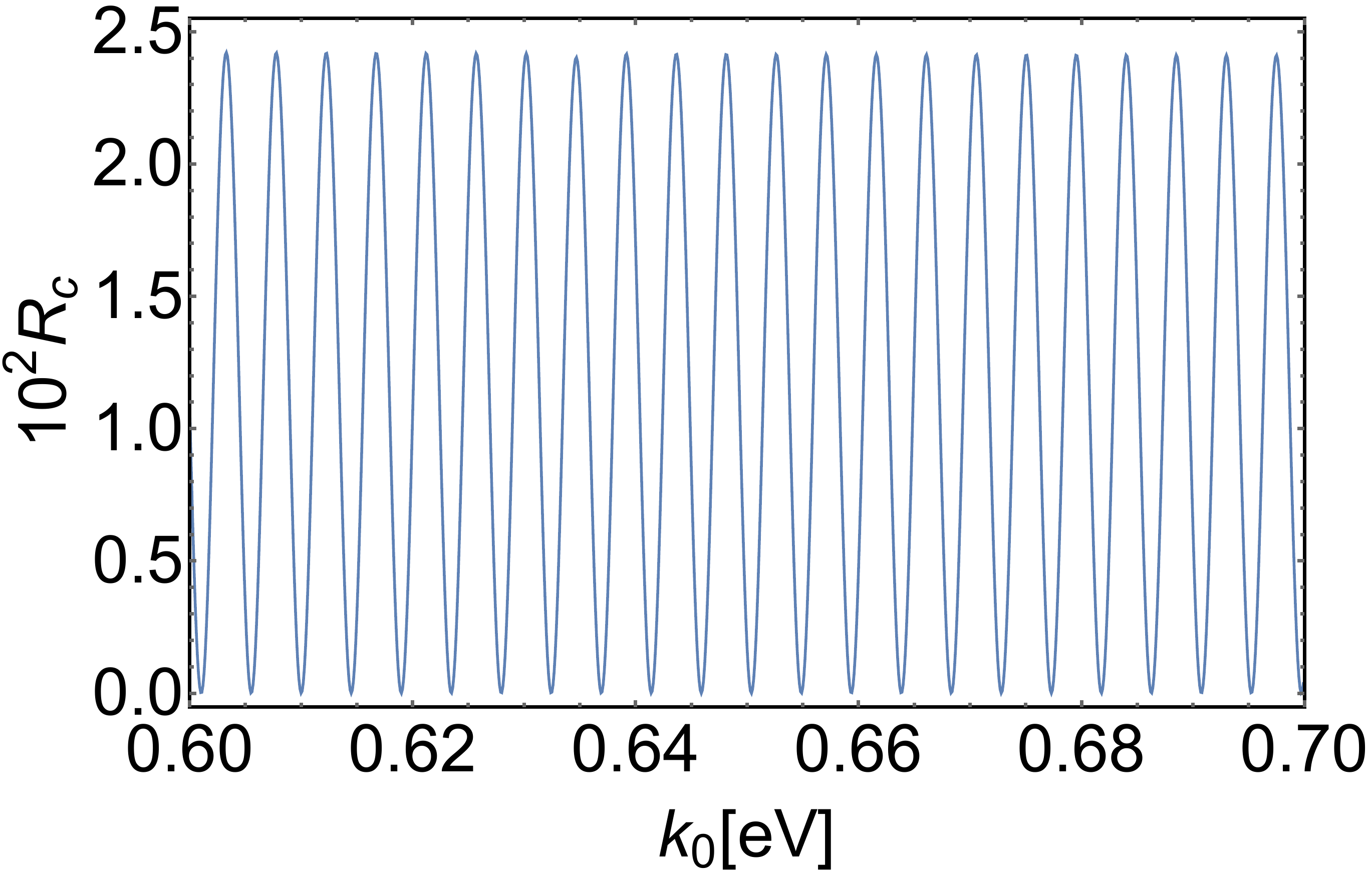}}\,
\raisebox{-0.5\height}{\includegraphics*[width=0.24\linewidth]{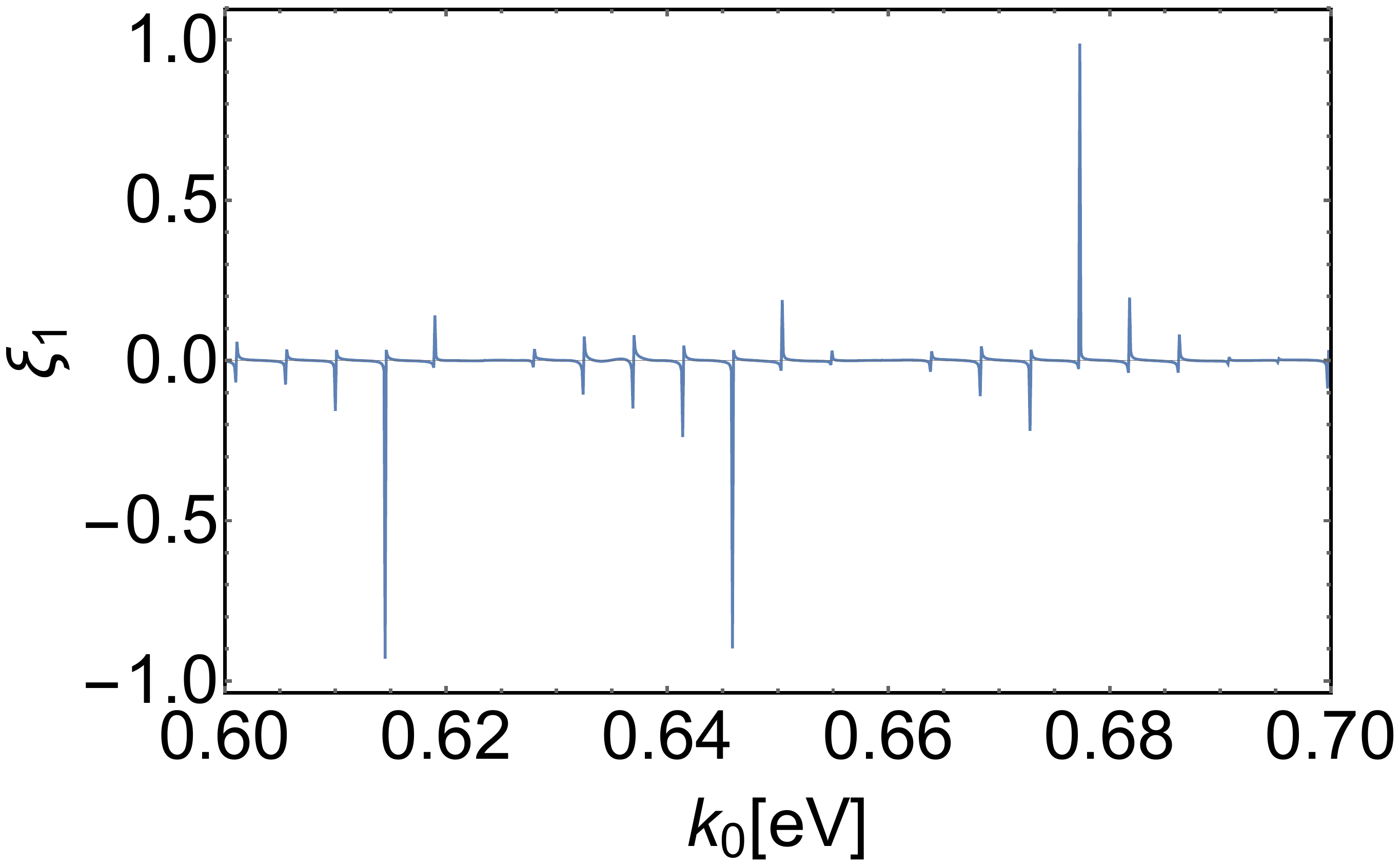}}\,
\raisebox{-0.5\height}{\includegraphics*[width=0.24\linewidth]{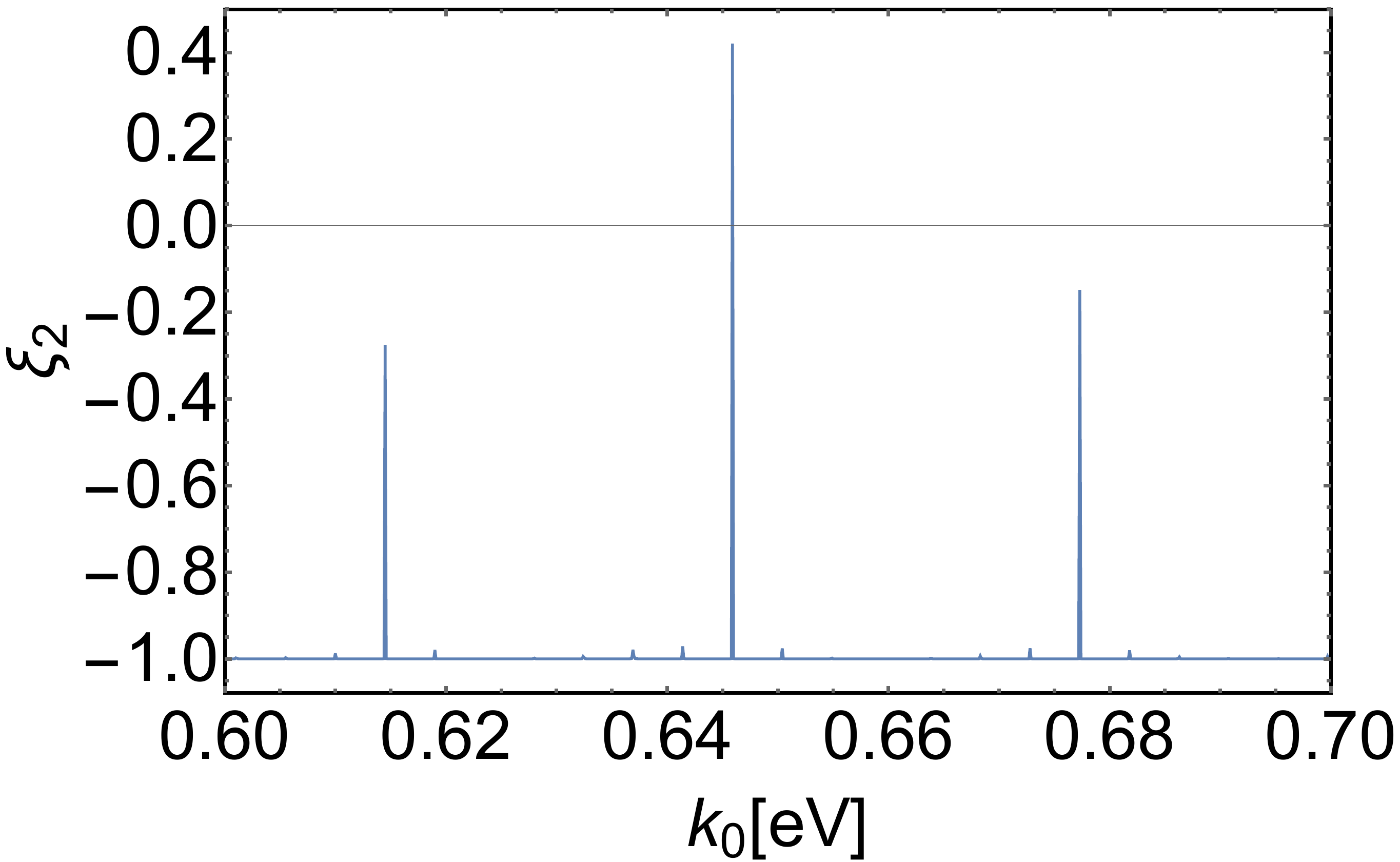}}\,
\raisebox{-0.5\height}{\includegraphics*[width=0.24\linewidth]{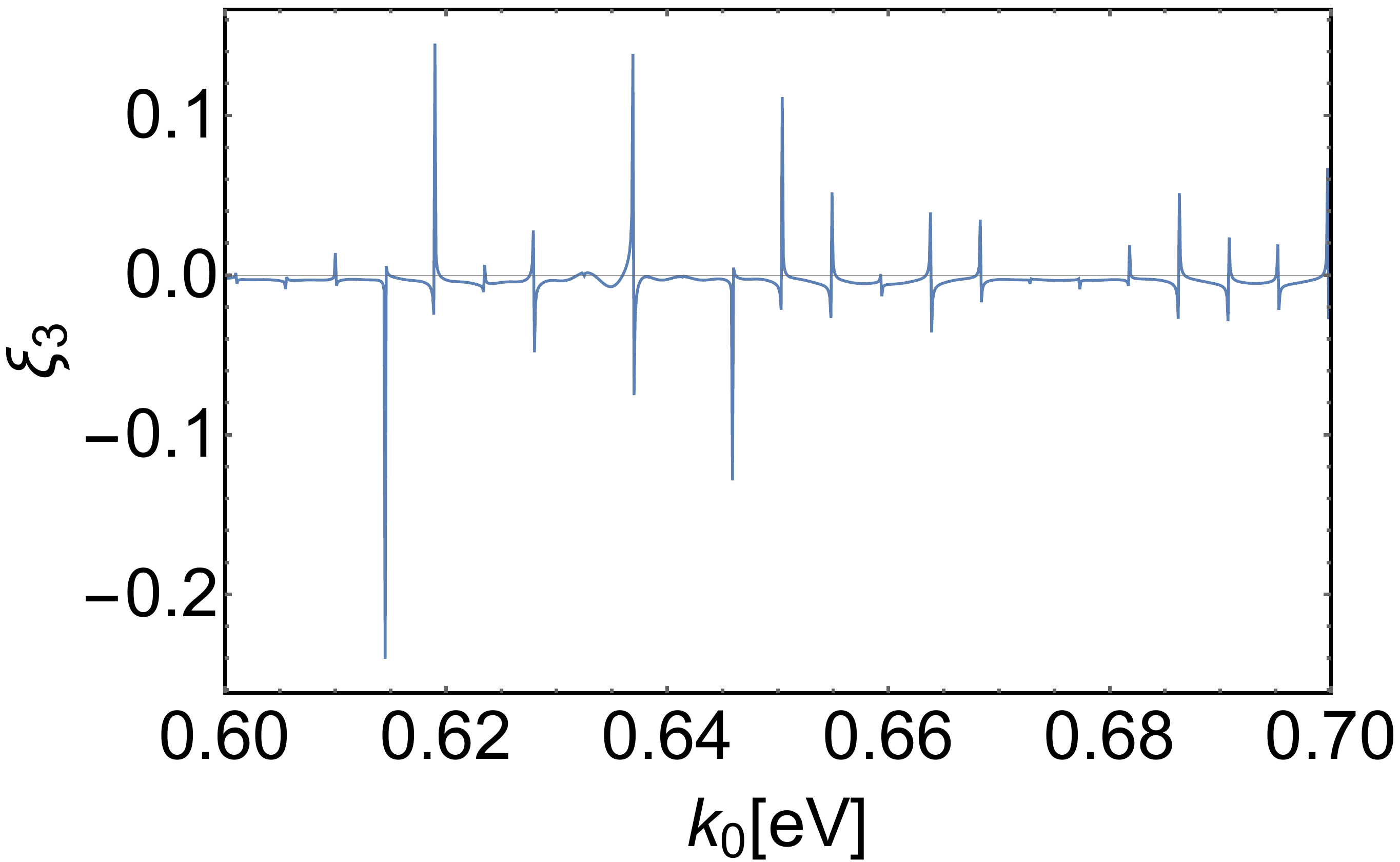}}\\
\raisebox{-0.5\height}{\includegraphics*[width=0.24\linewidth]{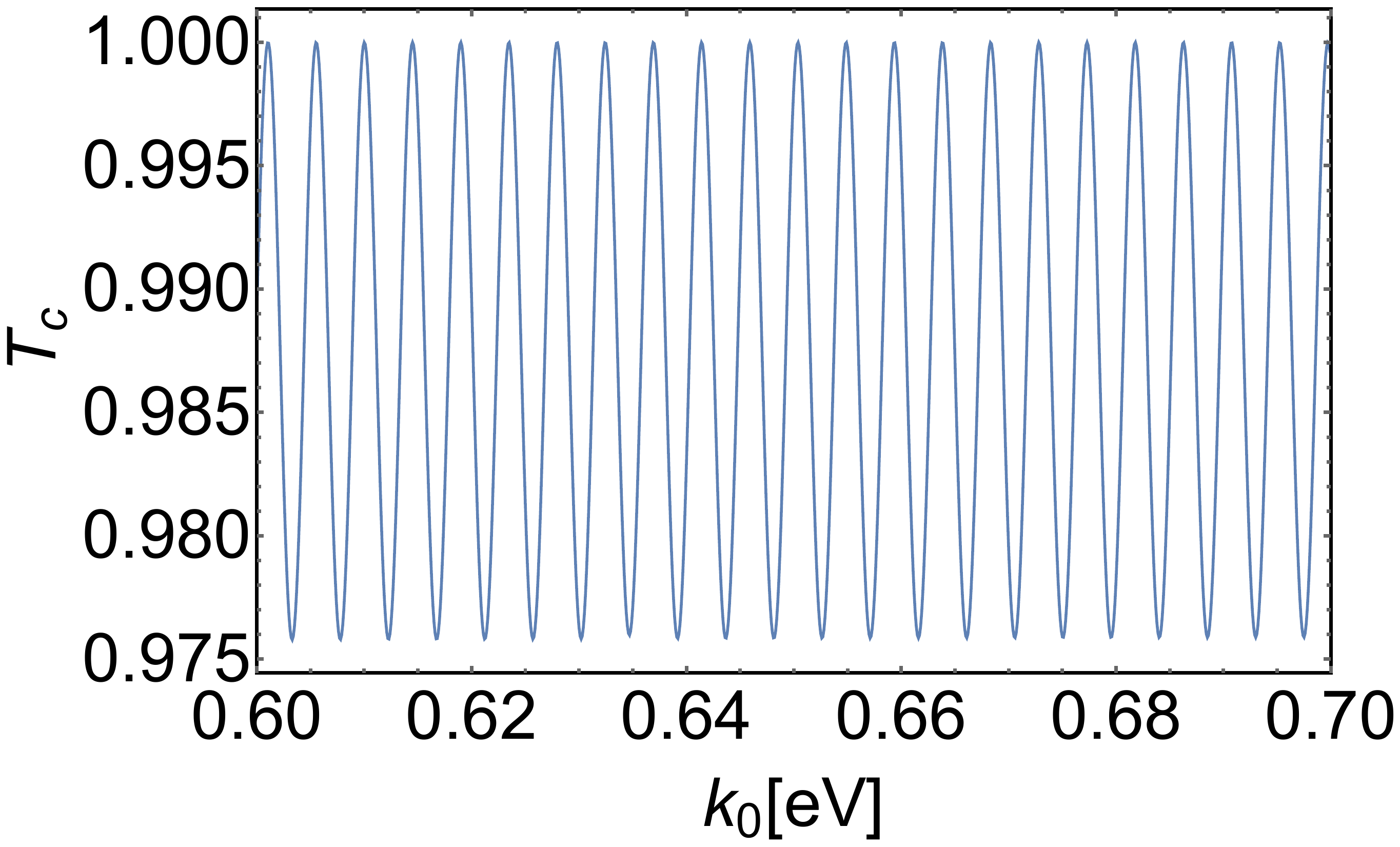}}\,
\raisebox{-0.5\height}{\includegraphics*[width=0.24\linewidth]{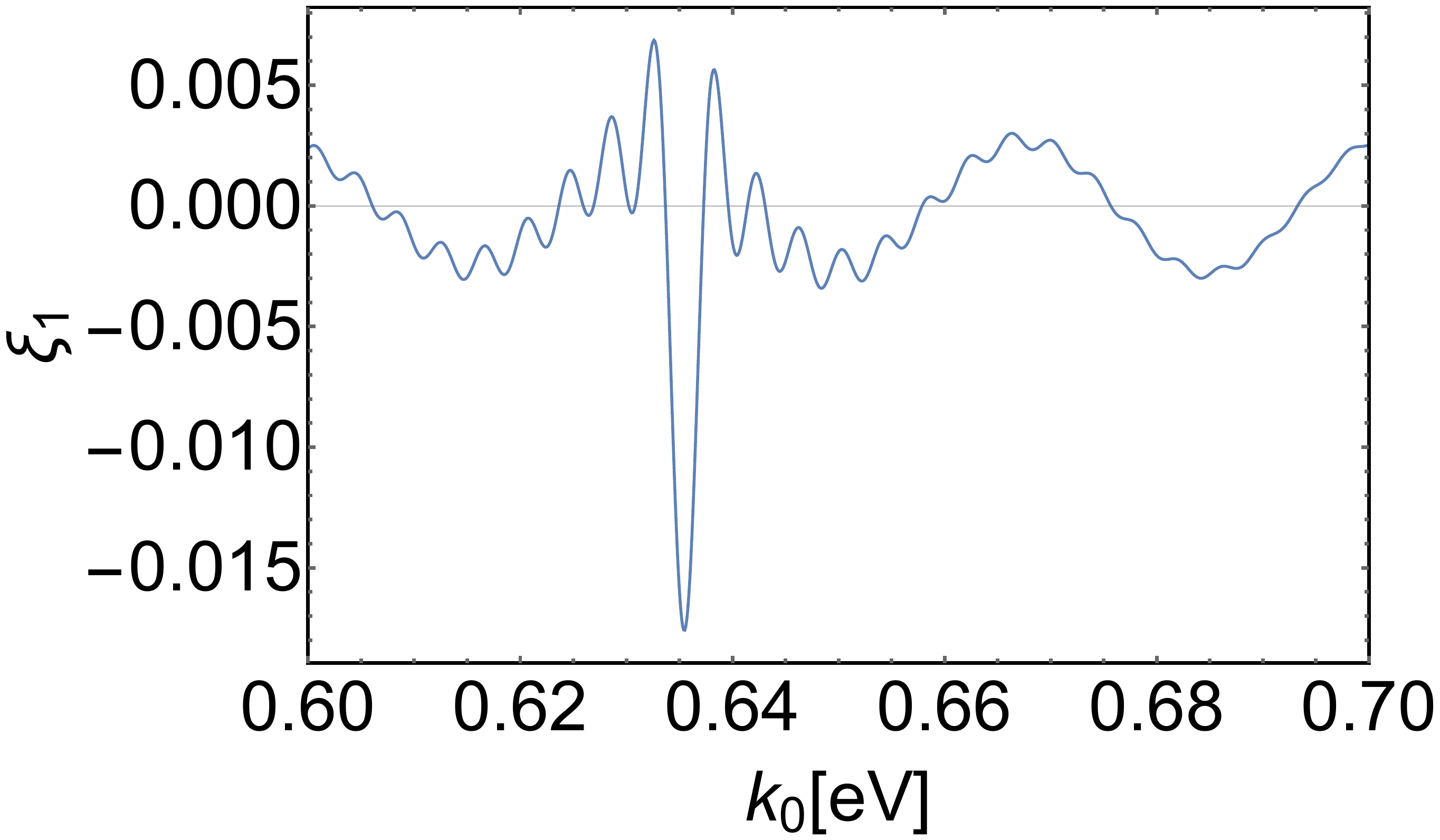}}\,
\raisebox{-0.5\height}{\includegraphics*[width=0.24\linewidth]{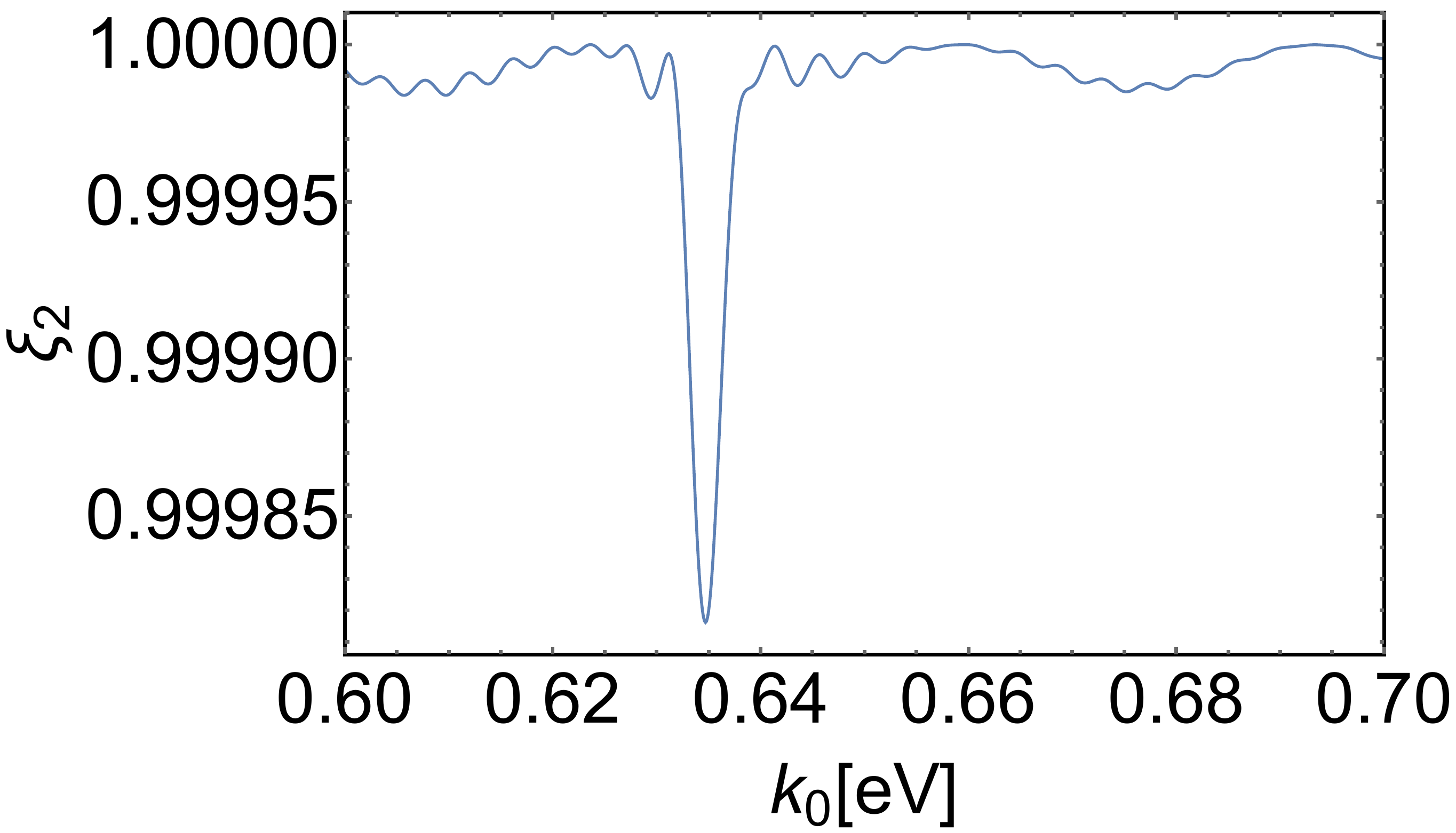}}\,
\raisebox{-0.5\height}{\includegraphics*[width=0.24\linewidth]{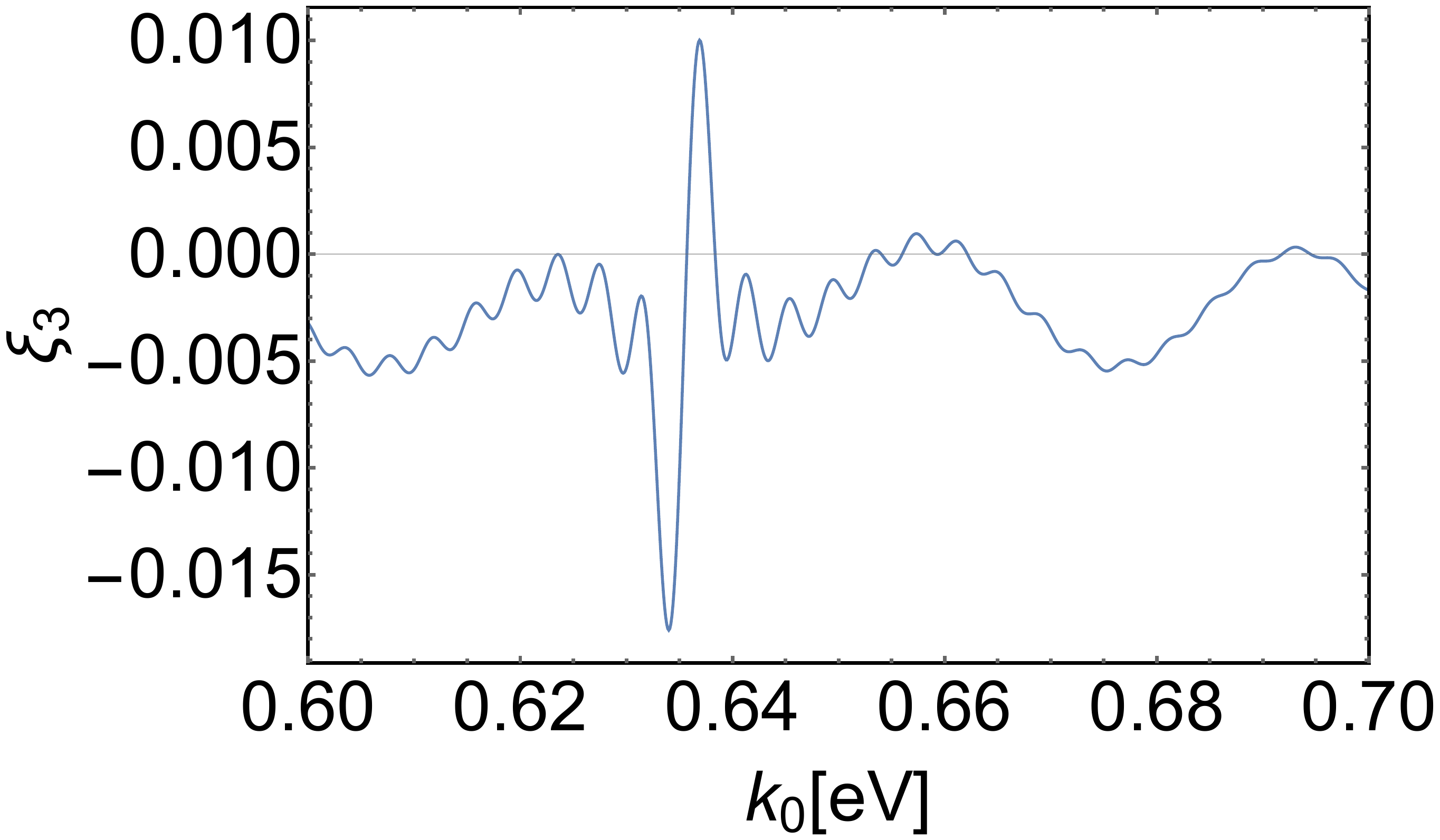}}\\
\raisebox{-0.5\height}{\includegraphics*[width=0.24\linewidth]{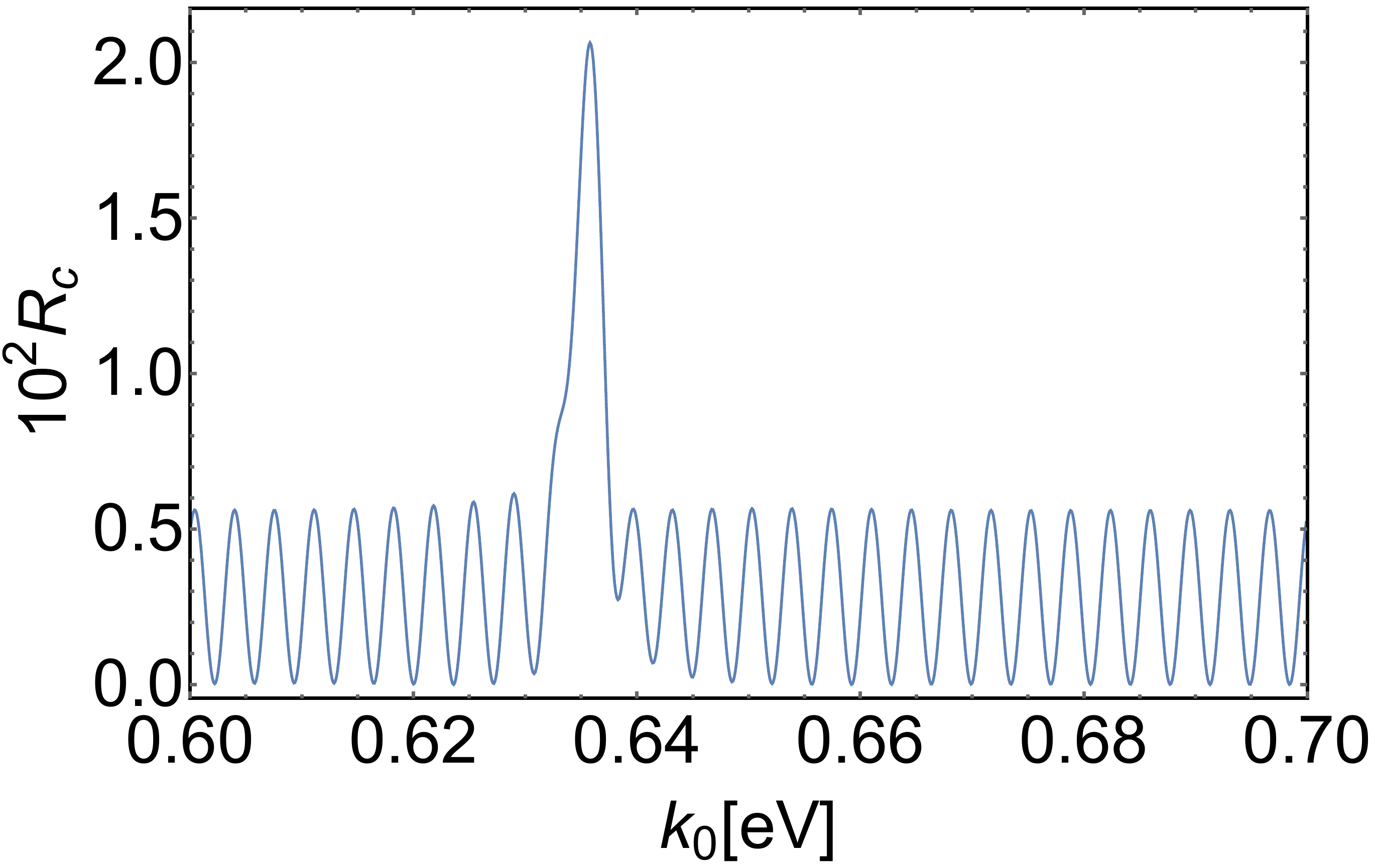}}\,
\raisebox{-0.5\height}{\includegraphics*[width=0.24\linewidth]{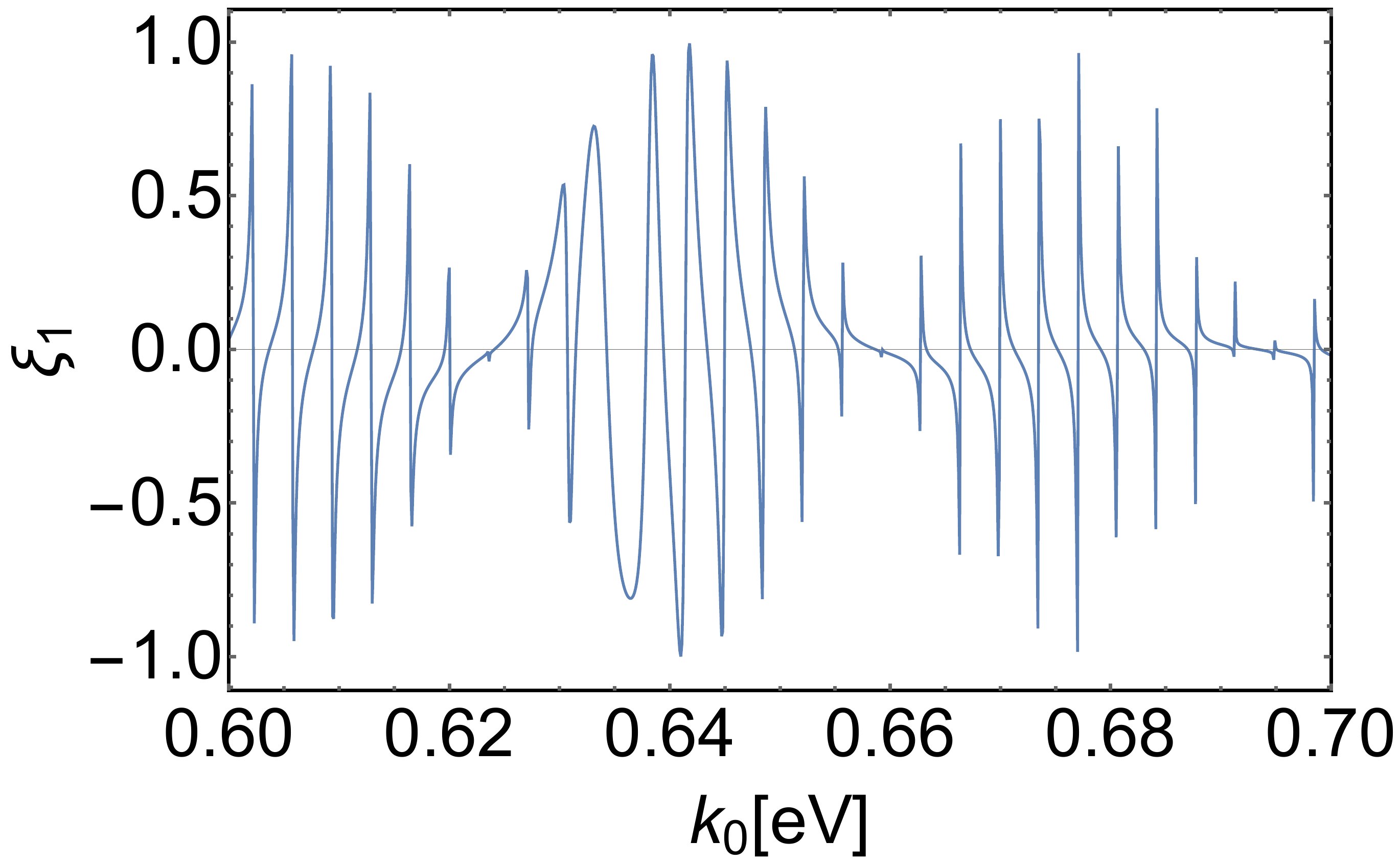}}\,
\raisebox{-0.5\height}{\includegraphics*[width=0.24\linewidth]{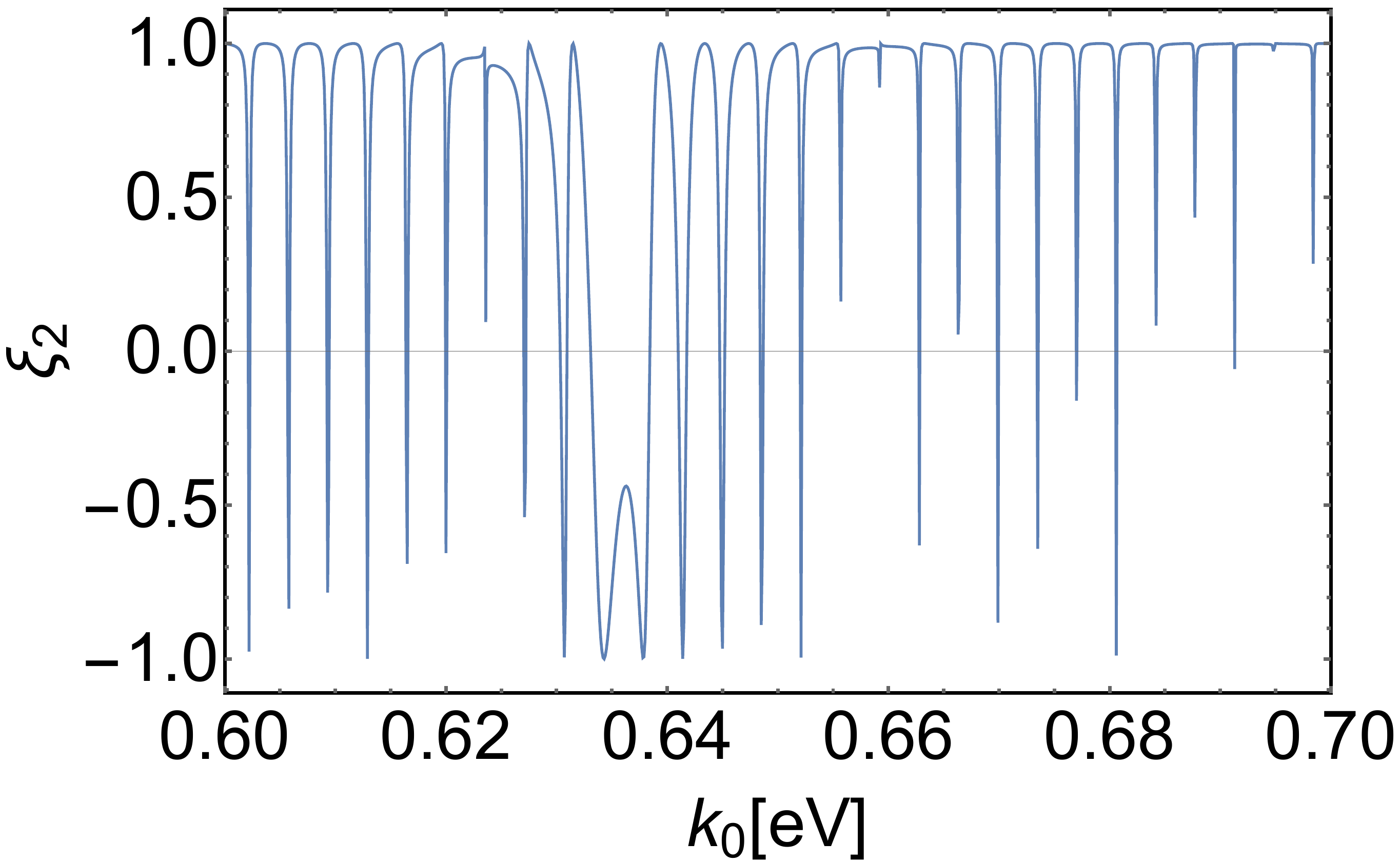}}\,
\raisebox{-0.5\height}{\includegraphics*[width=0.24\linewidth]{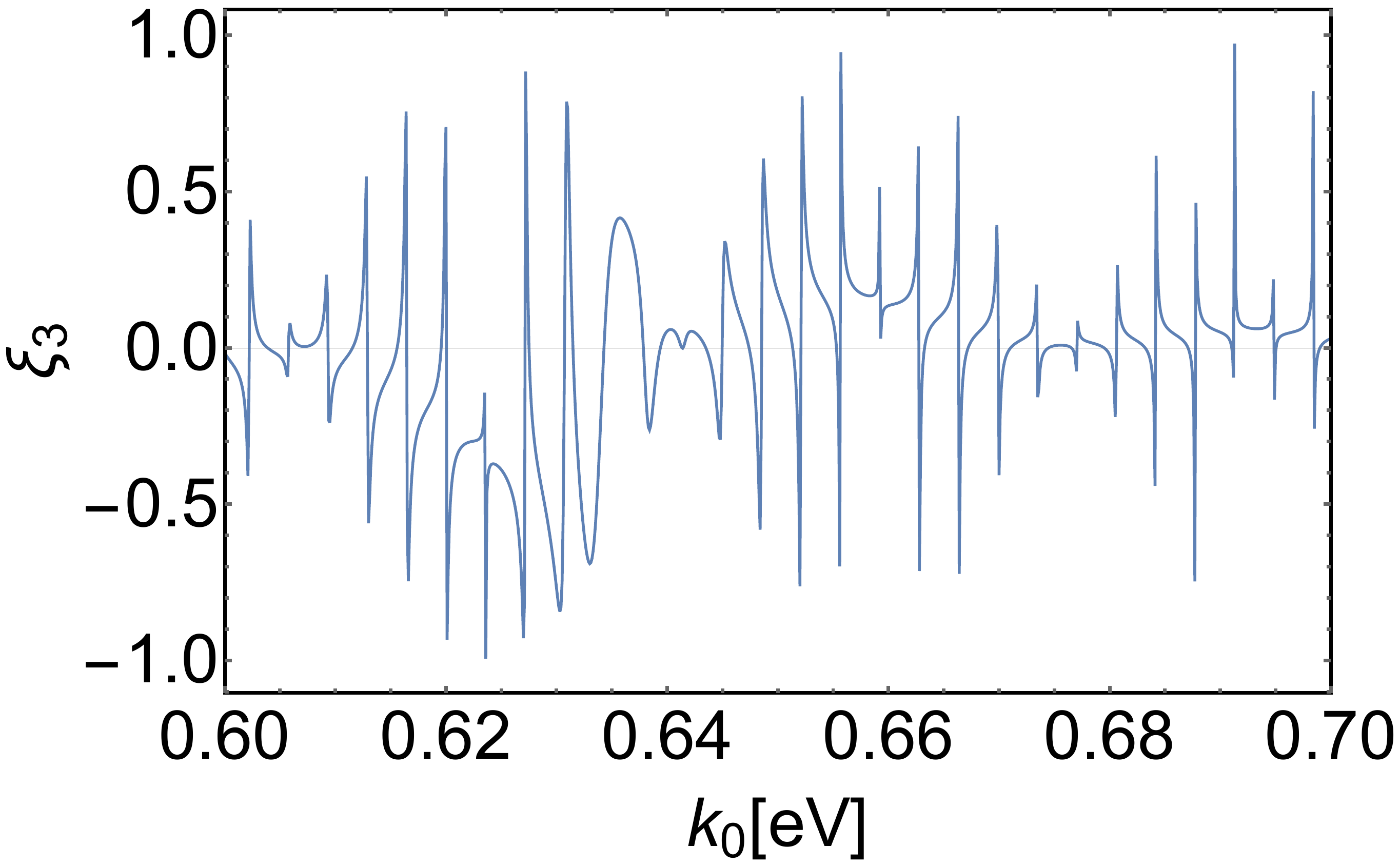}}\\
\raisebox{-0.5\height}{\includegraphics*[width=0.24\linewidth]{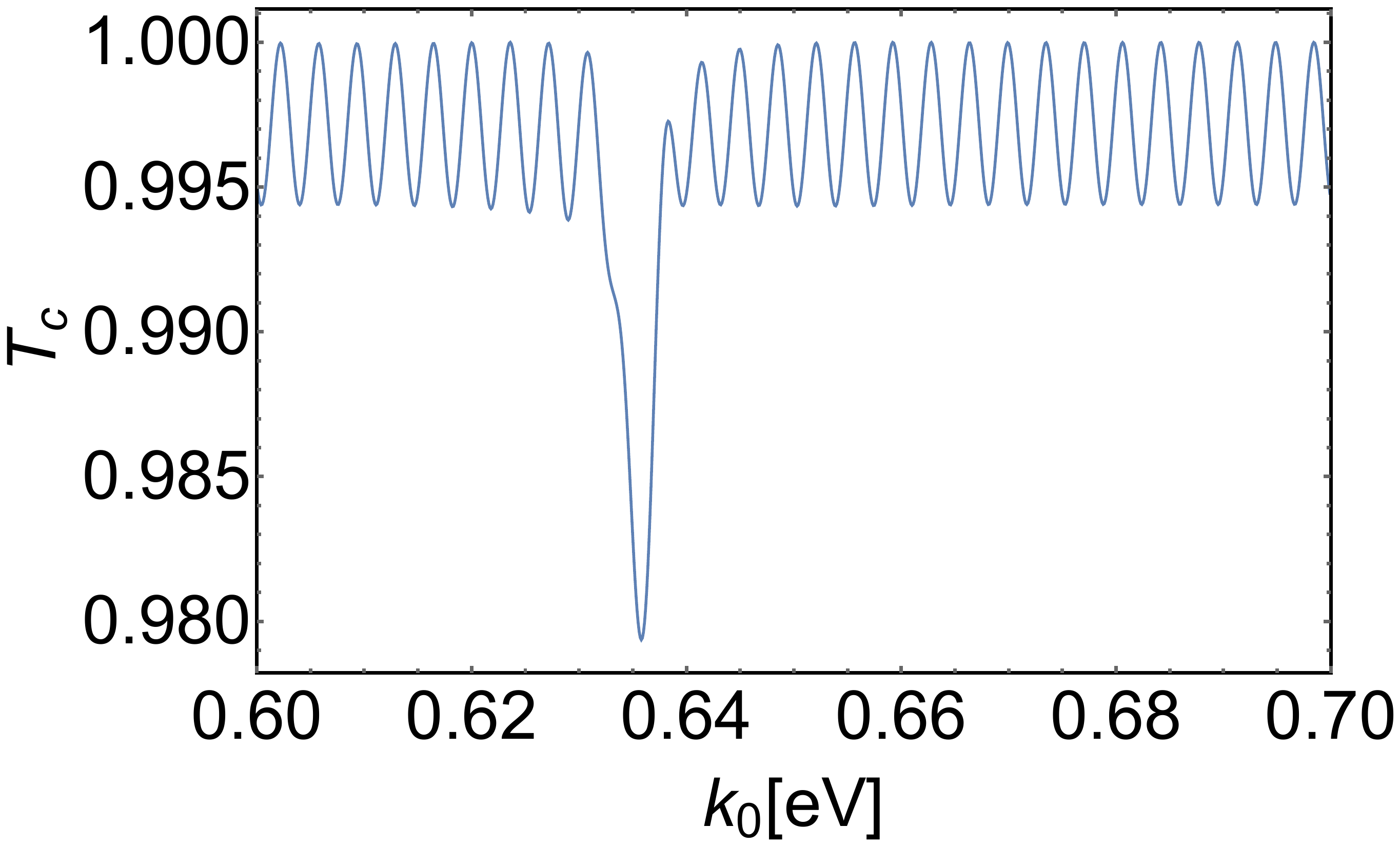}}\,
\raisebox{-0.5\height}{\includegraphics*[width=0.24\linewidth]{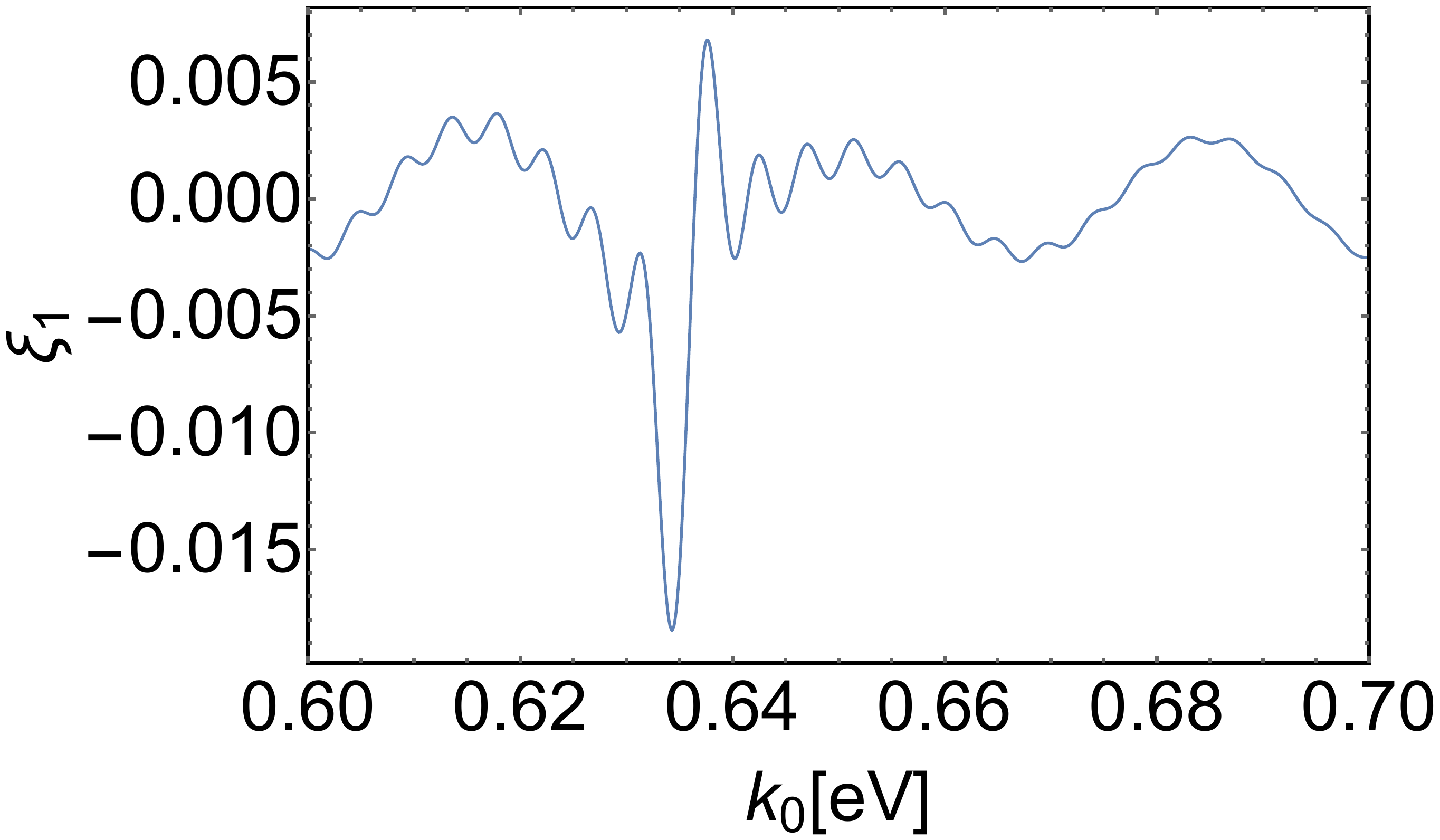}}\,
\raisebox{-0.5\height}{\includegraphics*[width=0.24\linewidth]{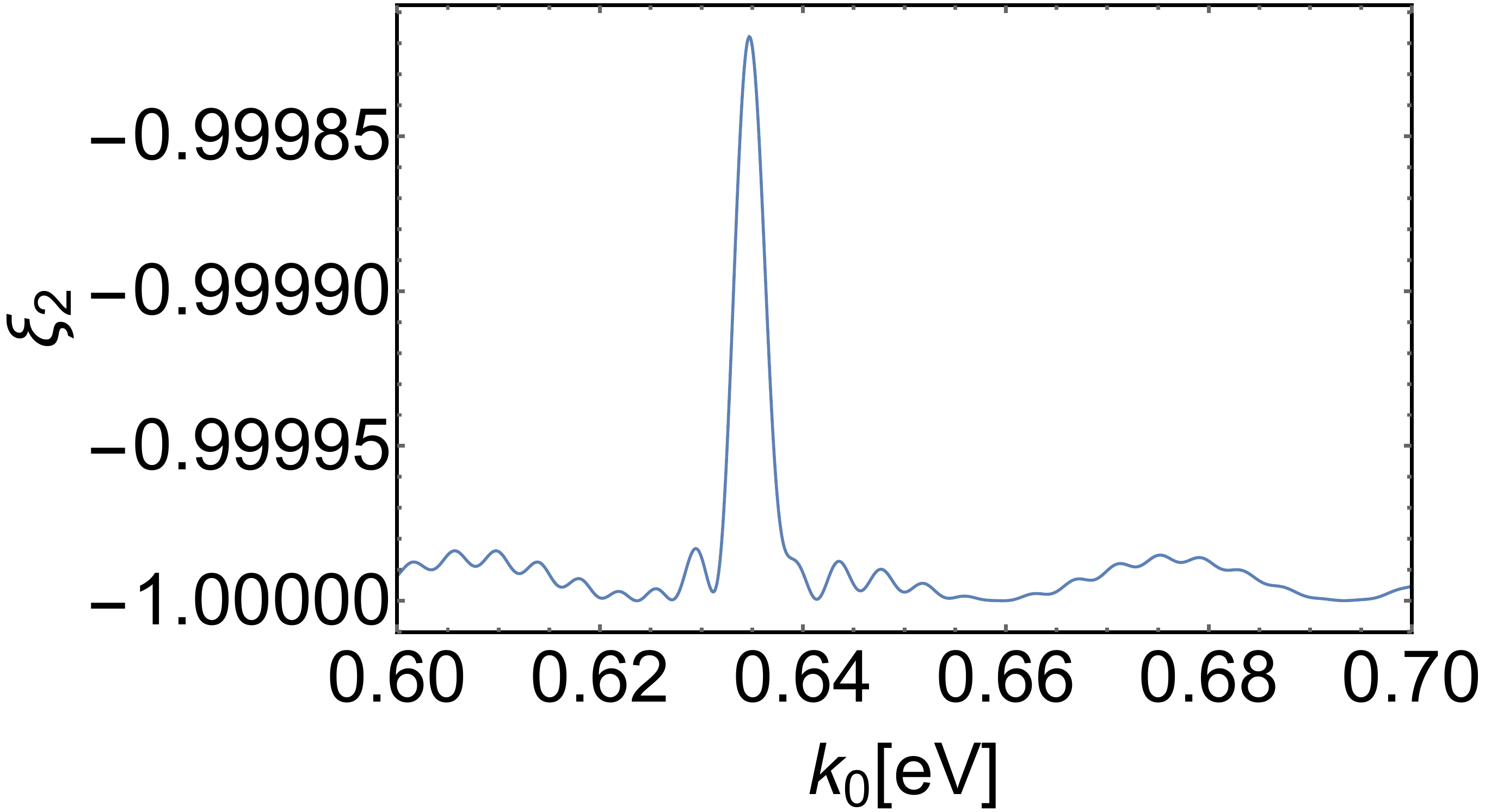}}\,
\raisebox{-0.5\height}{\includegraphics*[width=0.24\linewidth]{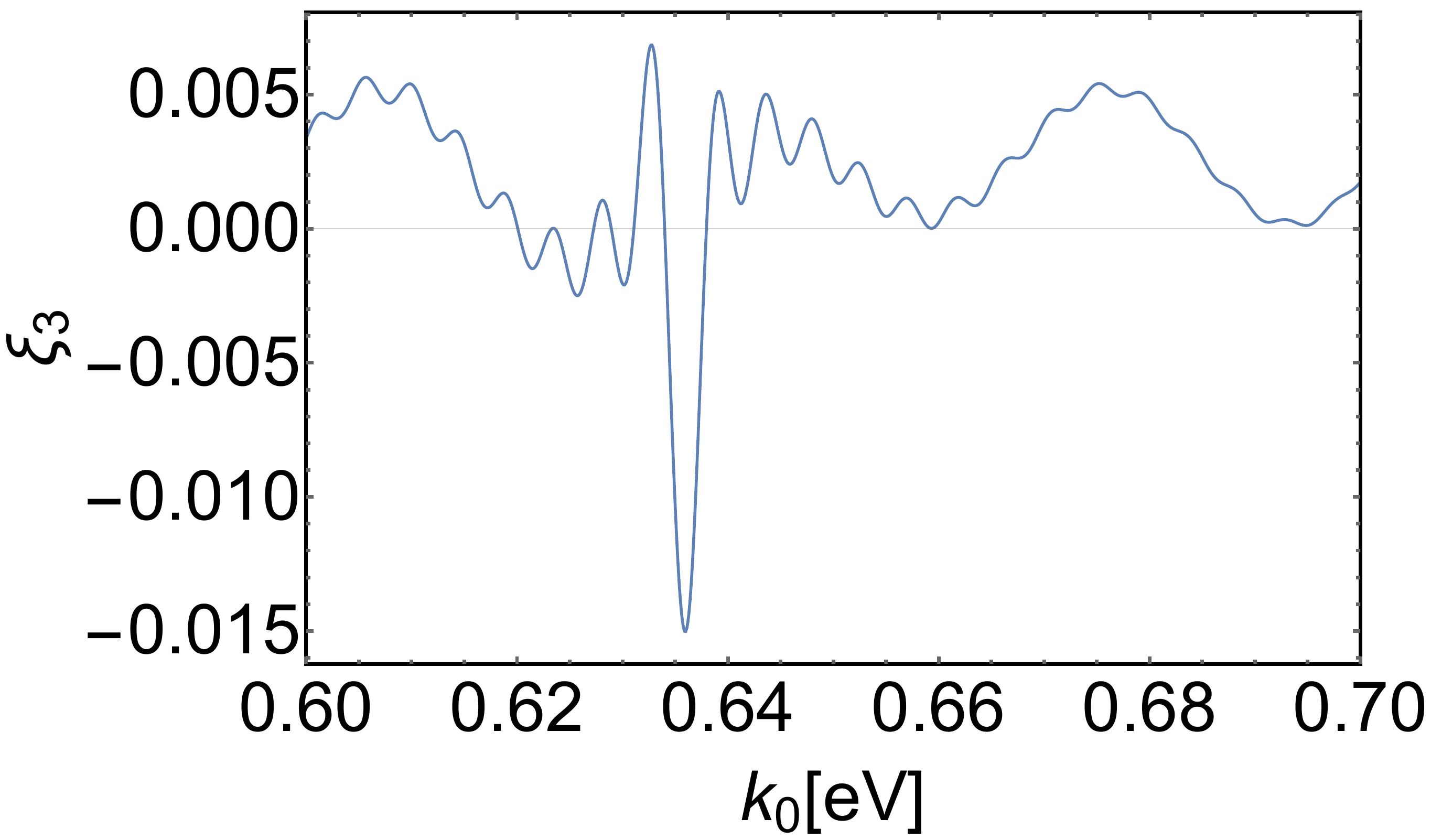}}\\
\raisebox{-0.5\height}{\includegraphics*[width=0.24\linewidth]{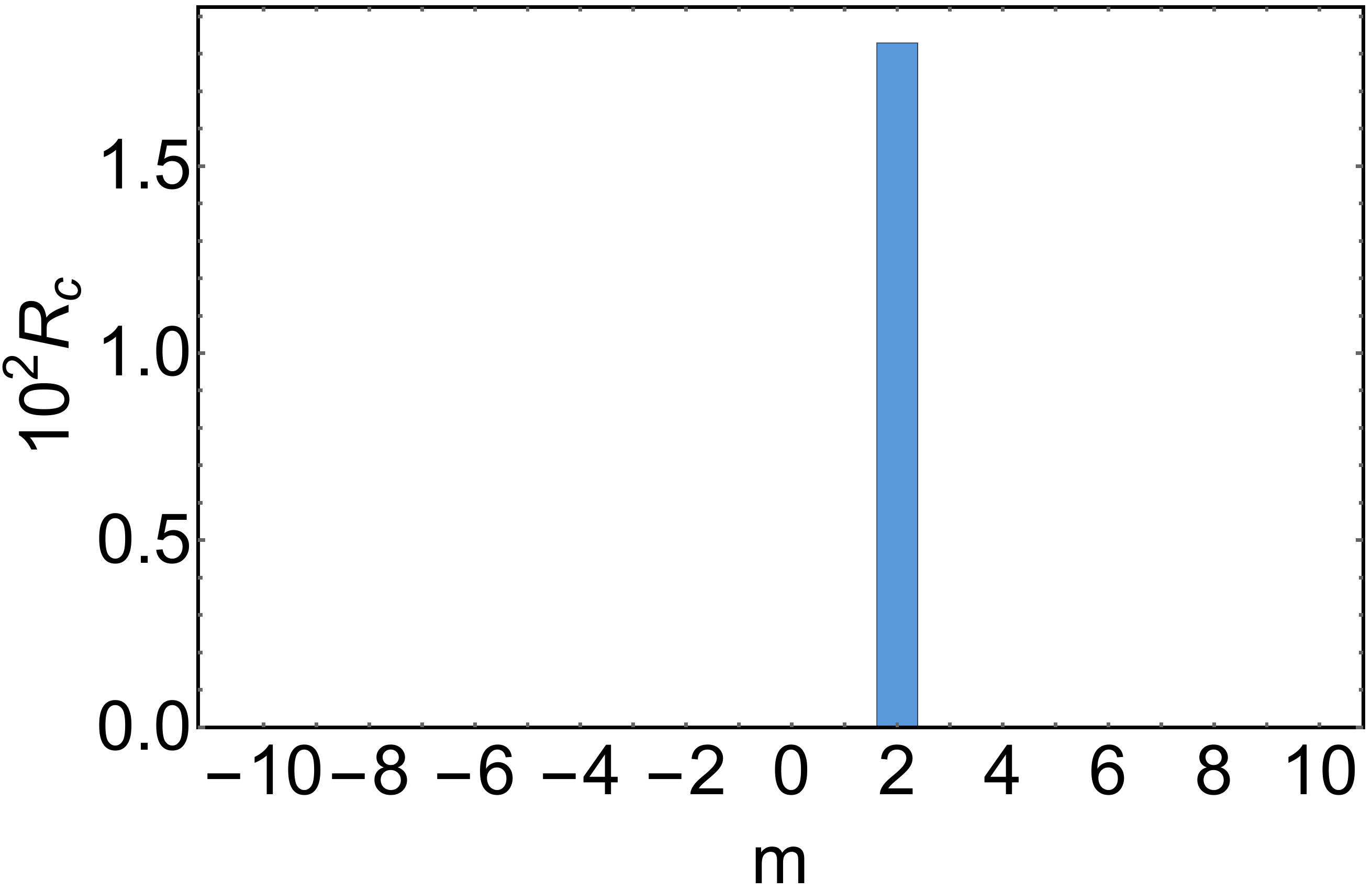}}\,
\raisebox{-0.5\height}{\includegraphics*[width=0.24\linewidth]{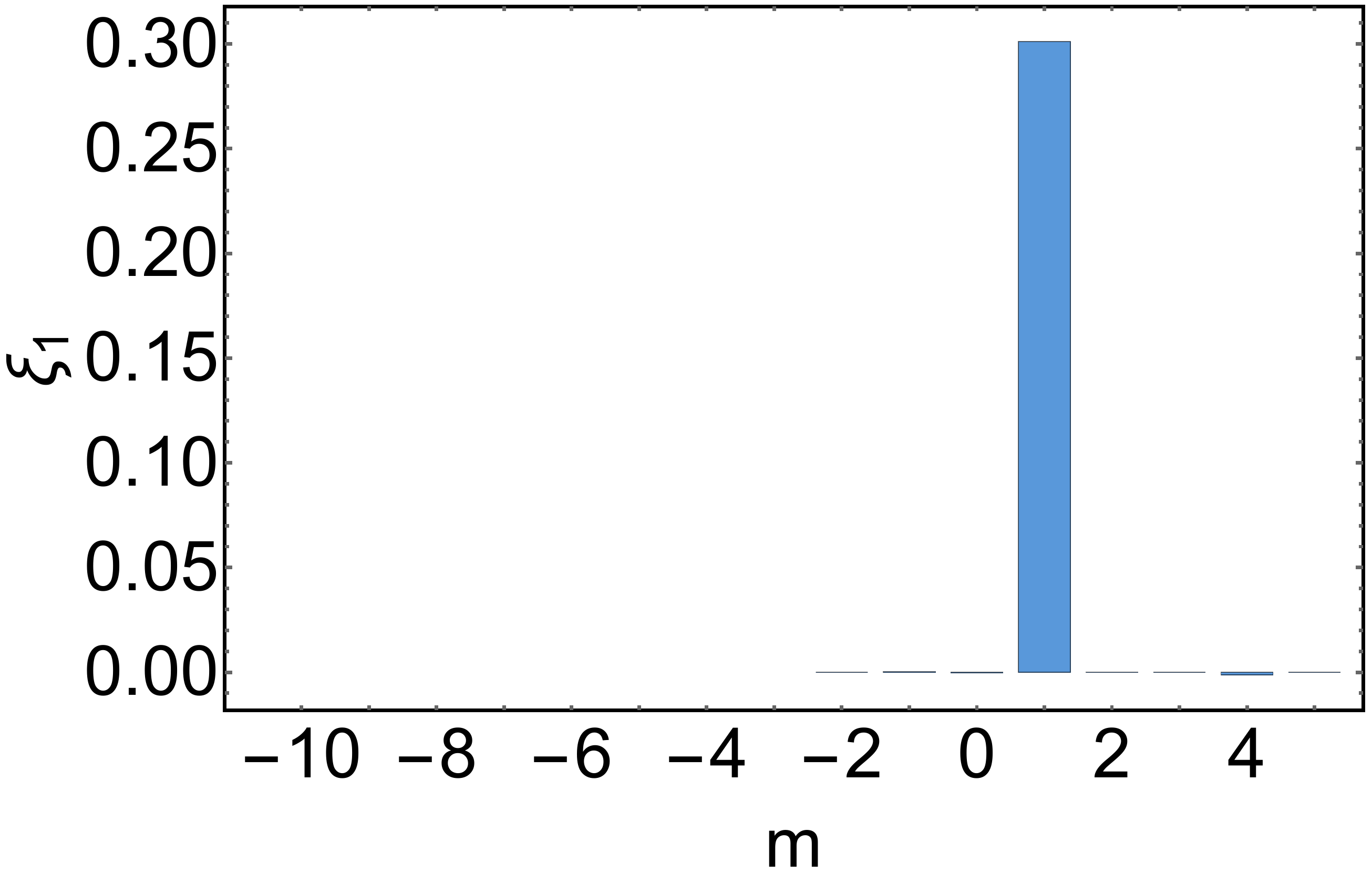}}\,
\raisebox{-0.5\height}{\includegraphics*[width=0.24\linewidth]{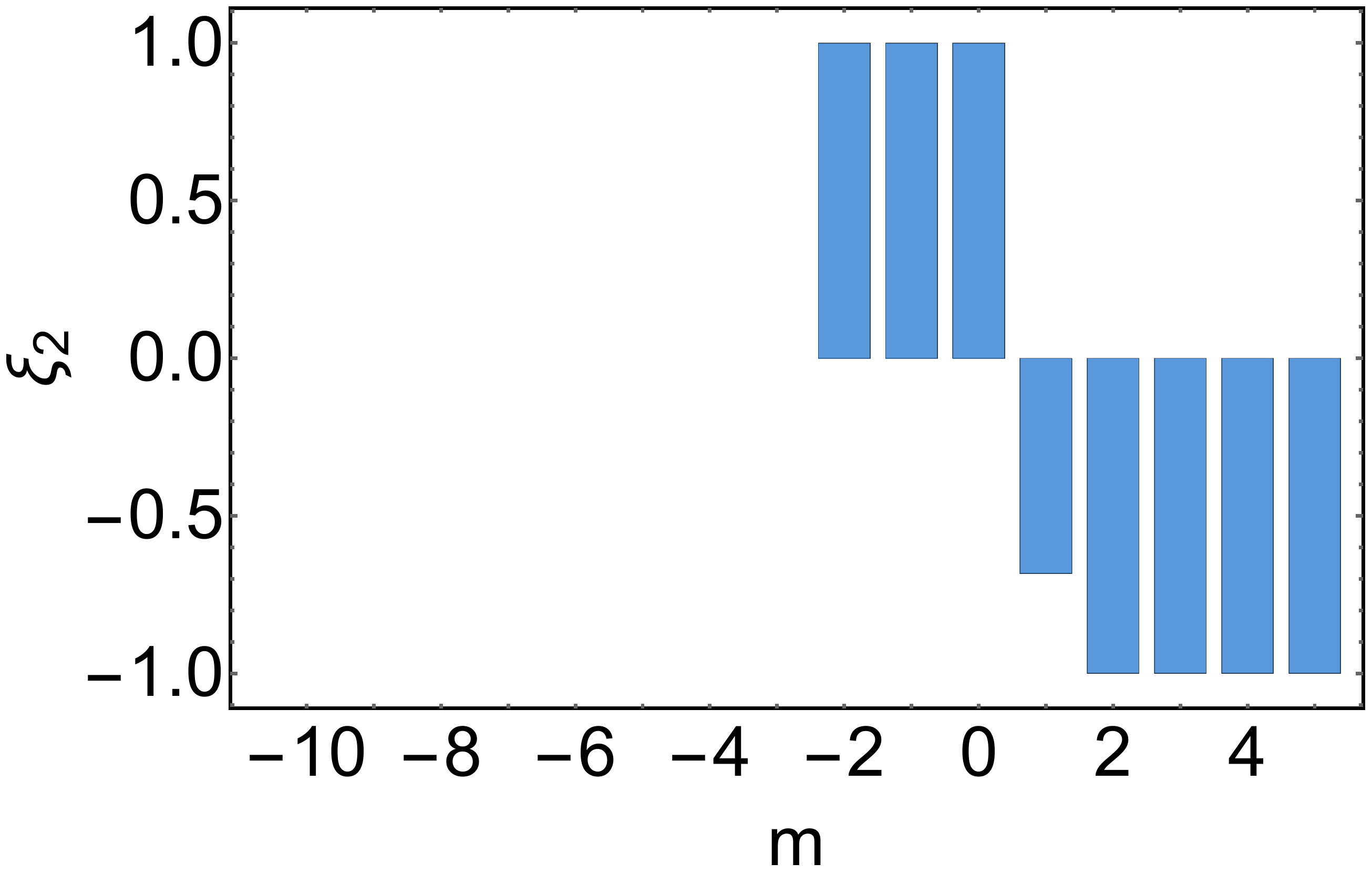}}\,
\raisebox{-0.5\height}{\includegraphics*[width=0.24\linewidth]{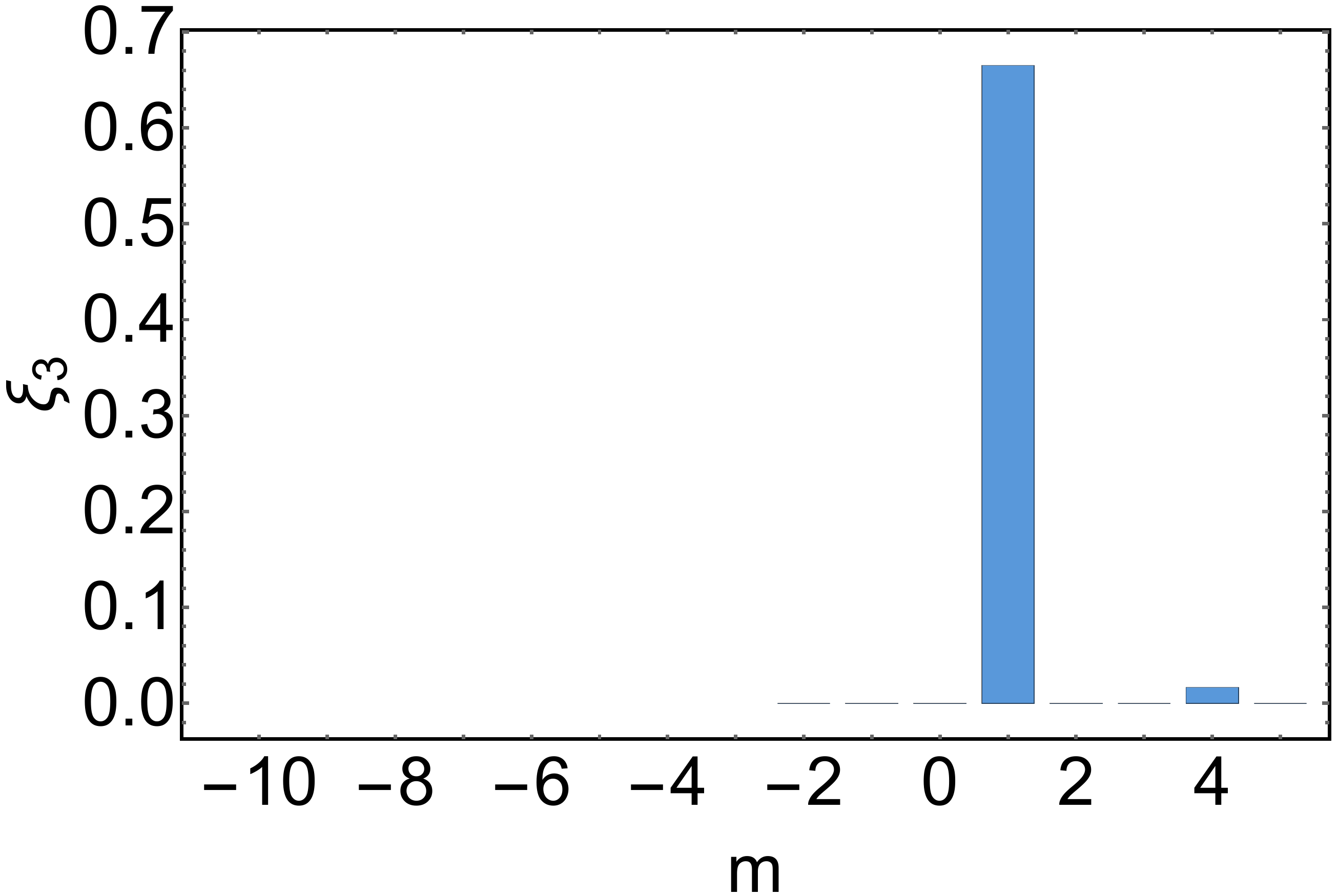}}\\
\raisebox{-0.5\height}{\includegraphics*[width=0.24\linewidth]{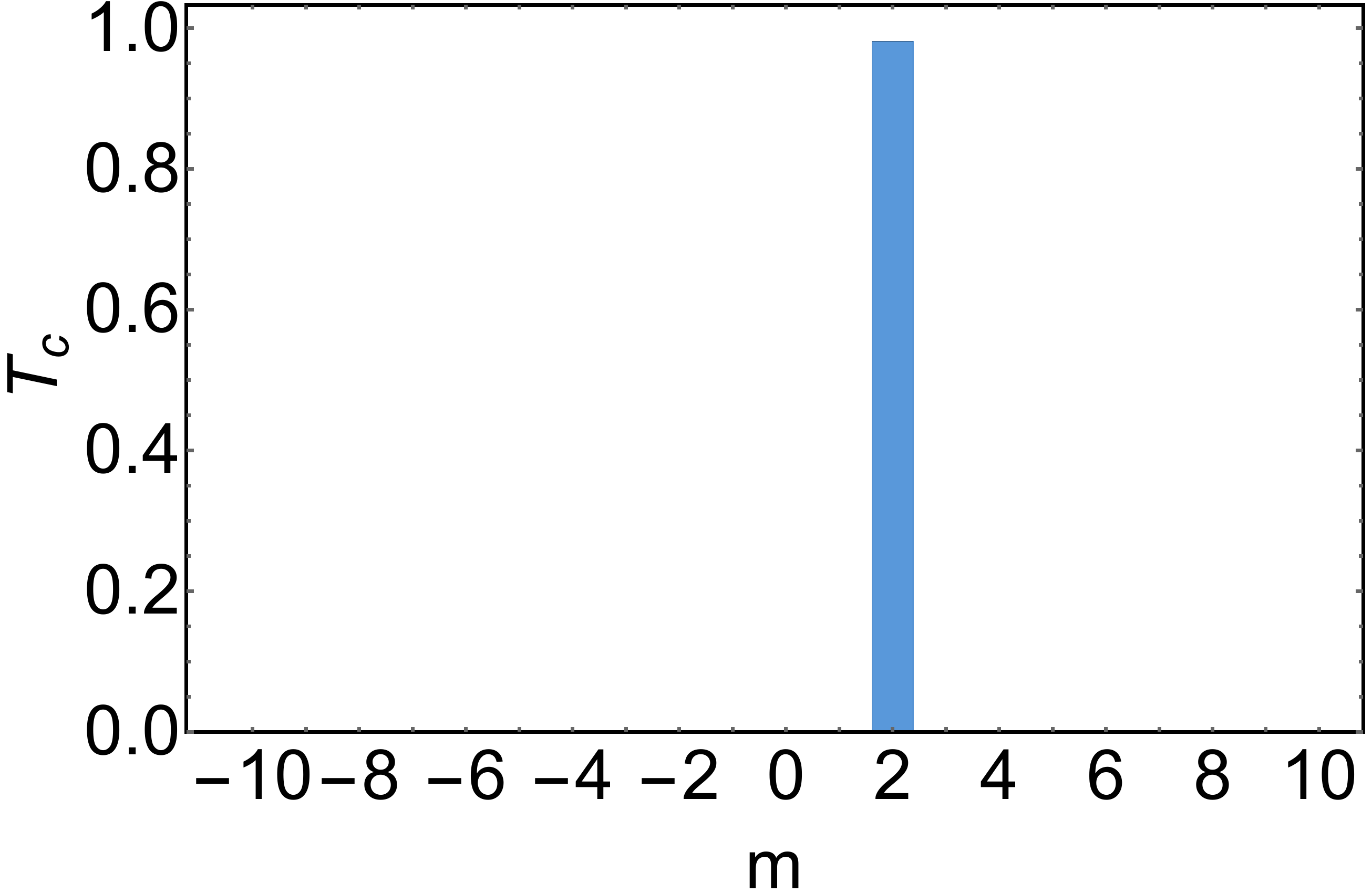}}\,
\raisebox{-0.5\height}{\includegraphics*[width=0.24\linewidth]{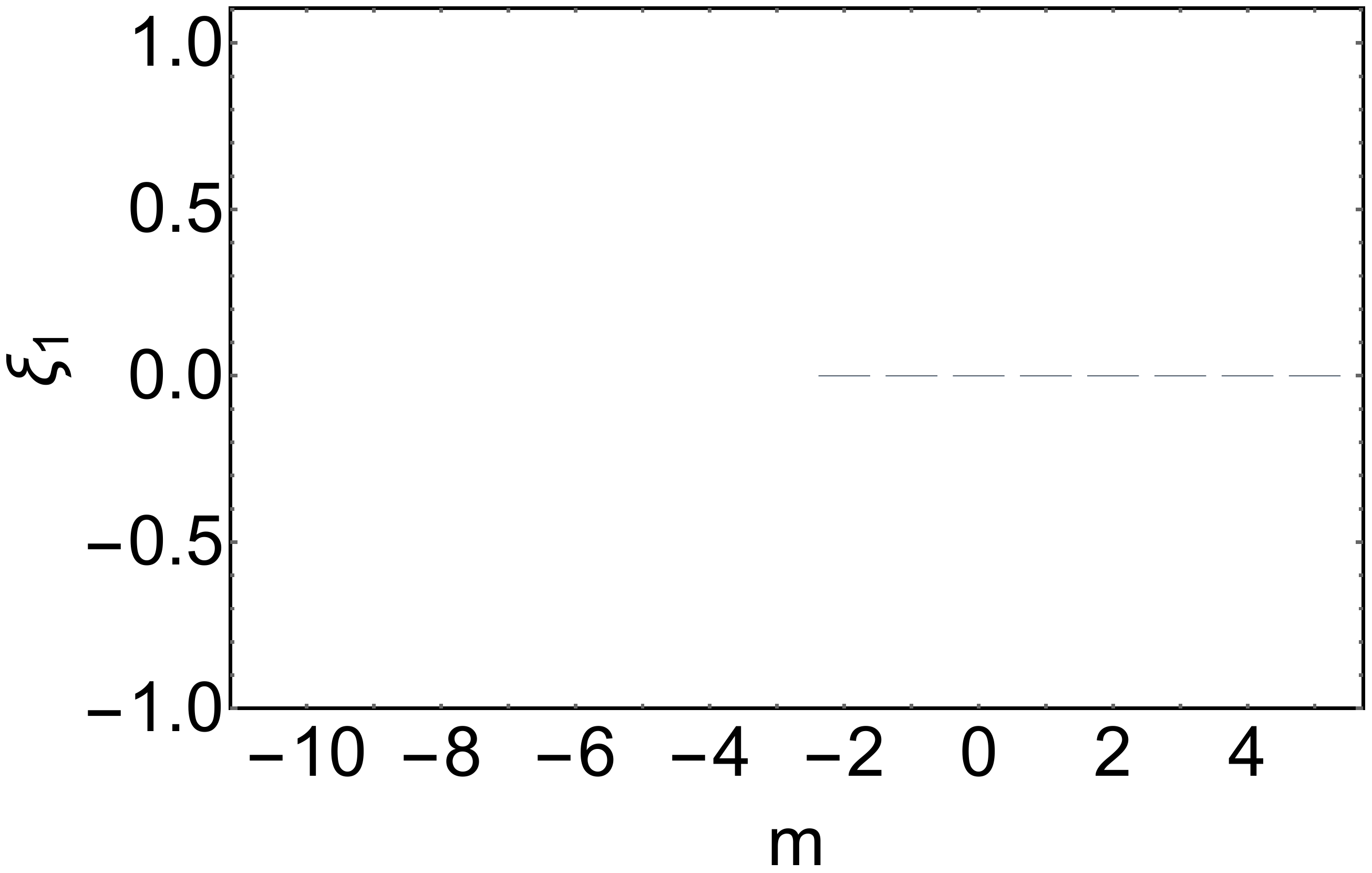}}\,
\raisebox{-0.5\height}{\includegraphics*[width=0.24\linewidth]{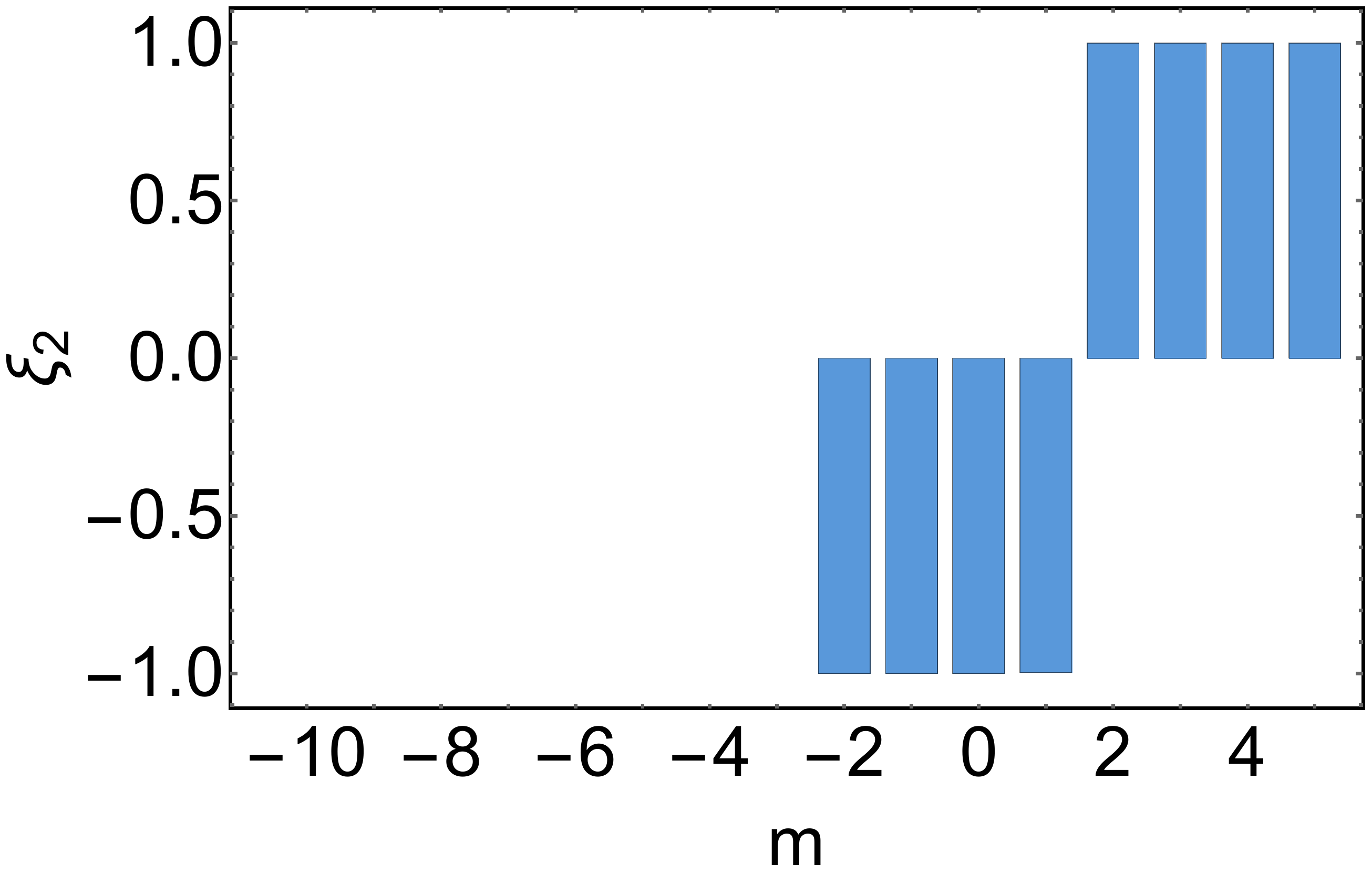}}\,
\raisebox{-0.5\height}{\includegraphics*[width=0.24\linewidth]{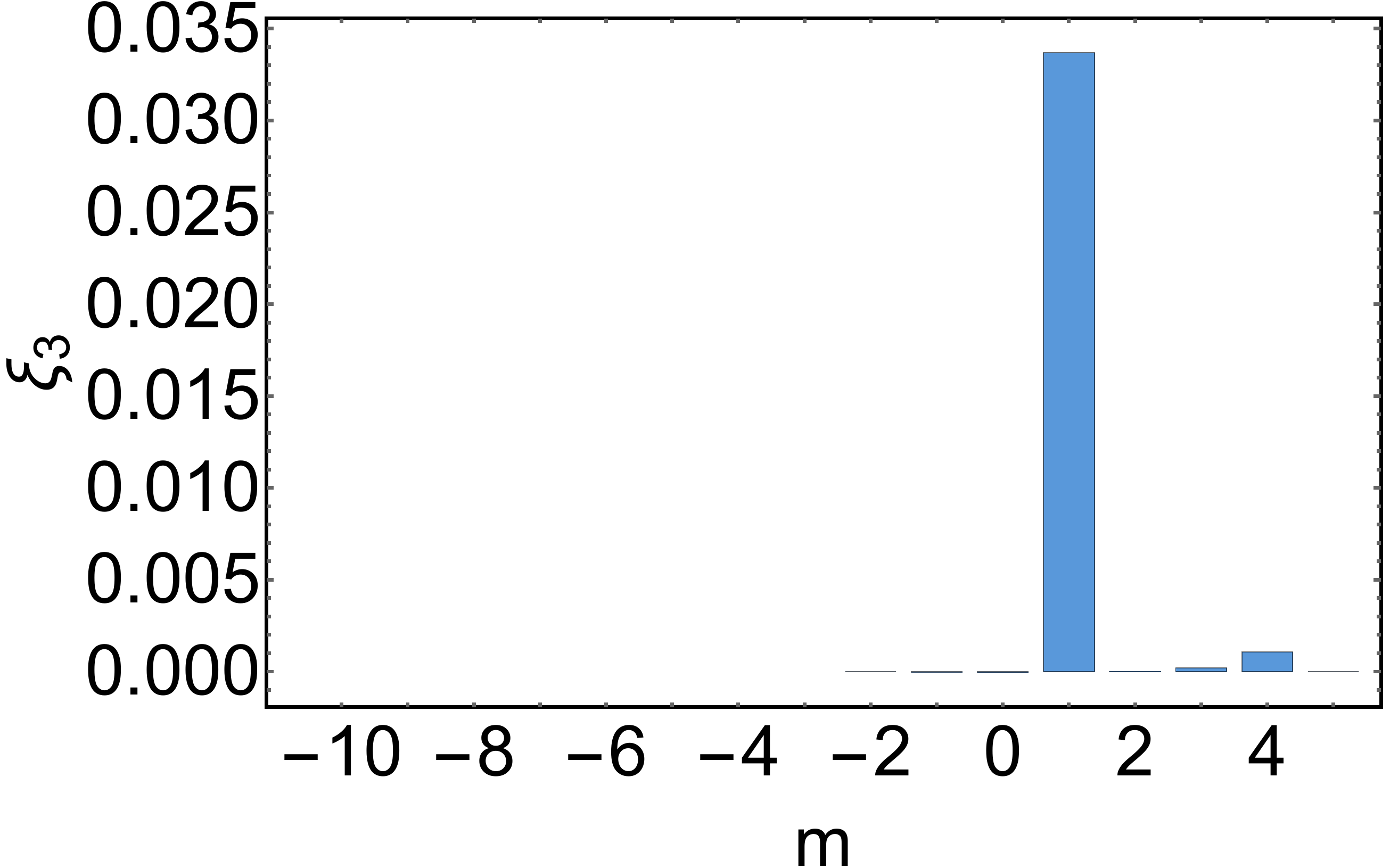}}\\
\raisebox{-0.5\height}{\includegraphics*[width=0.24\linewidth]{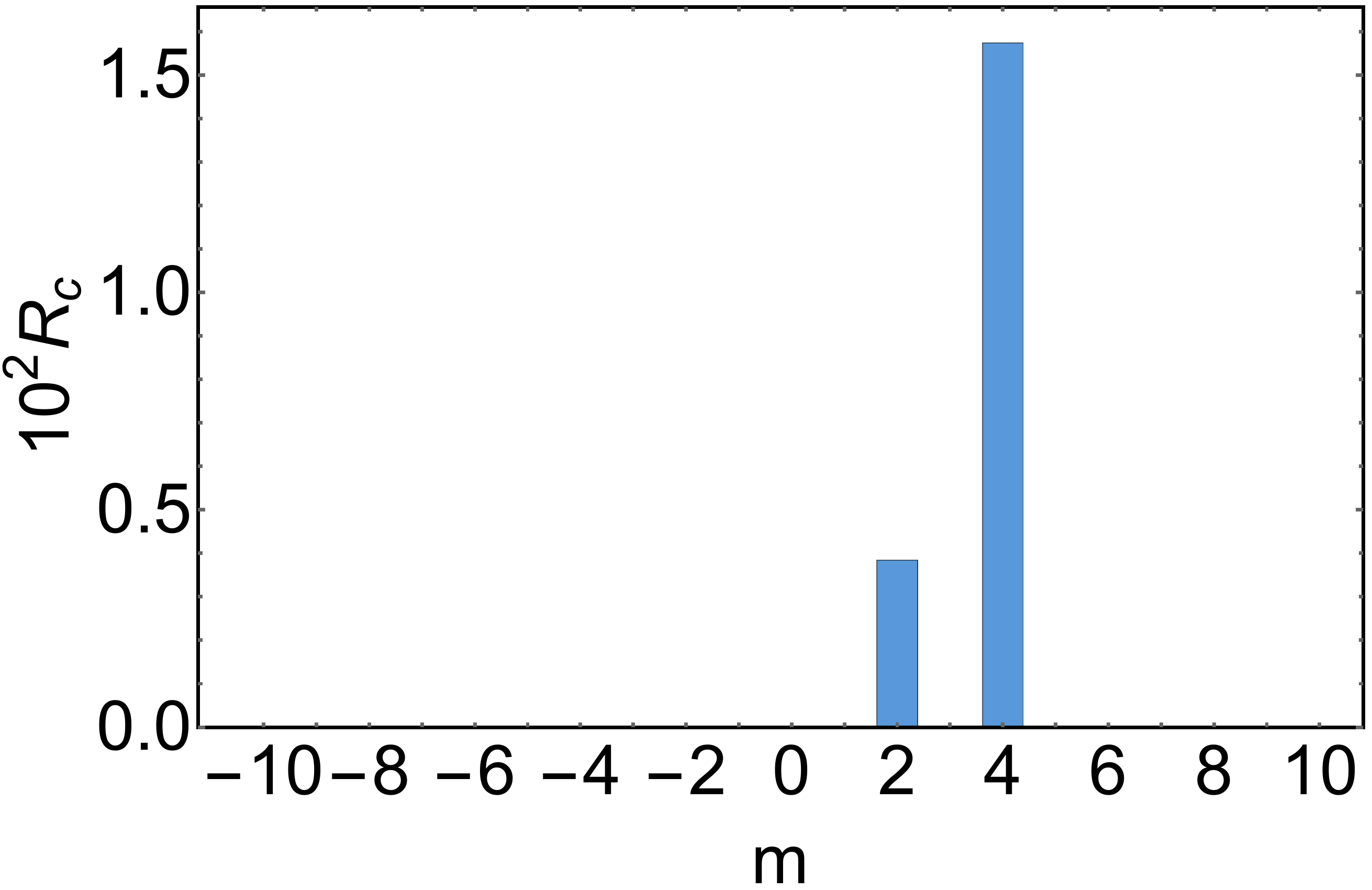}}\,
\raisebox{-0.5\height}{\includegraphics*[width=0.24\linewidth]{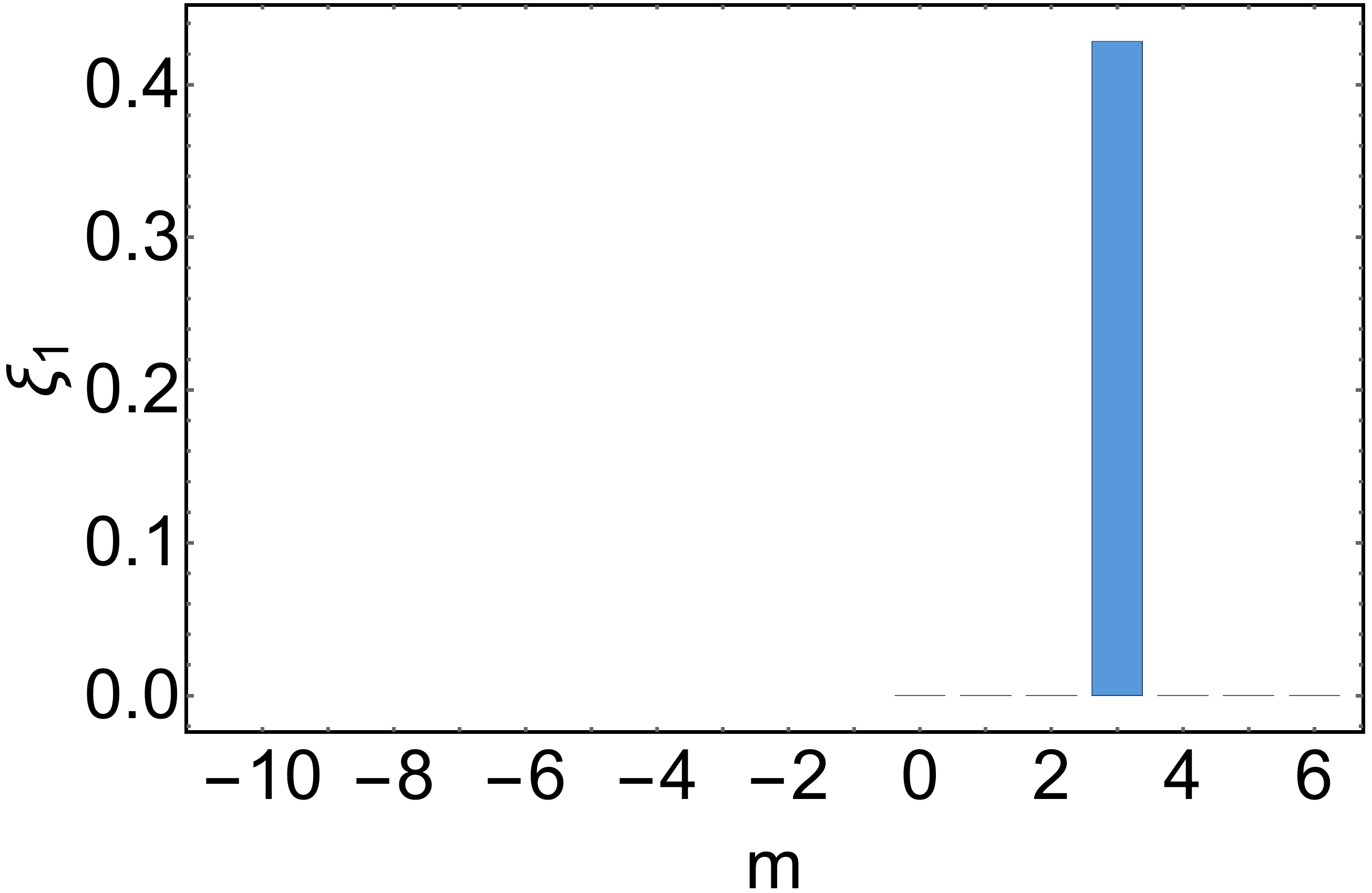}}\,
\raisebox{-0.5\height}{\includegraphics*[width=0.24\linewidth]{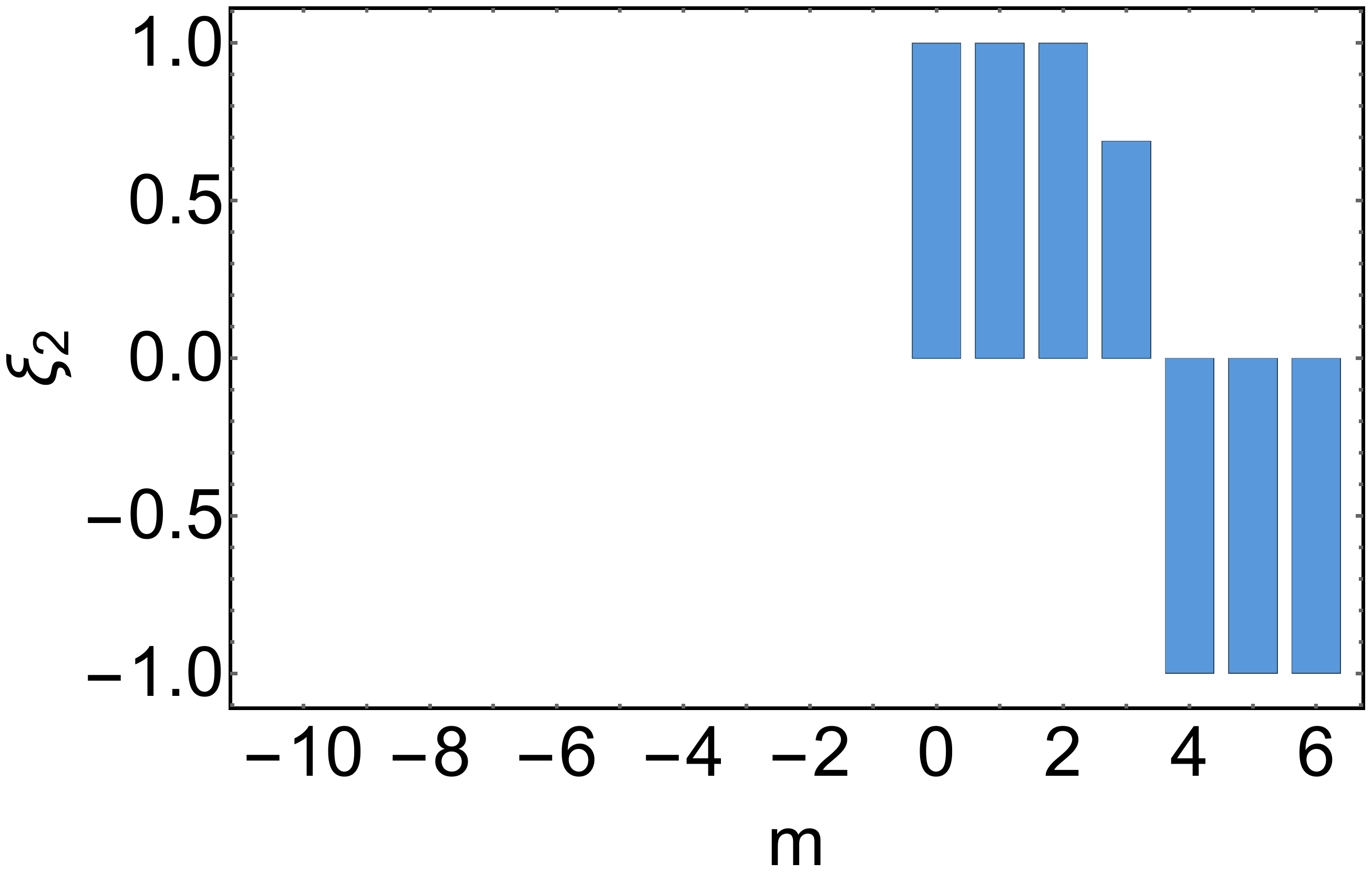}}\,
\raisebox{-0.5\height}{\includegraphics*[width=0.24\linewidth]{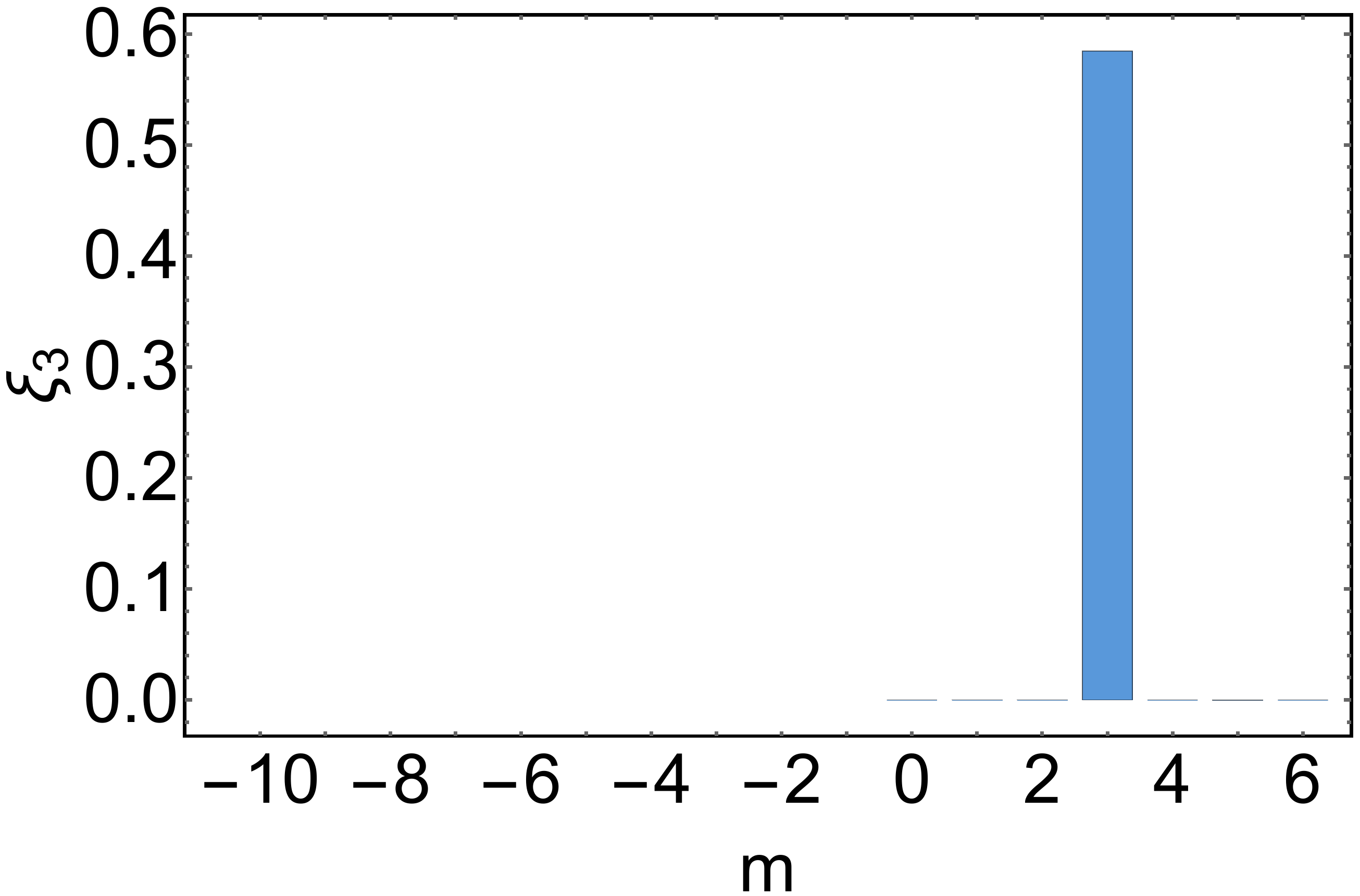}}\\
\raisebox{-0.5\height}{\includegraphics*[width=0.24\linewidth]{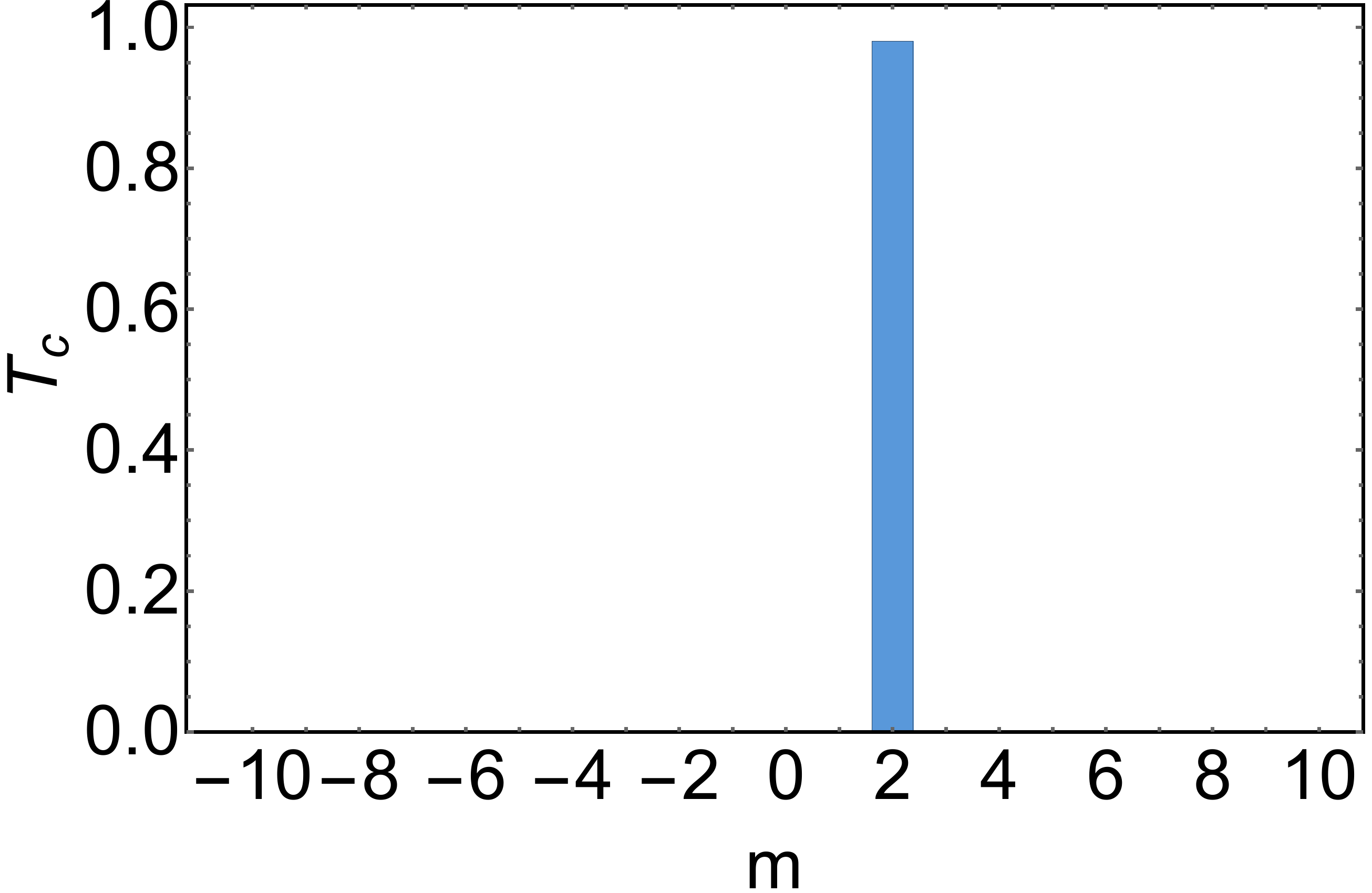}}\,
\raisebox{-0.5\height}{\includegraphics*[width=0.24\linewidth]{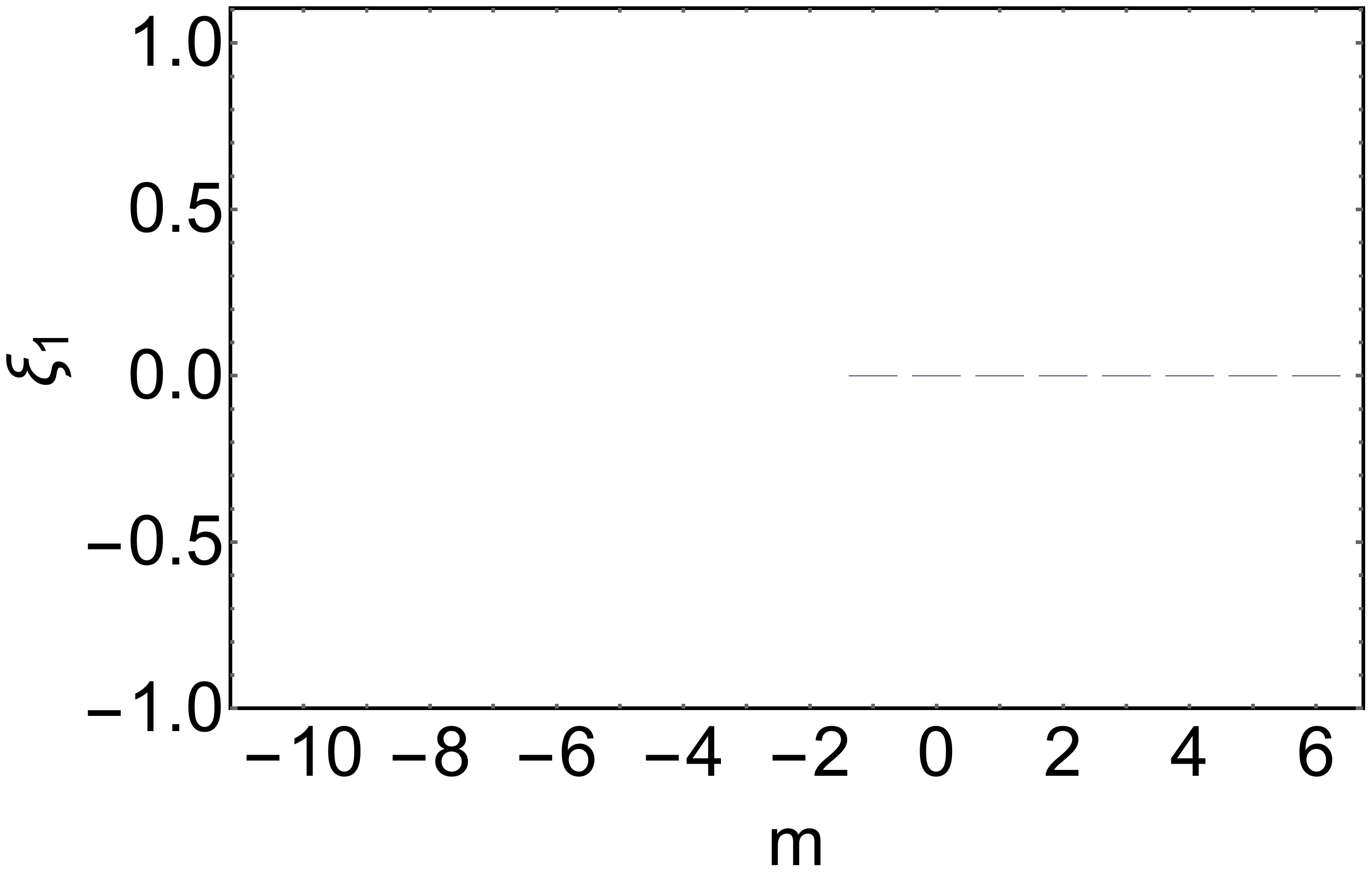}}\,
\raisebox{-0.5\height}{\includegraphics*[width=0.24\linewidth]{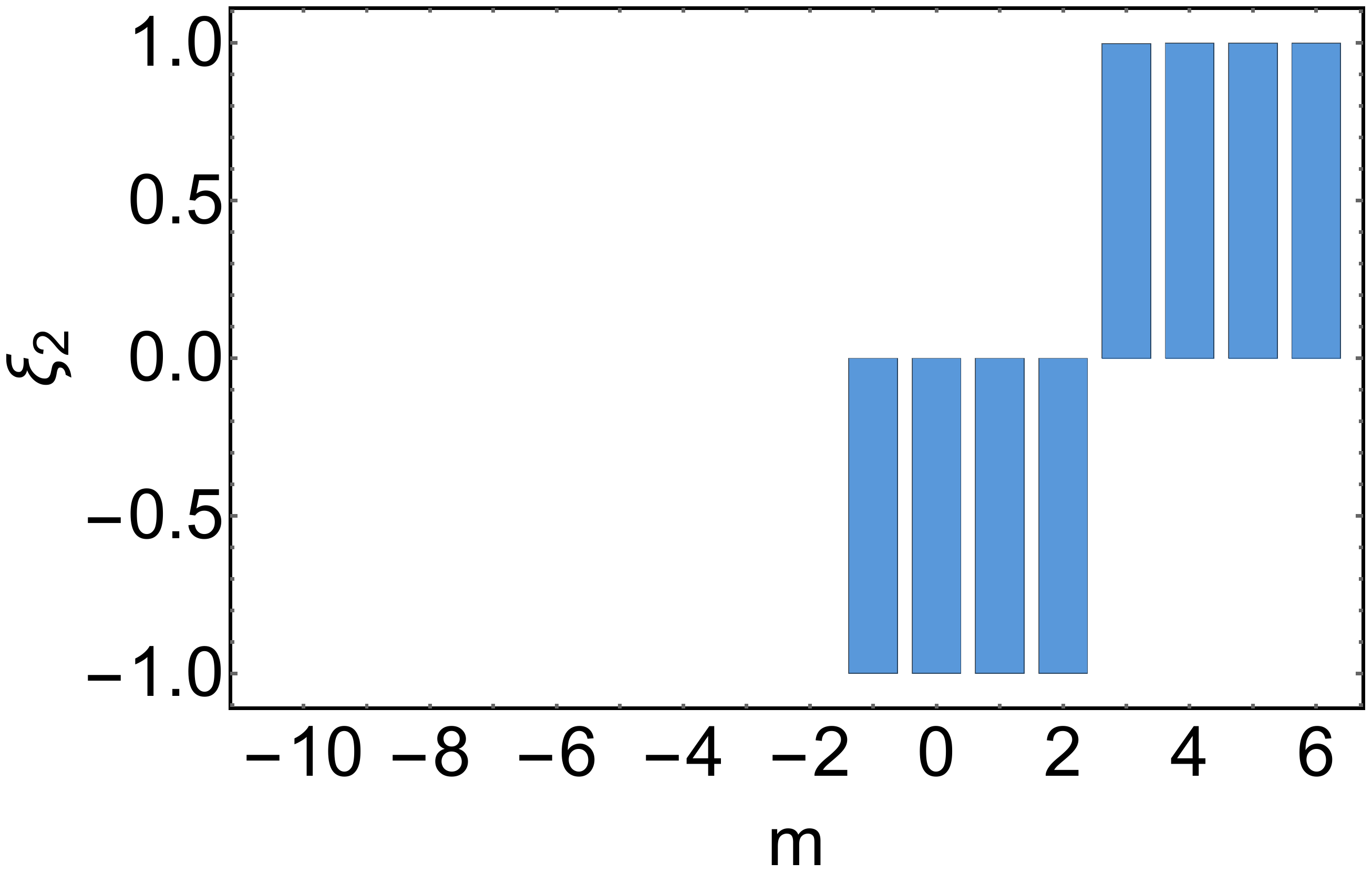}}\,
\raisebox{-0.5\height}{\includegraphics*[width=0.24\linewidth]{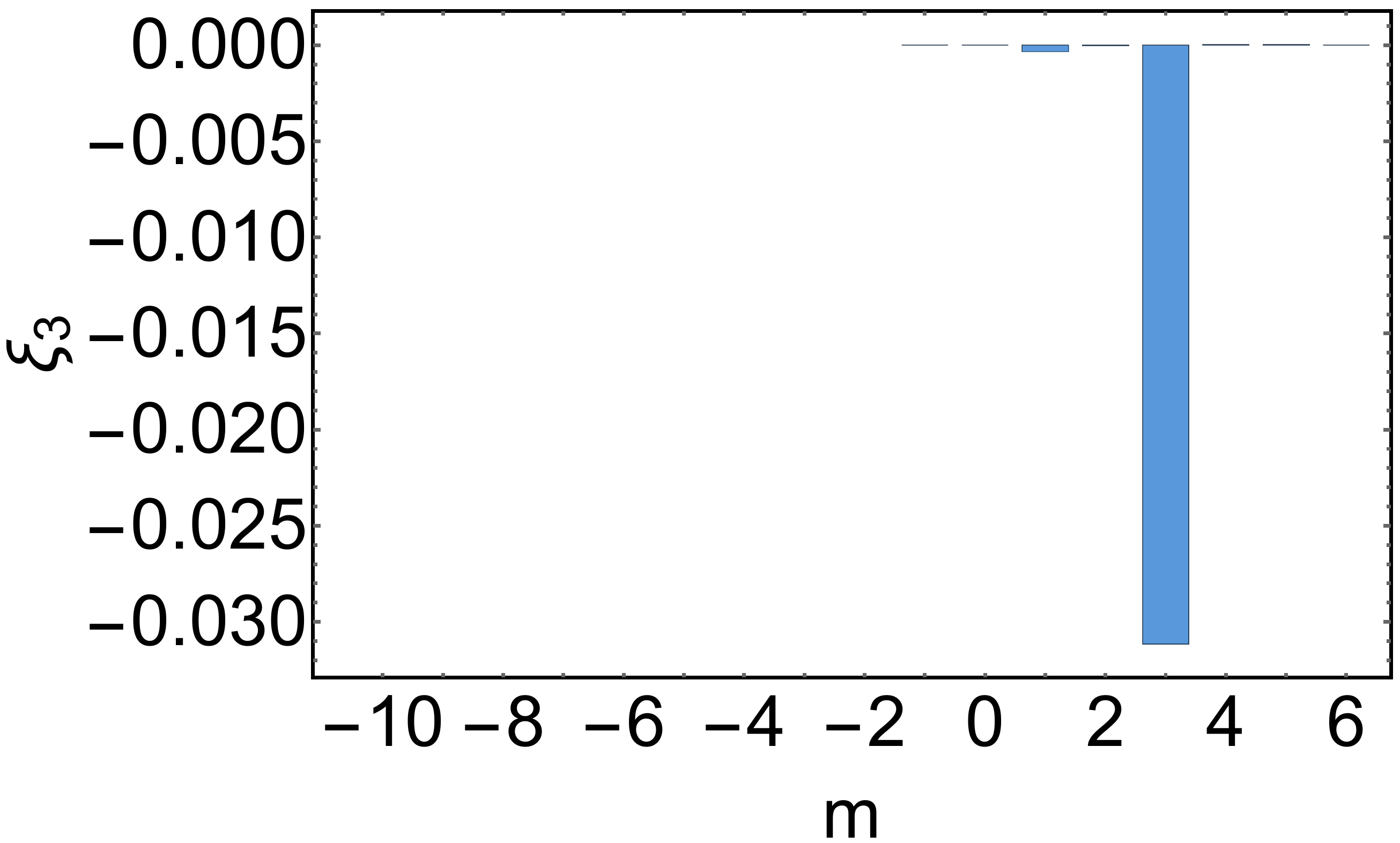}}
\caption{{\footnotesize The transmission and reflection coefficients and the Stokes parameters for the photons transmitted or reflected by the plate made of a helical medium near the real band gap corresponding to $s_h=\pm2$. The plate is immersed into the medium with permittivity $\e_v=2.65$. The parameters are the same as in Fig. \ref{Disp_Hels2_plots}. The lines $1$-$4$: The case of plane-wave photons scattered in the $(x,z)$ plane is considered. The lines $1$-$2$: The initial photon possesses the helicity $s=1$. The lines $3$-$4$: It has $s=-1$. The lines $5$-$8$: Scattering of the twisted photon with $m=2$ is considered at the energy $k_0=0.6354$ eV belonging to the band gap. The lines $5$-$6$ corresponds to $s=1$, whereas the lines $7$-$8$ are for $s=-1$. As is seen from the plots on the line $7$, the twisted photons with the same helicity as the helical medium, i.e., $qs<0$, acquire the additional projection of the total angular momentum as it follows from the selection rule \eqref{sel_rule}. However, as long as the gap parameter $|b|$ is small, the reflection coefficient is also small.}}
\label{Scatt_Hels2r_plots}
\end{figure}

%\newpage
\begin{figure}[tp]
\centering
\raisebox{-0.5\height}{\includegraphics*[width=0.24\linewidth]{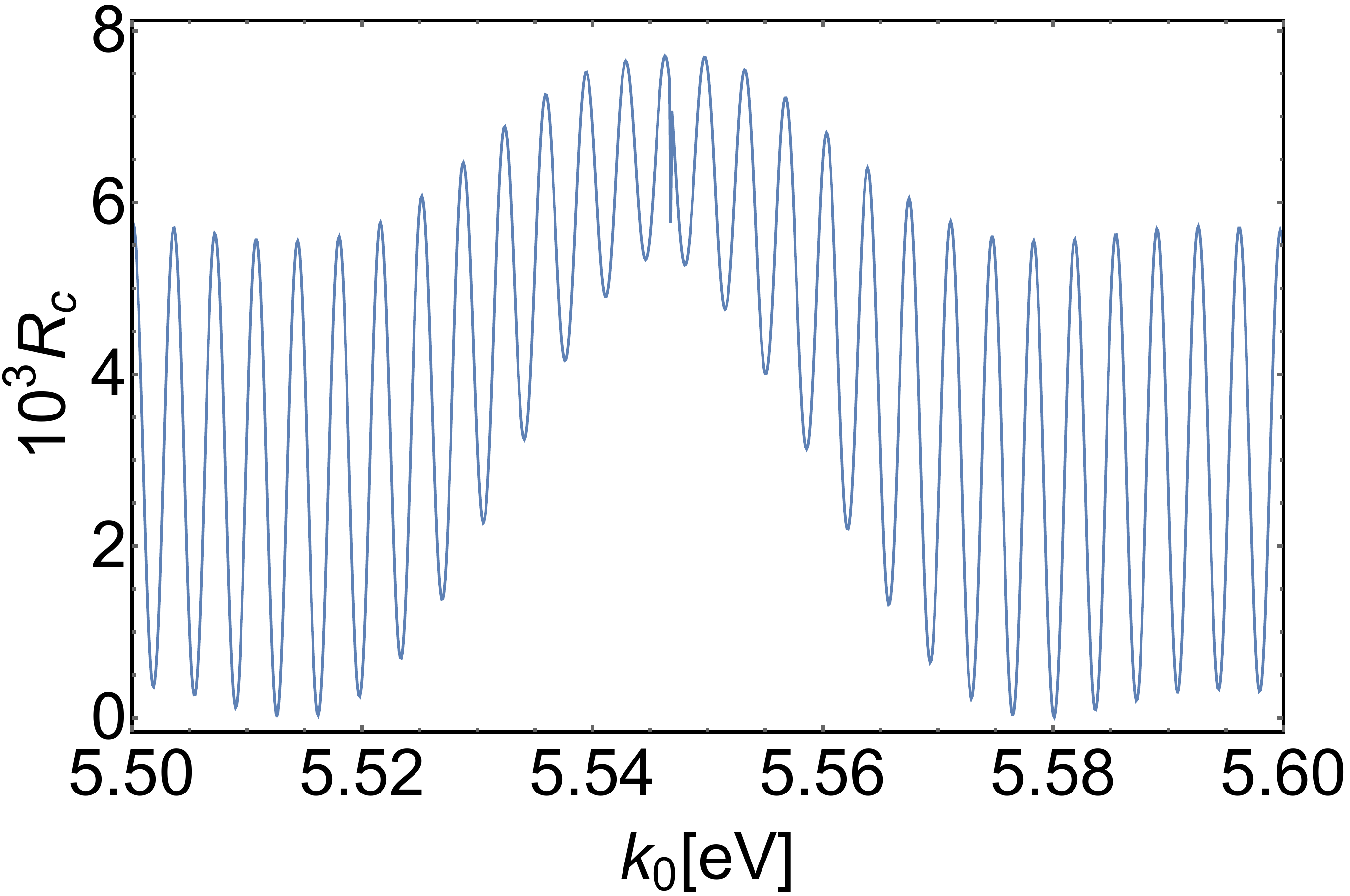}}\,
\raisebox{-0.5\height}{\includegraphics*[width=0.24\linewidth]{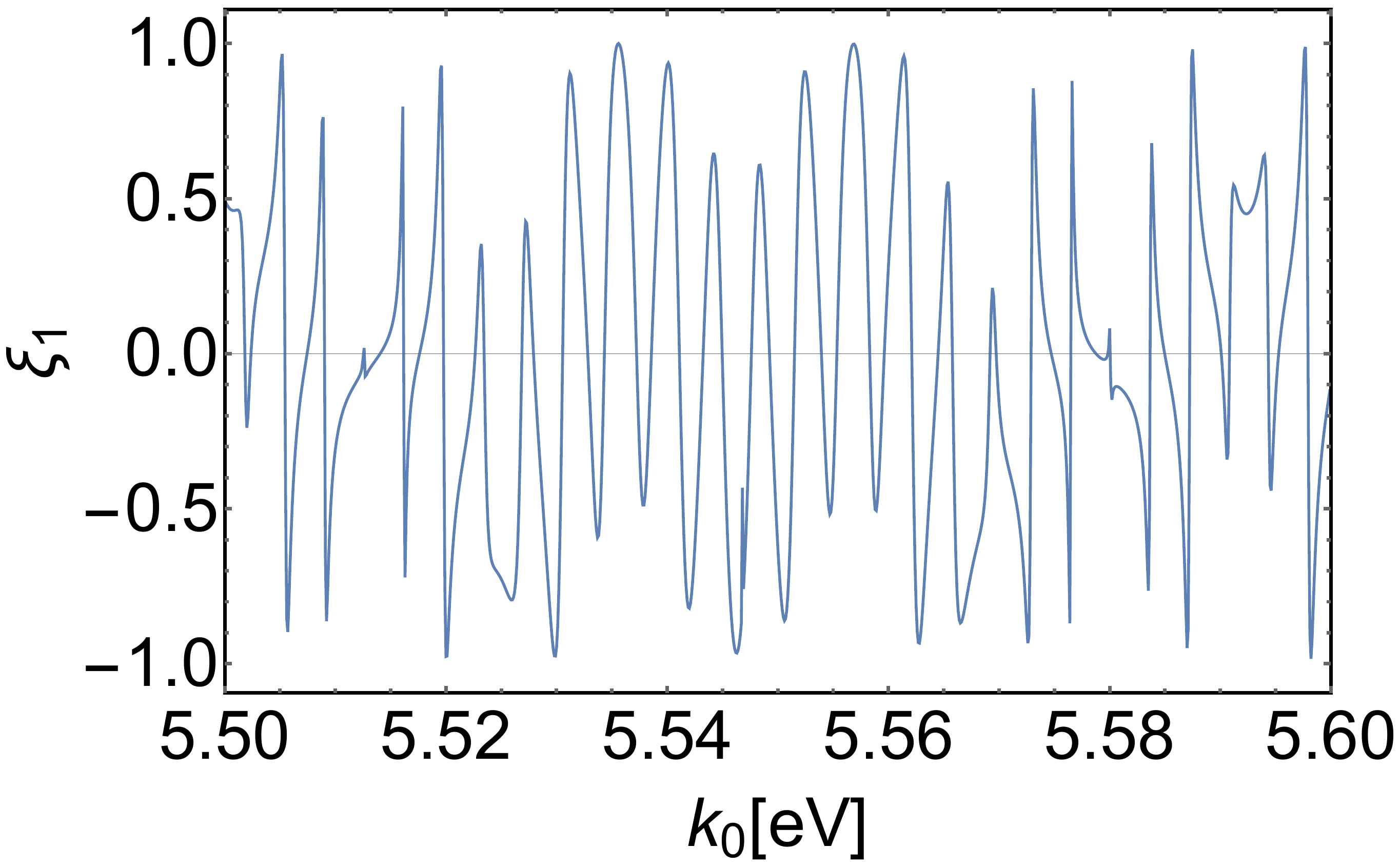}}\,
\raisebox{-0.5\height}{\includegraphics*[width=0.24\linewidth]{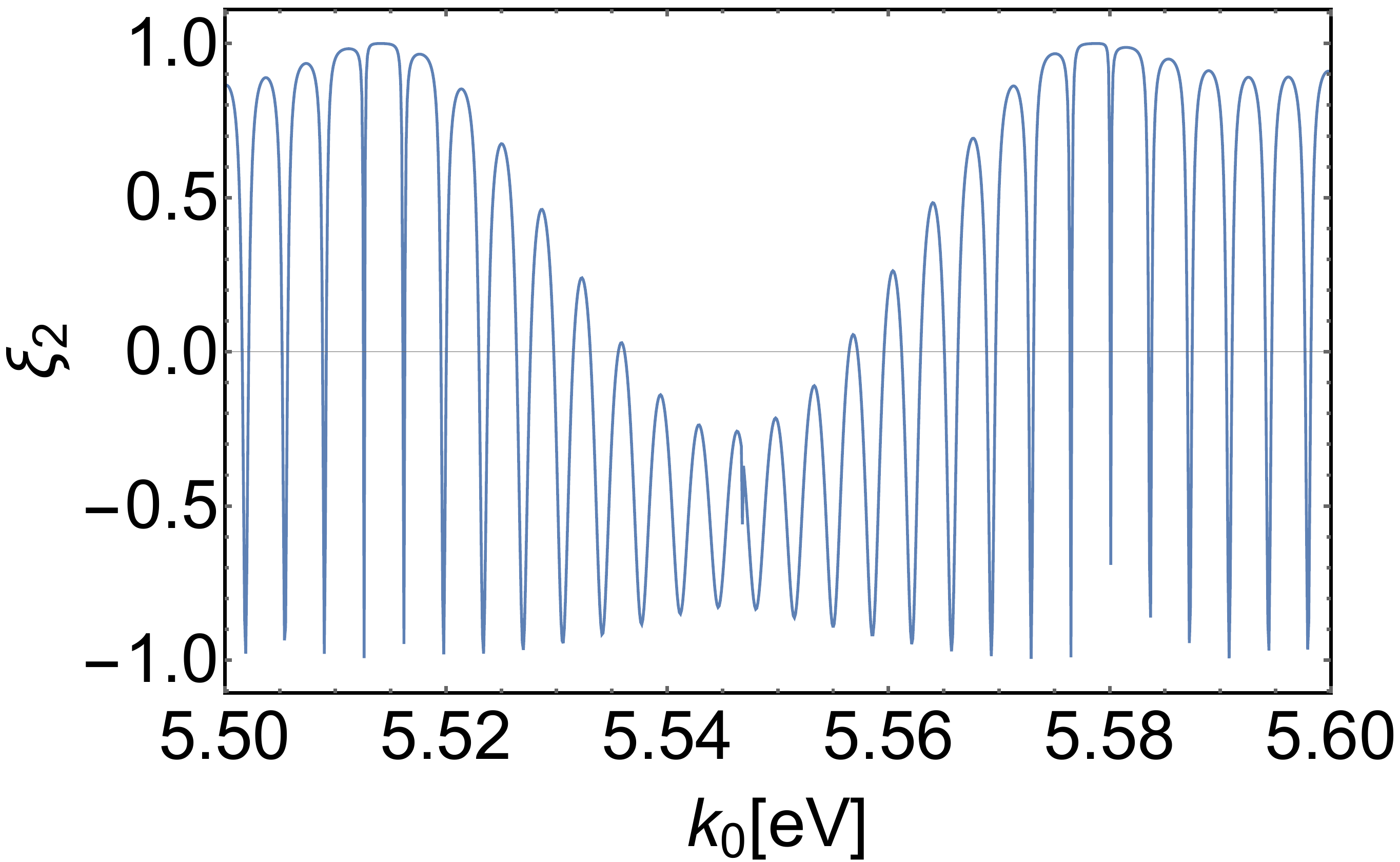}}\,
\raisebox{-0.5\height}{\includegraphics*[width=0.24\linewidth]{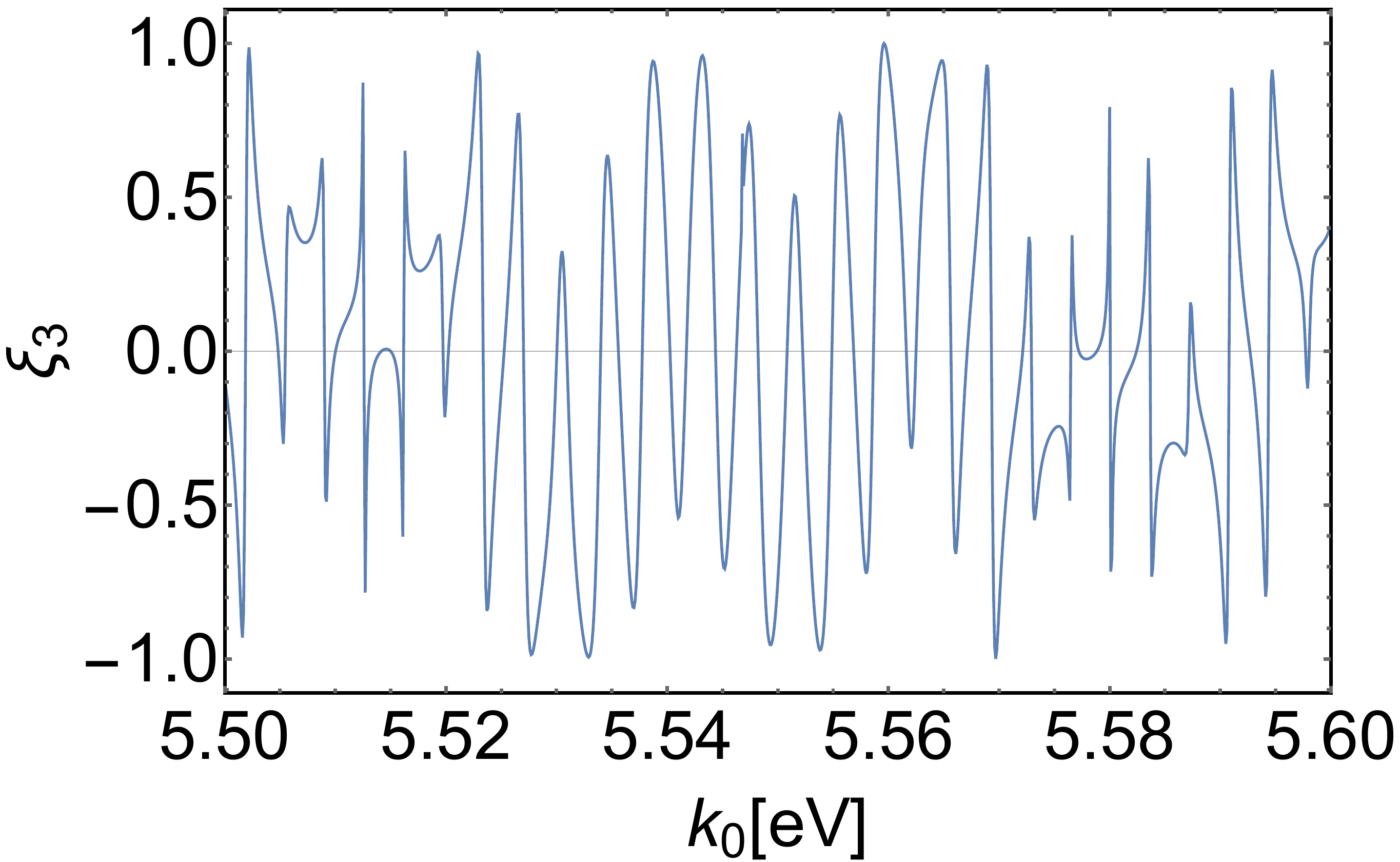}}\\
\raisebox{-0.5\height}{\includegraphics*[width=0.24\linewidth]{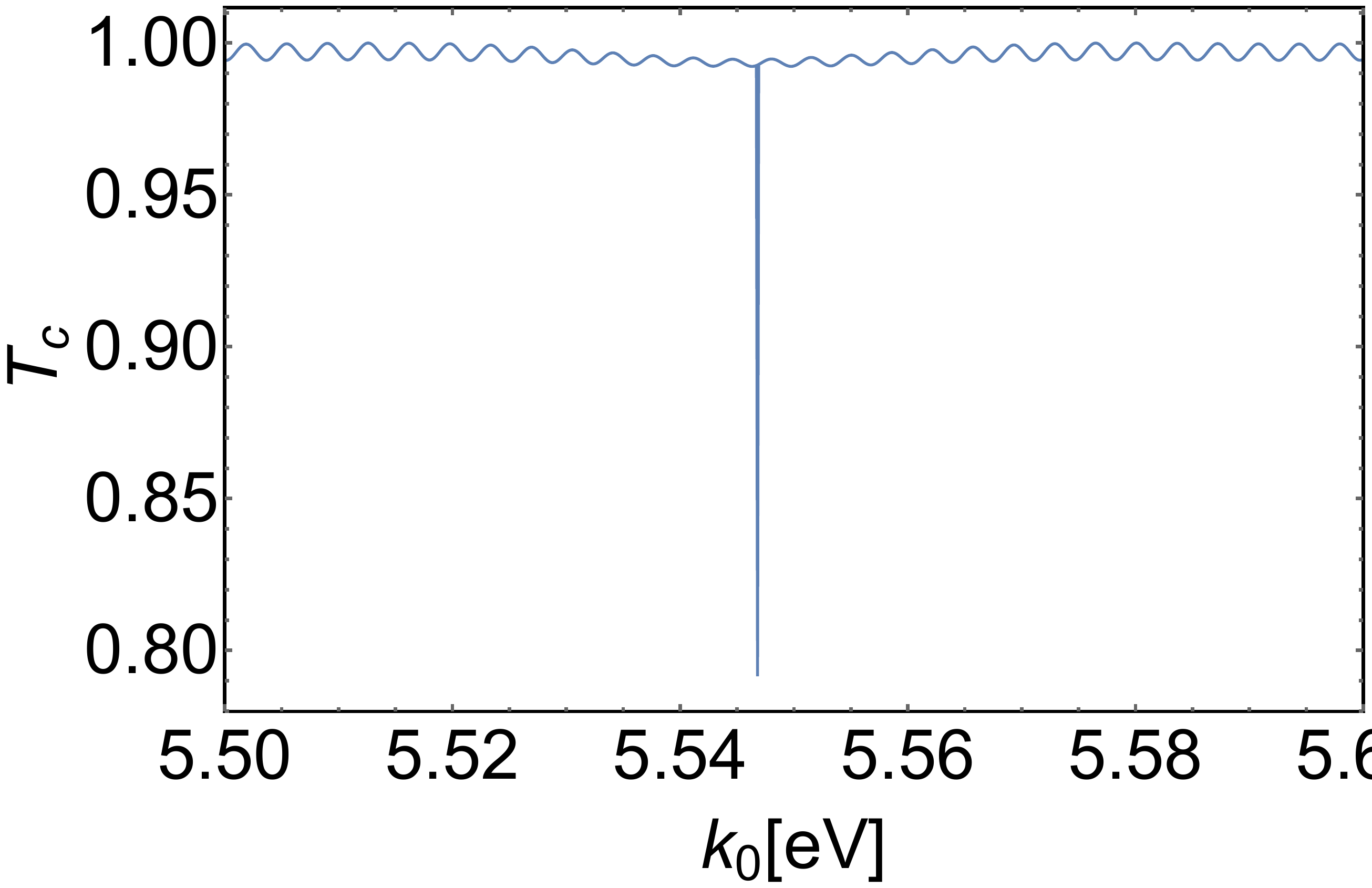}}\,
\raisebox{-0.5\height}{\includegraphics*[width=0.24\linewidth]{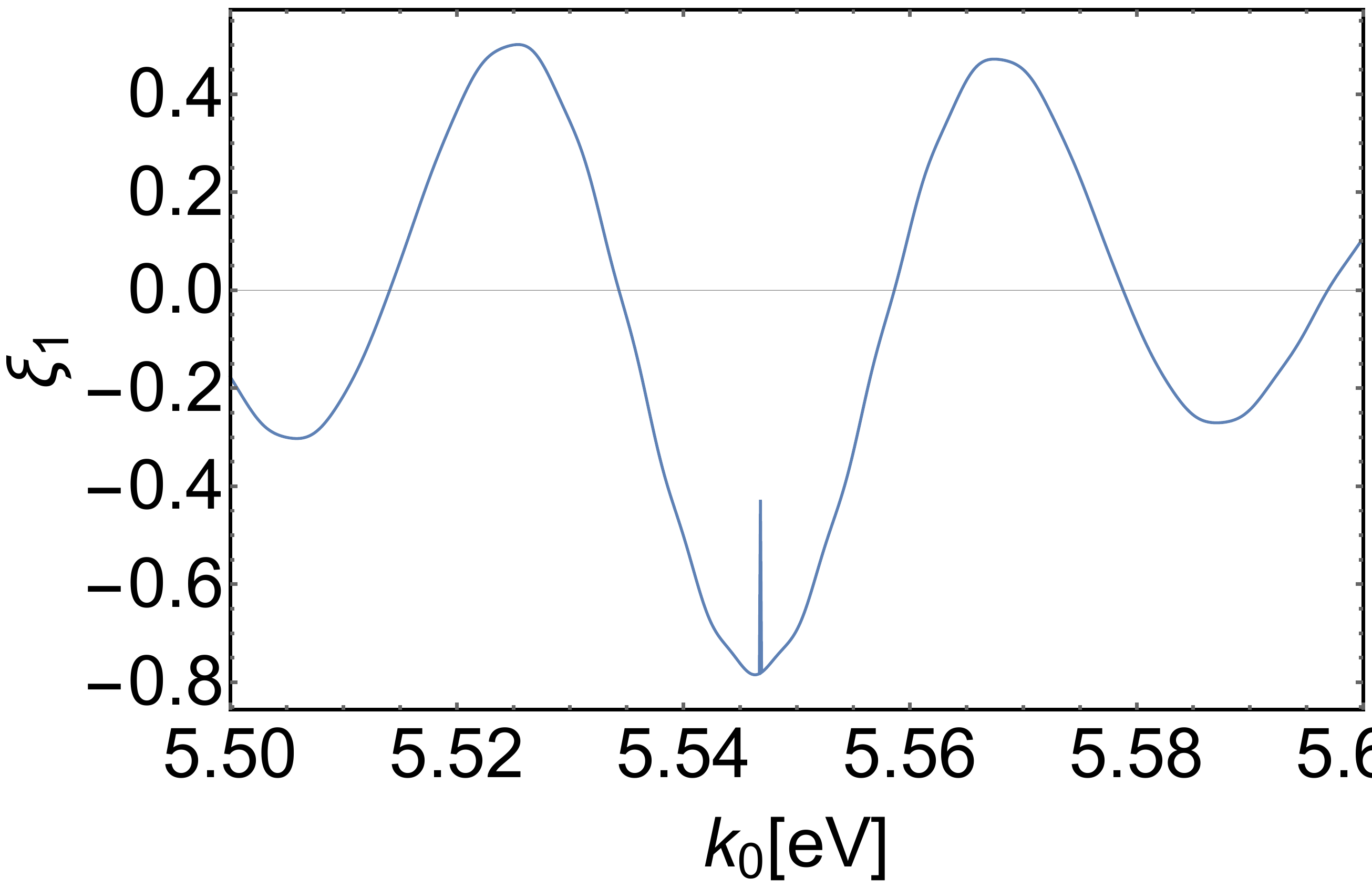}}\,
\raisebox{-0.5\height}{\includegraphics*[width=0.24\linewidth]{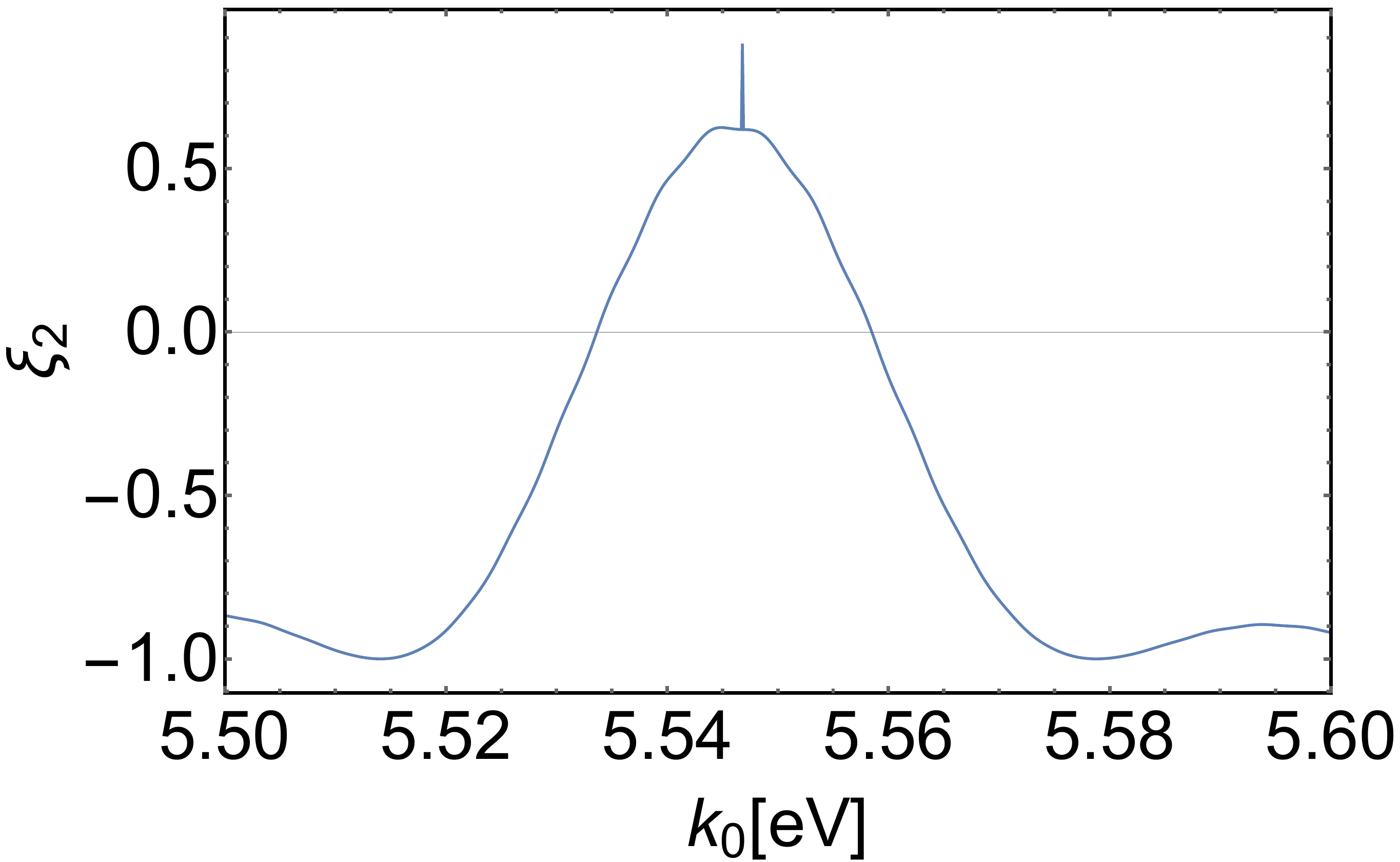}}\,
\raisebox{-0.5\height}{\includegraphics*[width=0.24\linewidth]{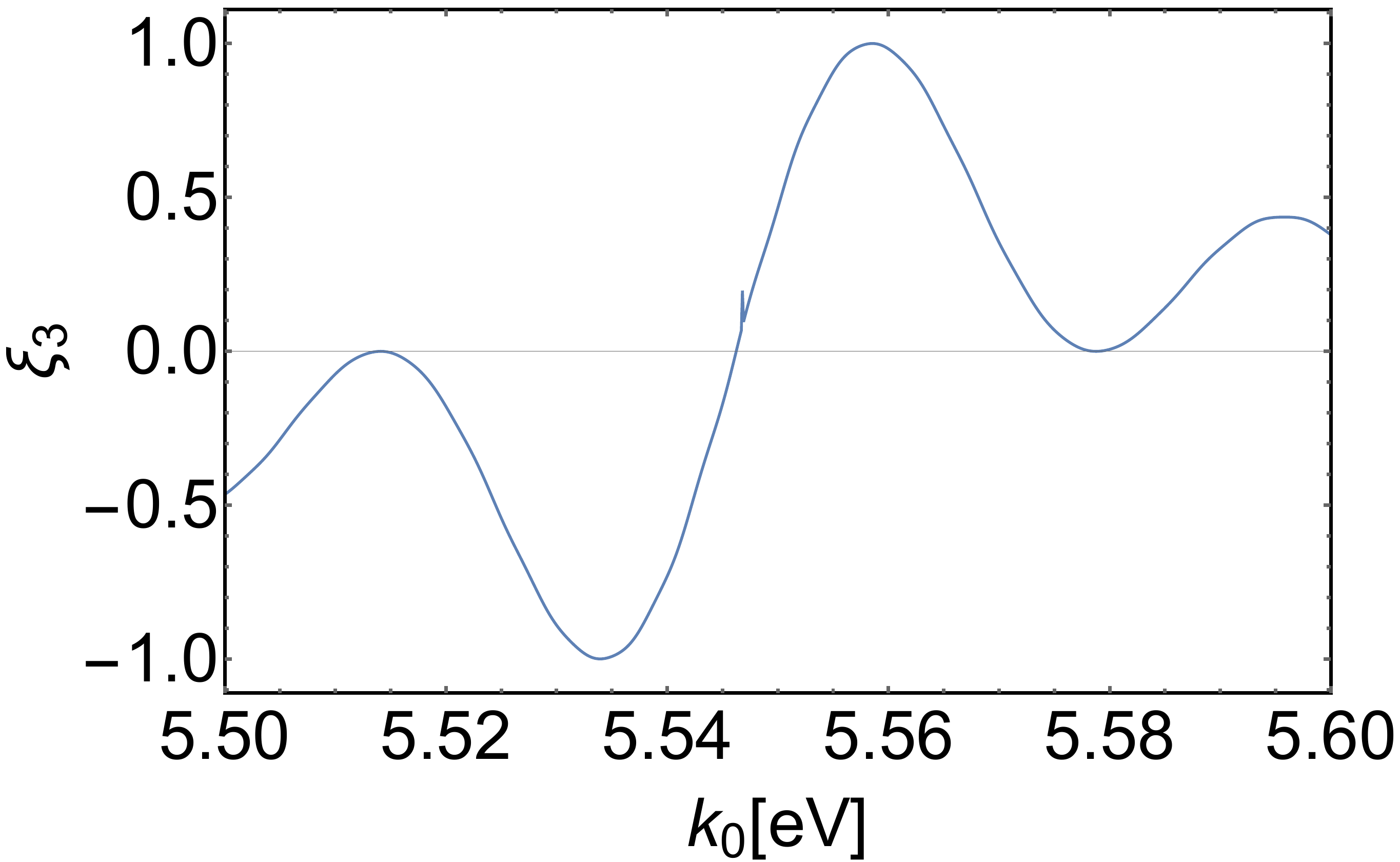}}\\
\raisebox{-0.5\height}{\includegraphics*[width=0.24\linewidth]{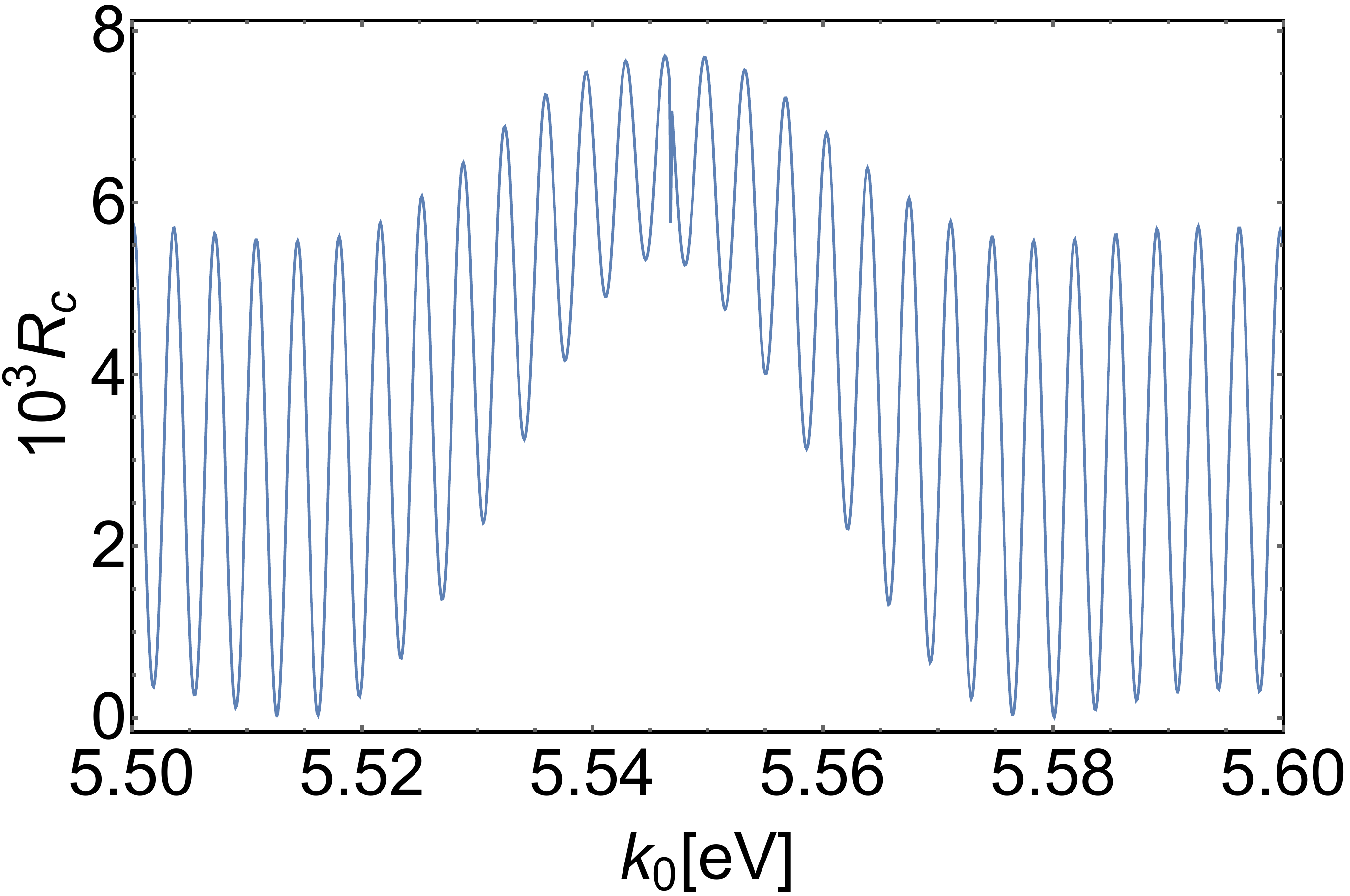}}\,
\raisebox{-0.5\height}{\includegraphics*[width=0.24\linewidth]{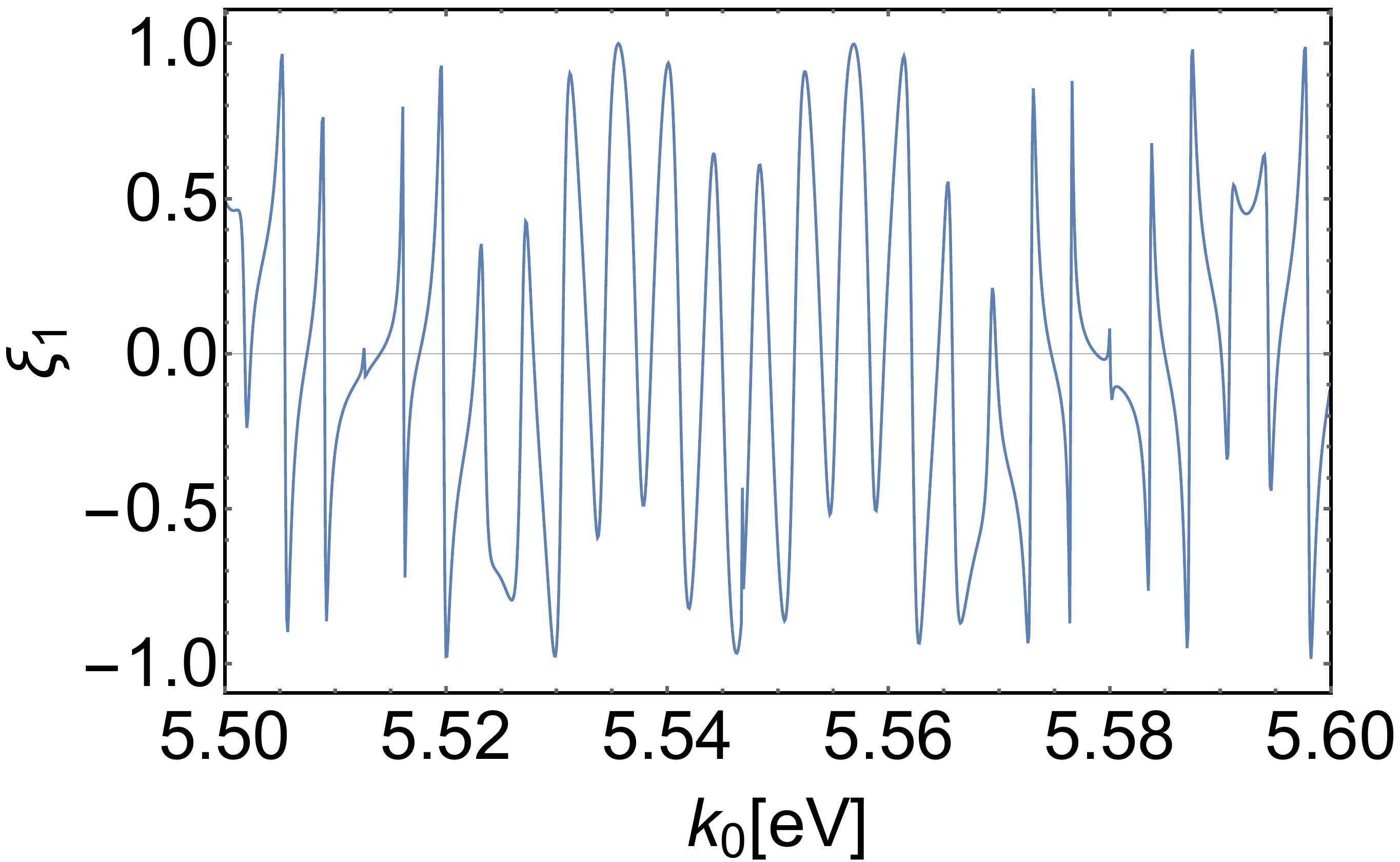}}\,
\raisebox{-0.5\height}{\includegraphics*[width=0.24\linewidth]{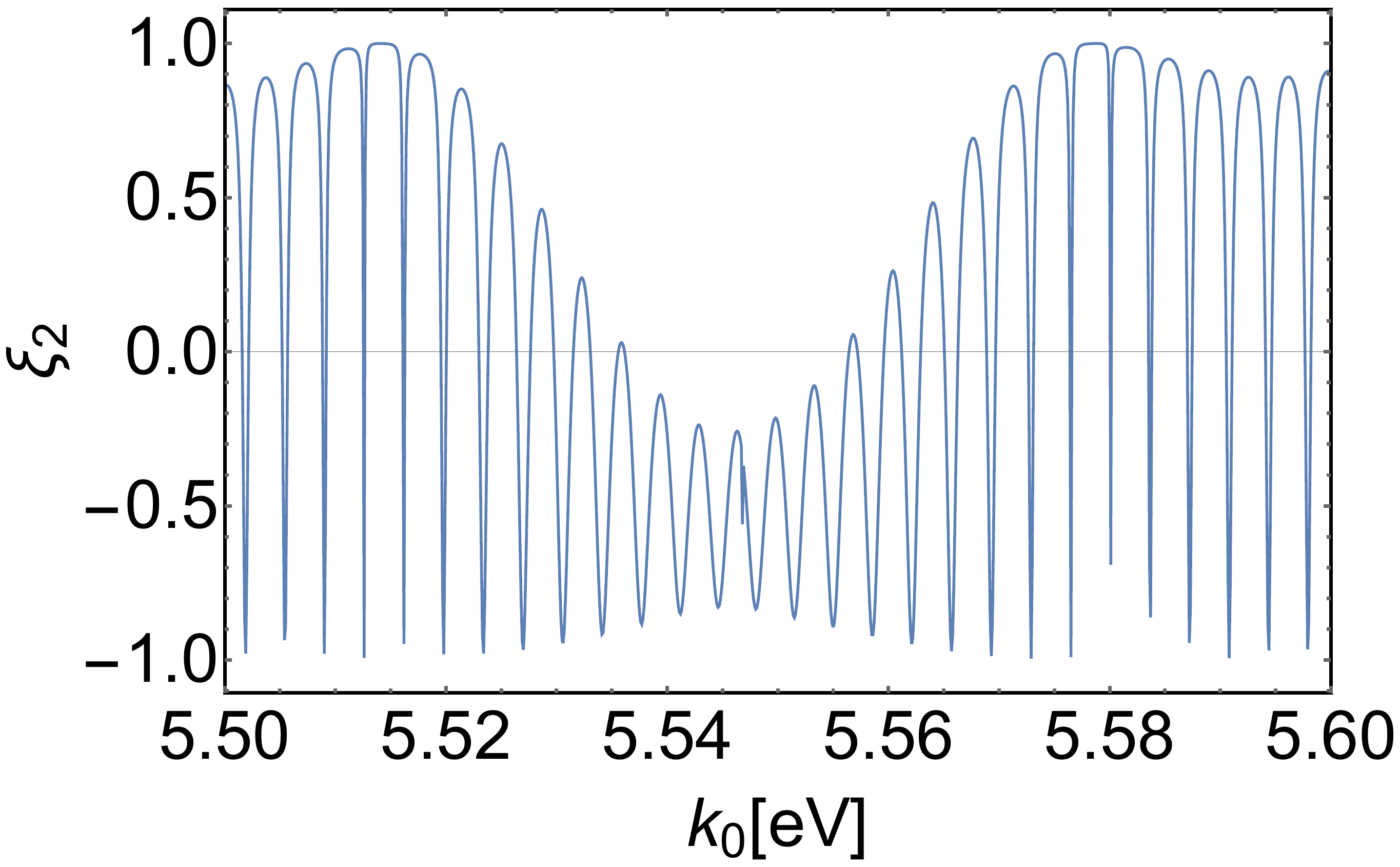}}\,
\raisebox{-0.5\height}{\includegraphics*[width=0.24\linewidth]{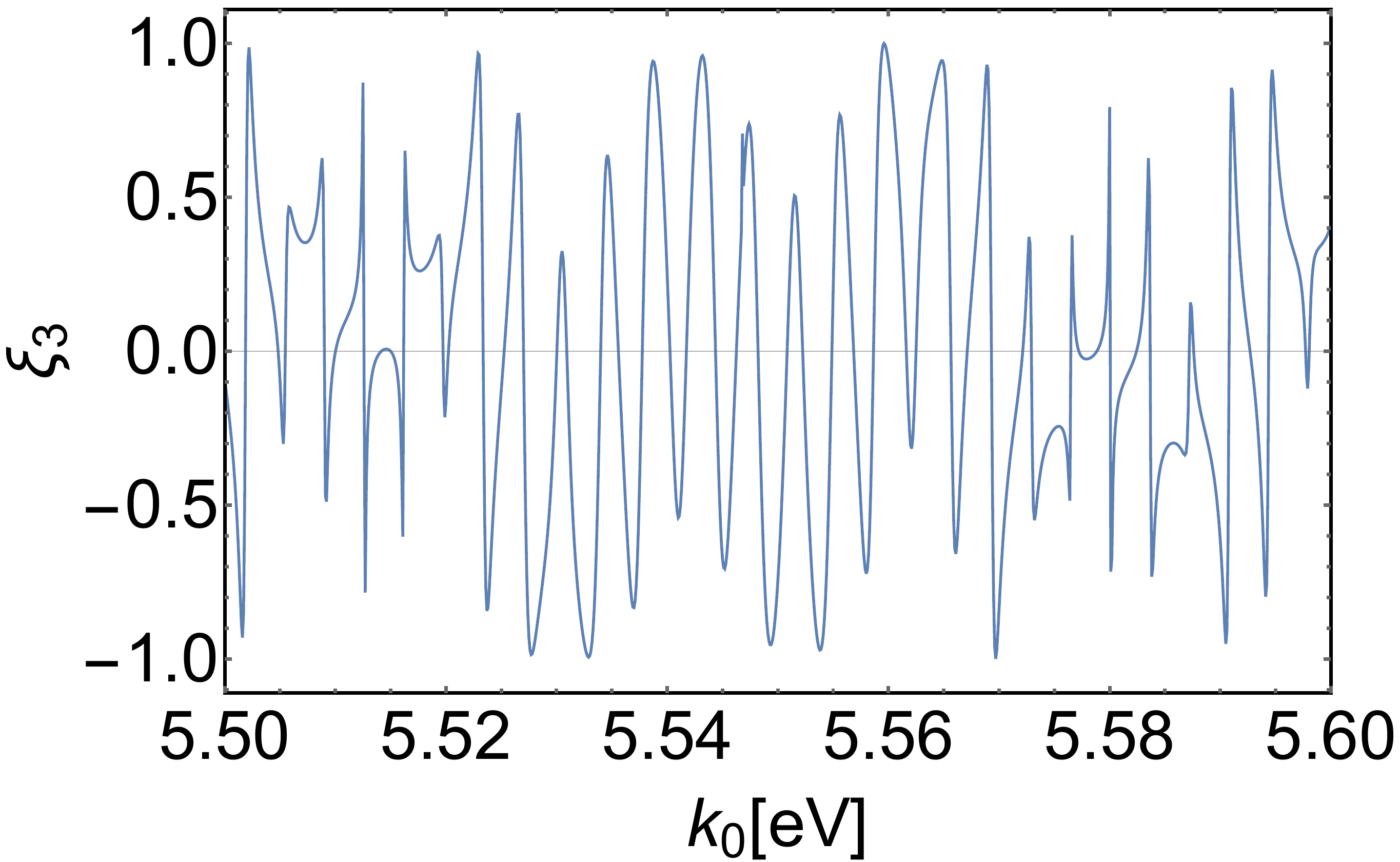}}\\
\raisebox{-0.5\height}{\includegraphics*[width=0.24\linewidth]{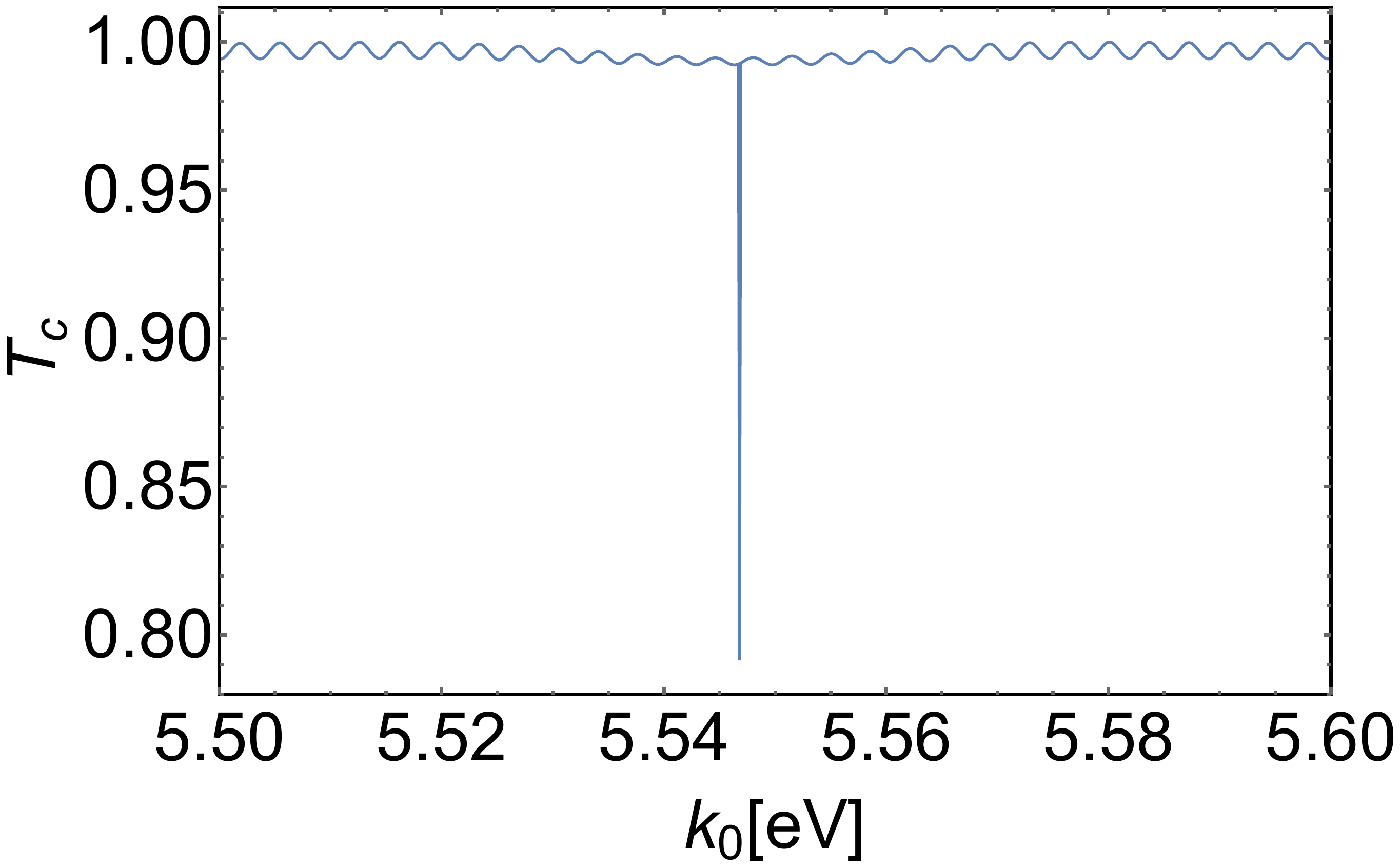}}\,
\raisebox{-0.5\height}{\includegraphics*[width=0.24\linewidth]{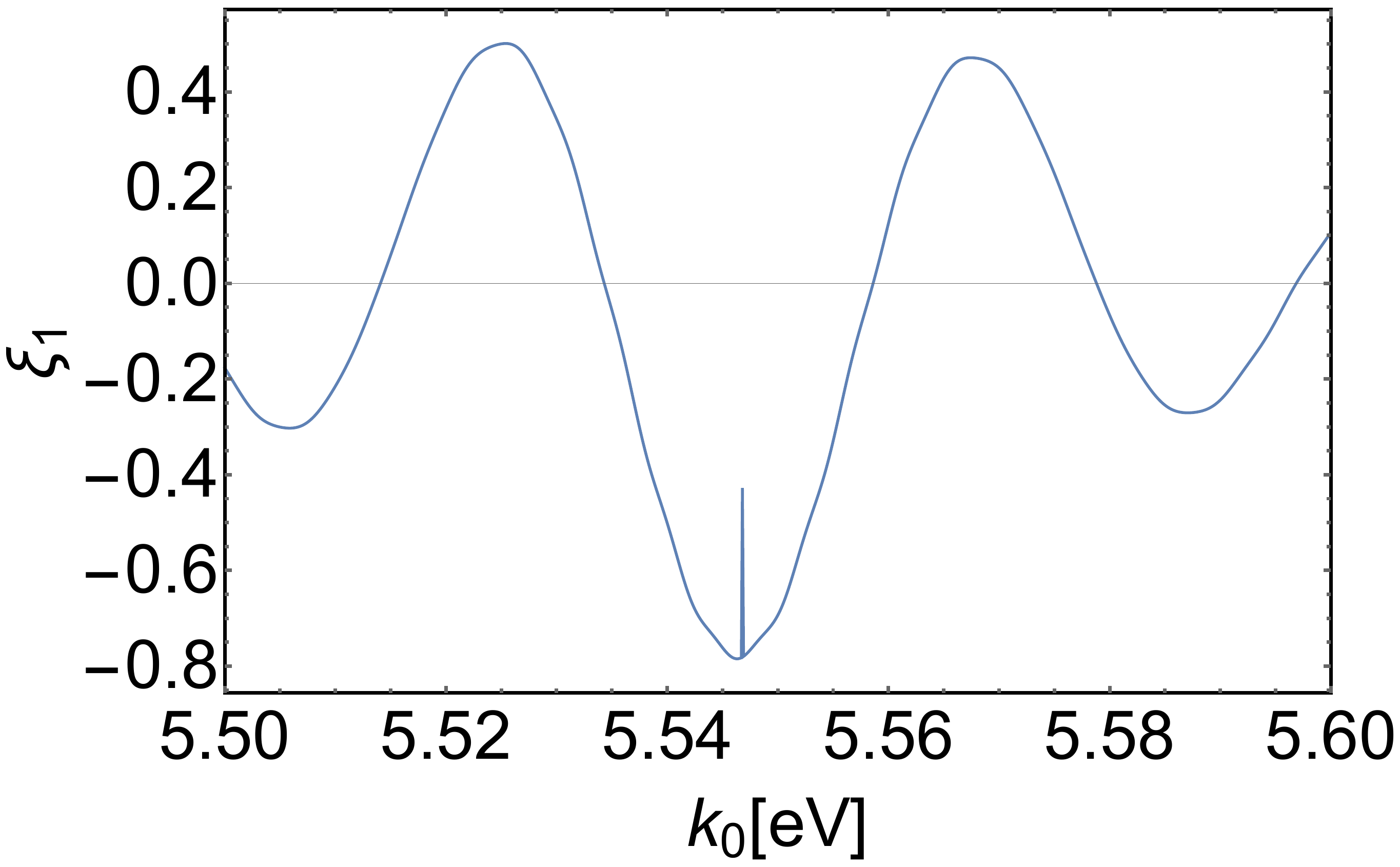}}\,
\raisebox{-0.5\height}{\includegraphics*[width=0.24\linewidth]{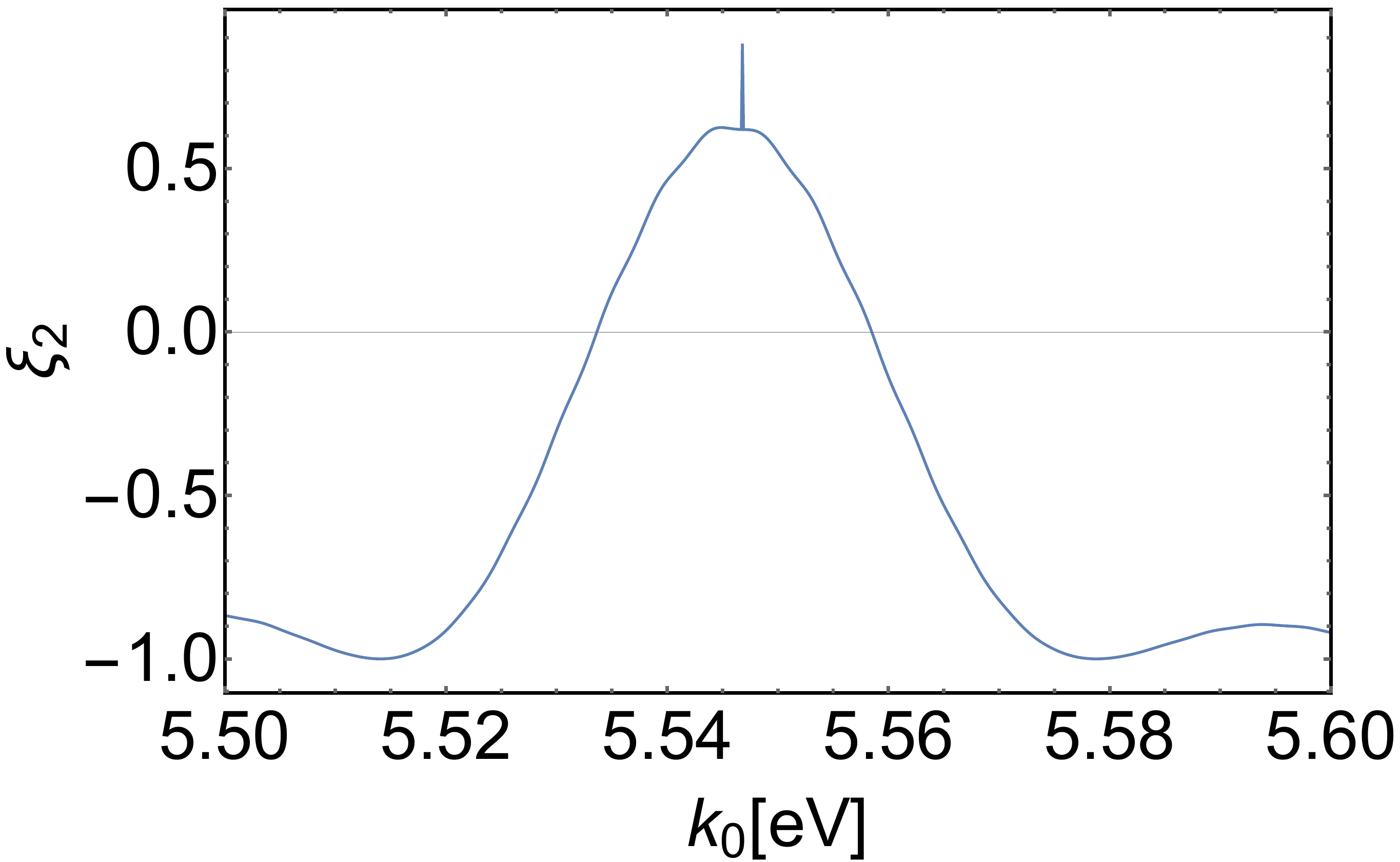}}\,
\raisebox{-0.5\height}{\includegraphics*[width=0.24\linewidth]{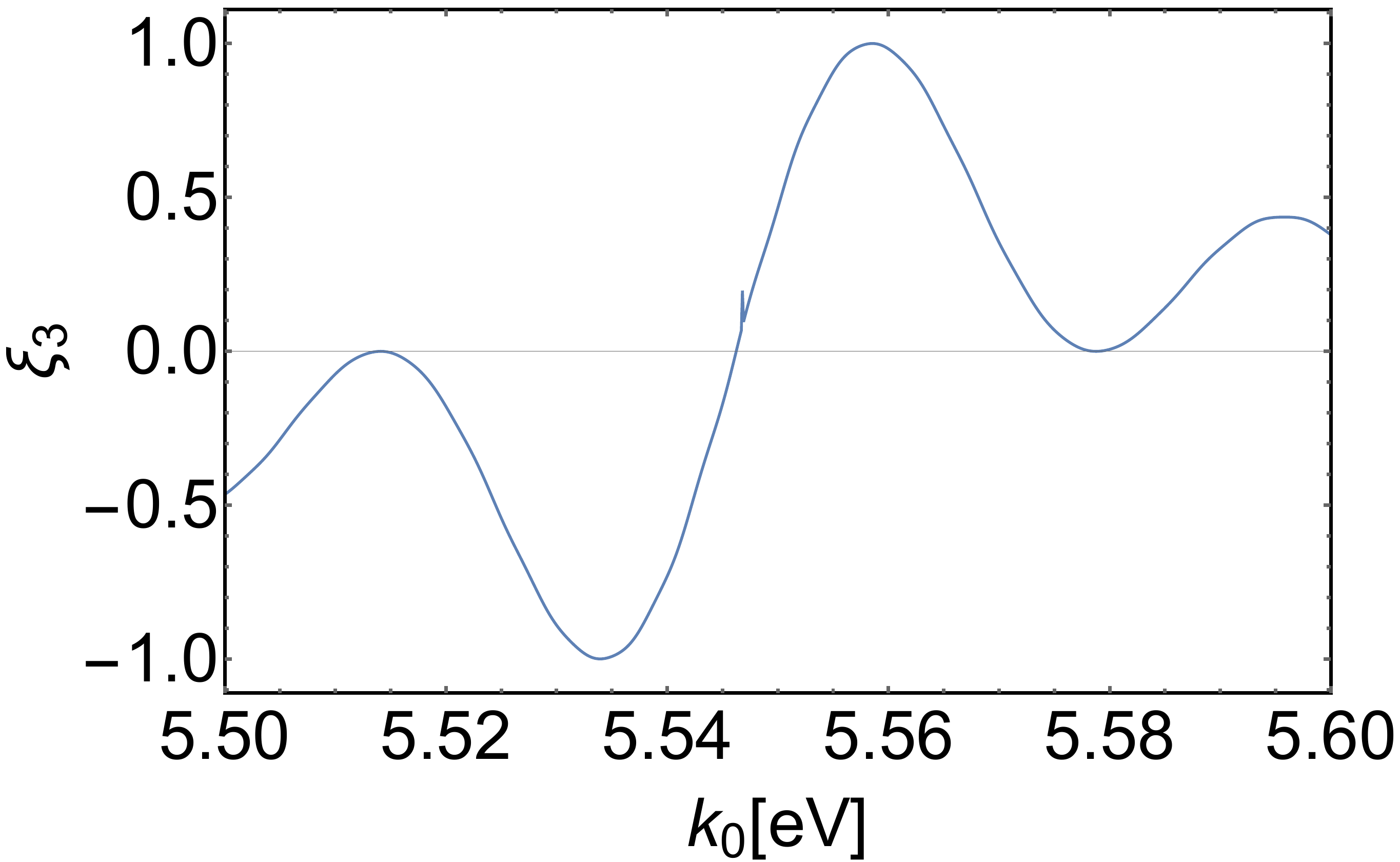}}\\
\raisebox{-0.5\height}{\includegraphics*[width=0.24\linewidth]{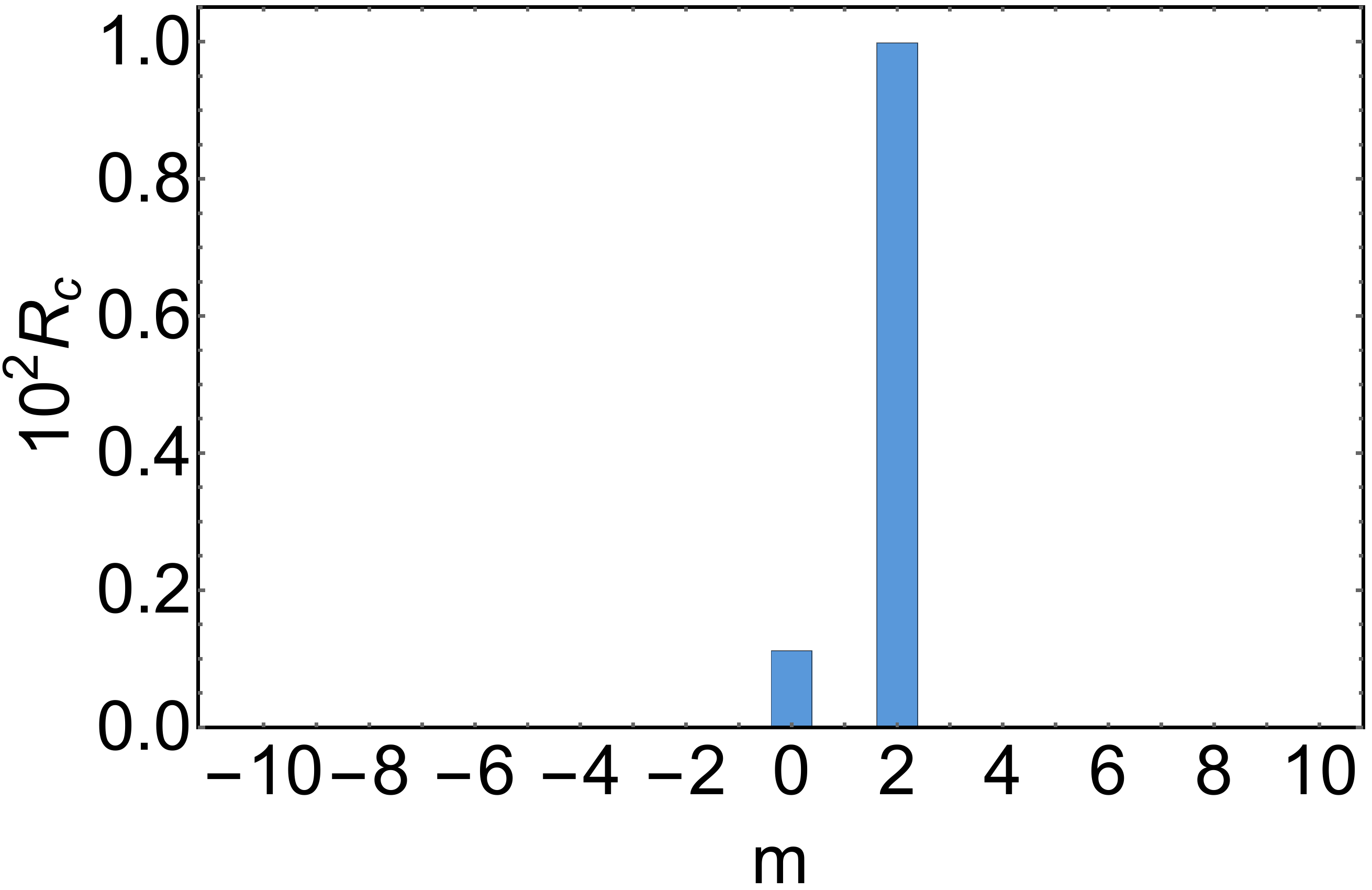}}\,
\raisebox{-0.5\height}{\includegraphics*[width=0.24\linewidth]{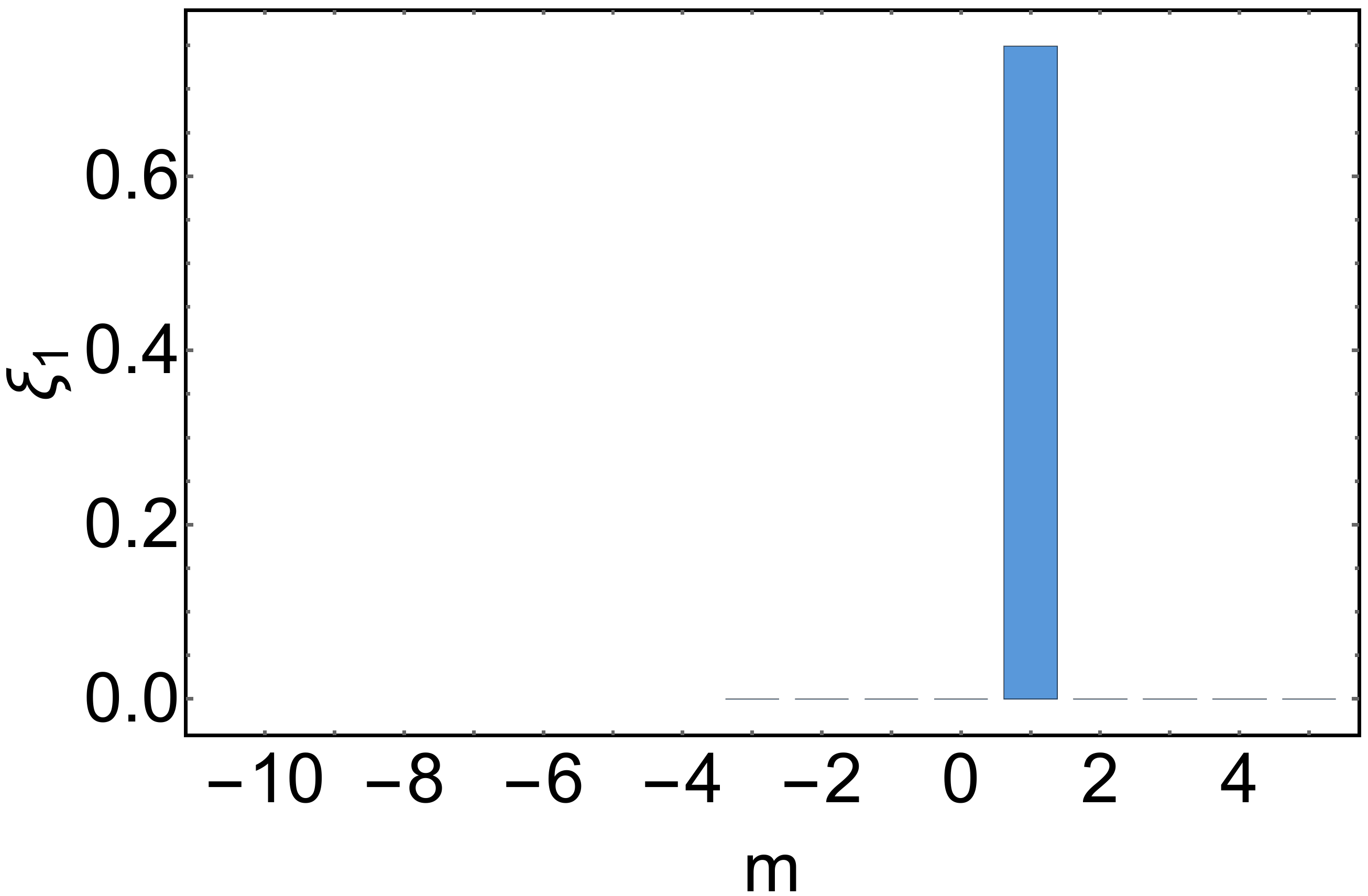}}\,
\raisebox{-0.5\height}{\includegraphics*[width=0.24\linewidth]{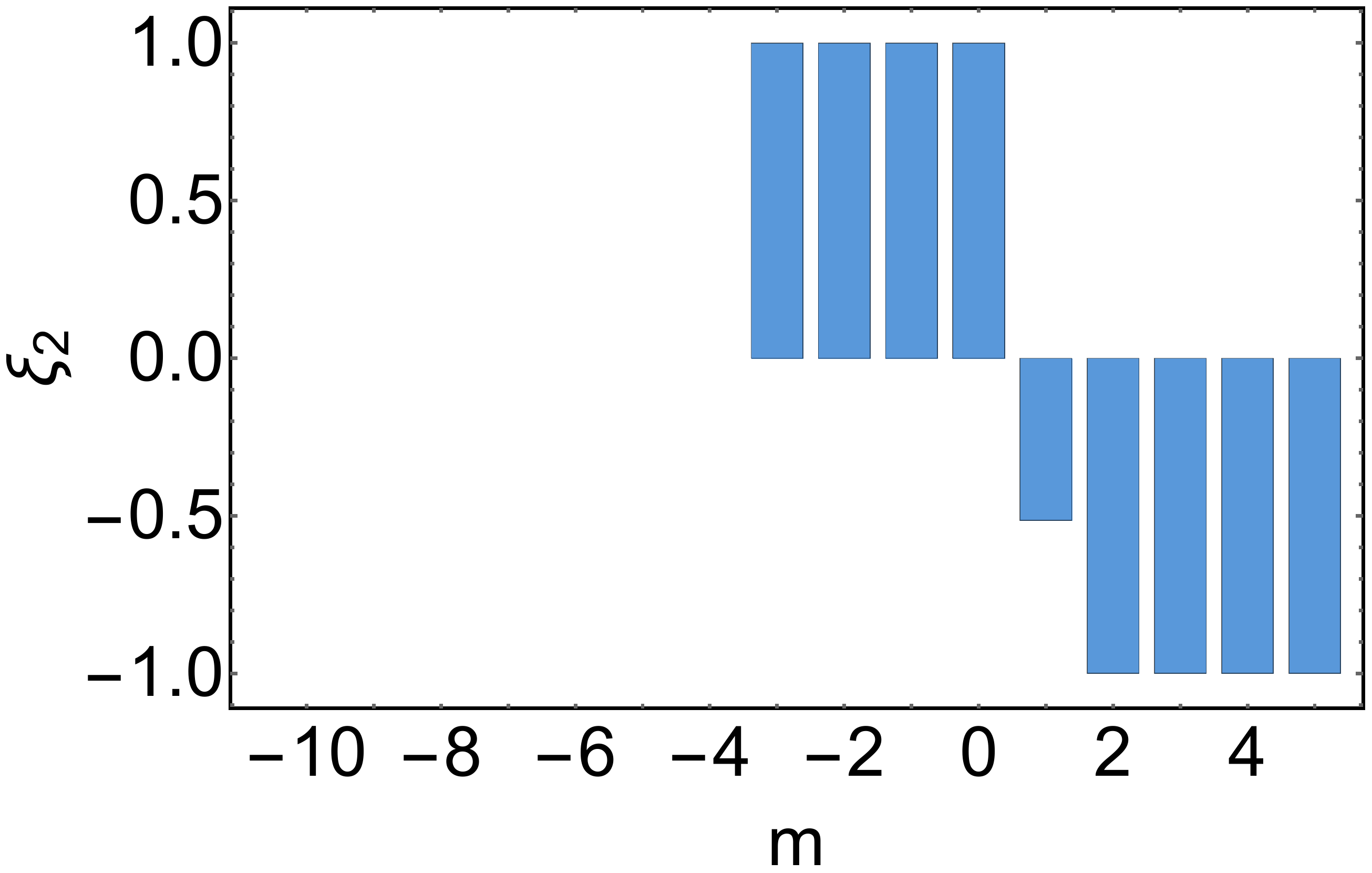}}\,
\raisebox{-0.5\height}{\includegraphics*[width=0.24\linewidth]{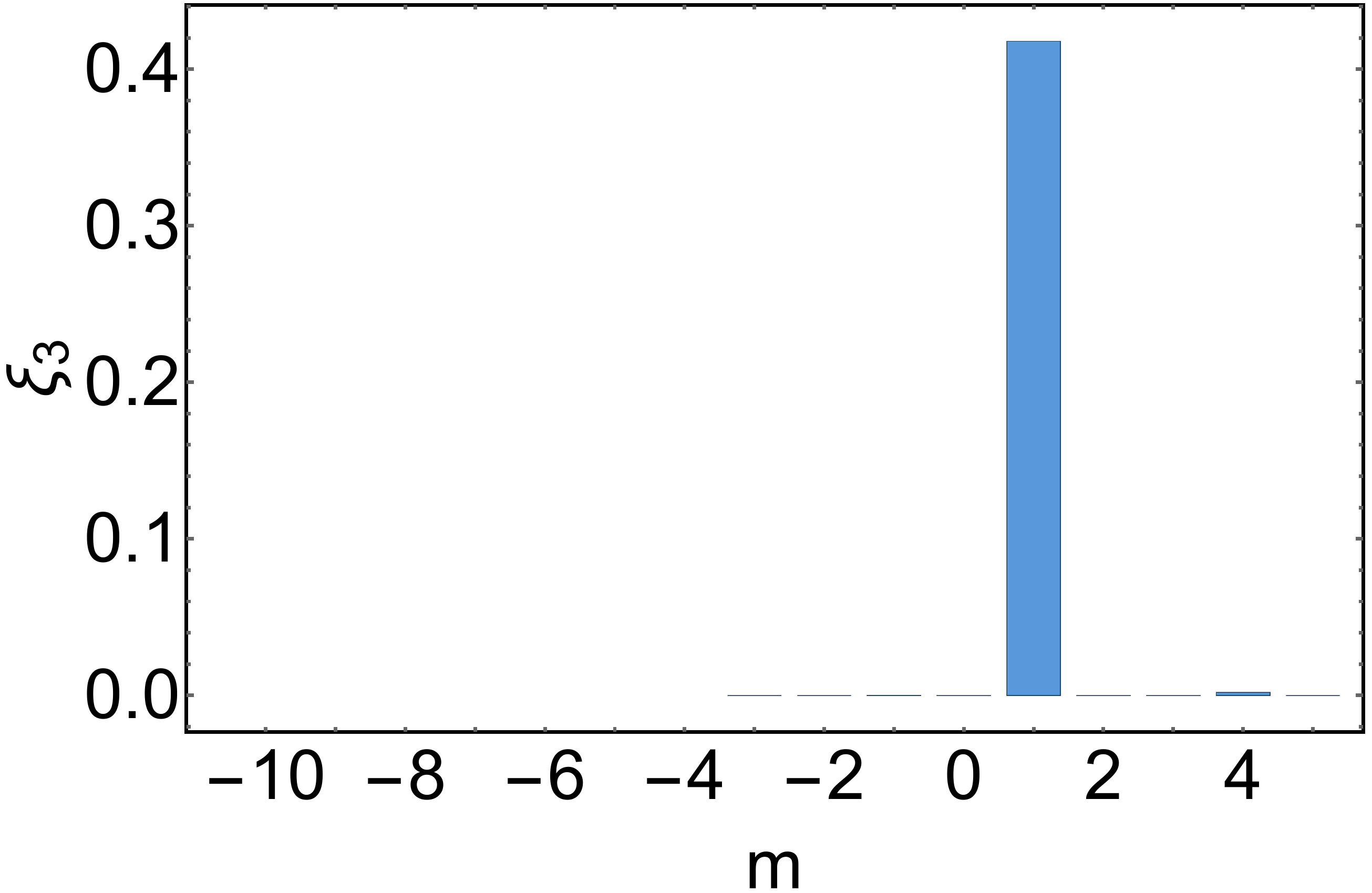}}\\
\raisebox{-0.5\height}{\includegraphics*[width=0.24\linewidth]{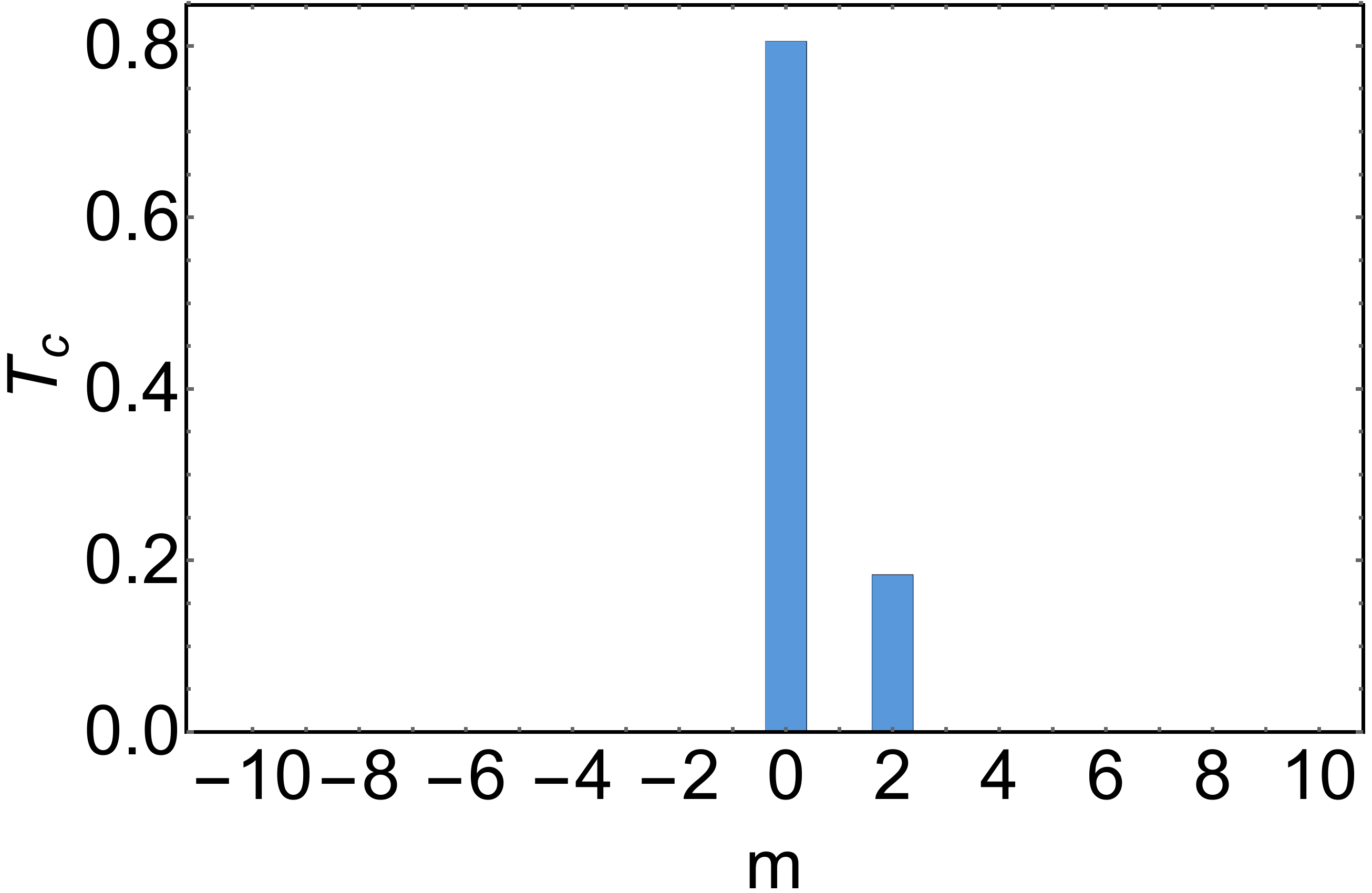}}\,
\raisebox{-0.5\height}{\includegraphics*[width=0.24\linewidth]{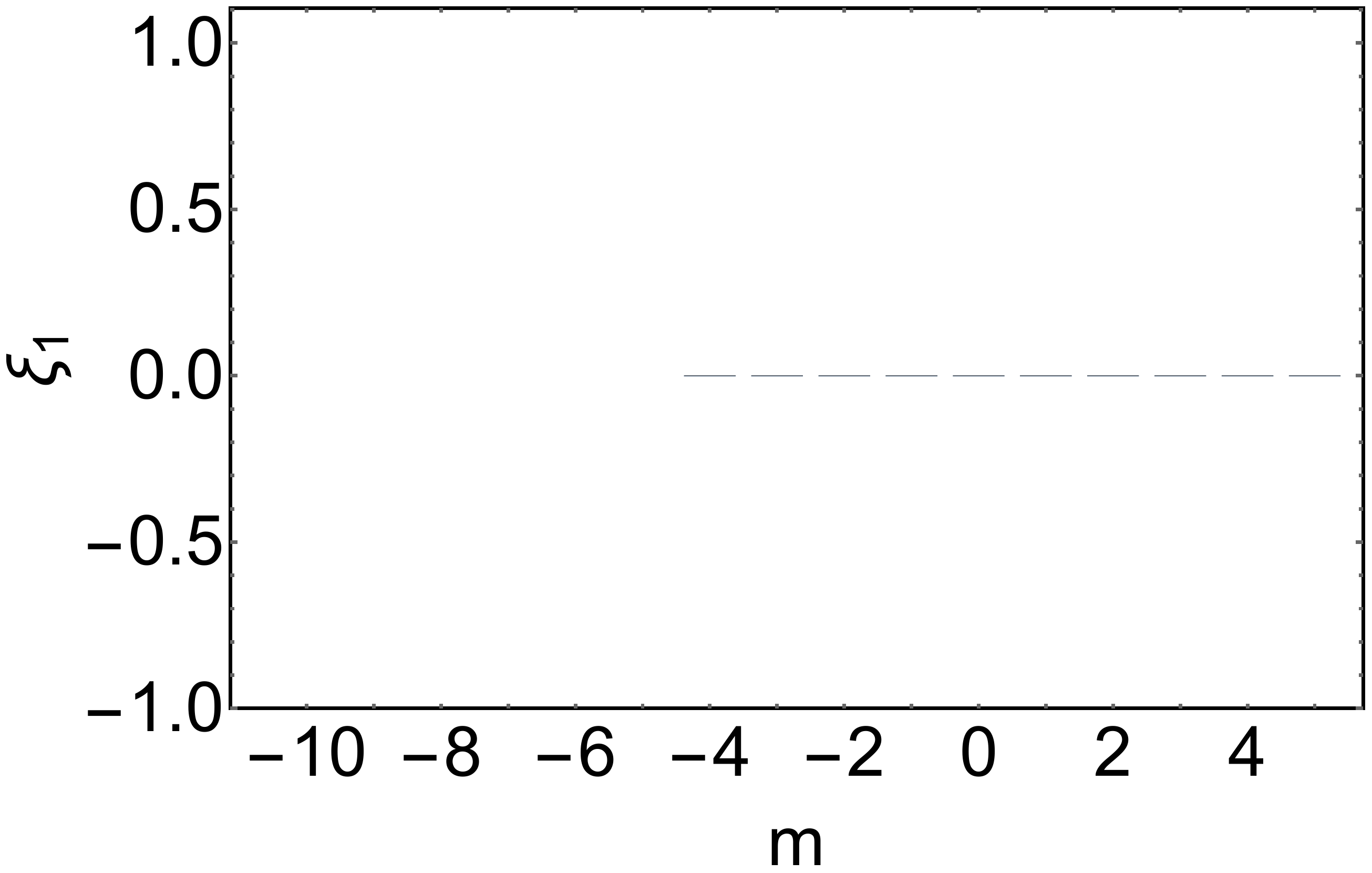}}\,
\raisebox{-0.5\height}{\includegraphics*[width=0.24\linewidth]{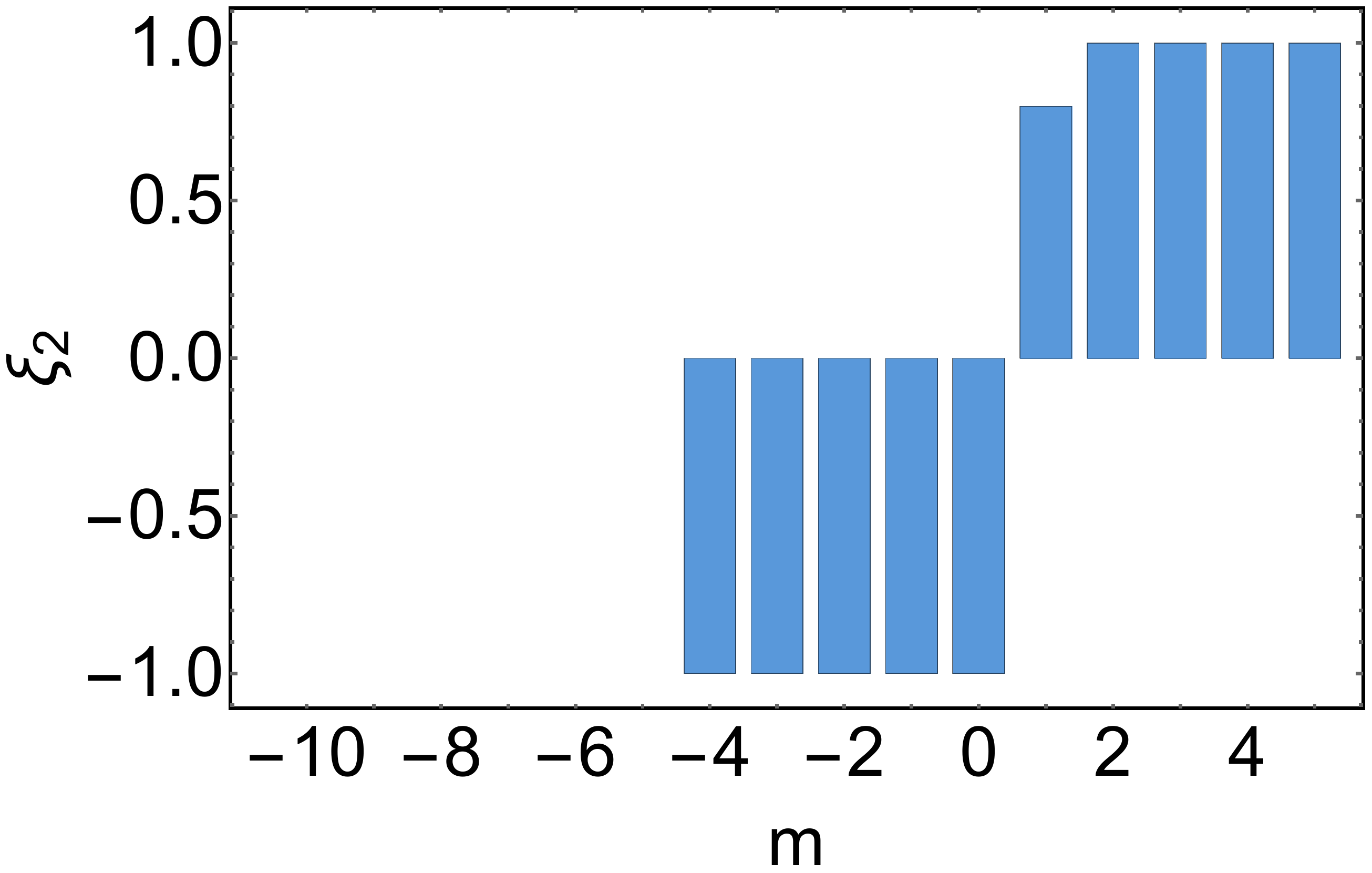}}\,
\raisebox{-0.5\height}{\includegraphics*[width=0.24\linewidth]{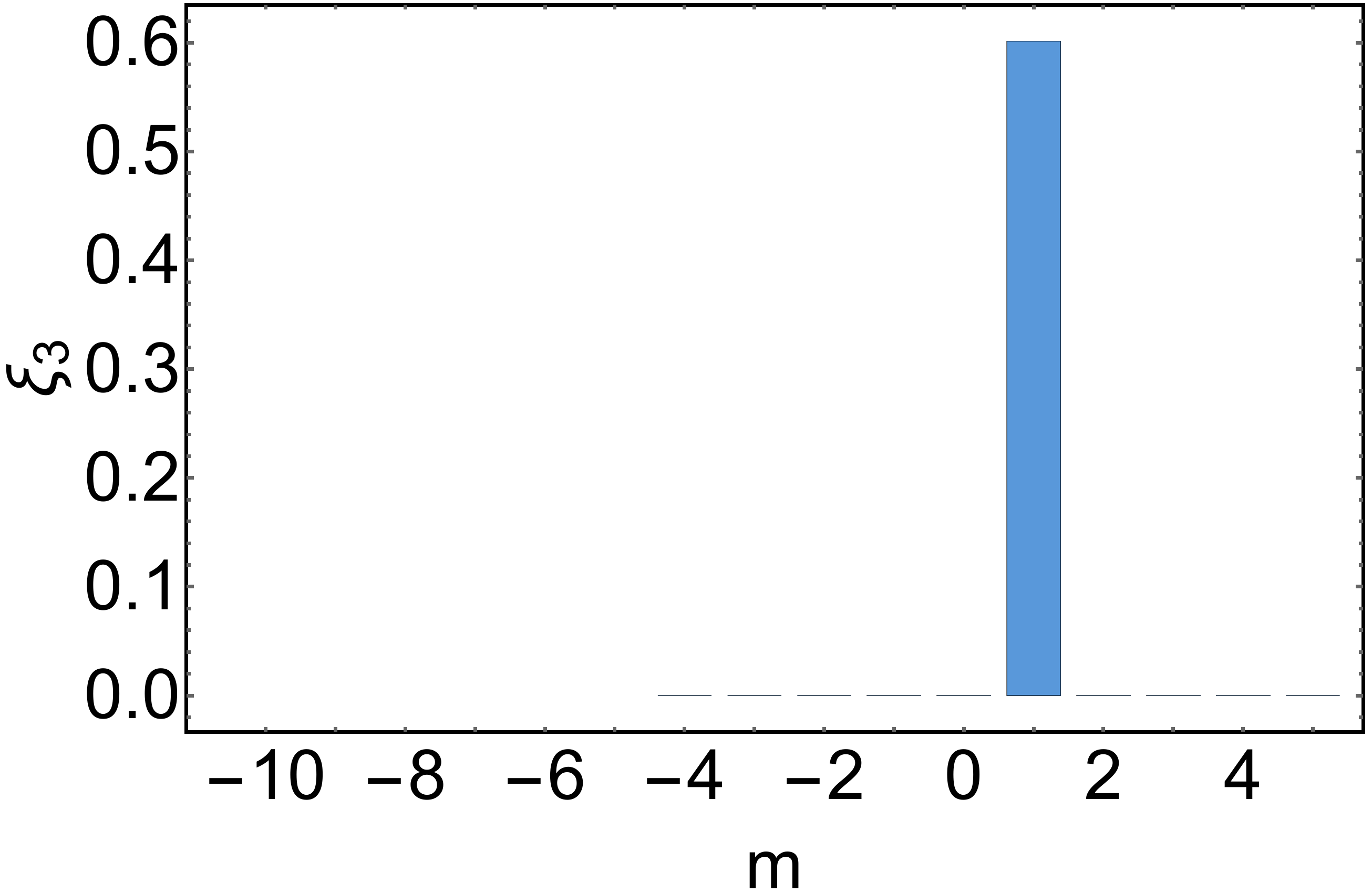}}\\
\raisebox{-0.5\height}{\includegraphics*[width=0.24\linewidth]{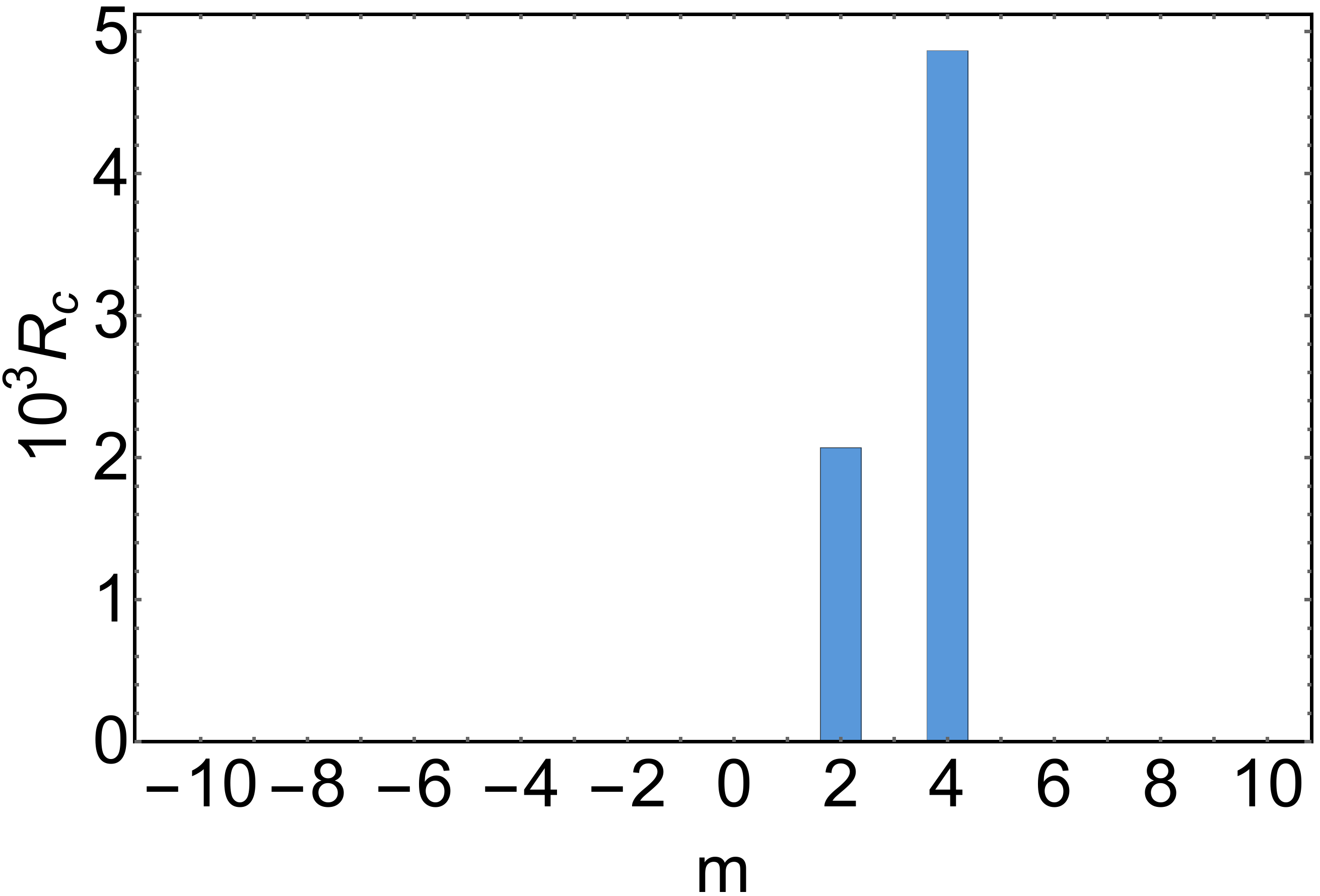}}\,
\raisebox{-0.5\height}{\includegraphics*[width=0.24\linewidth]{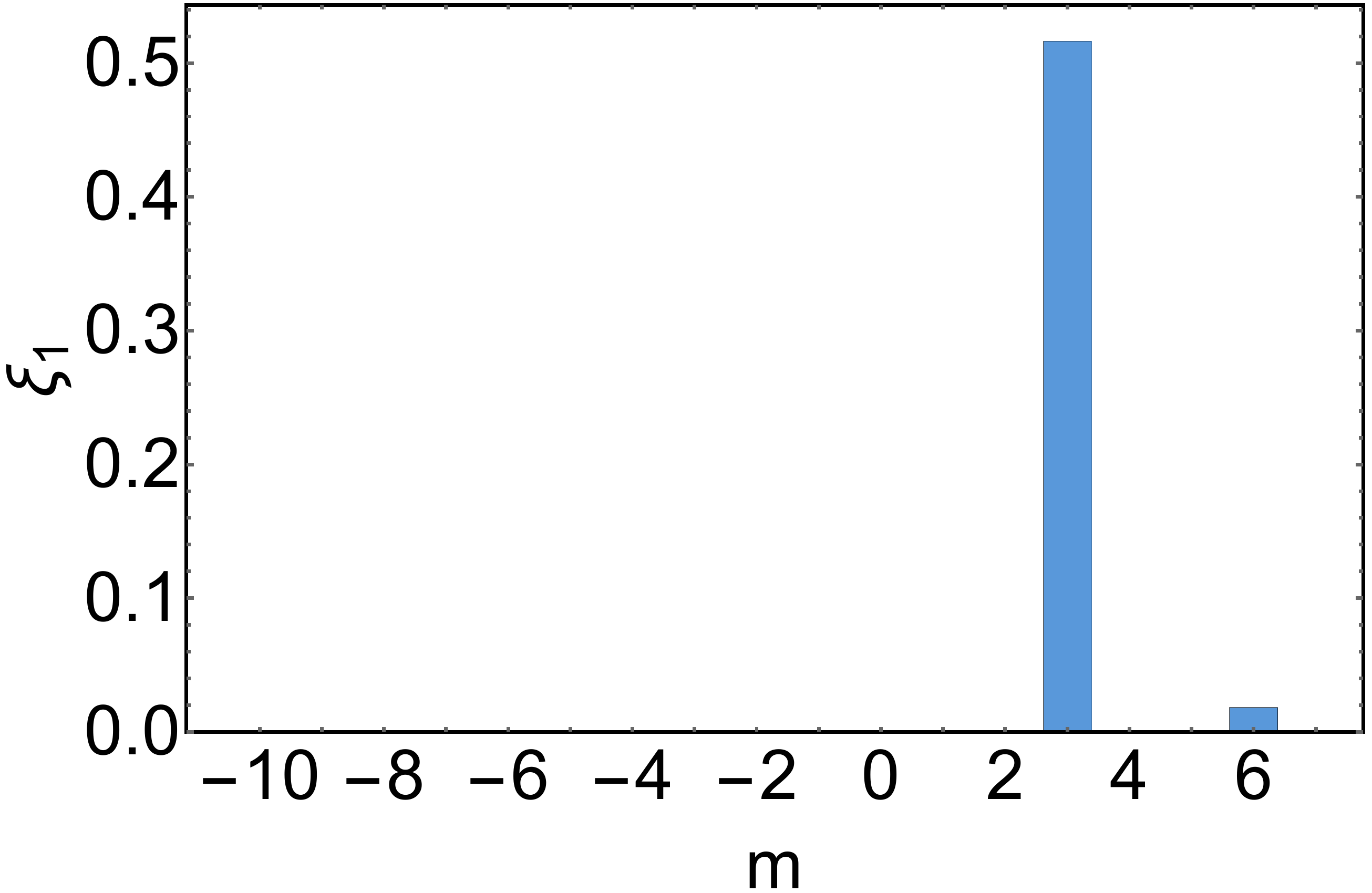}}\,
\raisebox{-0.5\height}{\includegraphics*[width=0.24\linewidth]{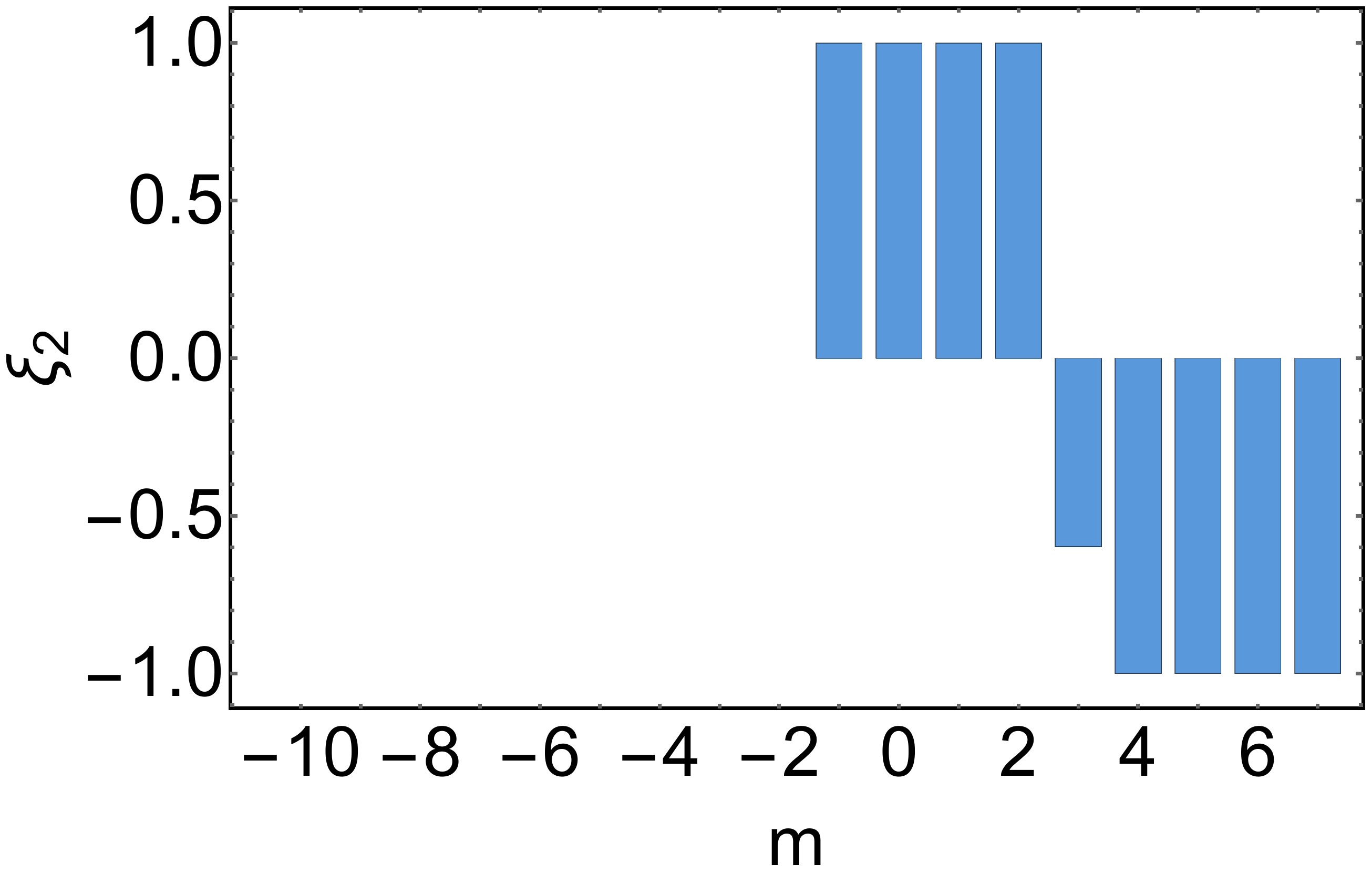}}\,
\raisebox{-0.5\height}{\includegraphics*[width=0.24\linewidth]{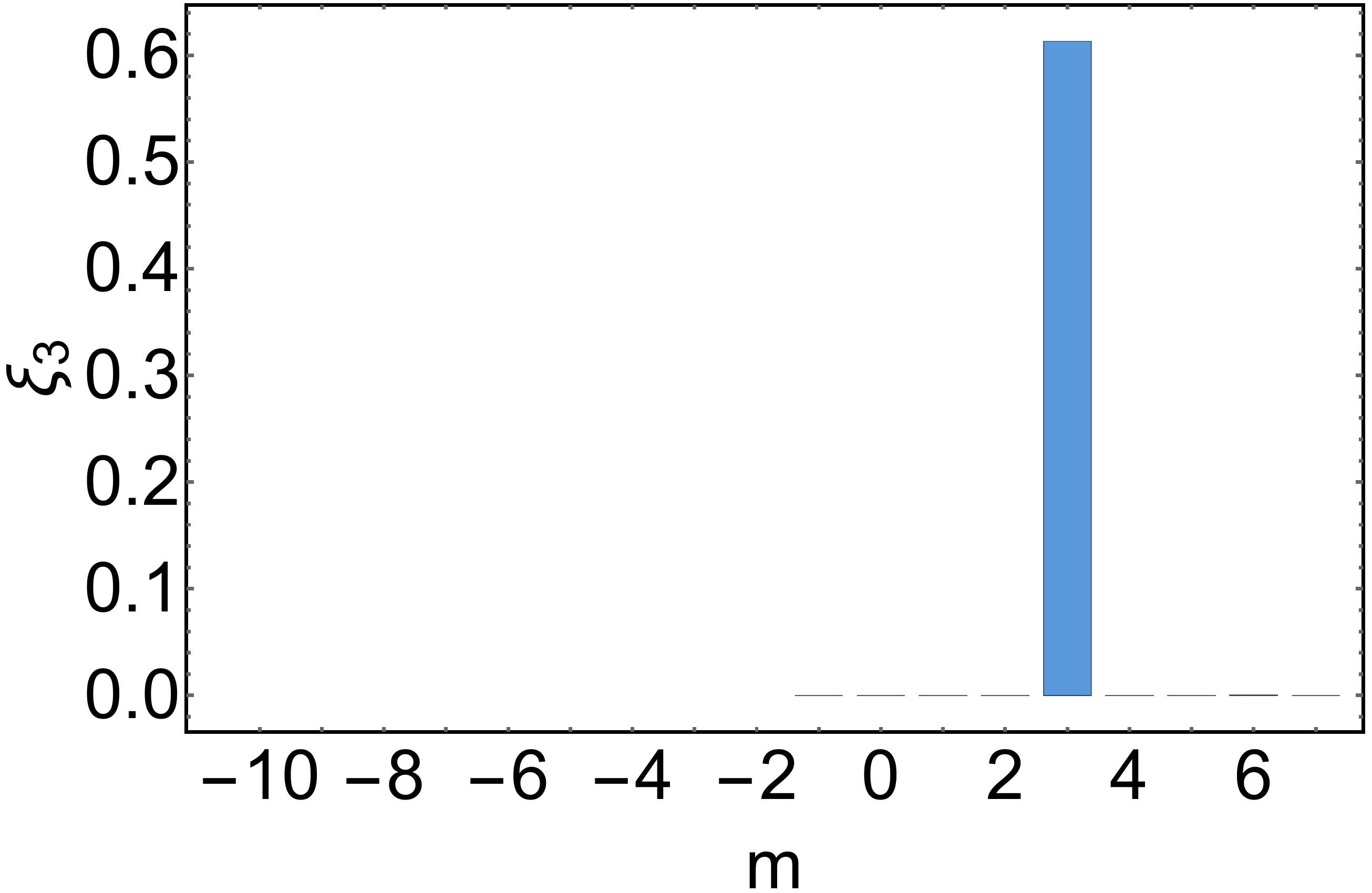}}\\
\raisebox{-0.5\height}{\includegraphics*[width=0.24\linewidth]{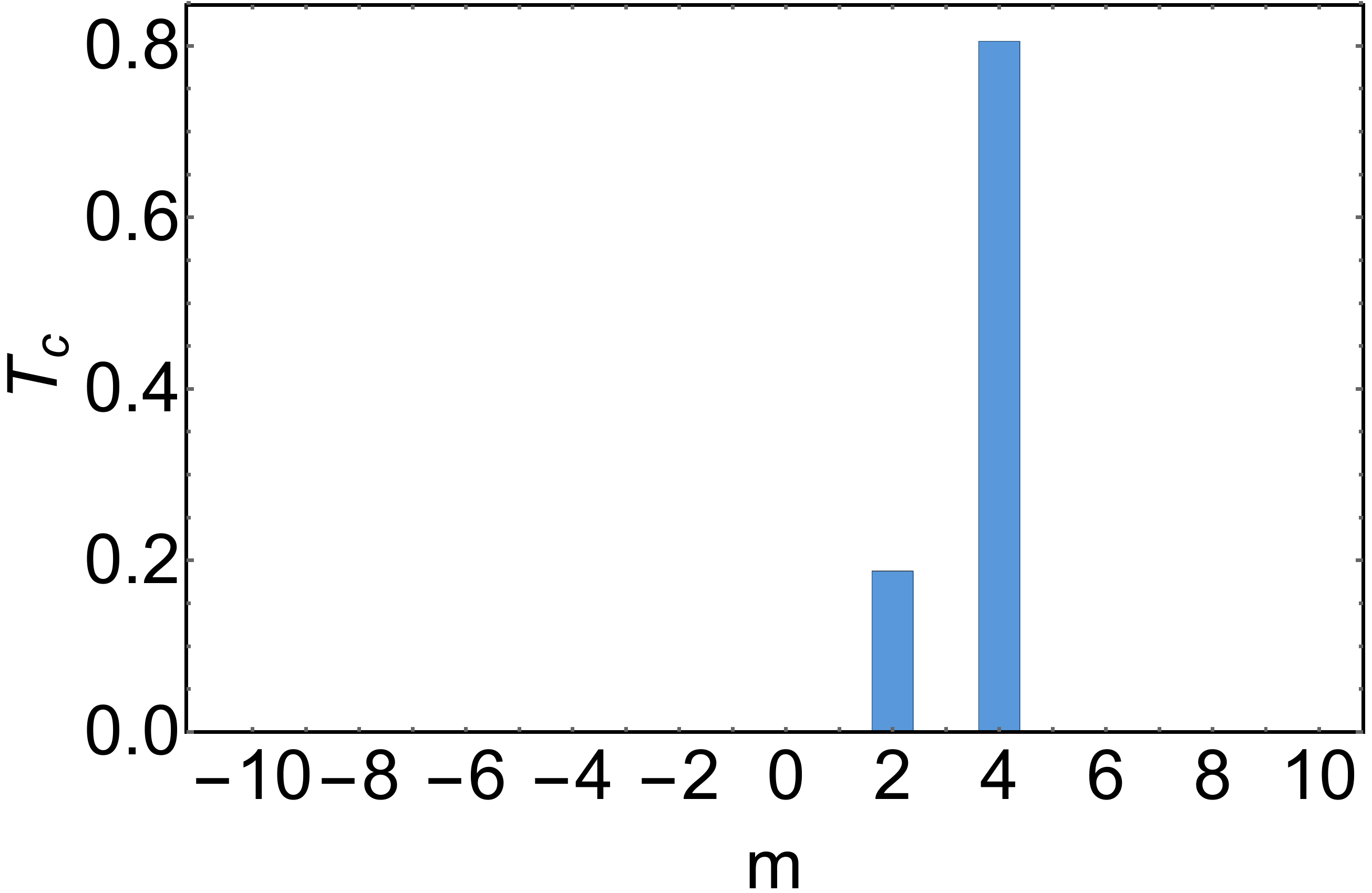}}\,
\raisebox{-0.5\height}{\includegraphics*[width=0.24\linewidth]{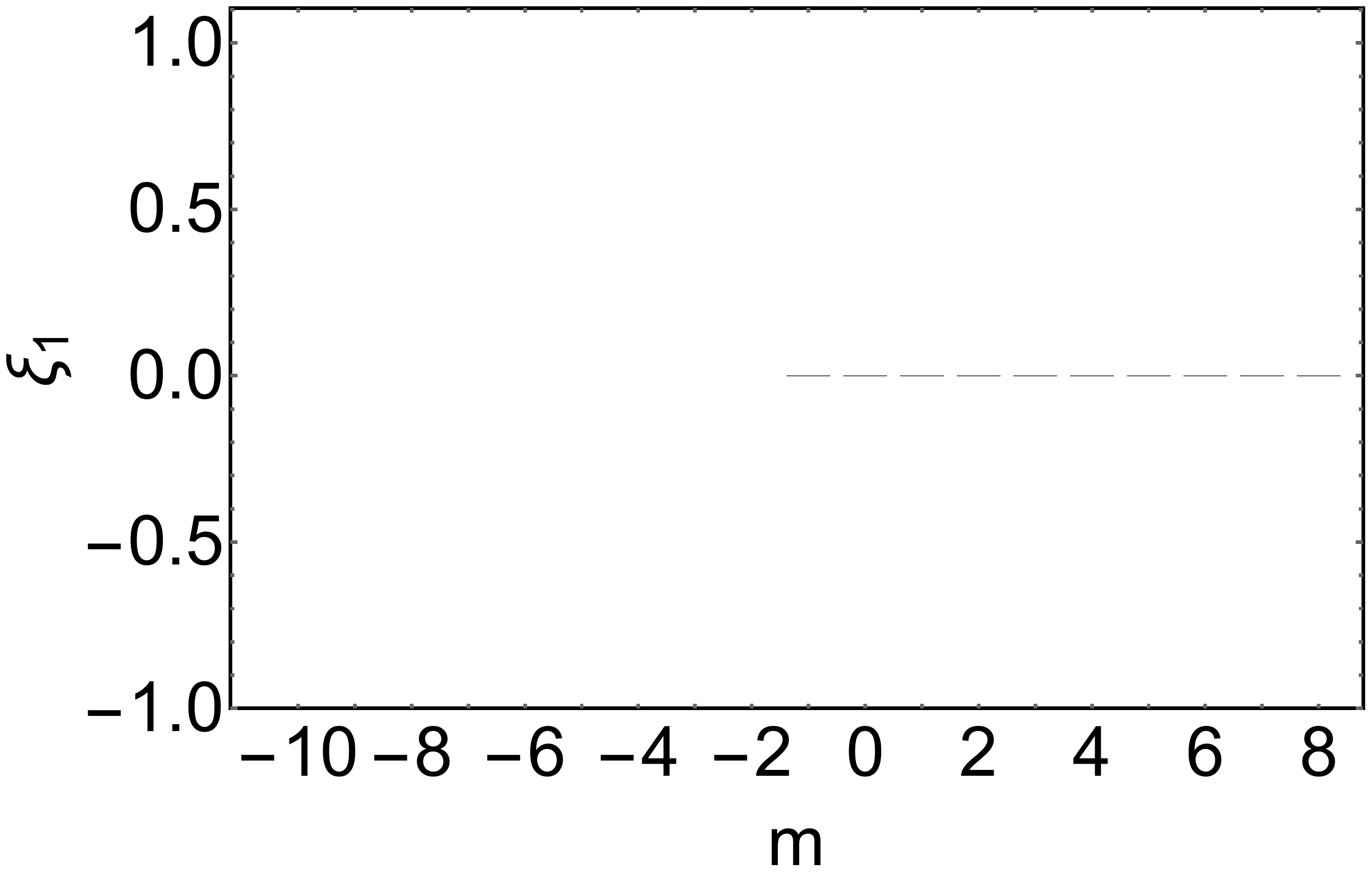}}\,
\raisebox{-0.5\height}{\includegraphics*[width=0.24\linewidth]{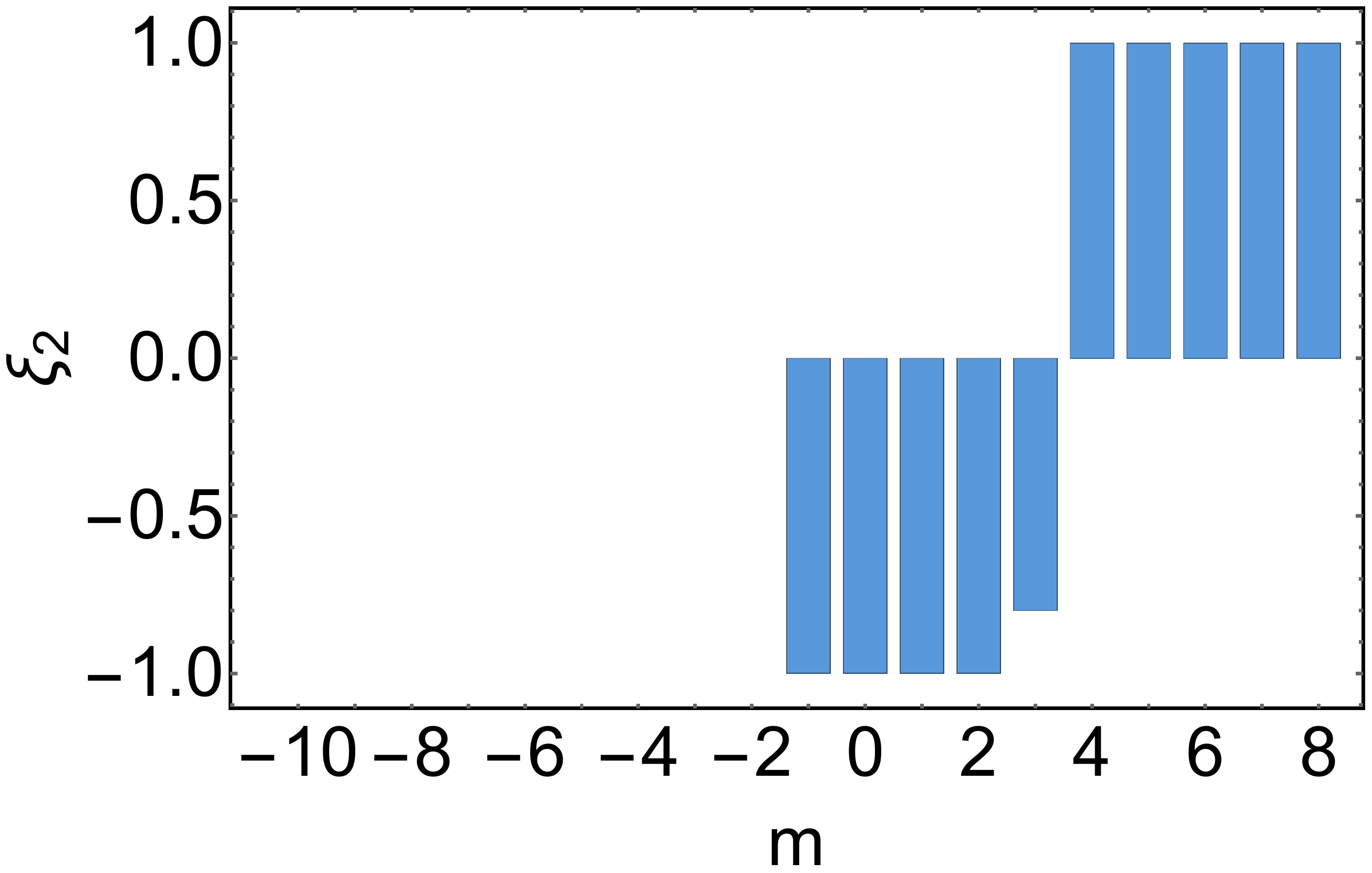}}\,
\raisebox{-0.5\height}{\includegraphics*[width=0.24\linewidth]{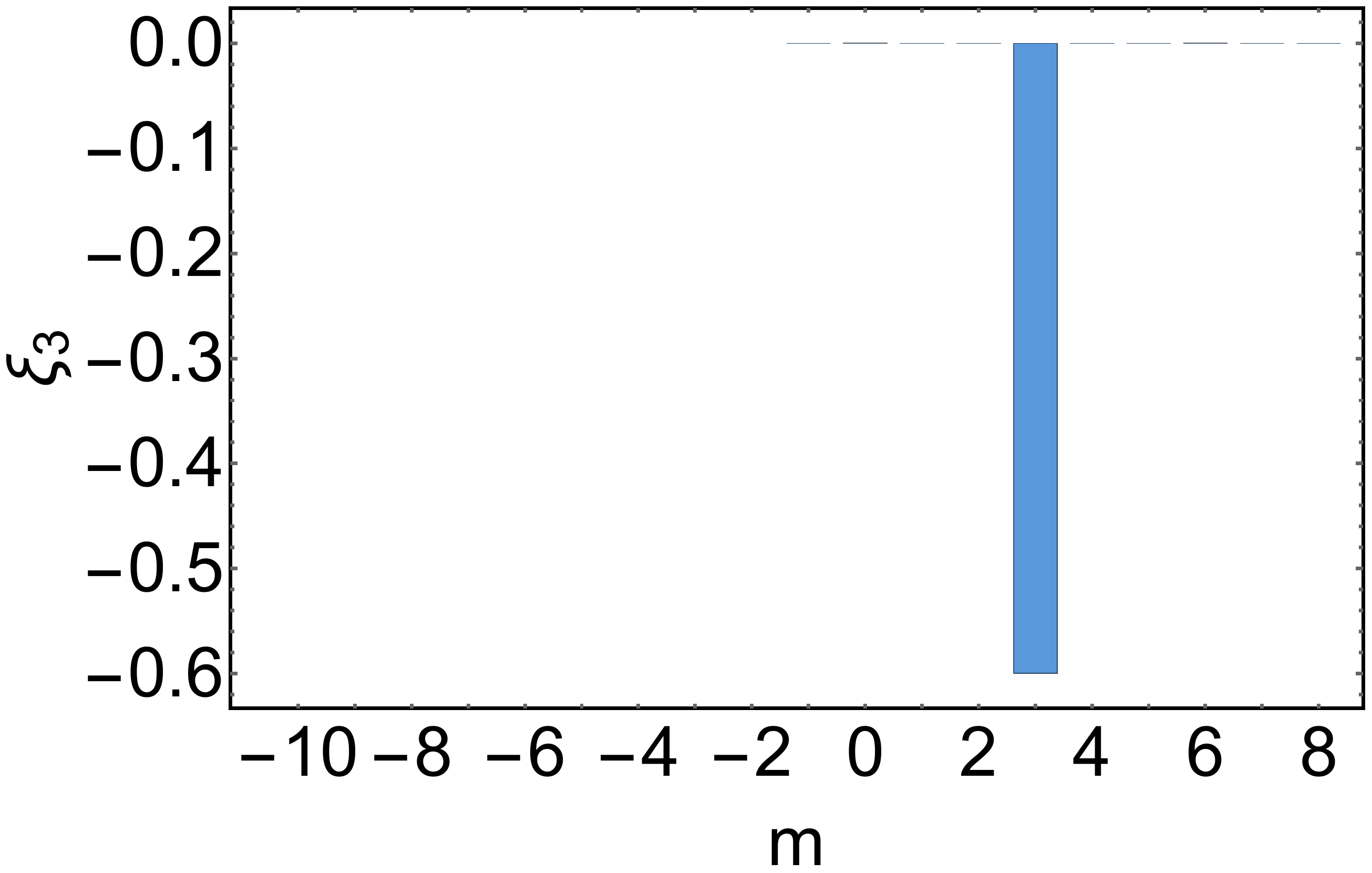}}
\caption{{\footnotesize The same as in Fig. \ref{Scatt_Hels2r_plots} but for the imaginary band gap corresponding to $s_h=\pm2$. The lines $1$-$4$: The case of plane-wave photons scattered in the $(x,z)$ plane is considered. The lines $1$-$2$: The initial photon possesses the helicity $s=1$. The lines $3$-$4$: It has $s=-1$. The lines $5$-$8$: Scattering of the twisted photon with $m=2$ is considered at the energy $k_0=5.547$ eV belonging to the band gap. The lines $5$-$6$ corresponds to $s=1$, whereas the lines $7$-$8$ are for $s=-1$.  As is seen from the $6$th and $8$th lines, the projection of the total angular momentum of a transmitted twisted photon is shifted by $-2s$ and the helicity sign is also flipped. Hence, the projection of the orbital angular momentum is not changed. As for the real $|s_h|=2$ gap, the reflected twisted photons with $qs<0$ acquire the additional projection of the total angular momentum.}}
\label{Scatt_Hels2c_plots}
\end{figure}

%\newpage
\begin{figure}[tp]
\centering
\raisebox{-0.5\height}{\includegraphics*[width=0.24\linewidth]{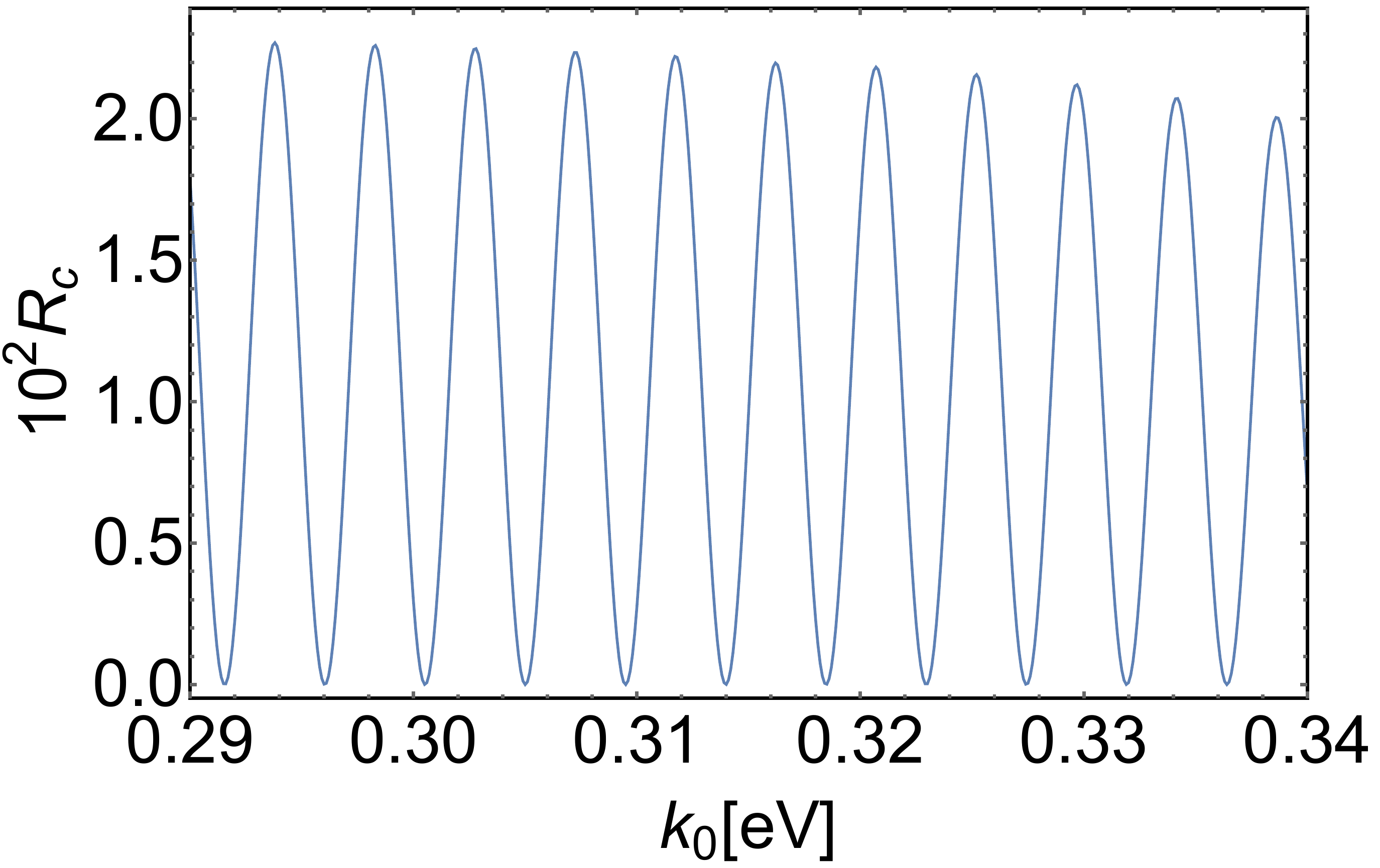}}\,
\raisebox{-0.5\height}{\includegraphics*[width=0.24\linewidth]{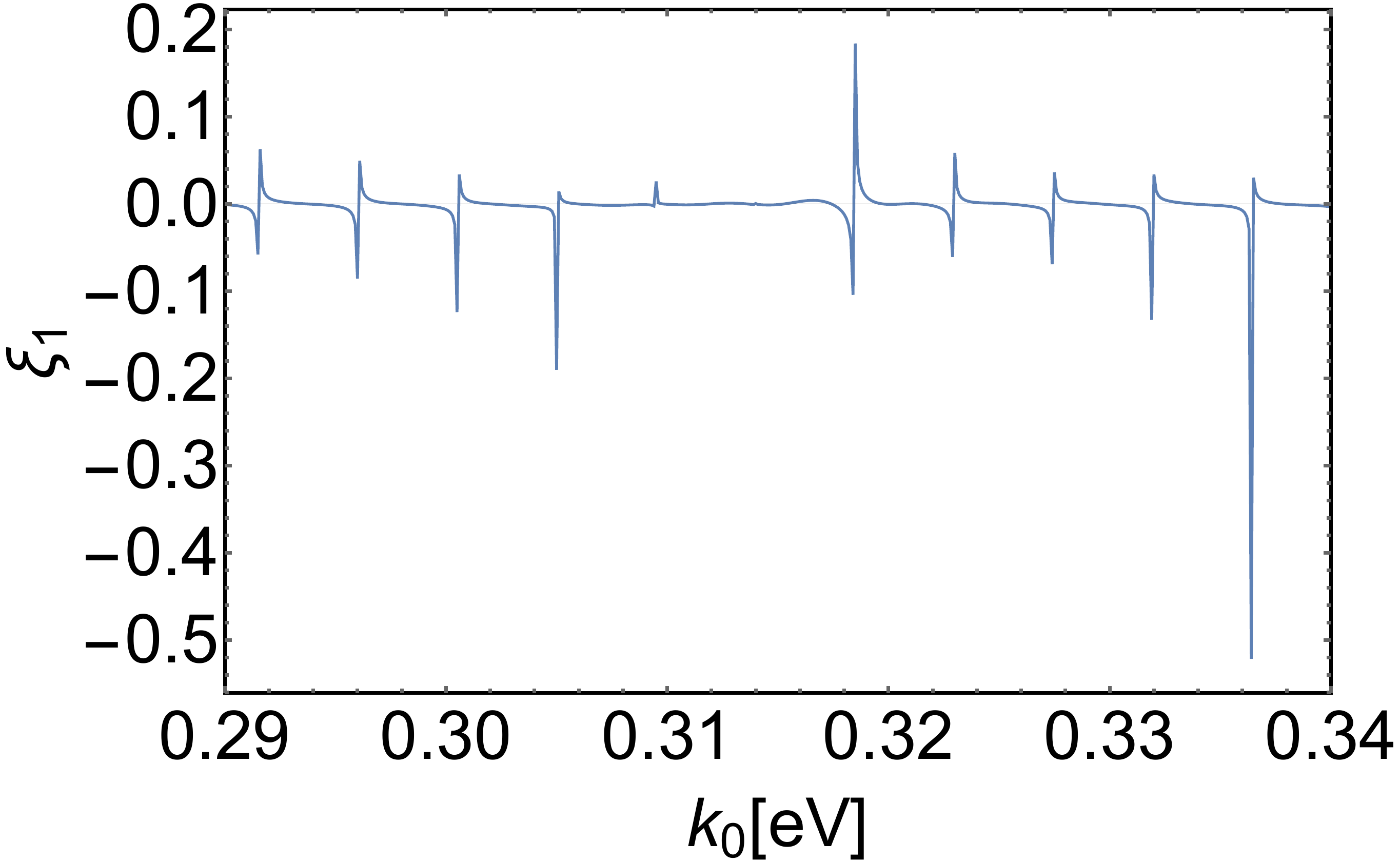}}\,
\raisebox{-0.5\height}{\includegraphics*[width=0.24\linewidth]{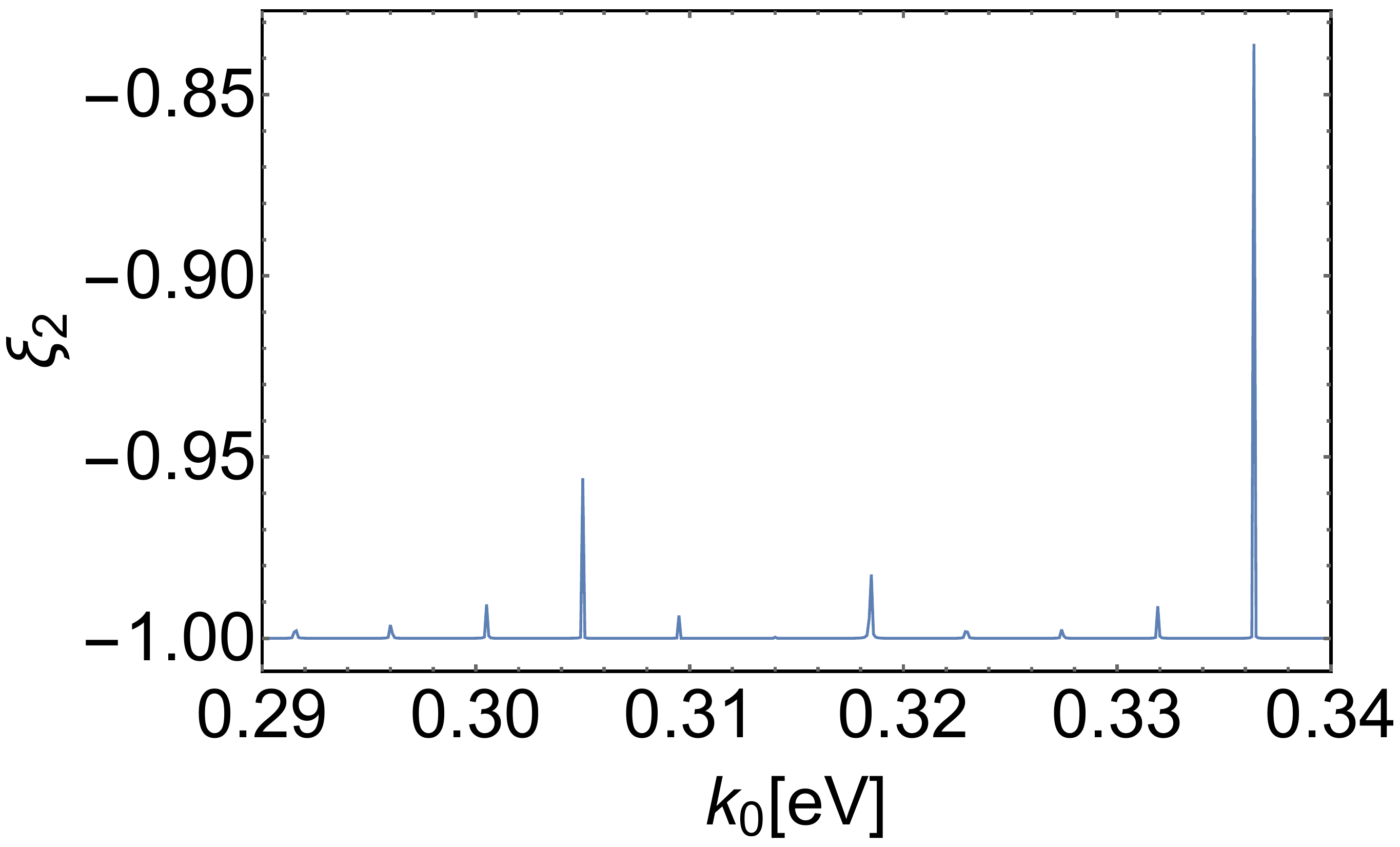}}\,
\raisebox{-0.5\height}{\includegraphics*[width=0.24\linewidth]{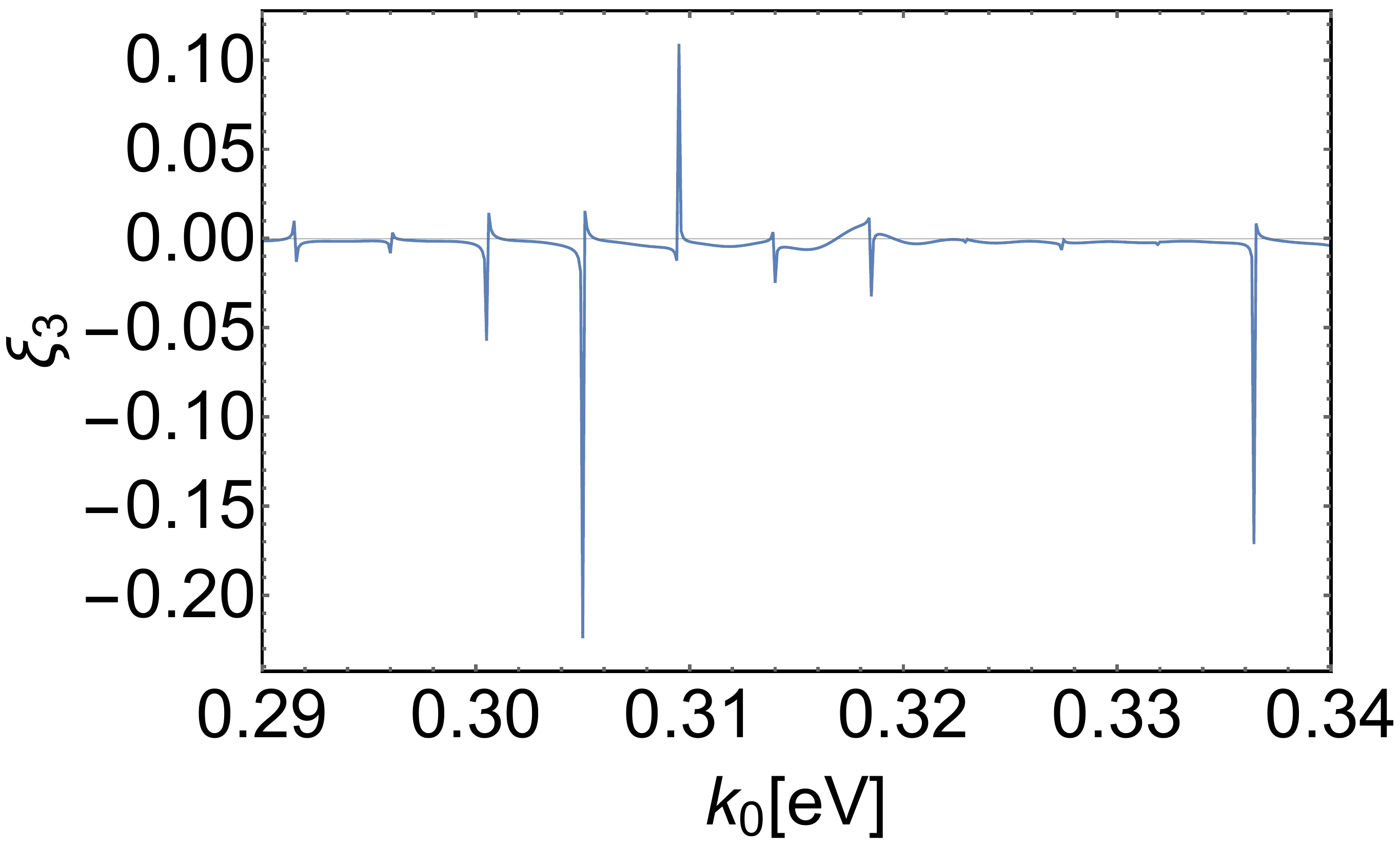}}\\
\raisebox{-0.5\height}{\includegraphics*[width=0.24\linewidth]{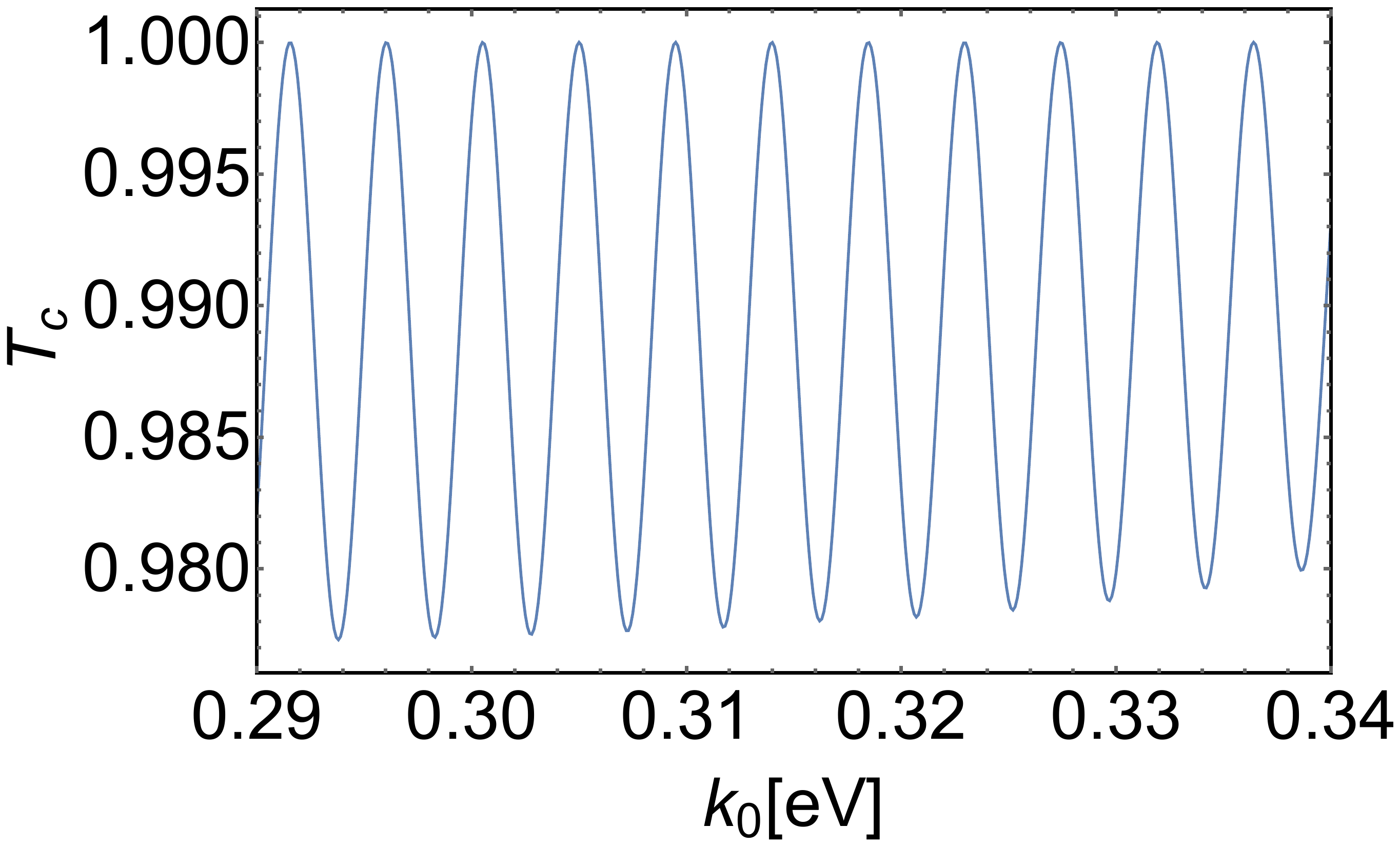}}\,
\raisebox{-0.5\height}{\includegraphics*[width=0.24\linewidth]{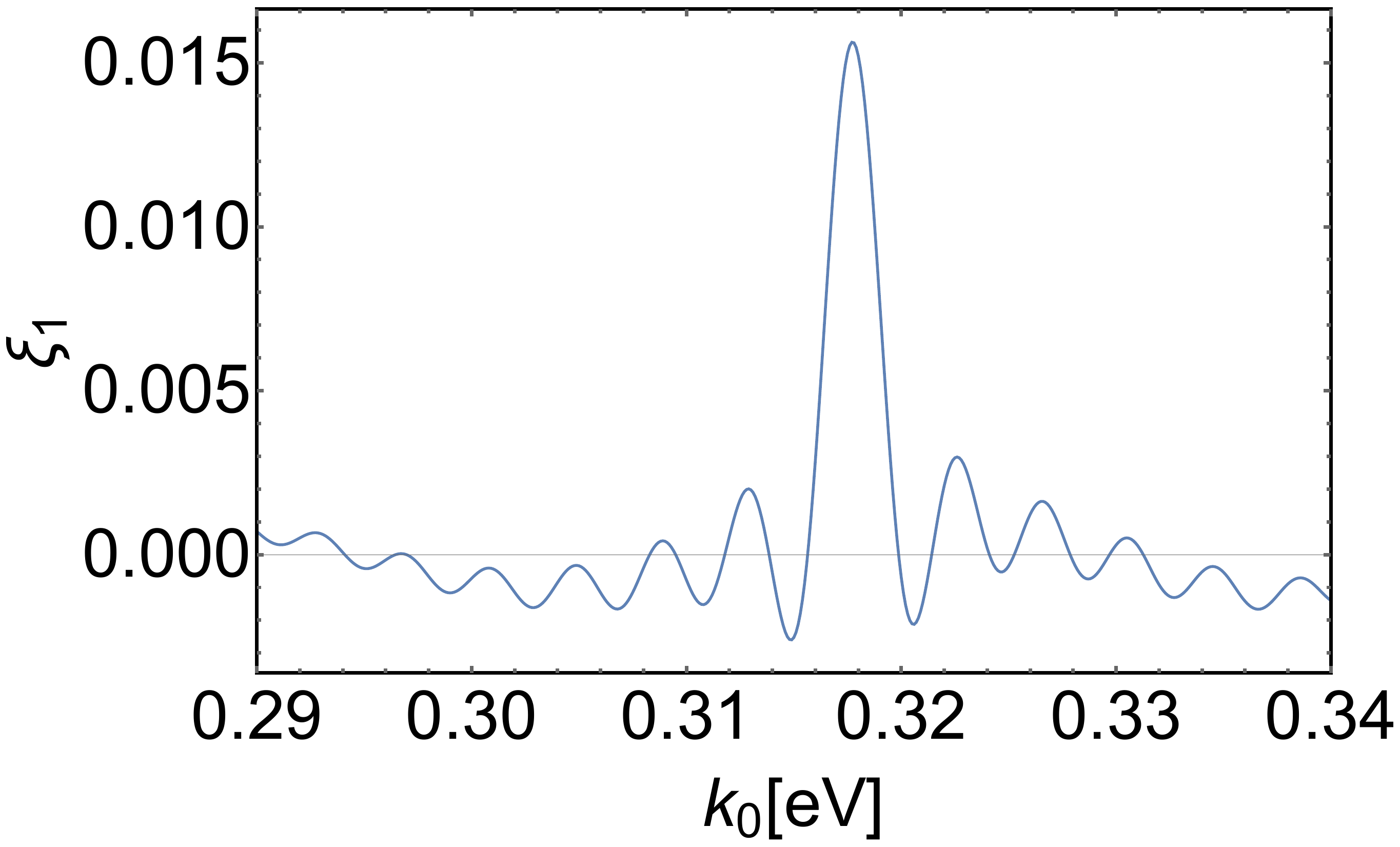}}\,
\raisebox{-0.5\height}{\includegraphics*[width=0.24\linewidth]{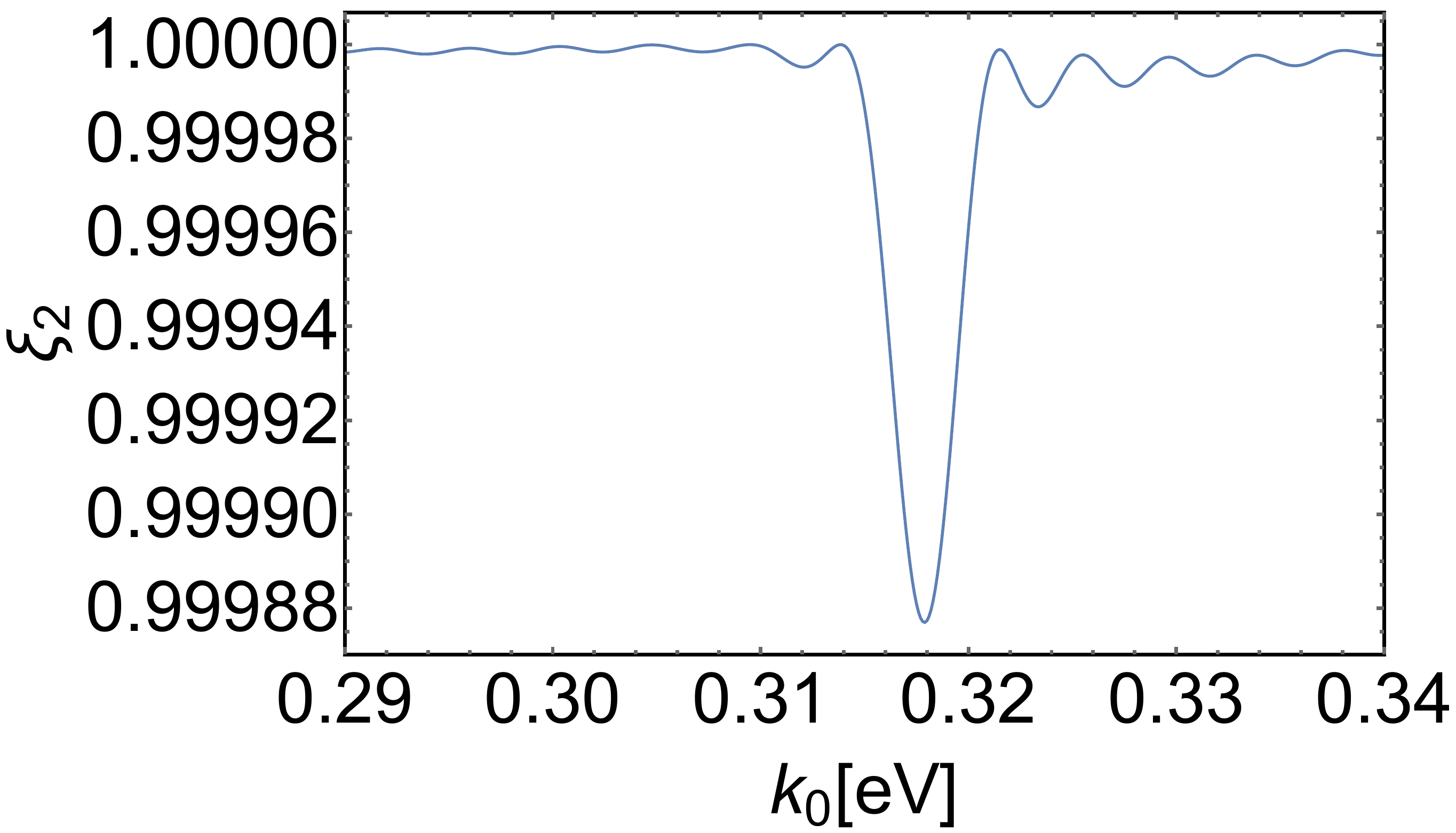}}\,
\raisebox{-0.5\height}{\includegraphics*[width=0.24\linewidth]{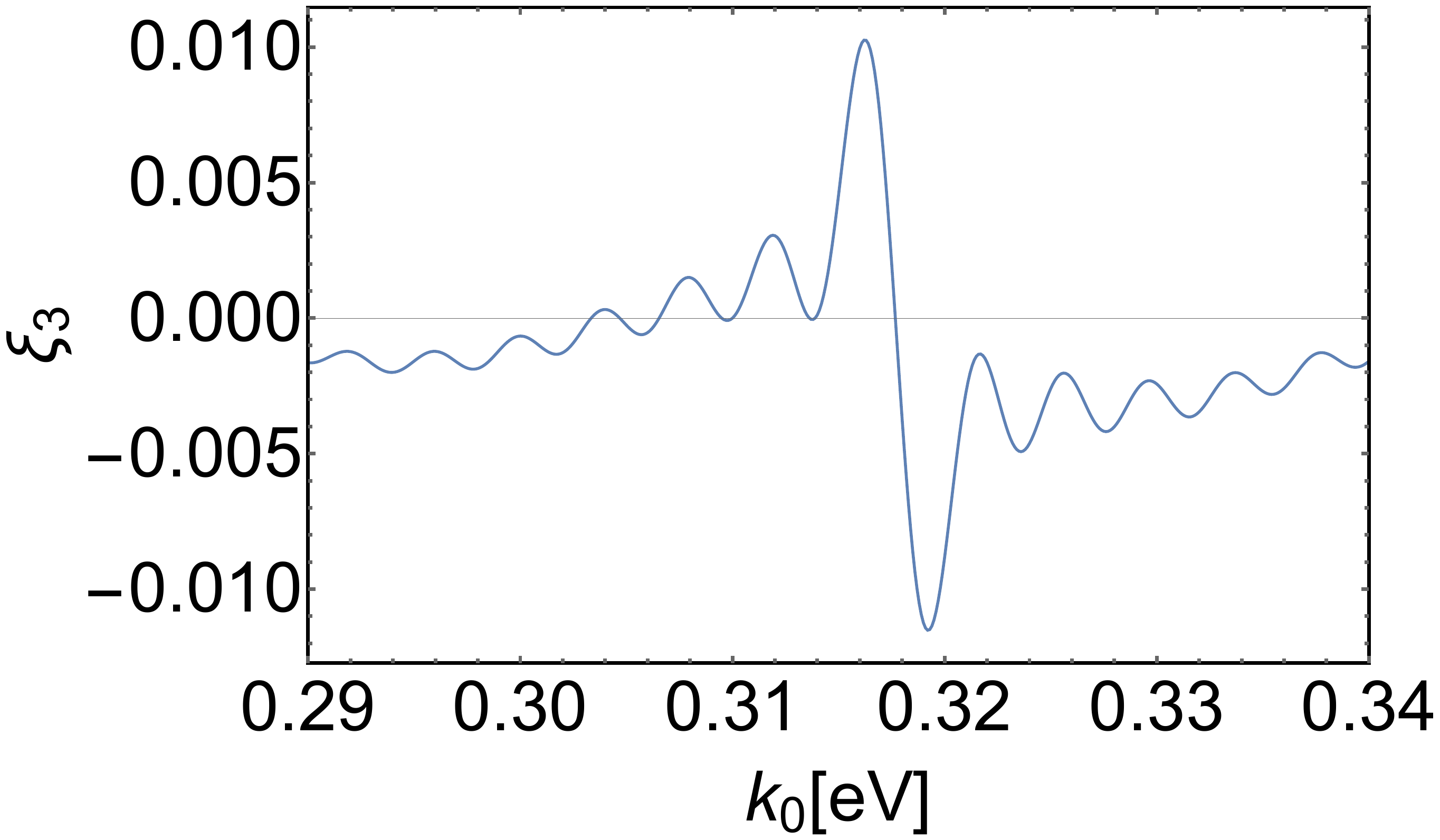}}\\
\raisebox{-0.5\height}{\includegraphics*[width=0.24\linewidth]{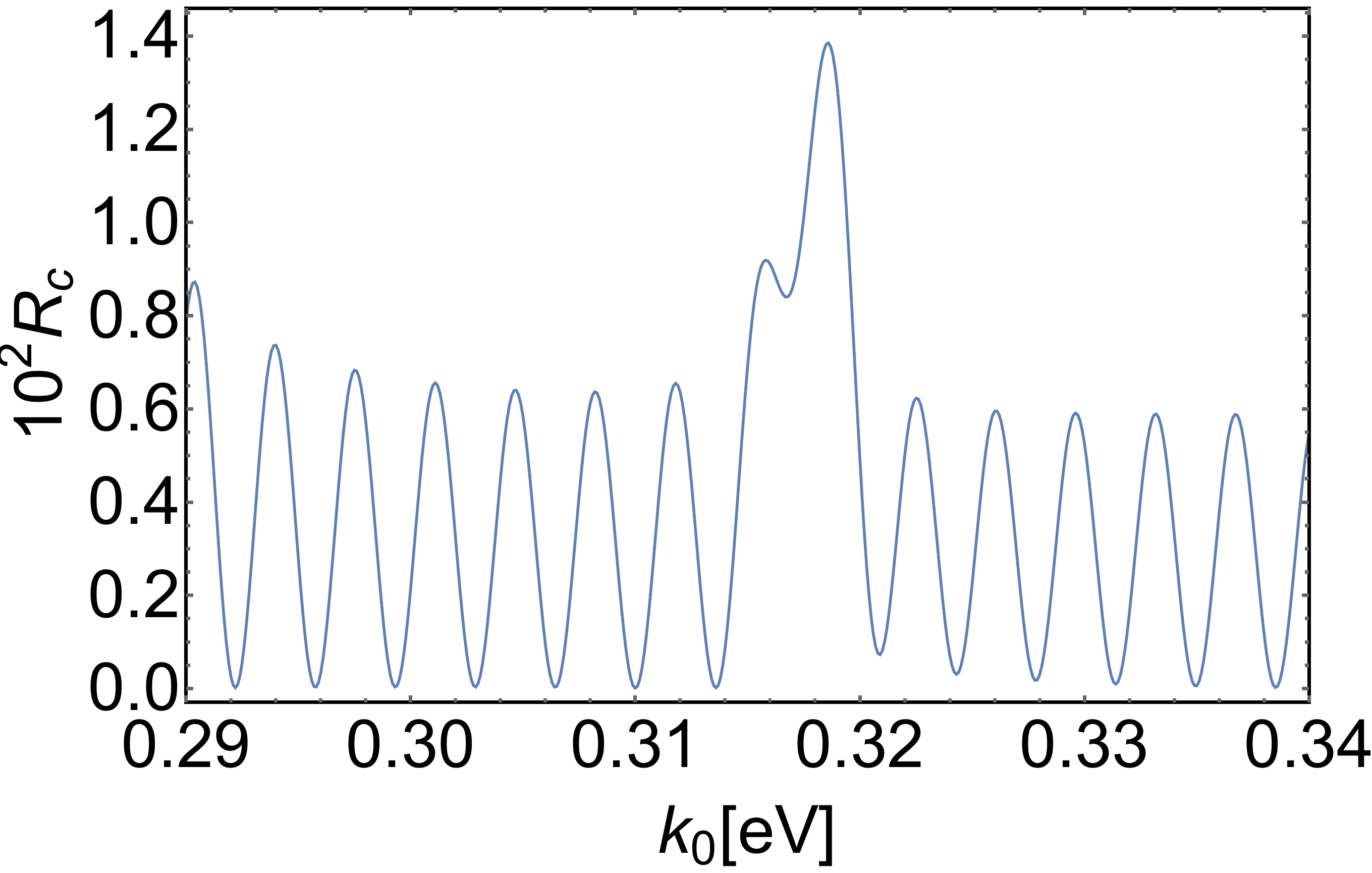}}\,
\raisebox{-0.5\height}{\includegraphics*[width=0.24\linewidth]{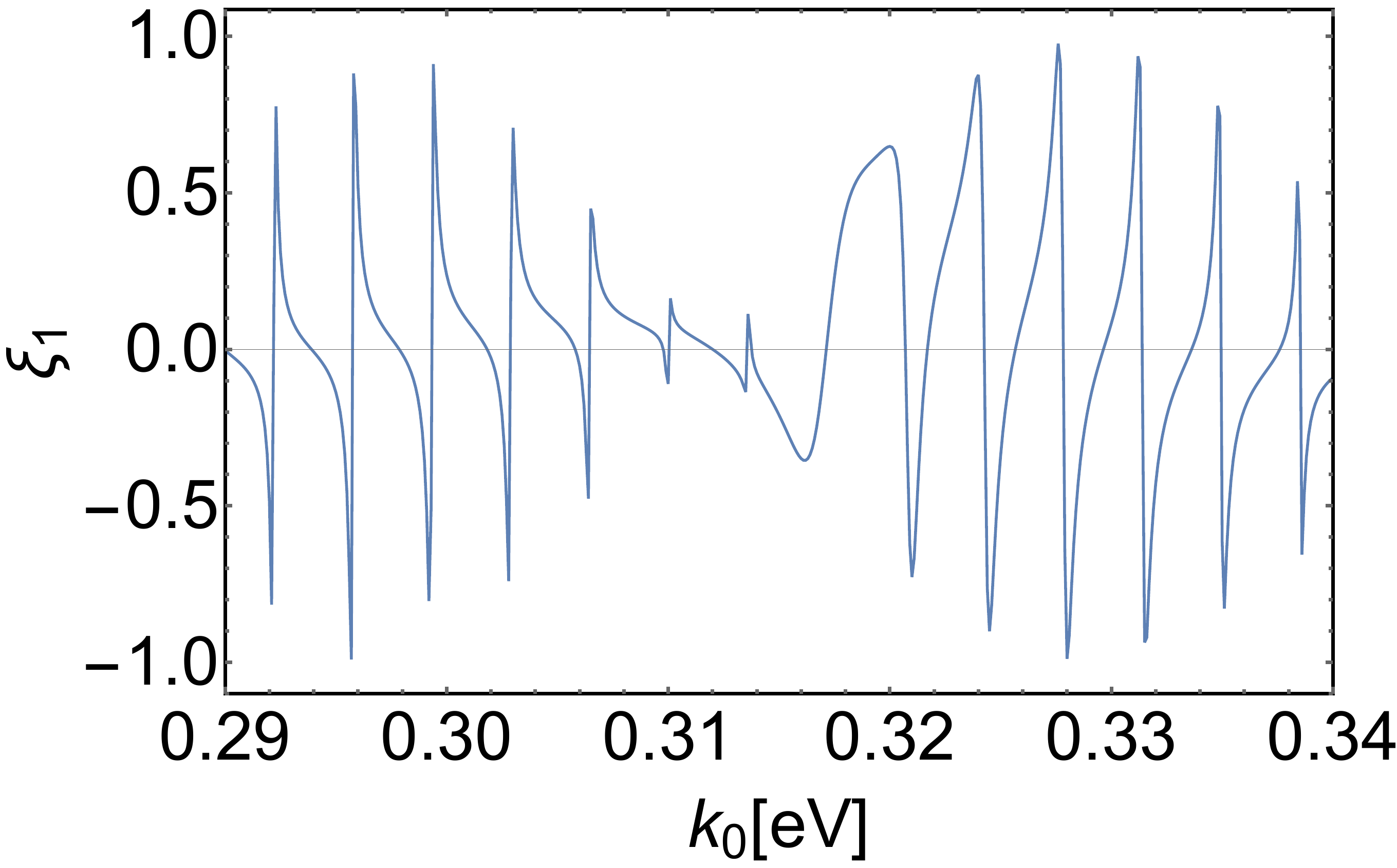}}\,
\raisebox{-0.5\height}{\includegraphics*[width=0.24\linewidth]{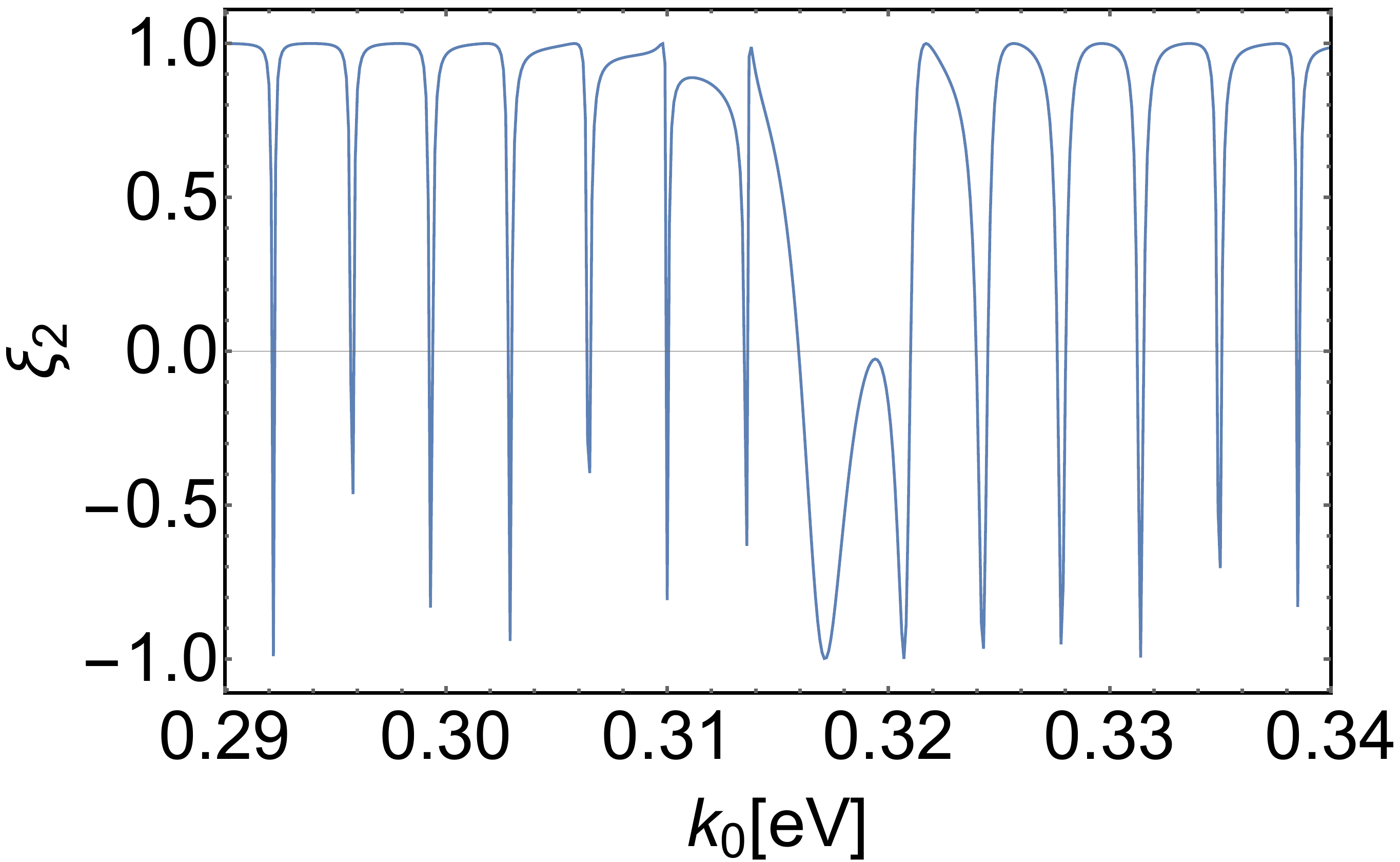}}\,
\raisebox{-0.5\height}{\includegraphics*[width=0.24\linewidth]{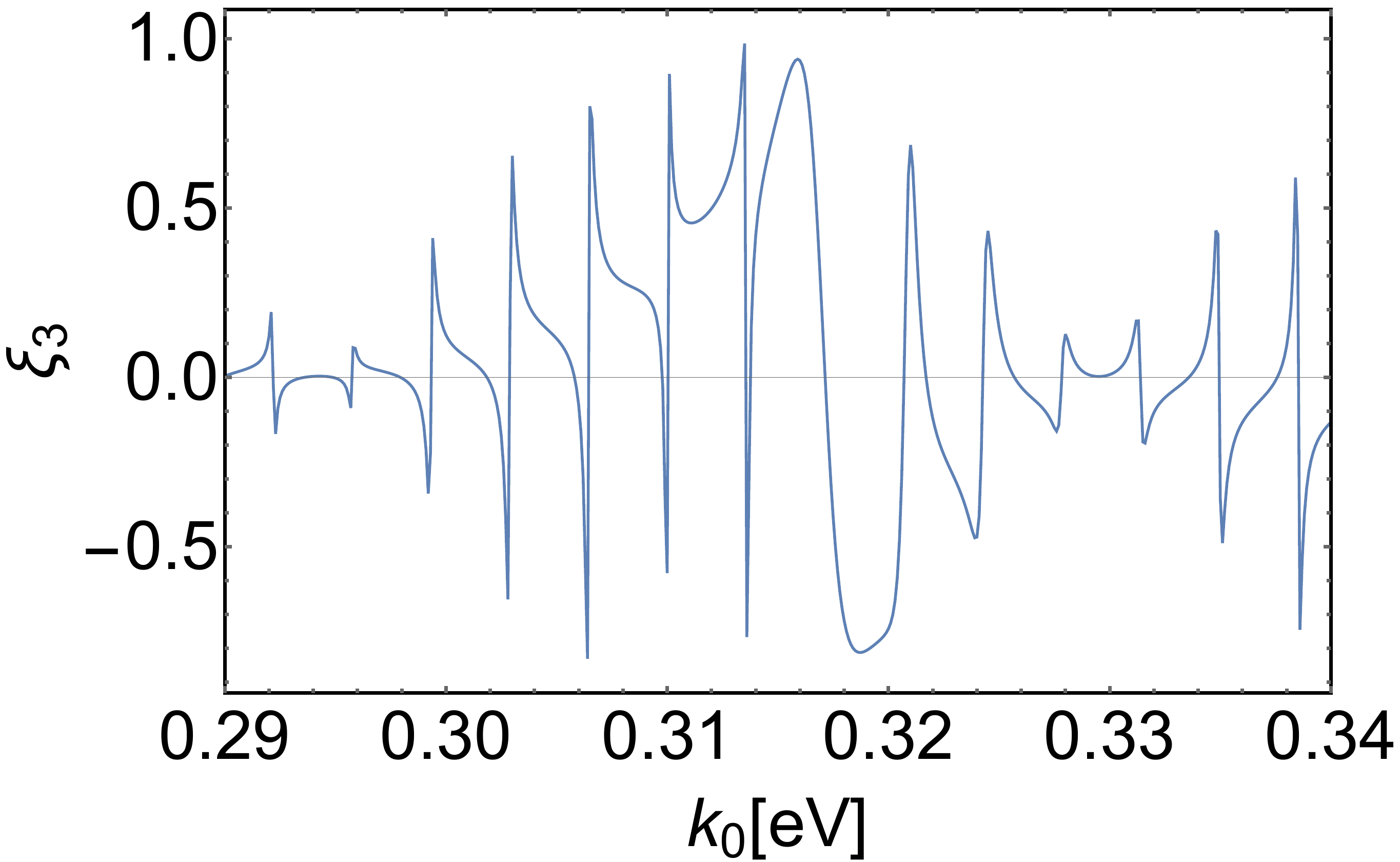}}\\
\raisebox{-0.5\height}{\includegraphics*[width=0.24\linewidth]{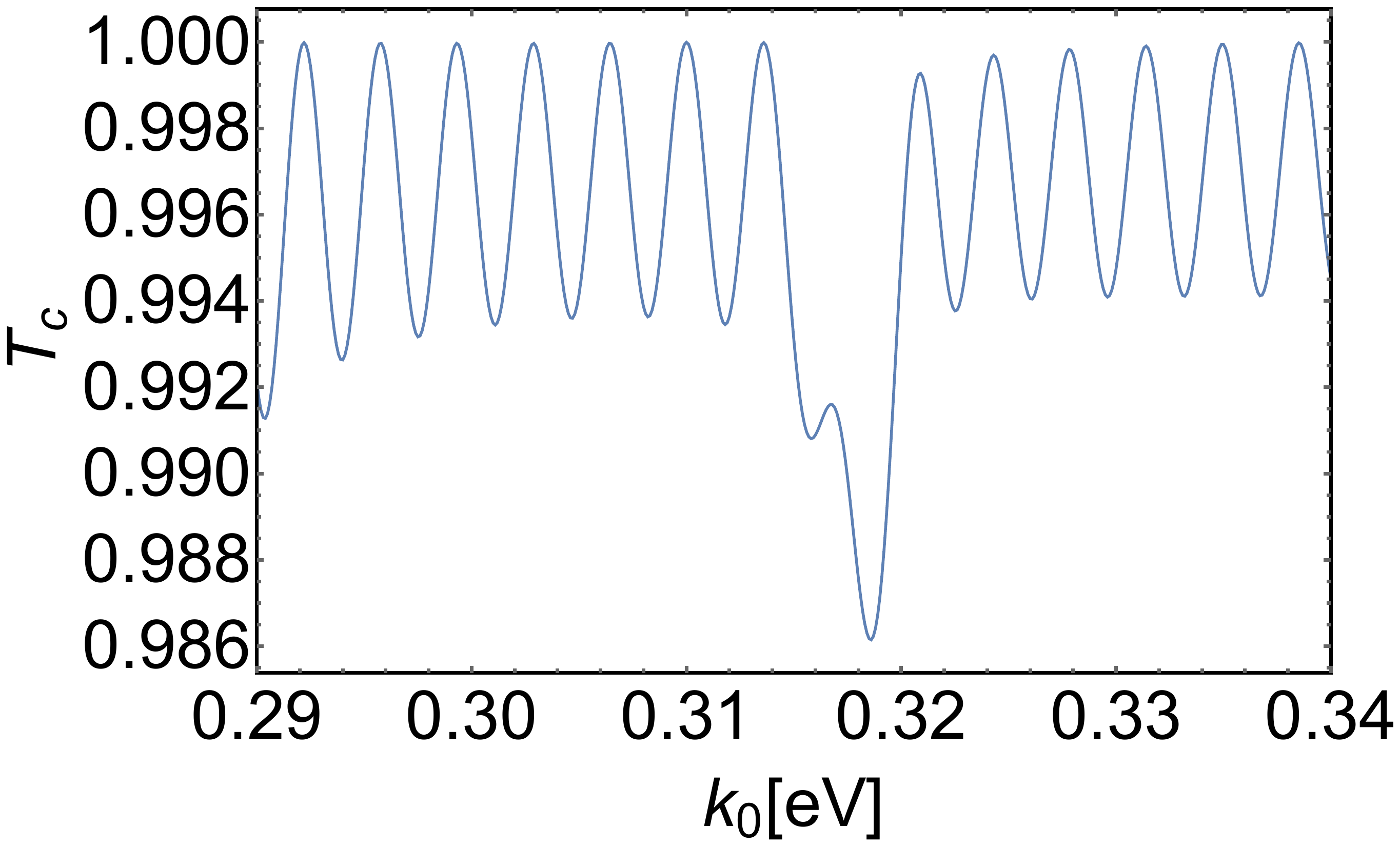}}\,
\raisebox{-0.5\height}{\includegraphics*[width=0.24\linewidth]{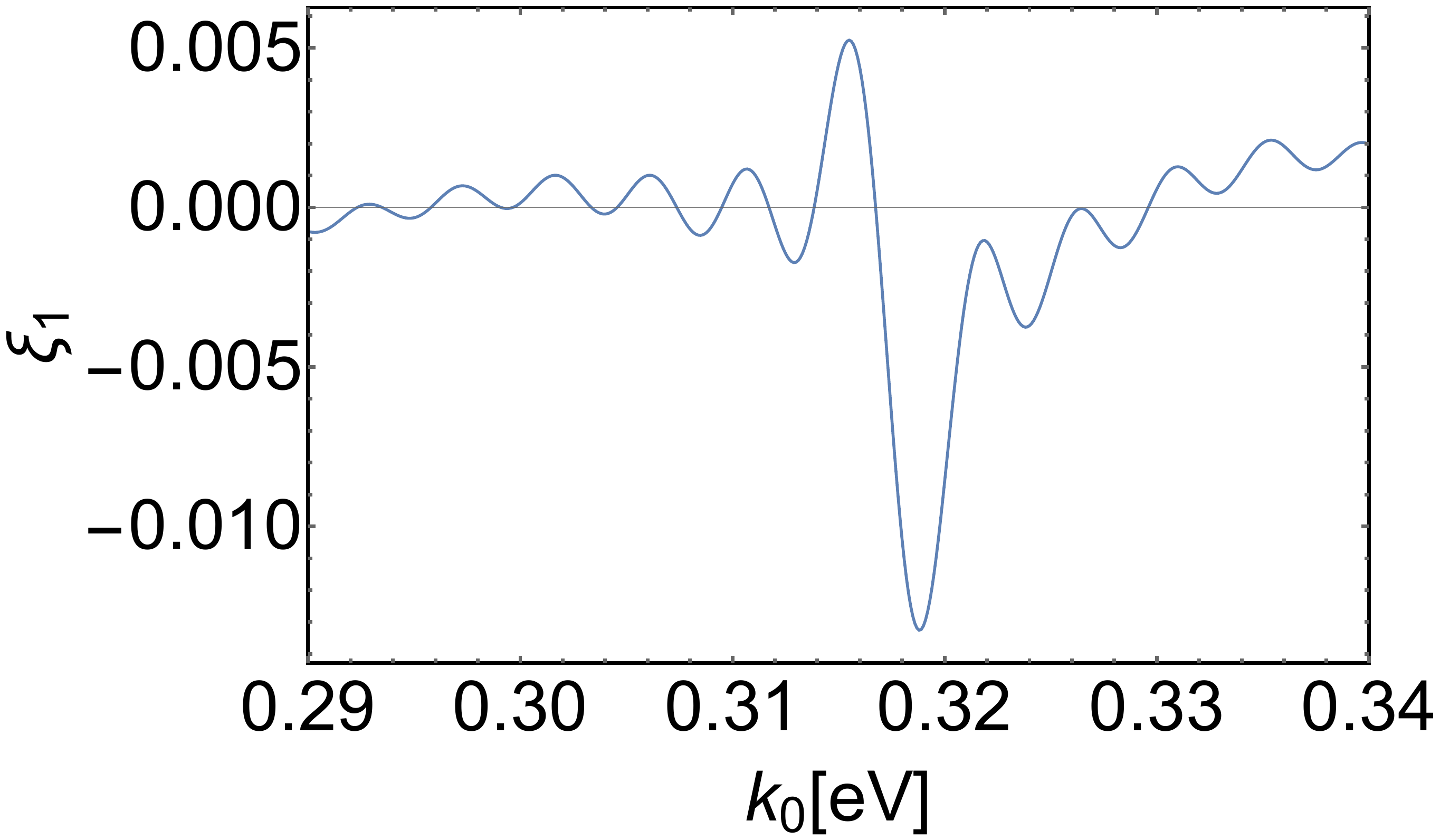}}\,
\raisebox{-0.5\height}{\includegraphics*[width=0.24\linewidth]{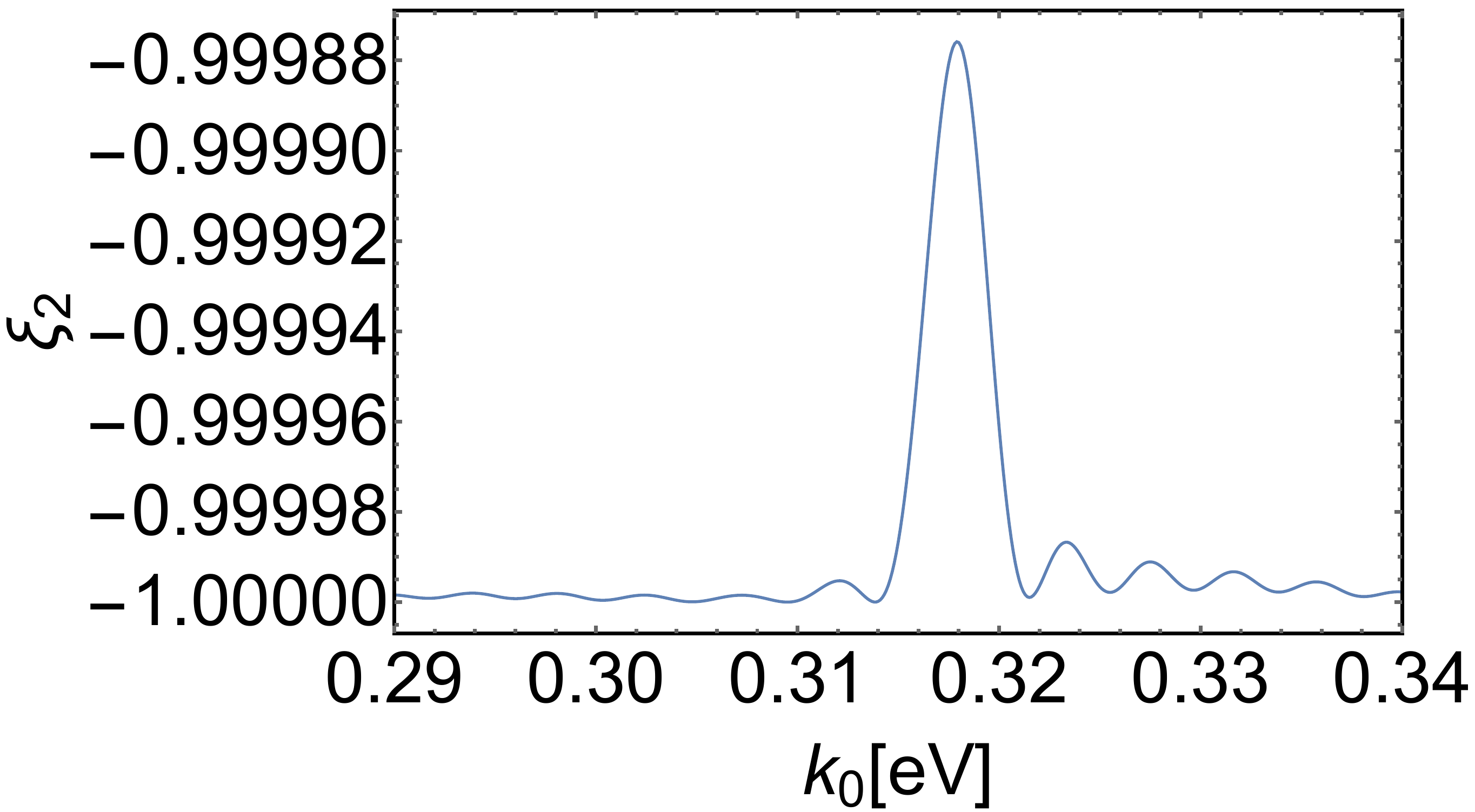}}\,
\raisebox{-0.5\height}{\includegraphics*[width=0.24\linewidth]{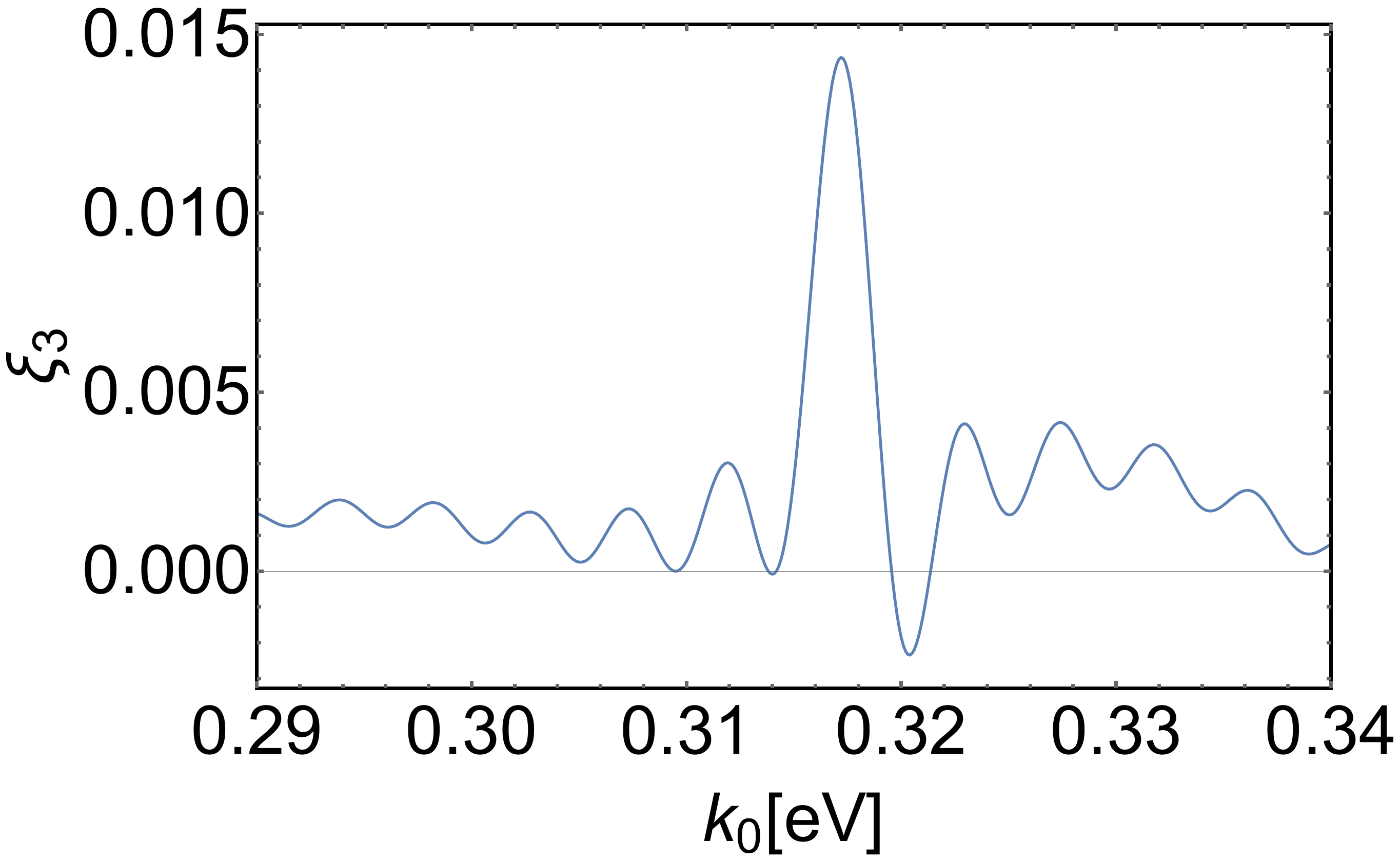}}\\
\raisebox{-0.5\height}{\includegraphics*[width=0.24\linewidth]{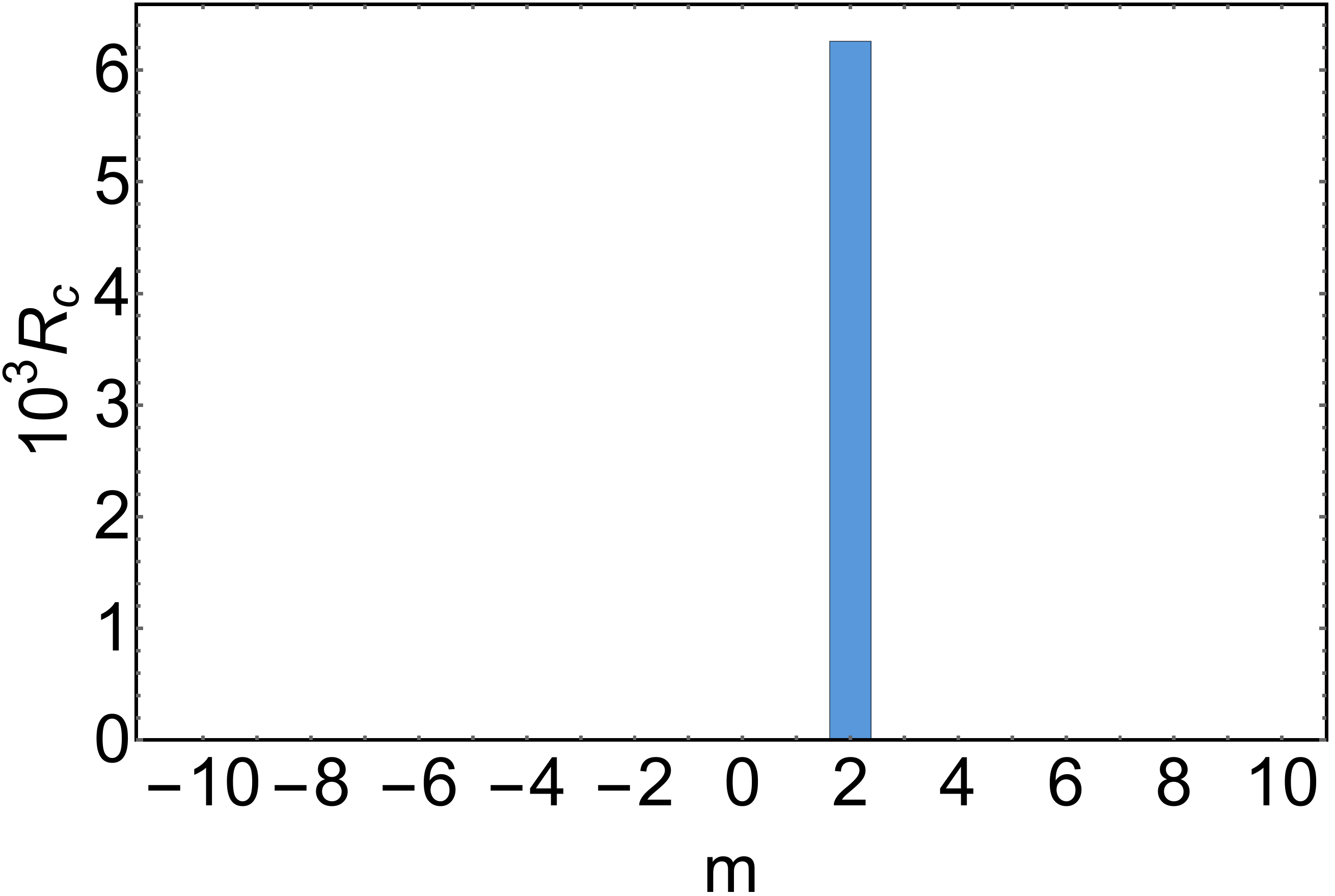}}\,
\raisebox{-0.5\height}{\includegraphics*[width=0.24\linewidth]{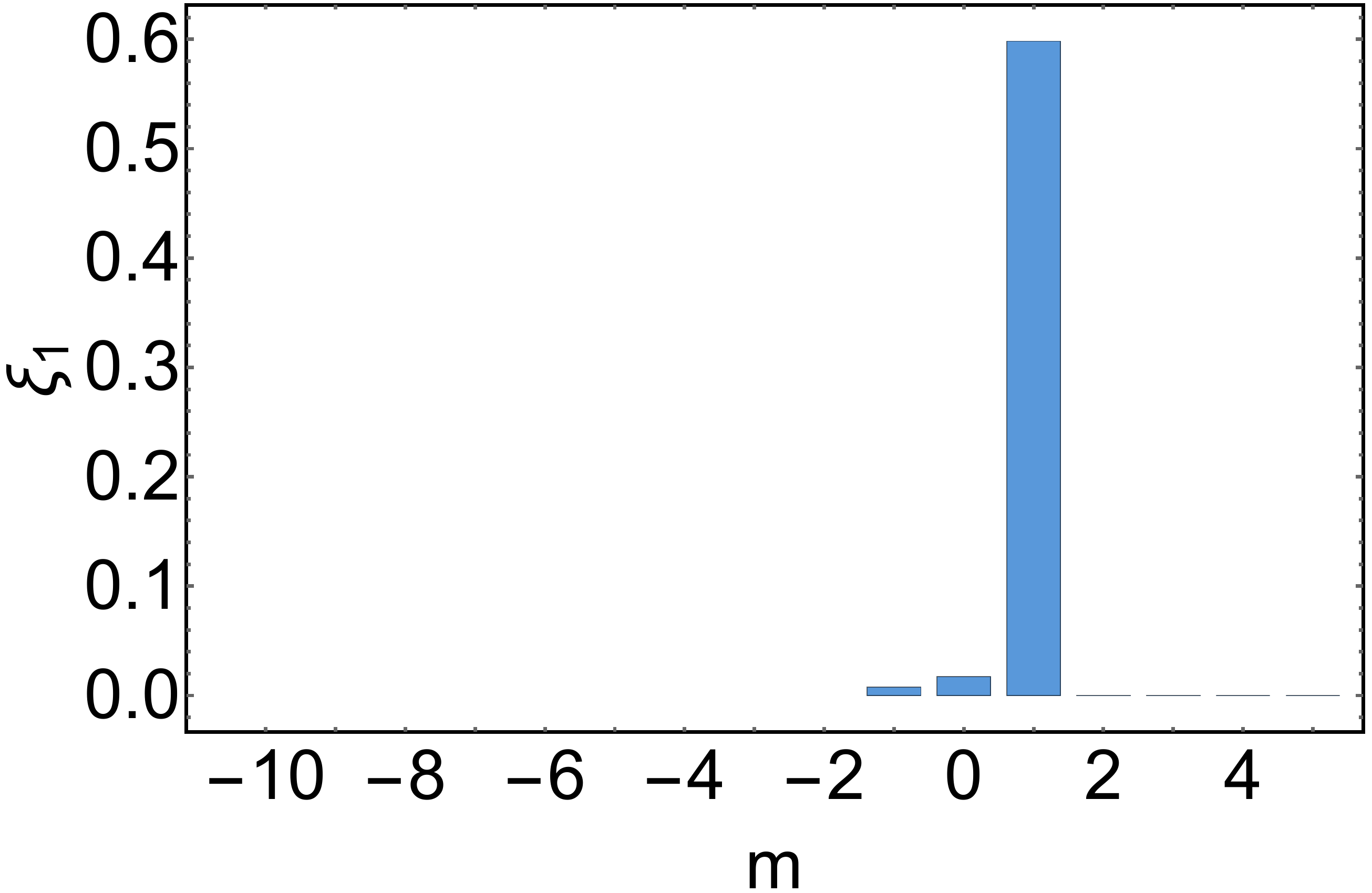}}\,
\raisebox{-0.5\height}{\includegraphics*[width=0.24\linewidth]{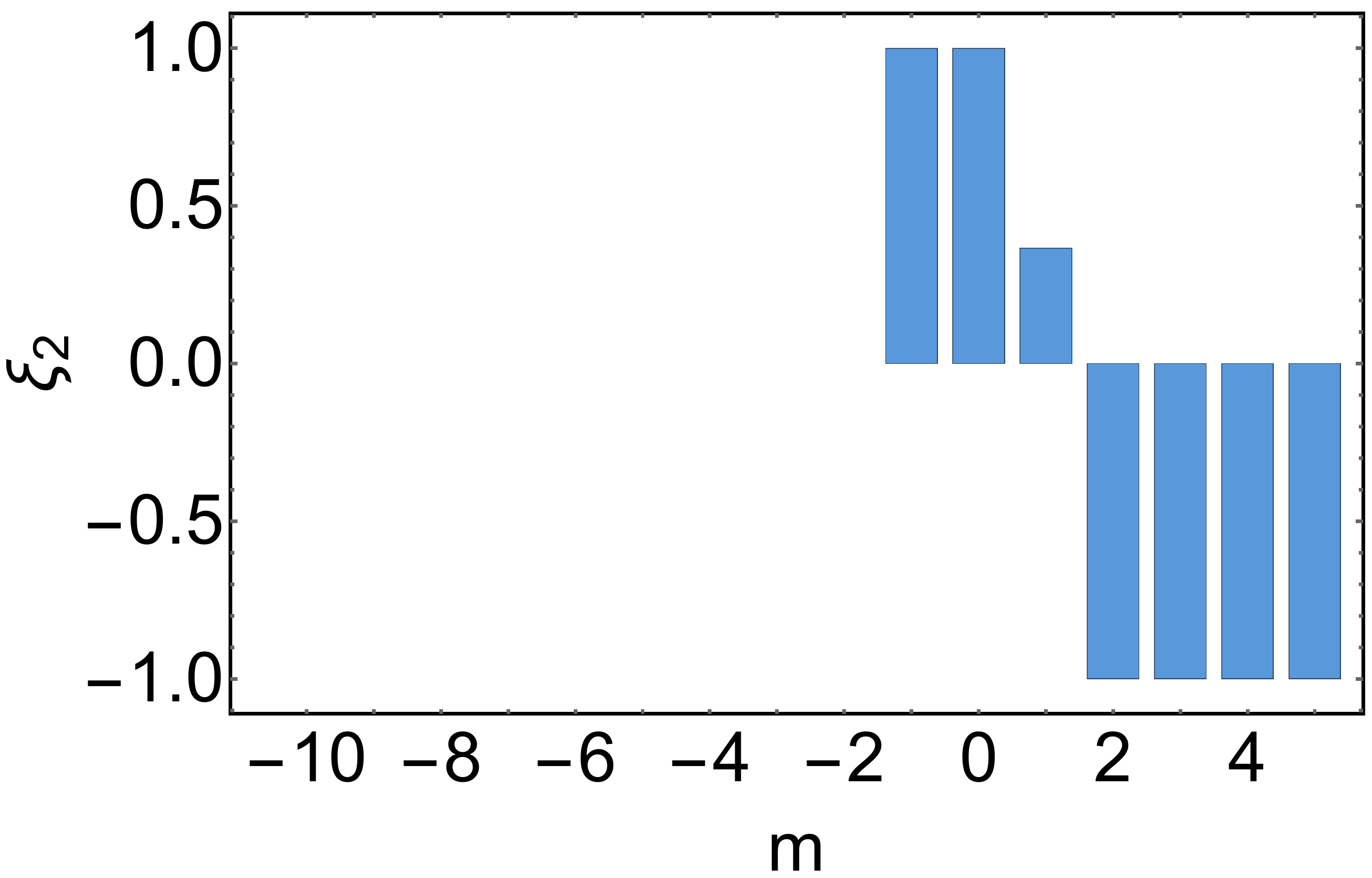}}\,
\raisebox{-0.5\height}{\includegraphics*[width=0.24\linewidth]{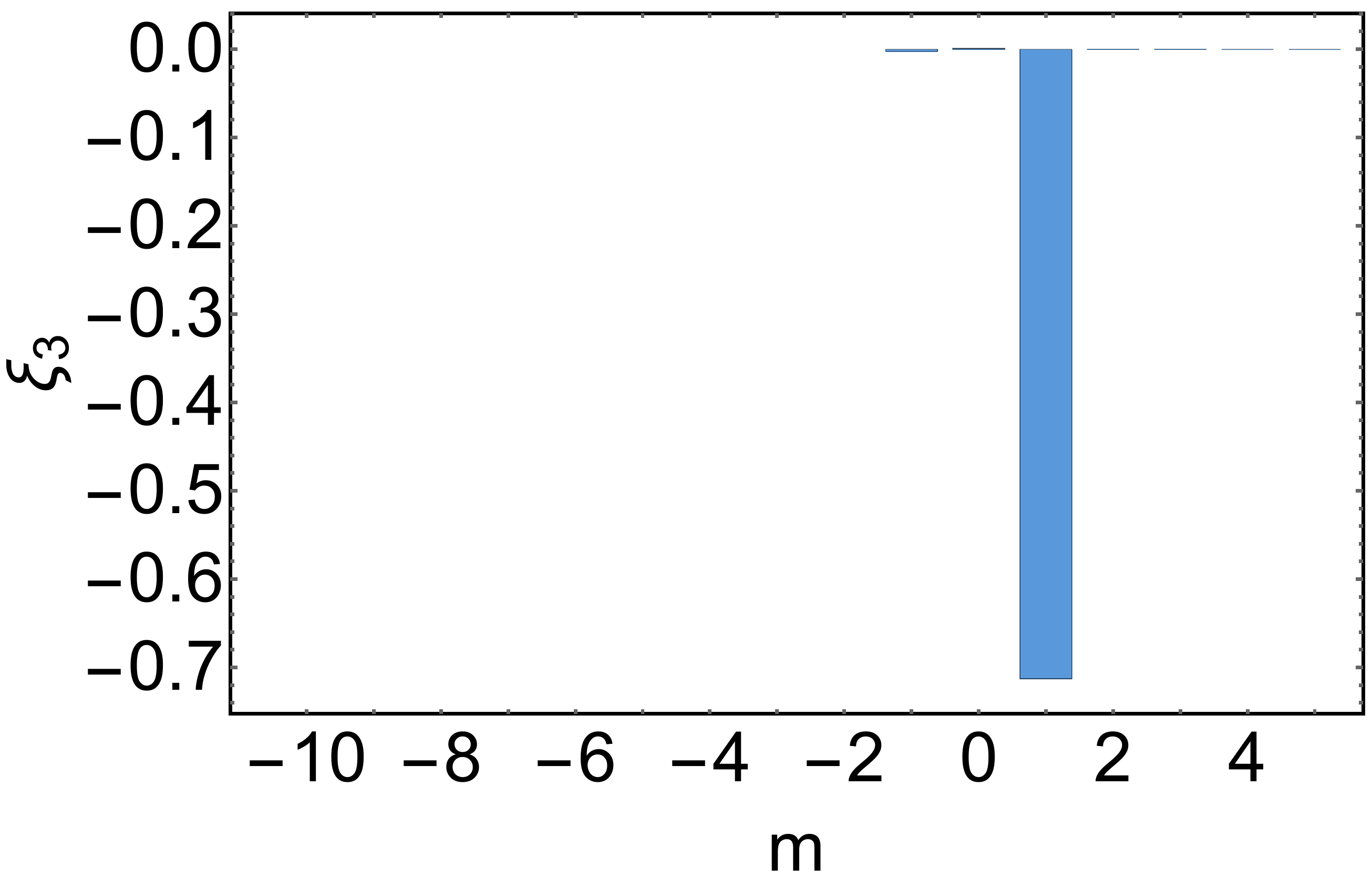}}\\
\raisebox{-0.5\height}{\includegraphics*[width=0.24\linewidth]{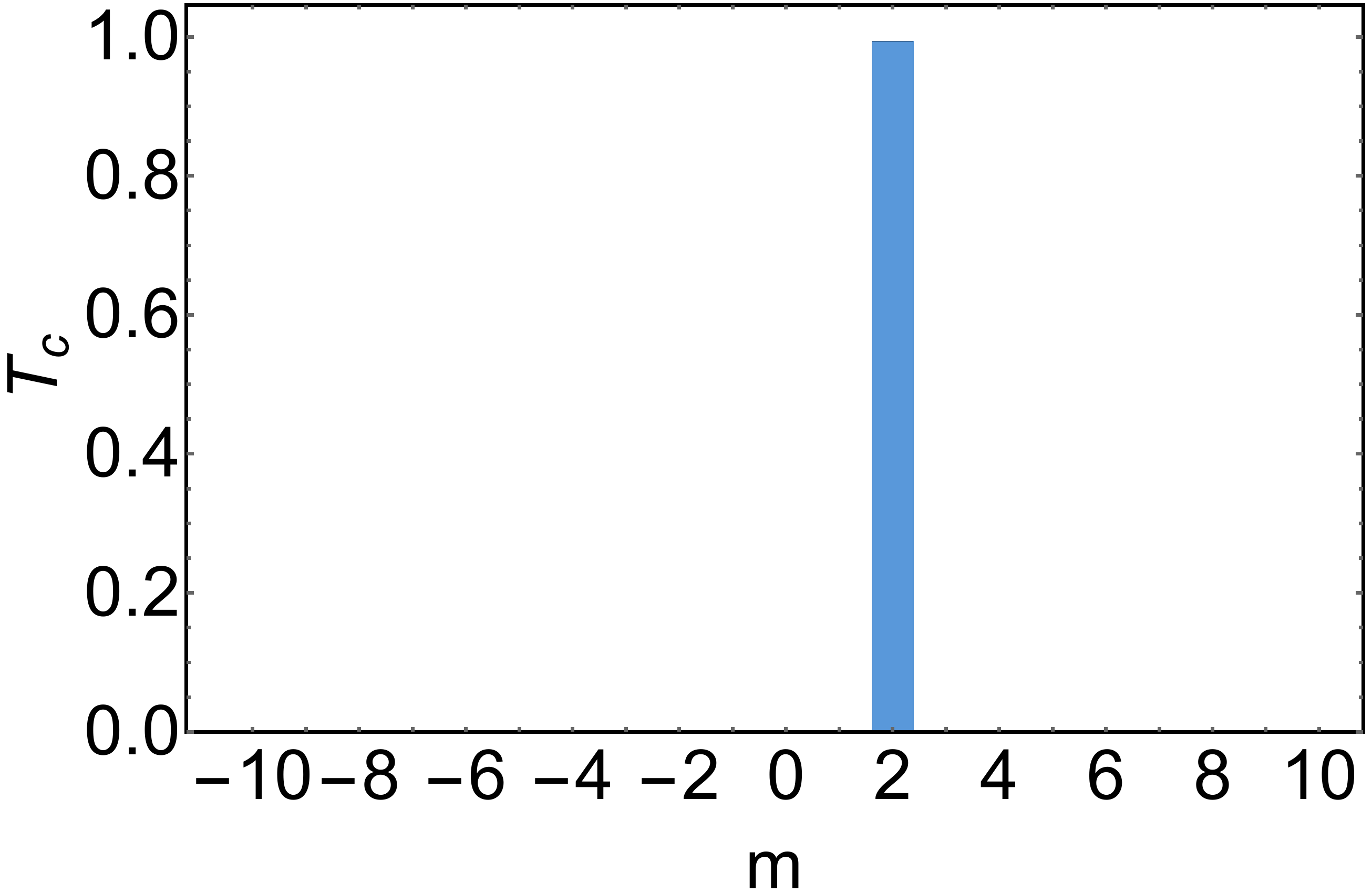}}\,
\raisebox{-0.5\height}{\includegraphics*[width=0.24\linewidth]{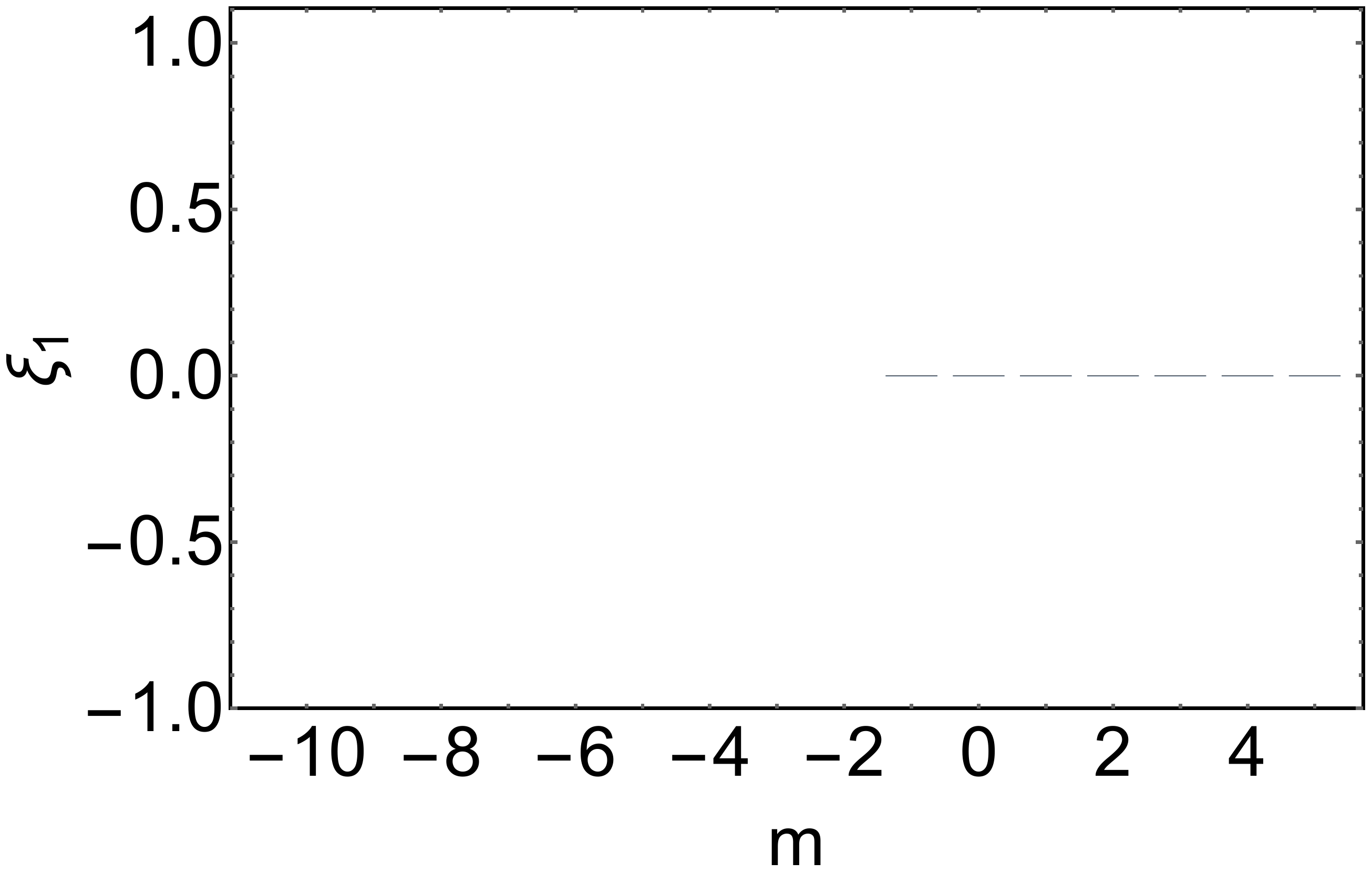}}\,
\raisebox{-0.5\height}{\includegraphics*[width=0.24\linewidth]{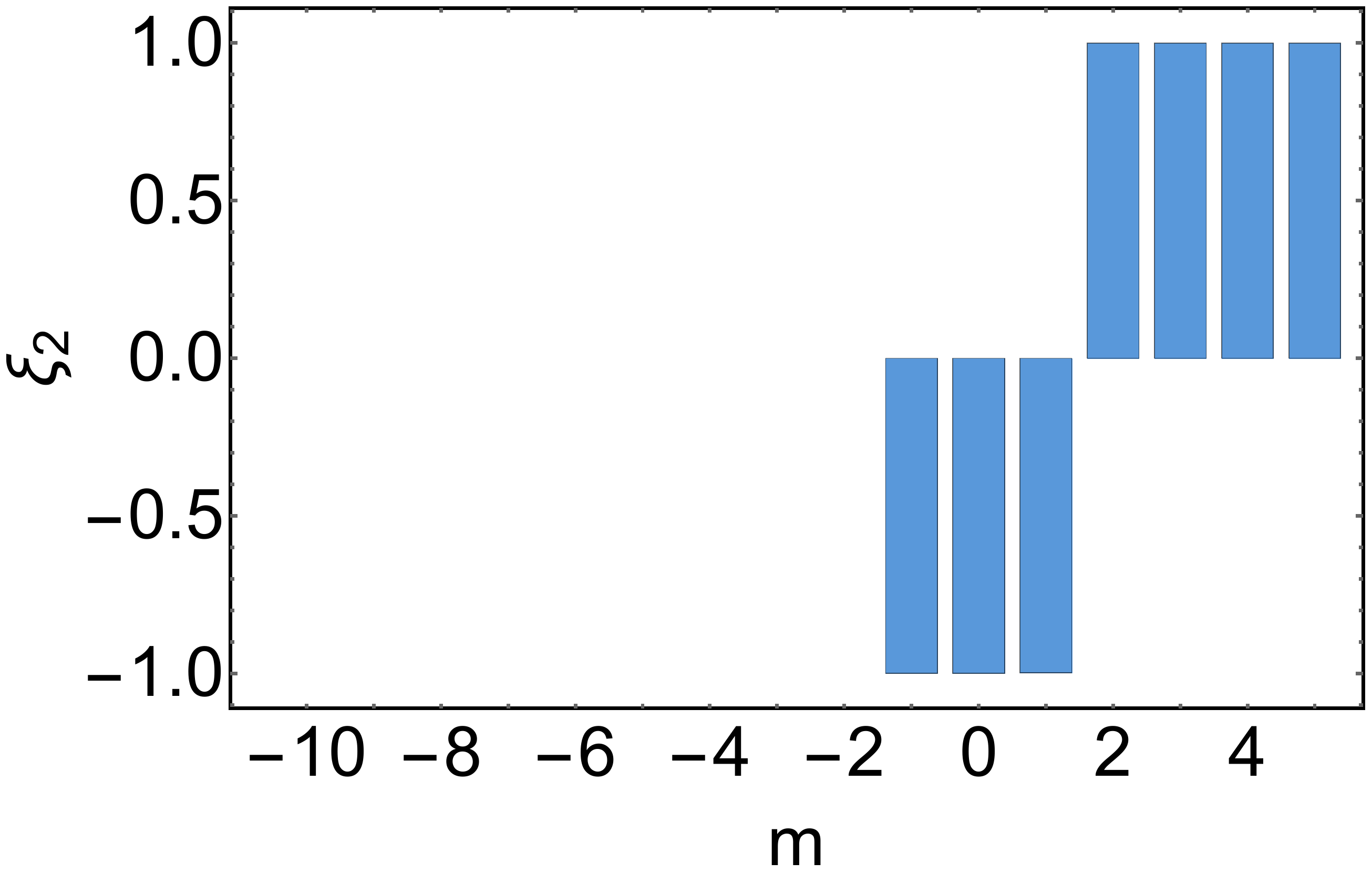}}\,
\raisebox{-0.5\height}{\includegraphics*[width=0.24\linewidth]{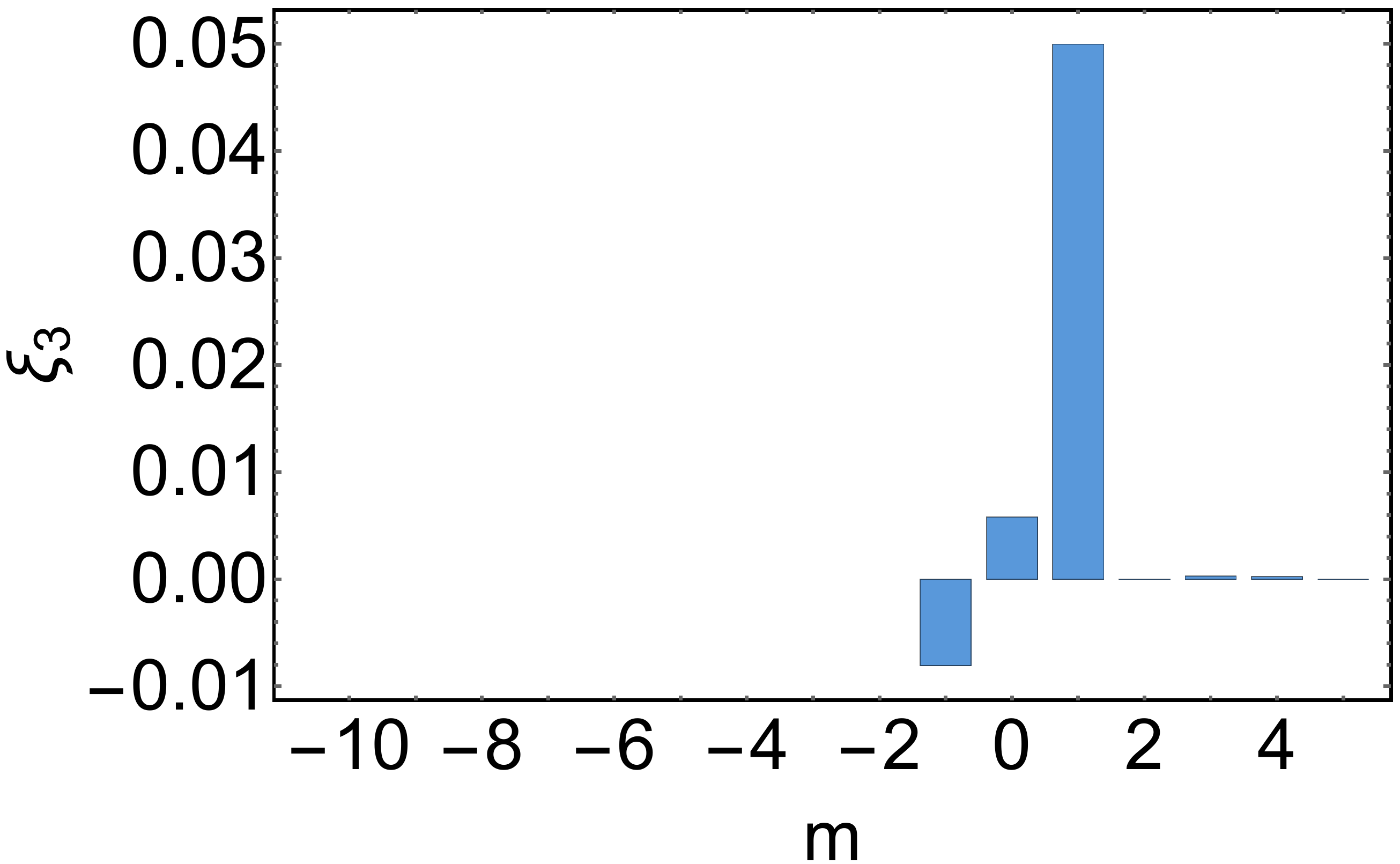}}\\
\raisebox{-0.5\height}{\includegraphics*[width=0.24\linewidth]{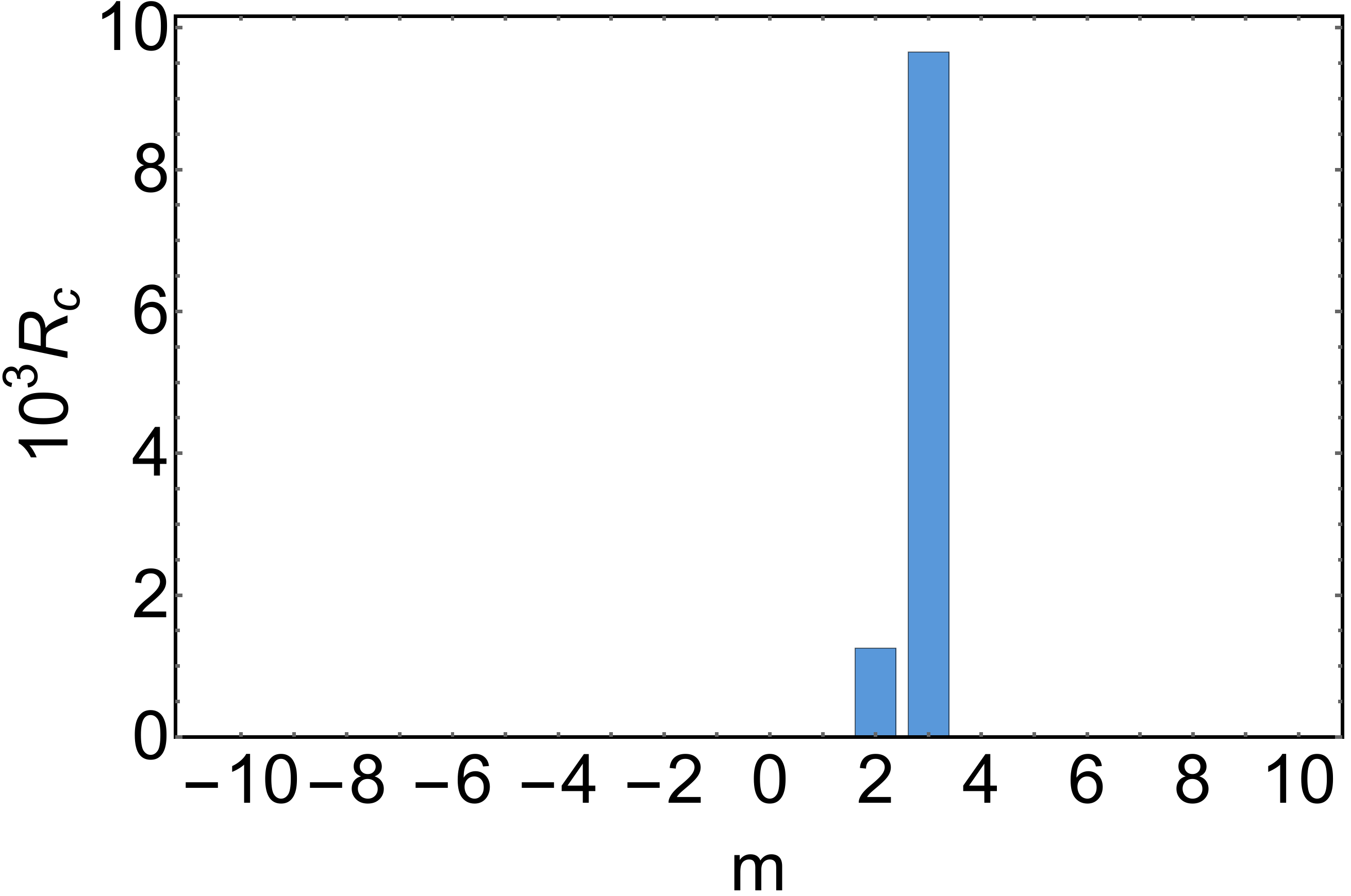}}\,
\raisebox{-0.5\height}{\includegraphics*[width=0.24\linewidth]{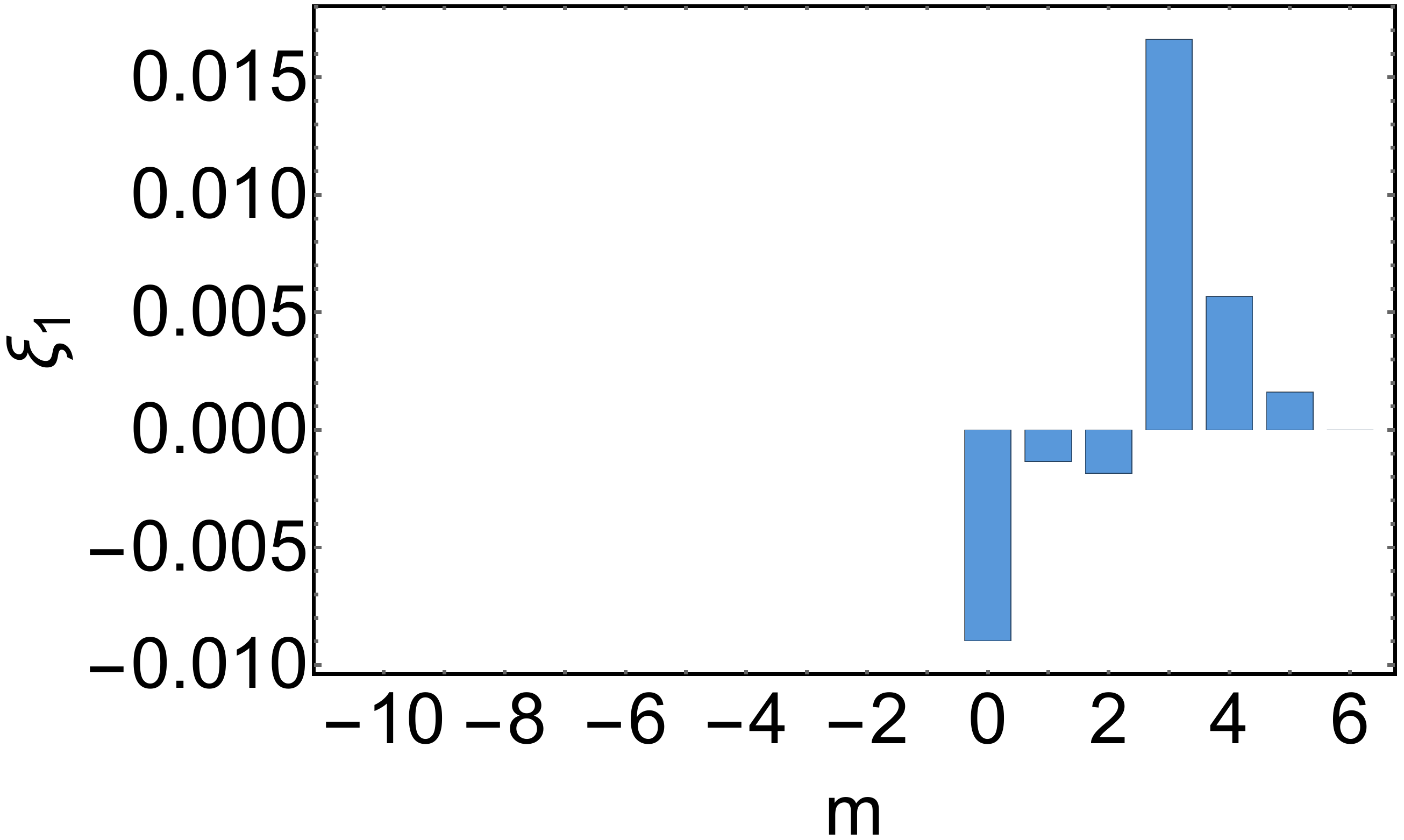}}\,
\raisebox{-0.5\height}{\includegraphics*[width=0.24\linewidth]{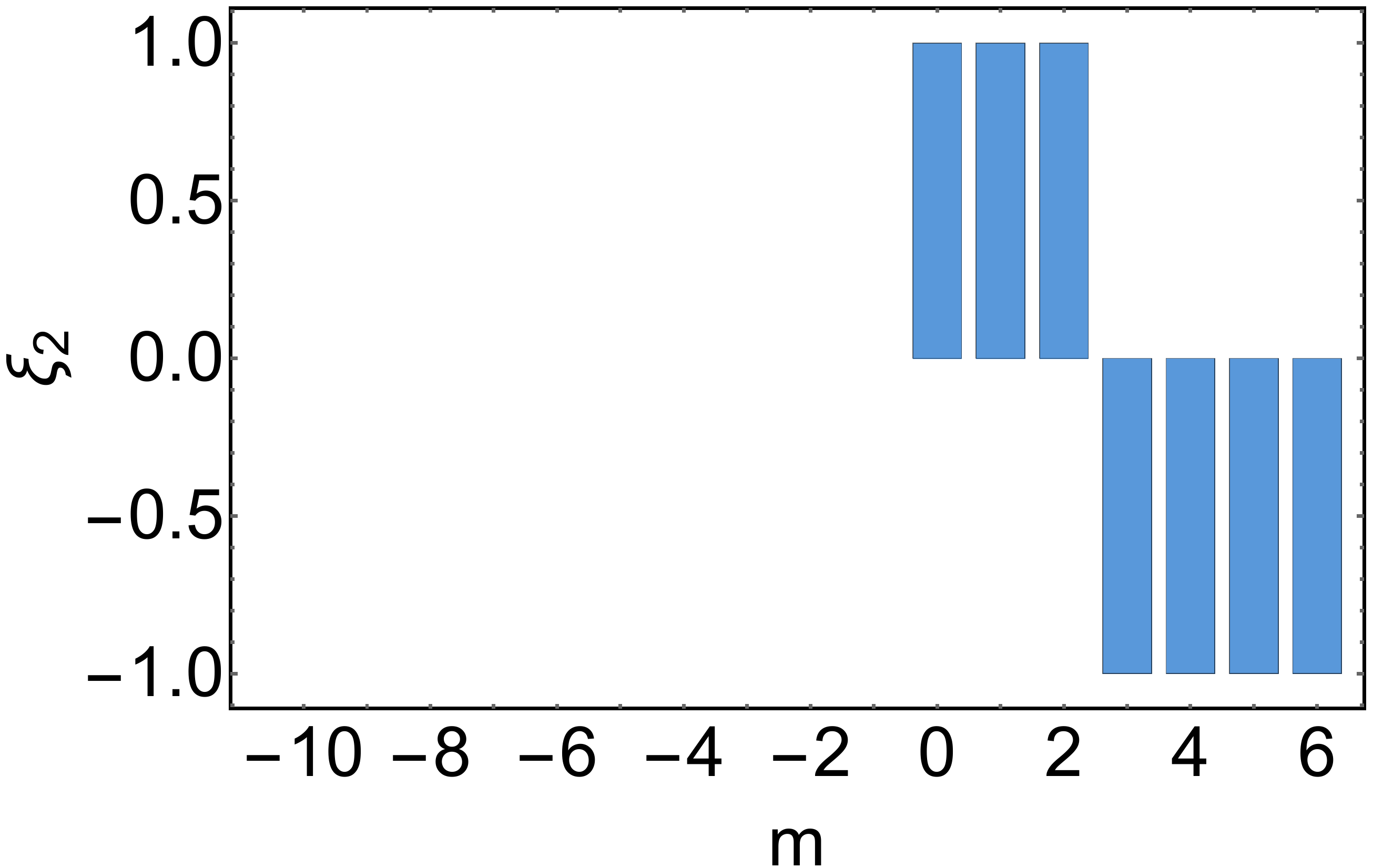}}\,
\raisebox{-0.5\height}{\includegraphics*[width=0.24\linewidth]{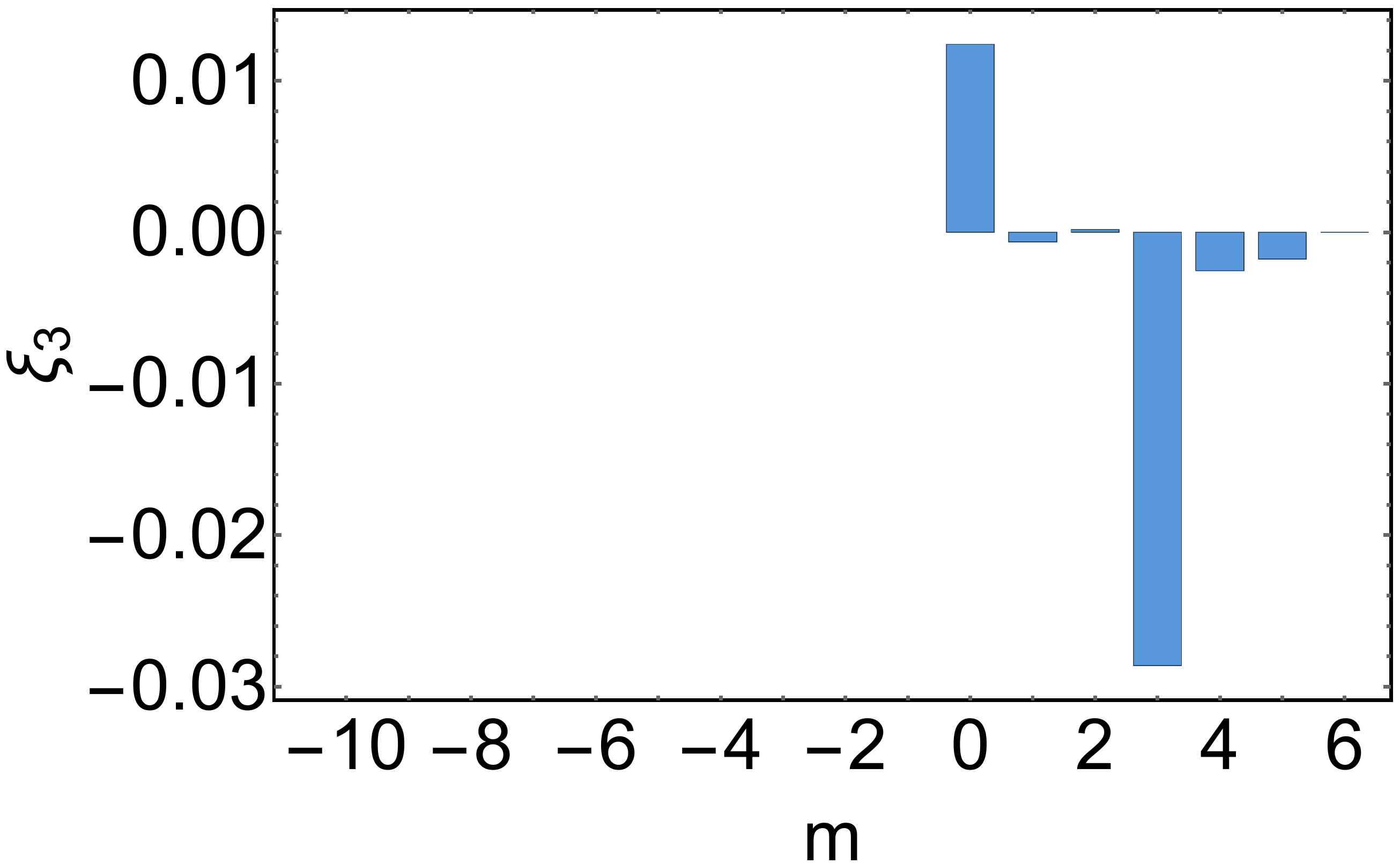}}\\
\raisebox{-0.5\height}{\includegraphics*[width=0.24\linewidth]{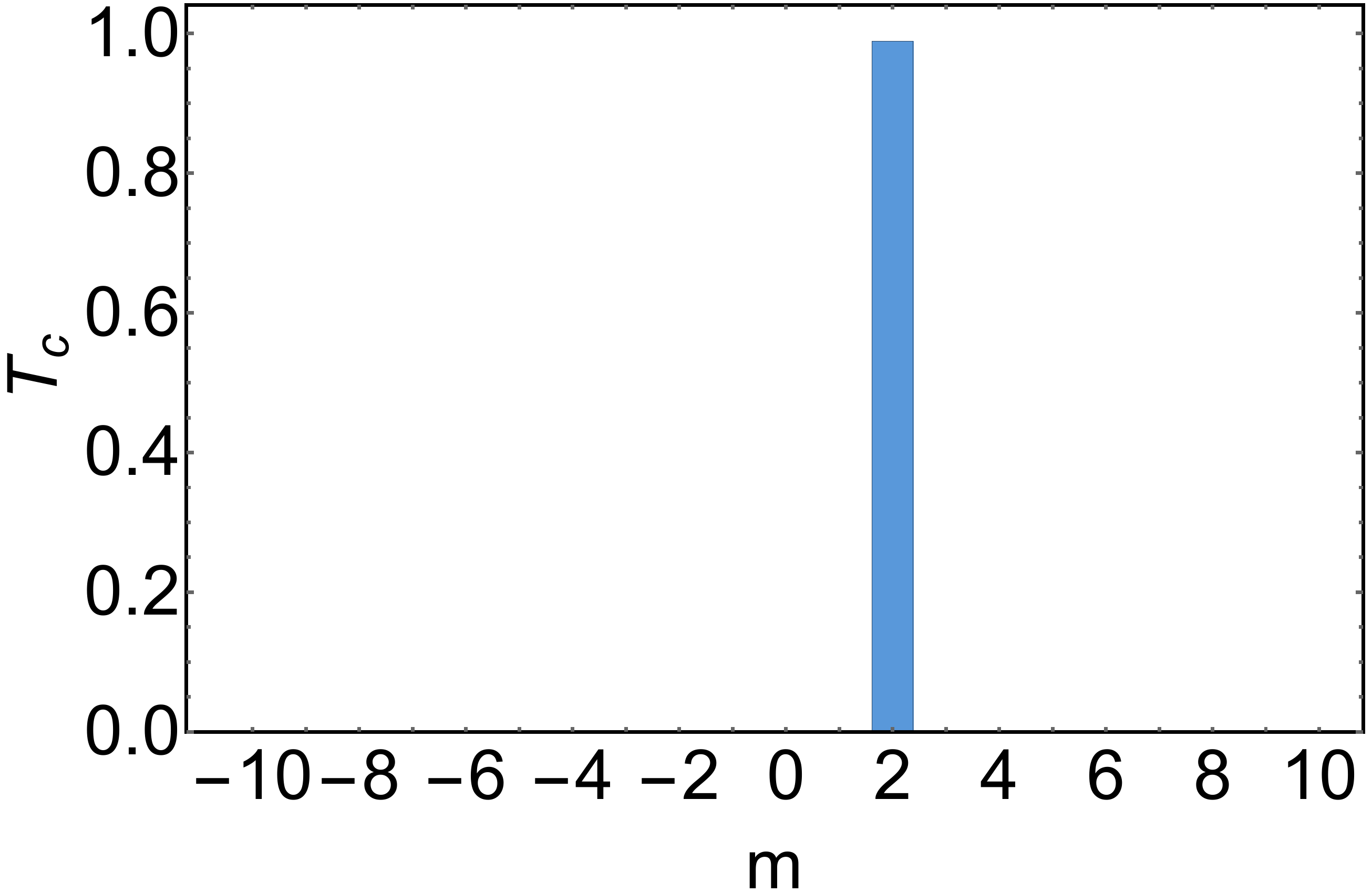}}\,
\raisebox{-0.5\height}{\includegraphics*[width=0.24\linewidth]{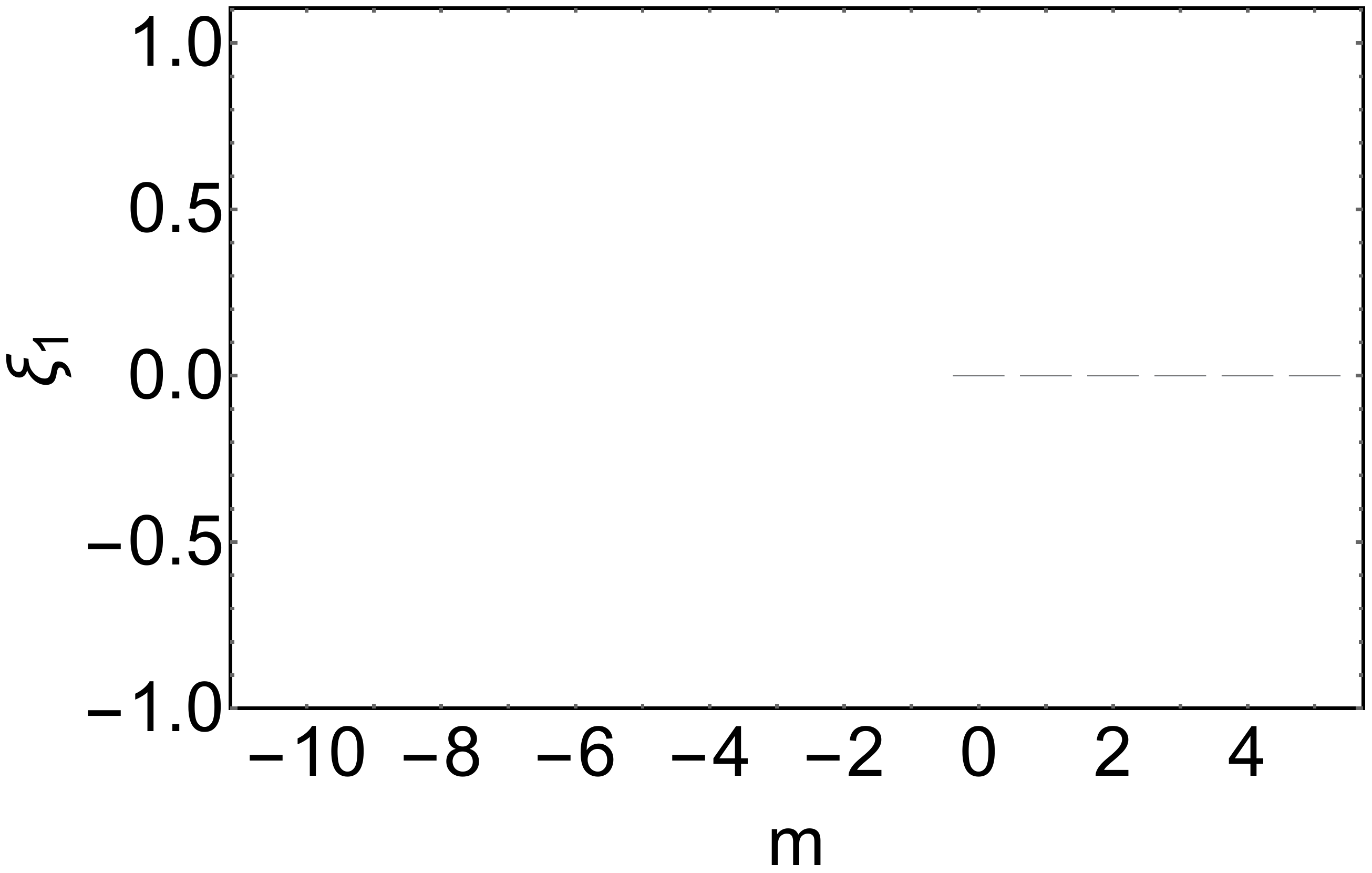}}\,
\raisebox{-0.5\height}{\includegraphics*[width=0.24\linewidth]{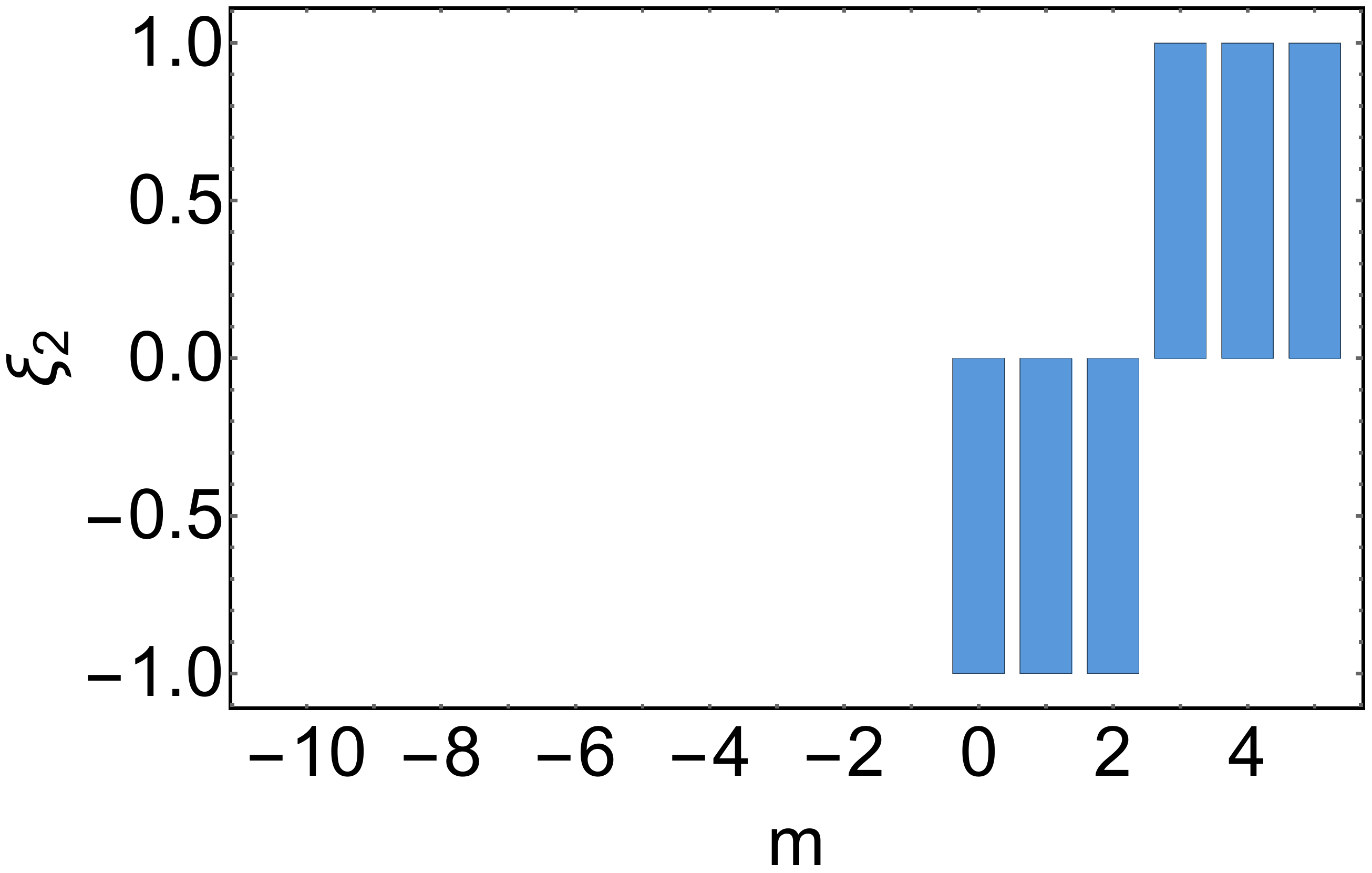}}\,
\raisebox{-0.5\height}{\includegraphics*[width=0.24\linewidth]{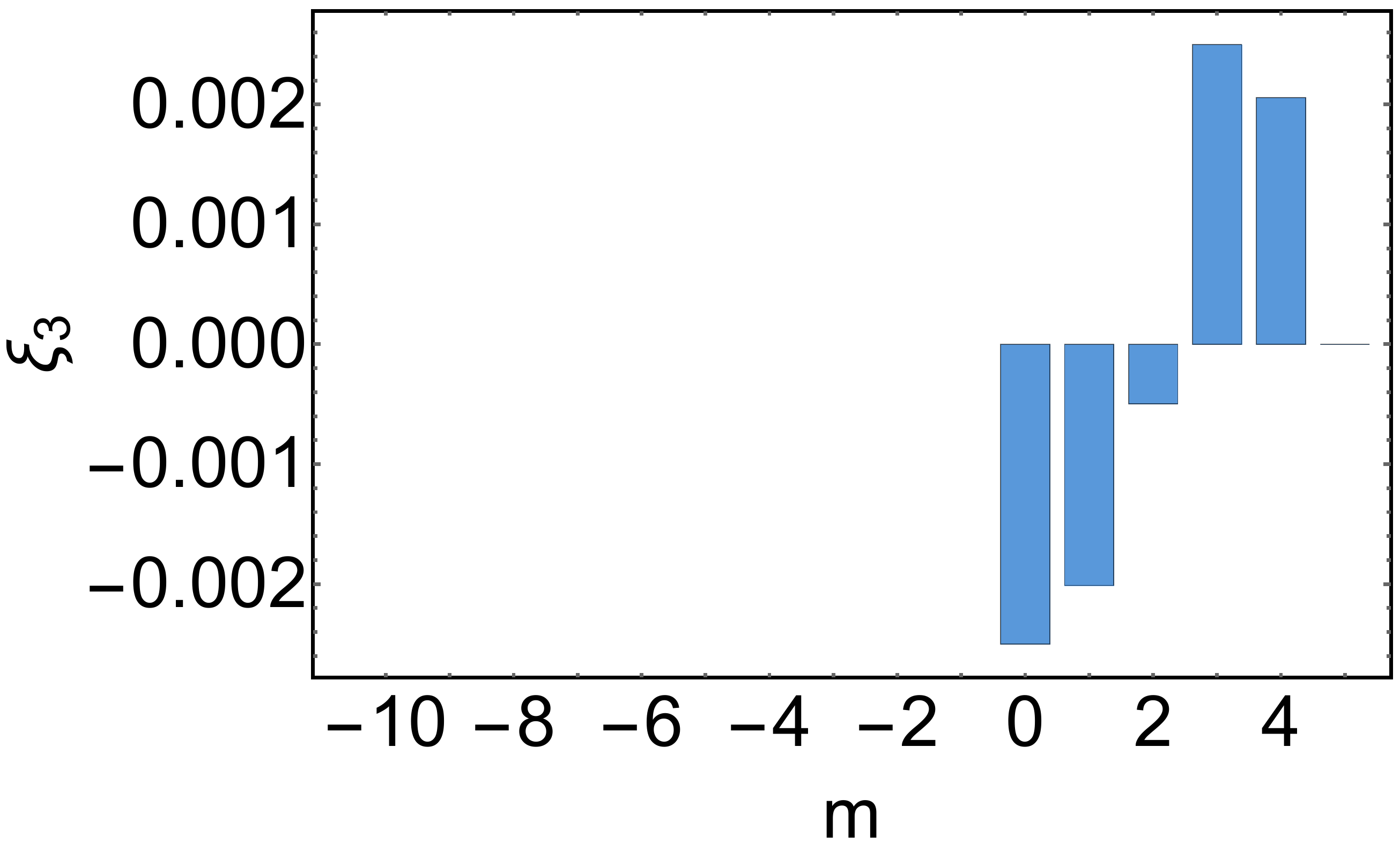}}
\caption{{\footnotesize The same as in Fig. \ref{Scatt_Hels2r_plots} but for the real band gap corresponding to $s_h=\pm1$. The lines $1$-$4$: The case of plane-wave photons scattered in the $(x,z)$ plane is considered. The lines $1$-$2$: The initial photon possesses the helicity $s=1$. The lines $3$-$4$: It has $s=-1$. The lines $5$-$8$: Scattering of the twisted photon with $m=2$ is considered at the energy $k_0=0.3177$ eV belonging to the band gap. The lines $5$-$6$ corresponds to $s=1$, whereas the lines $7$-$8$ are for $s=-1$.  As is seen from the plots on the line $7$, for $qs<0$, the projection of the total angular momentum of reflected twisted photons is shifted by $1$ in accordance with the selection rule \eqref{sel_rule}. In virtue of the fact that the gap parameter $|b|$ is small, the reflection coefficient is also small. It can be made to be almost equal to unity by increasing $|b|$ and/or by increasing the number of periods $N_h$.}}
\label{Scatt_Hels1r_plots}
\end{figure}

%\newpage
\begin{figure}[tp]
\centering
\raisebox{-0.5\height}{\includegraphics*[width=0.24\linewidth]{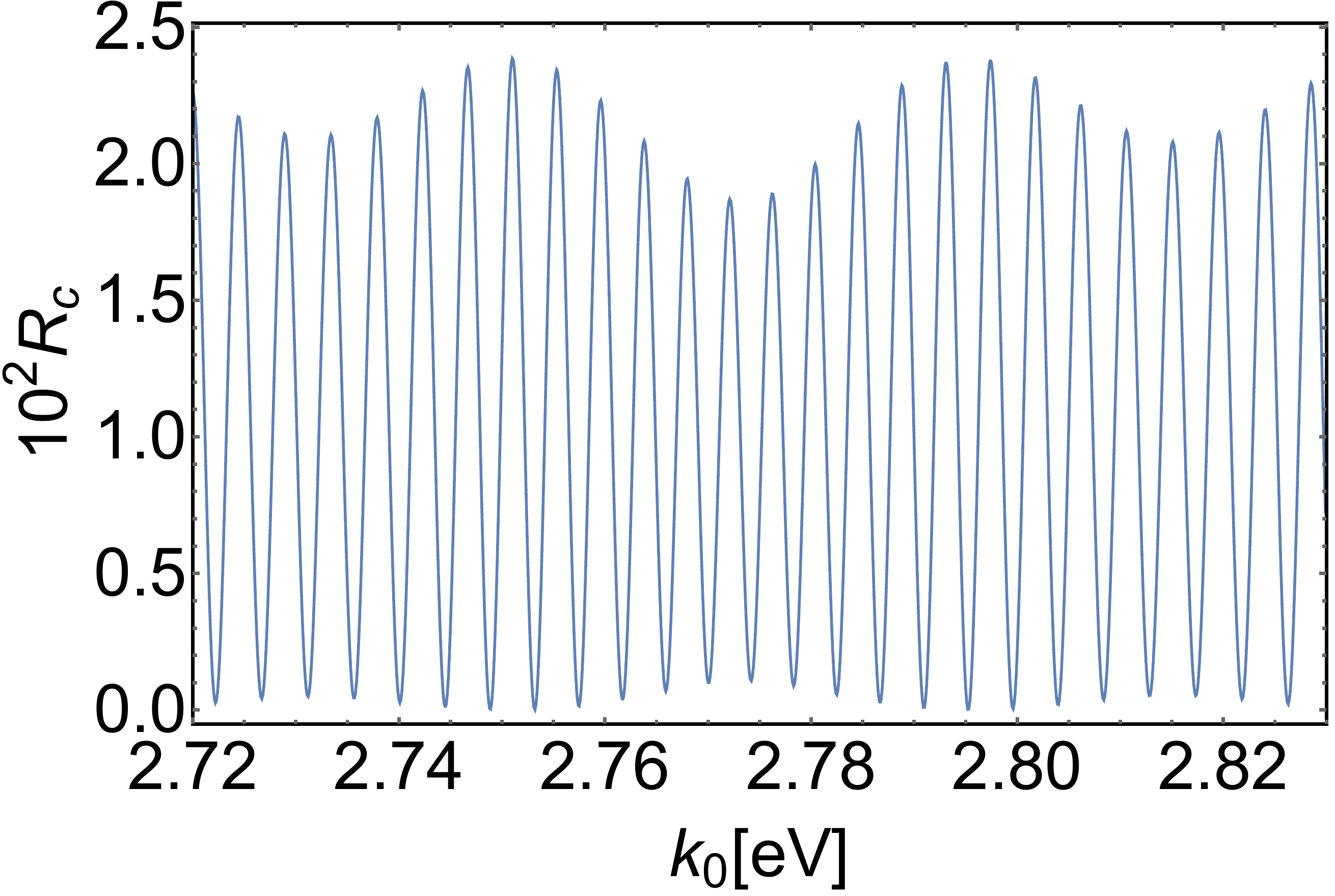}}\,
\raisebox{-0.5\height}{\includegraphics*[width=0.24\linewidth]{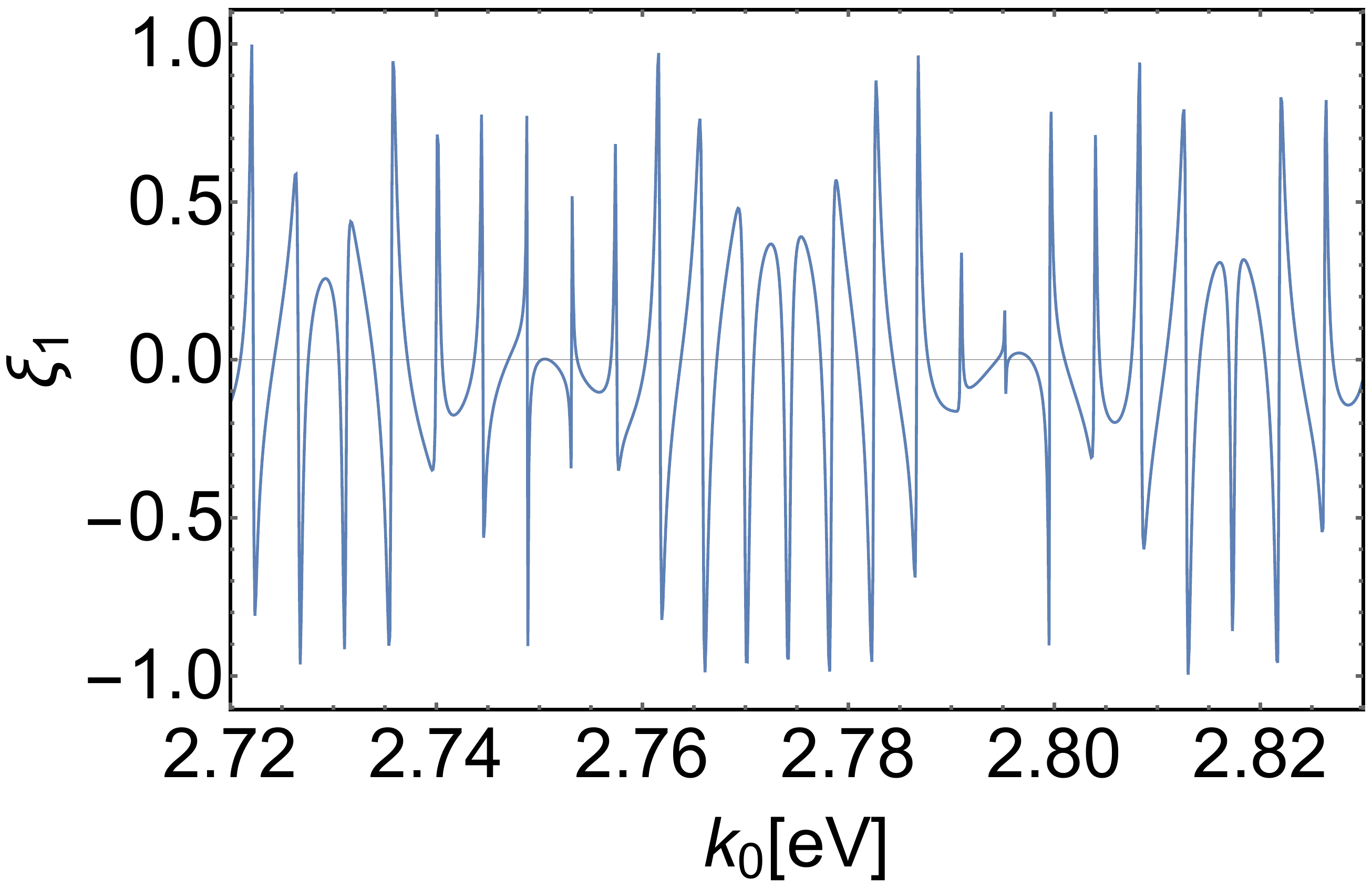}}\,
\raisebox{-0.5\height}{\includegraphics*[width=0.24\linewidth]{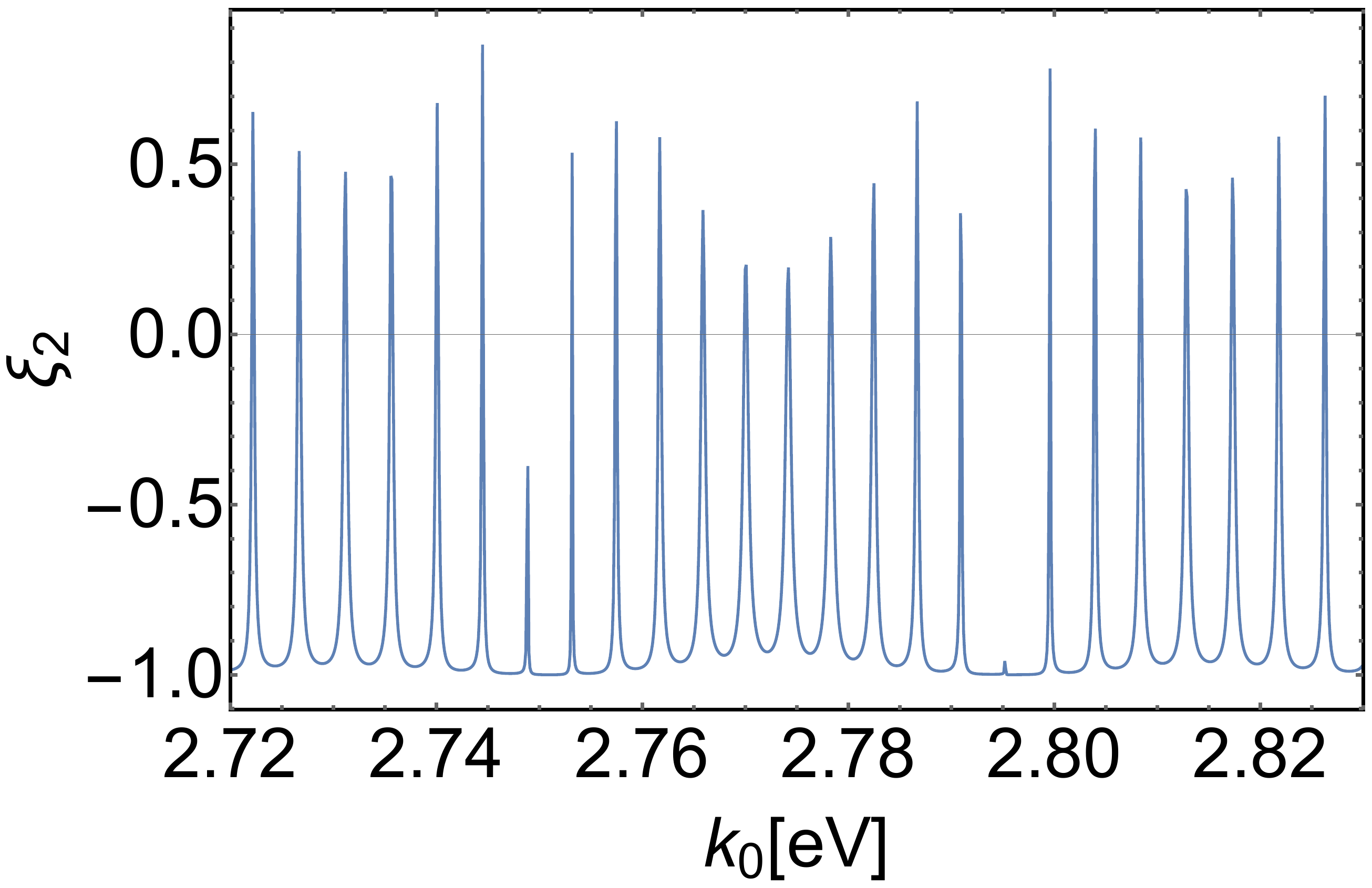}}\,
\raisebox{-0.5\height}{\includegraphics*[width=0.24\linewidth]{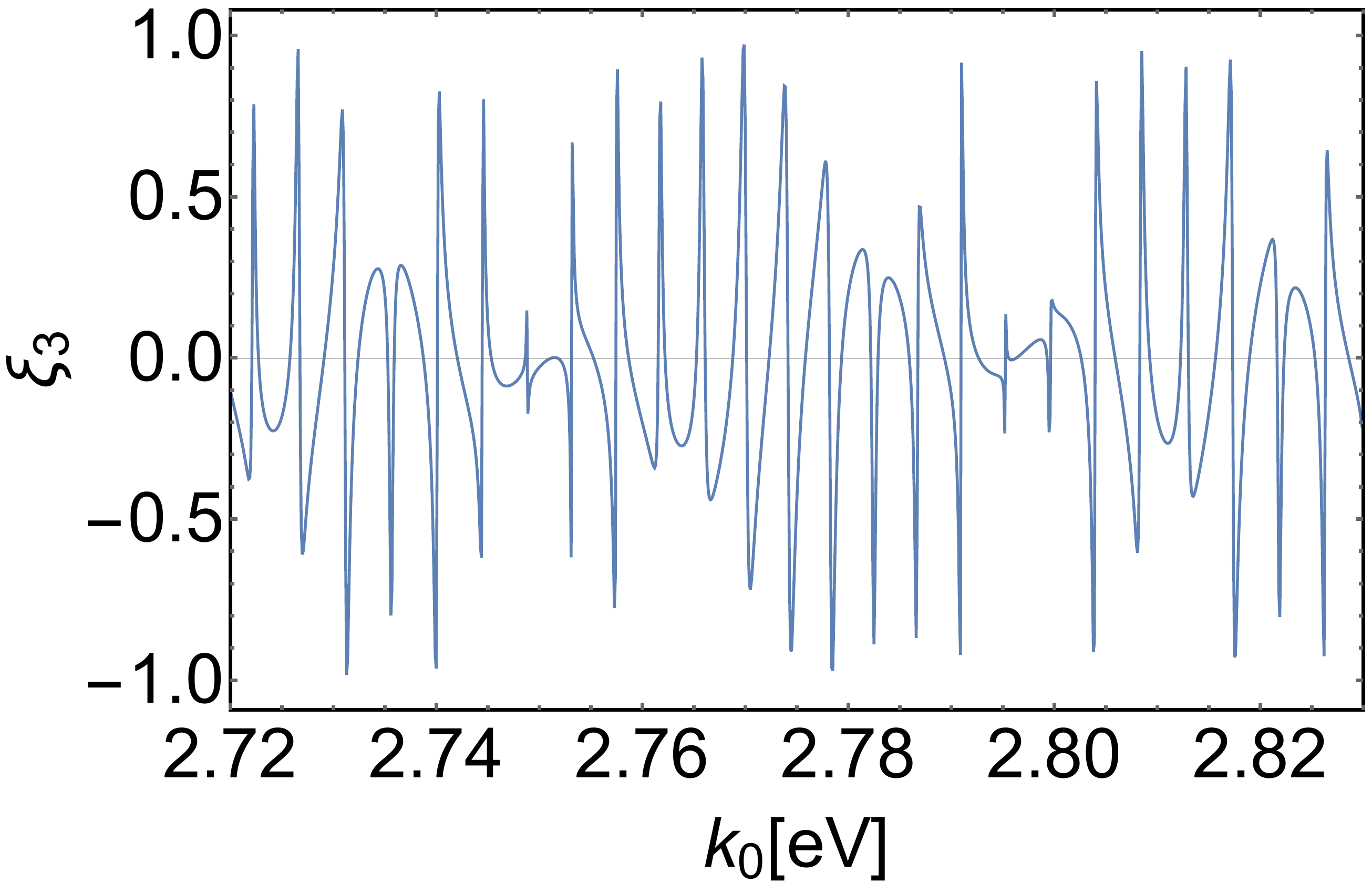}}\\
\raisebox{-0.5\height}{\includegraphics*[width=0.24\linewidth]{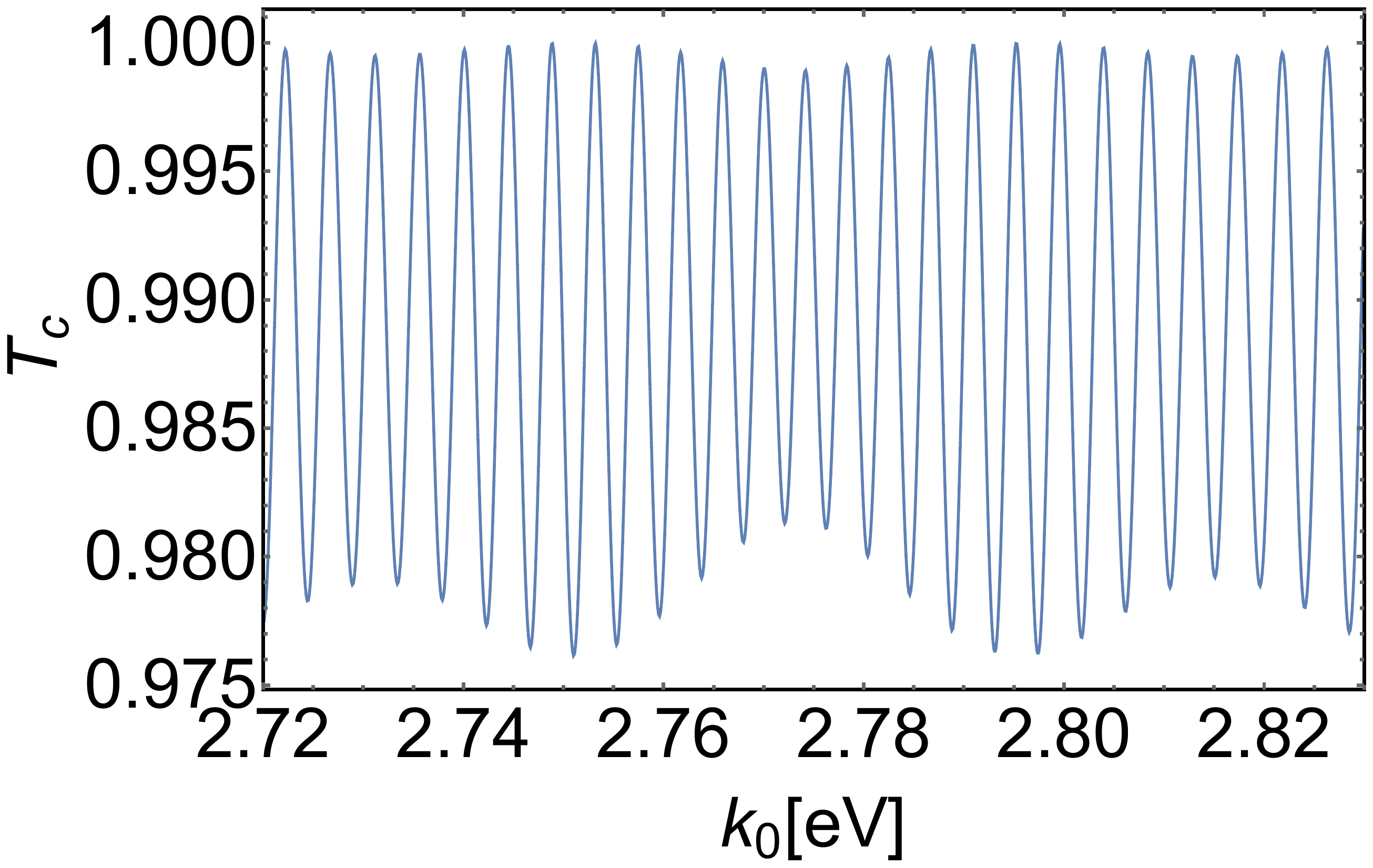}}\,
\raisebox{-0.5\height}{\includegraphics*[width=0.24\linewidth]{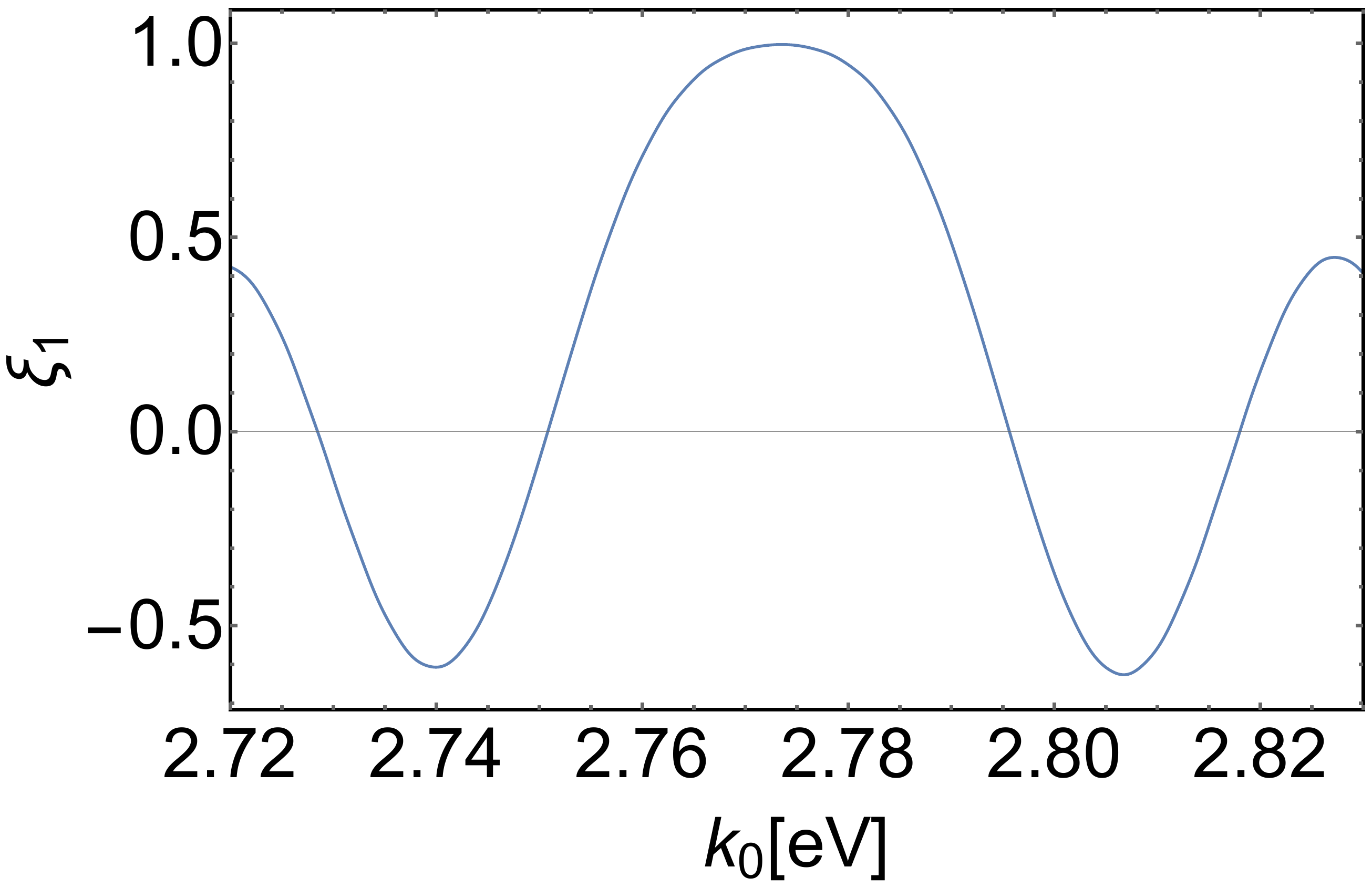}}\,
\raisebox{-0.5\height}{\includegraphics*[width=0.24\linewidth]{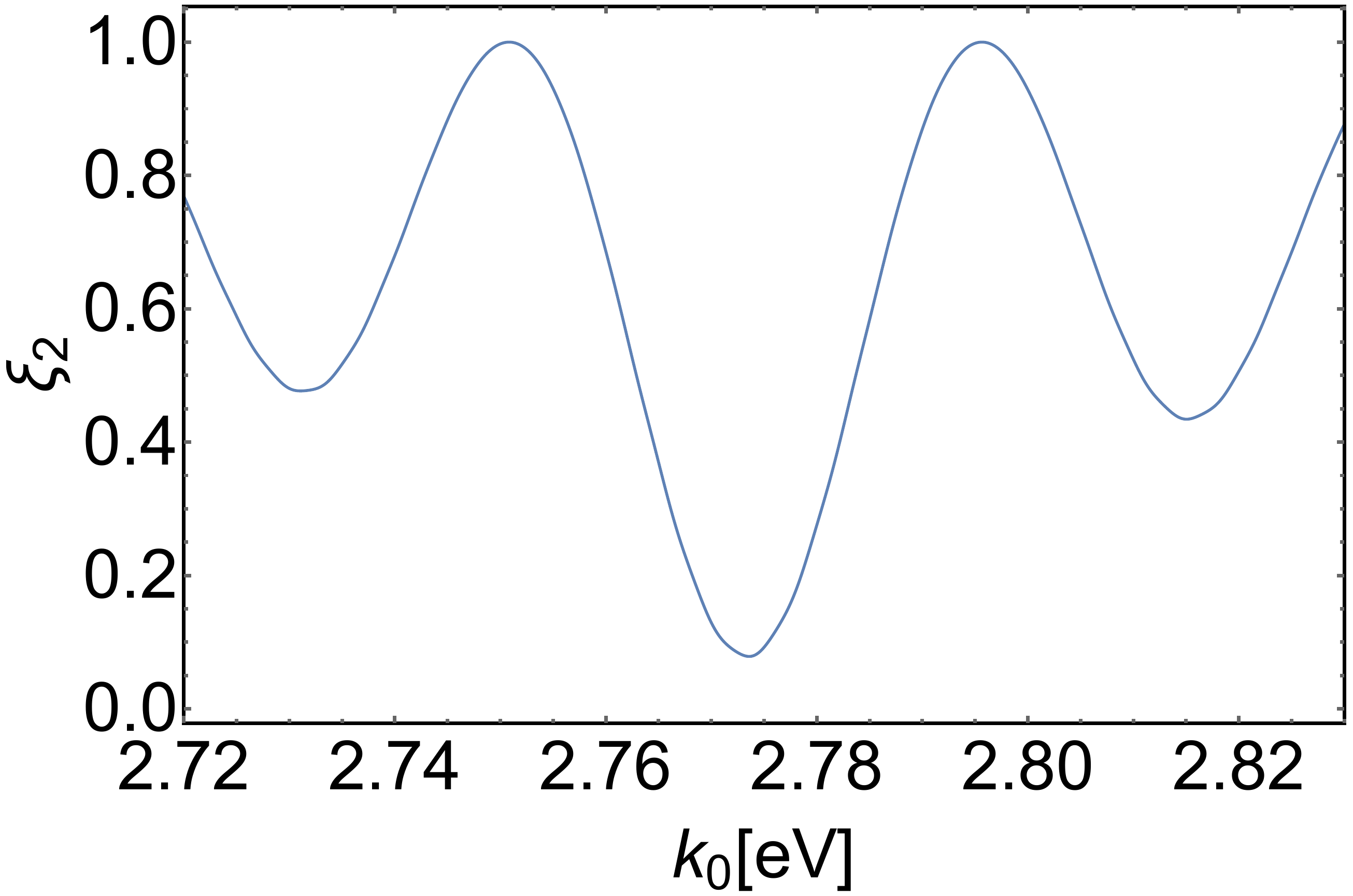}}\,
\raisebox{-0.5\height}{\includegraphics*[width=0.24\linewidth]{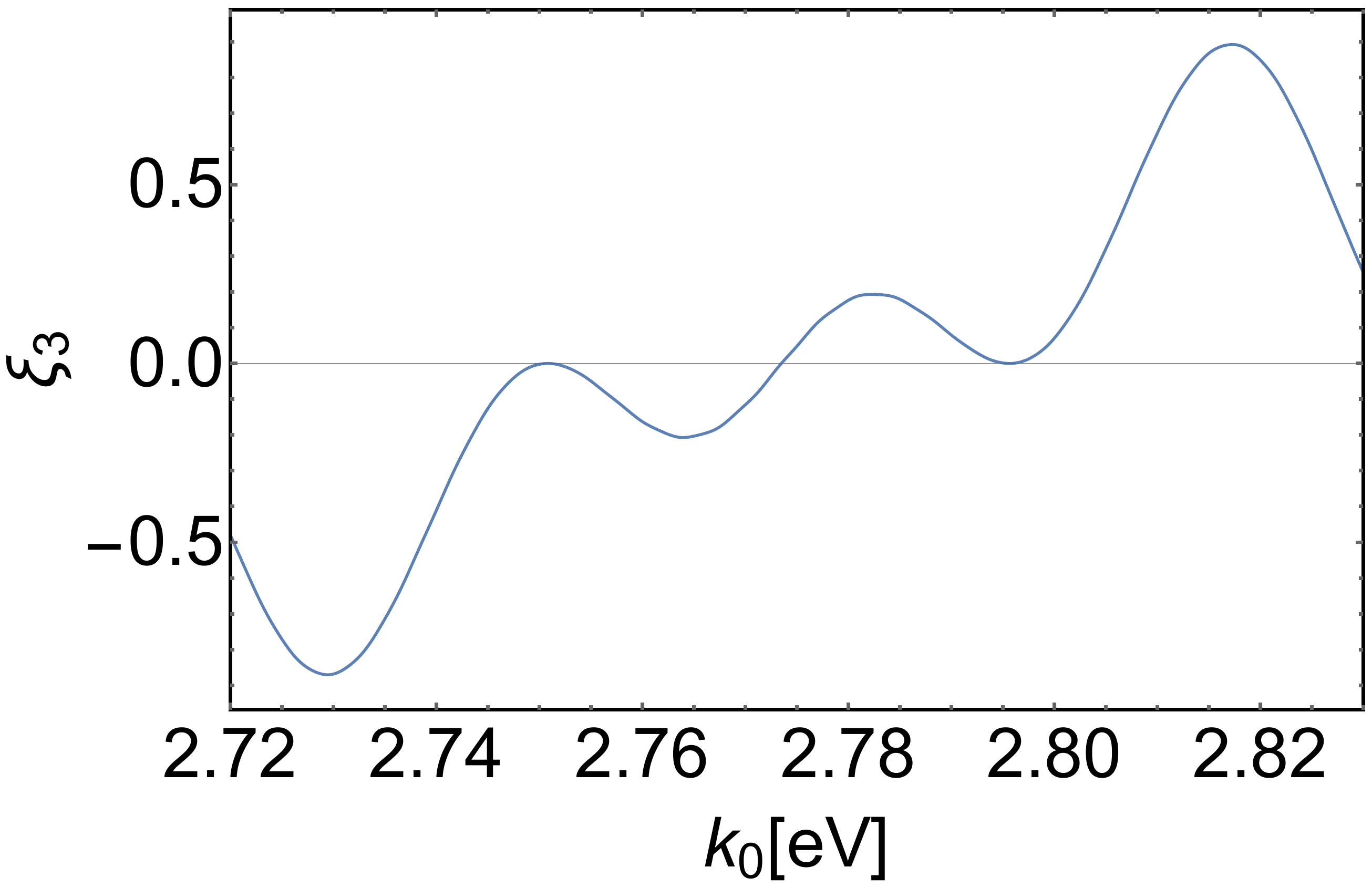}}\\
\raisebox{-0.5\height}{\includegraphics*[width=0.24\linewidth]{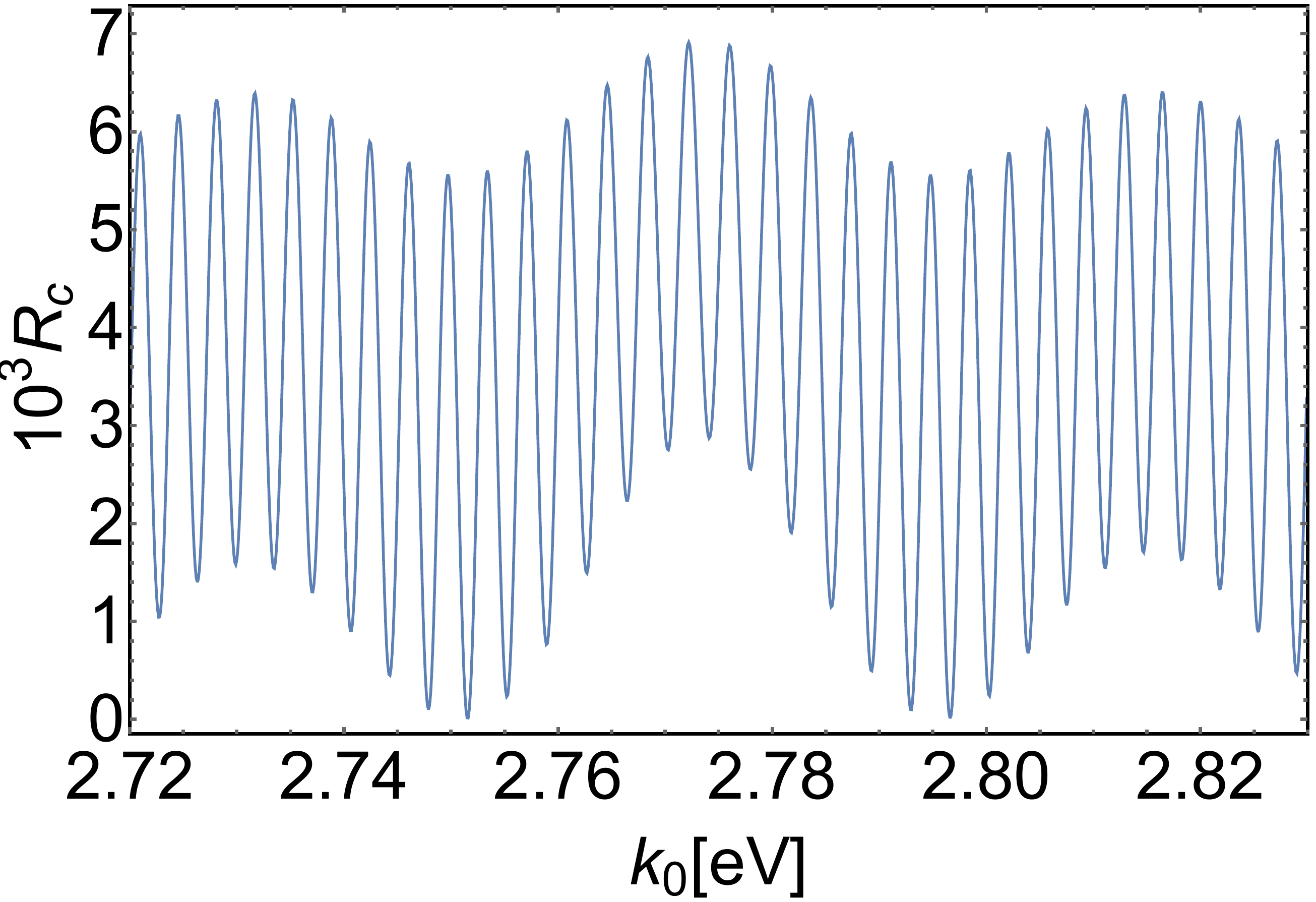}}\,
\raisebox{-0.5\height}{\includegraphics*[width=0.24\linewidth]{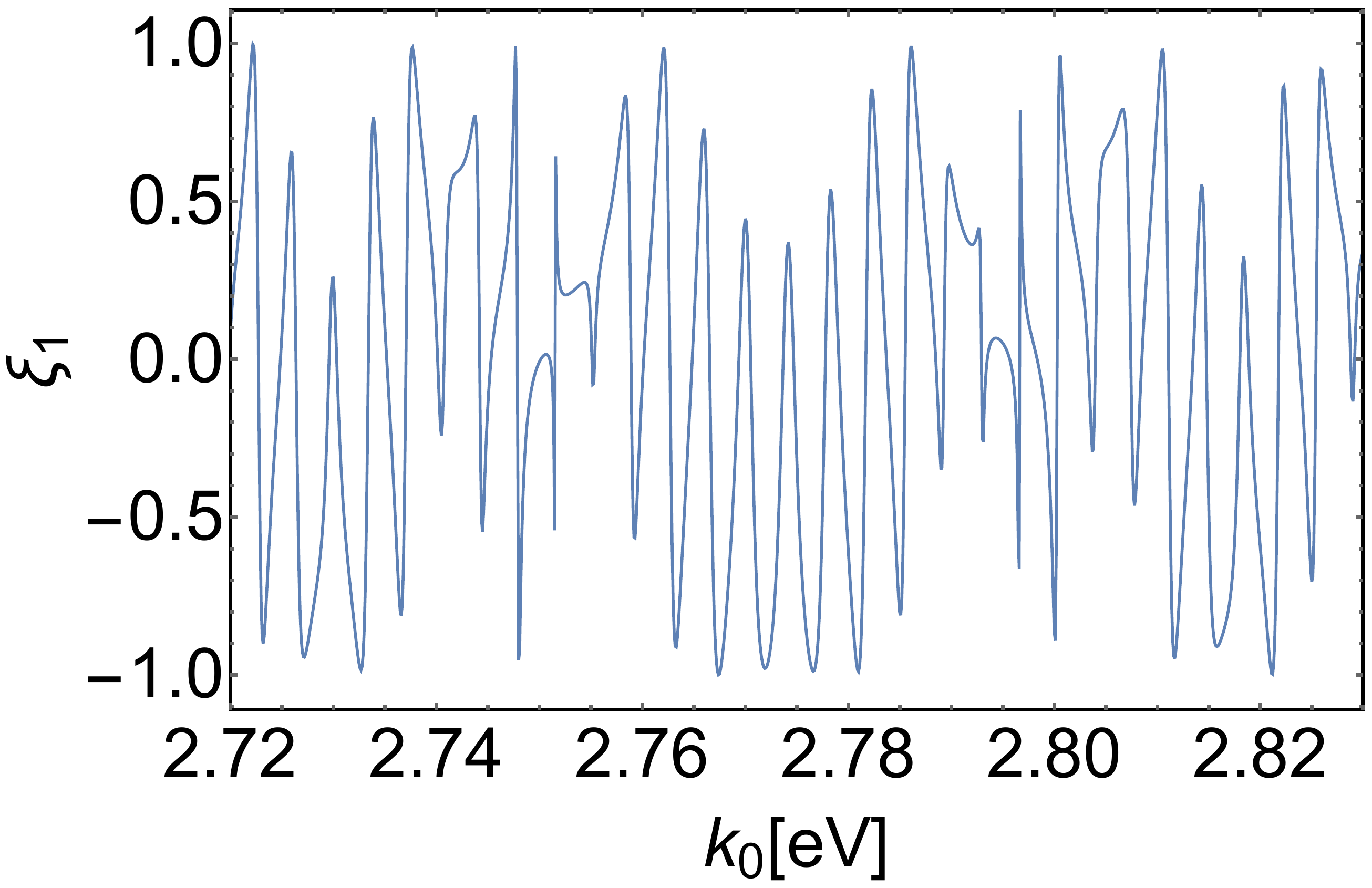}}\,
\raisebox{-0.5\height}{\includegraphics*[width=0.24\linewidth]{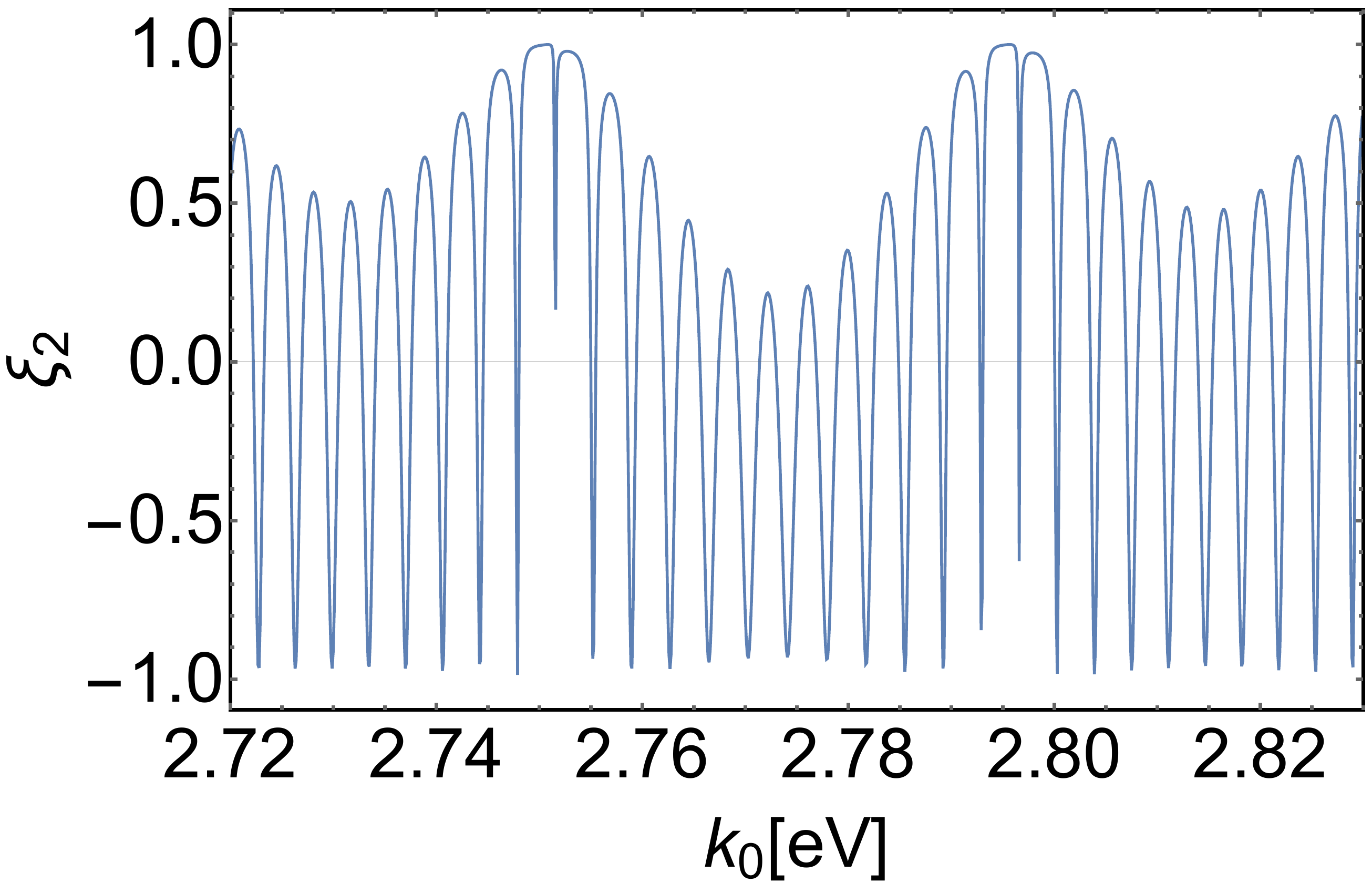}}\,
\raisebox{-0.5\height}{\includegraphics*[width=0.24\linewidth]{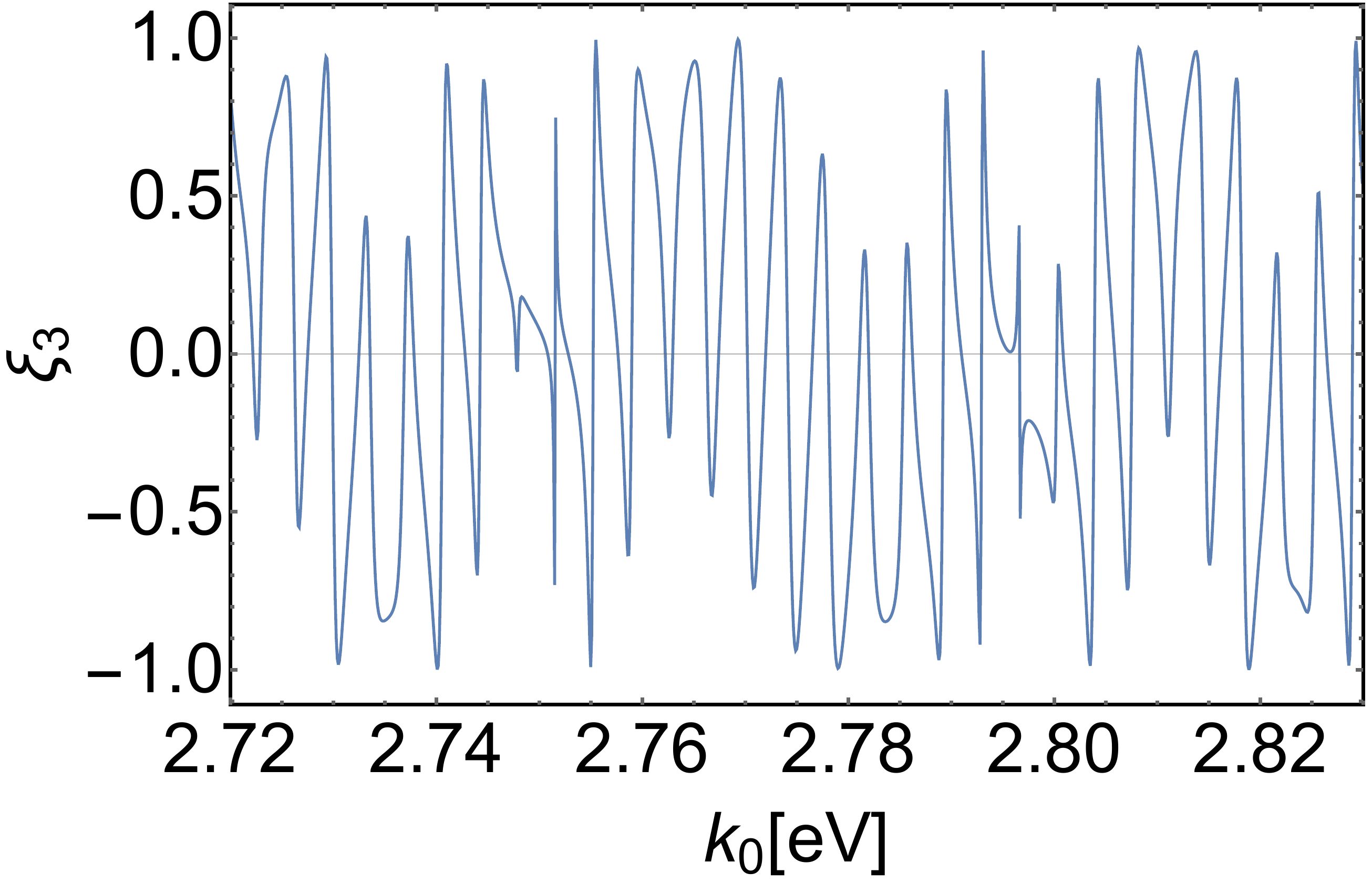}}\\
\raisebox{-0.5\height}{\includegraphics*[width=0.24\linewidth]{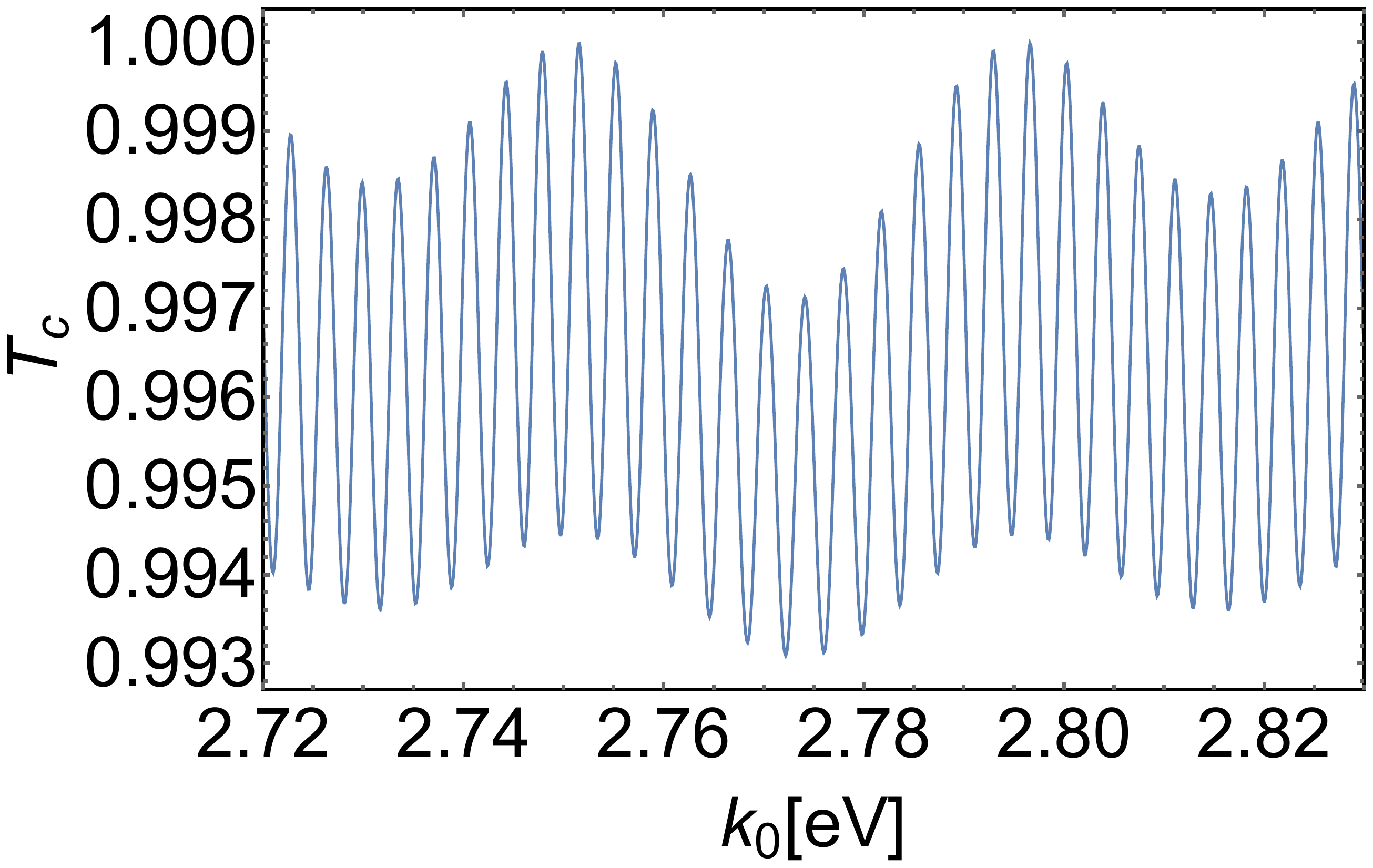}}\,
\raisebox{-0.5\height}{\includegraphics*[width=0.24\linewidth]{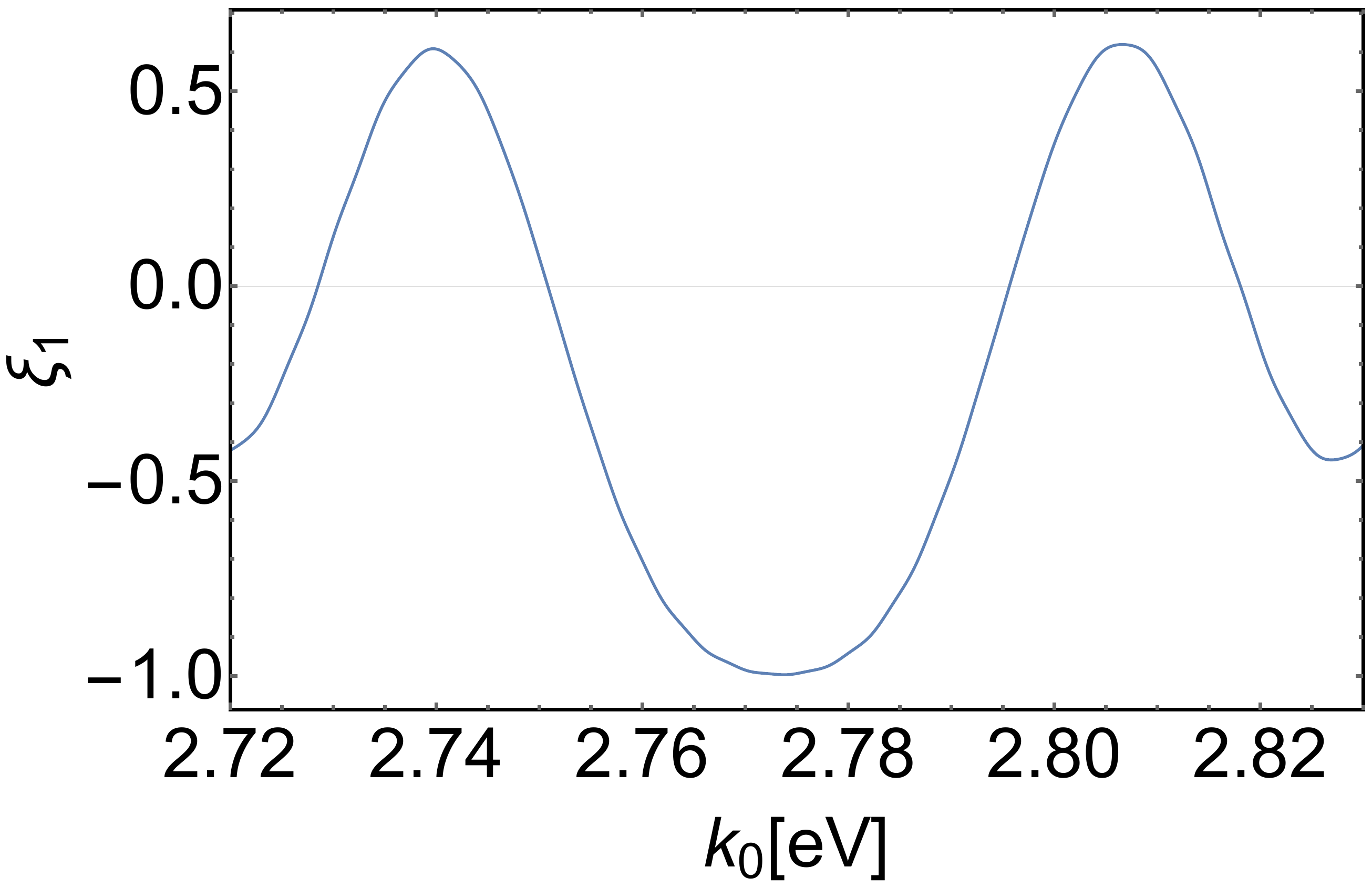}}\,
\raisebox{-0.5\height}{\includegraphics*[width=0.24\linewidth]{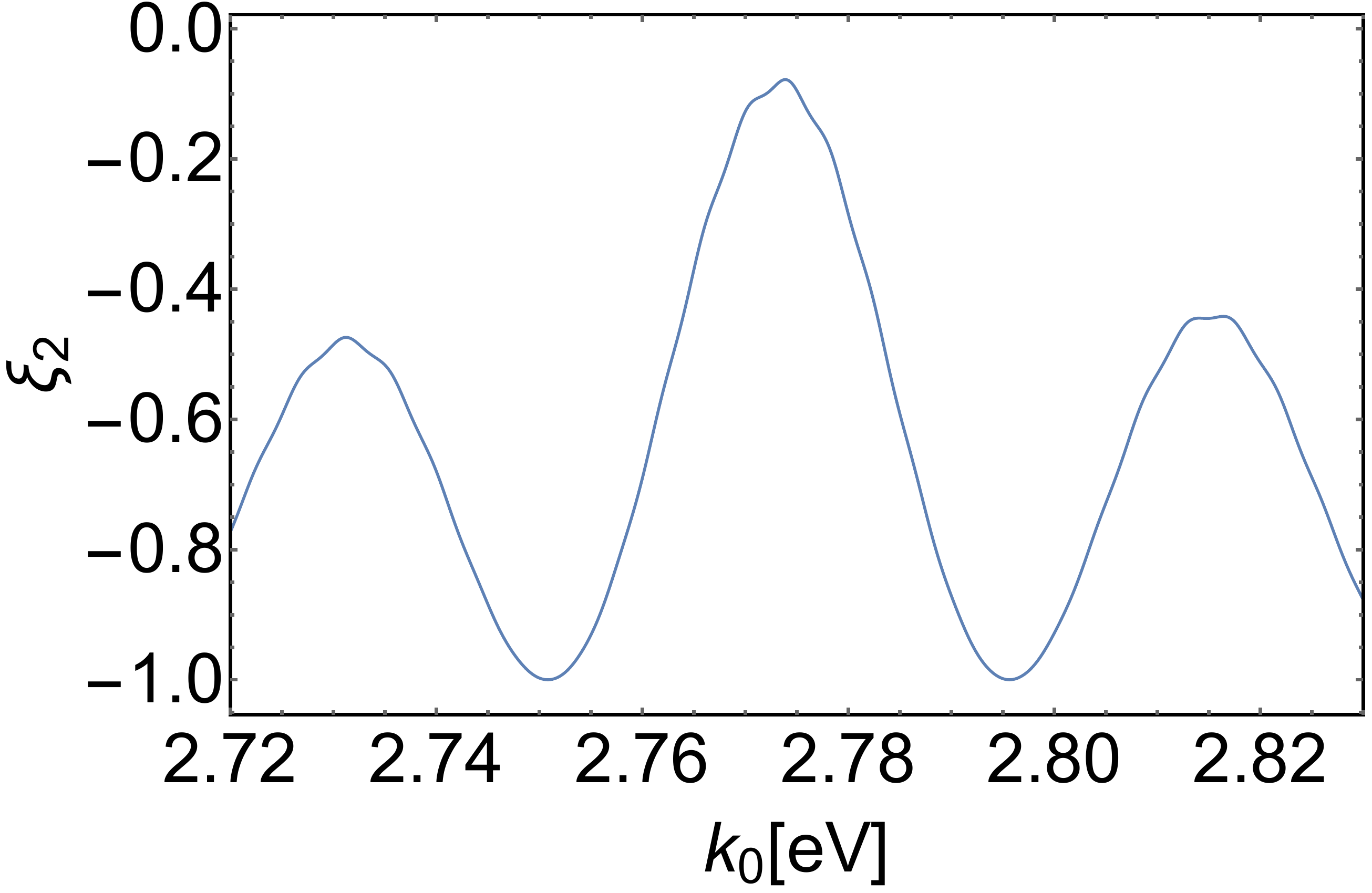}}\,
\raisebox{-0.5\height}{\includegraphics*[width=0.24\linewidth]{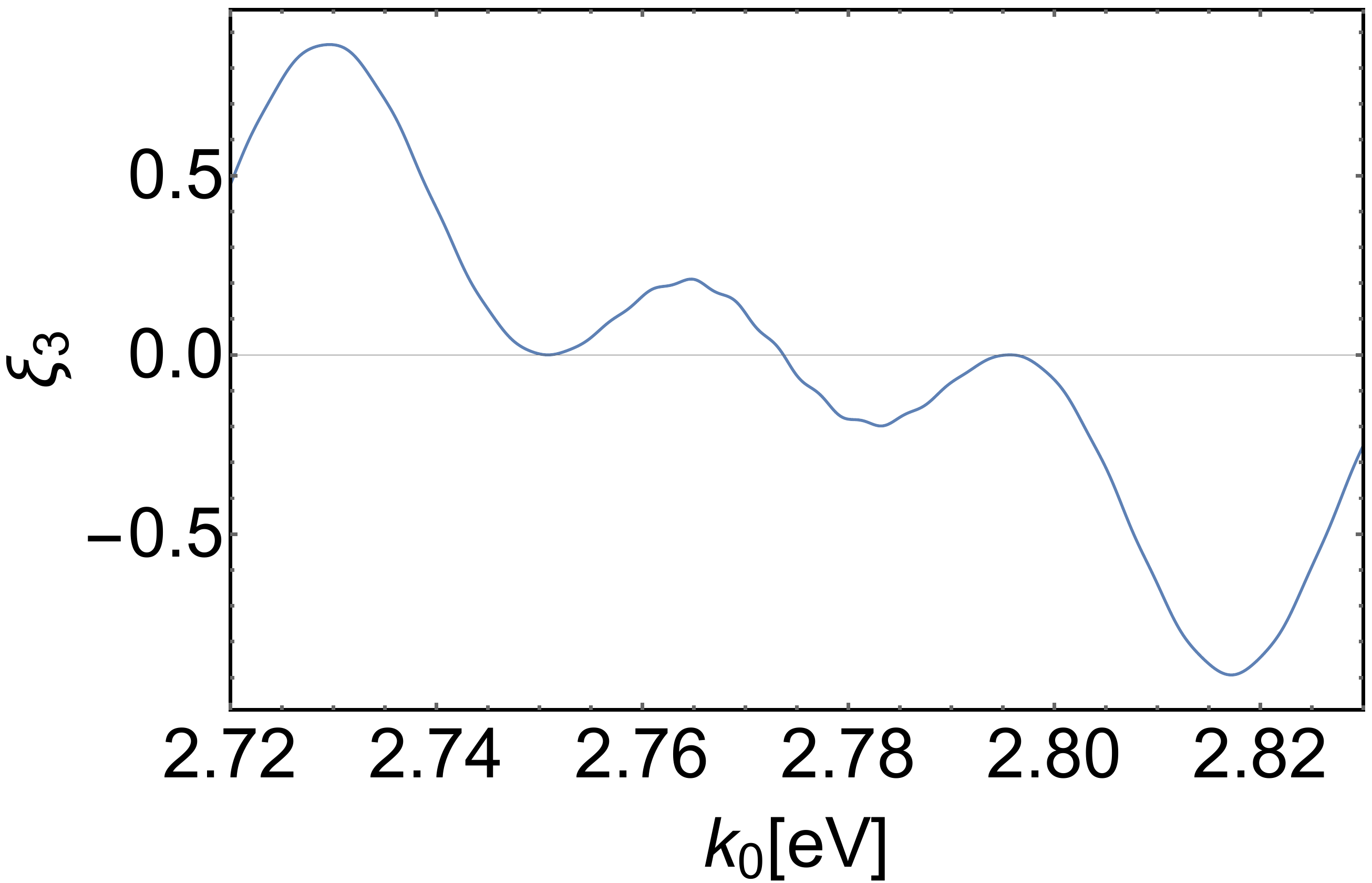}}\\
\raisebox{-0.5\height}{\includegraphics*[width=0.24\linewidth]{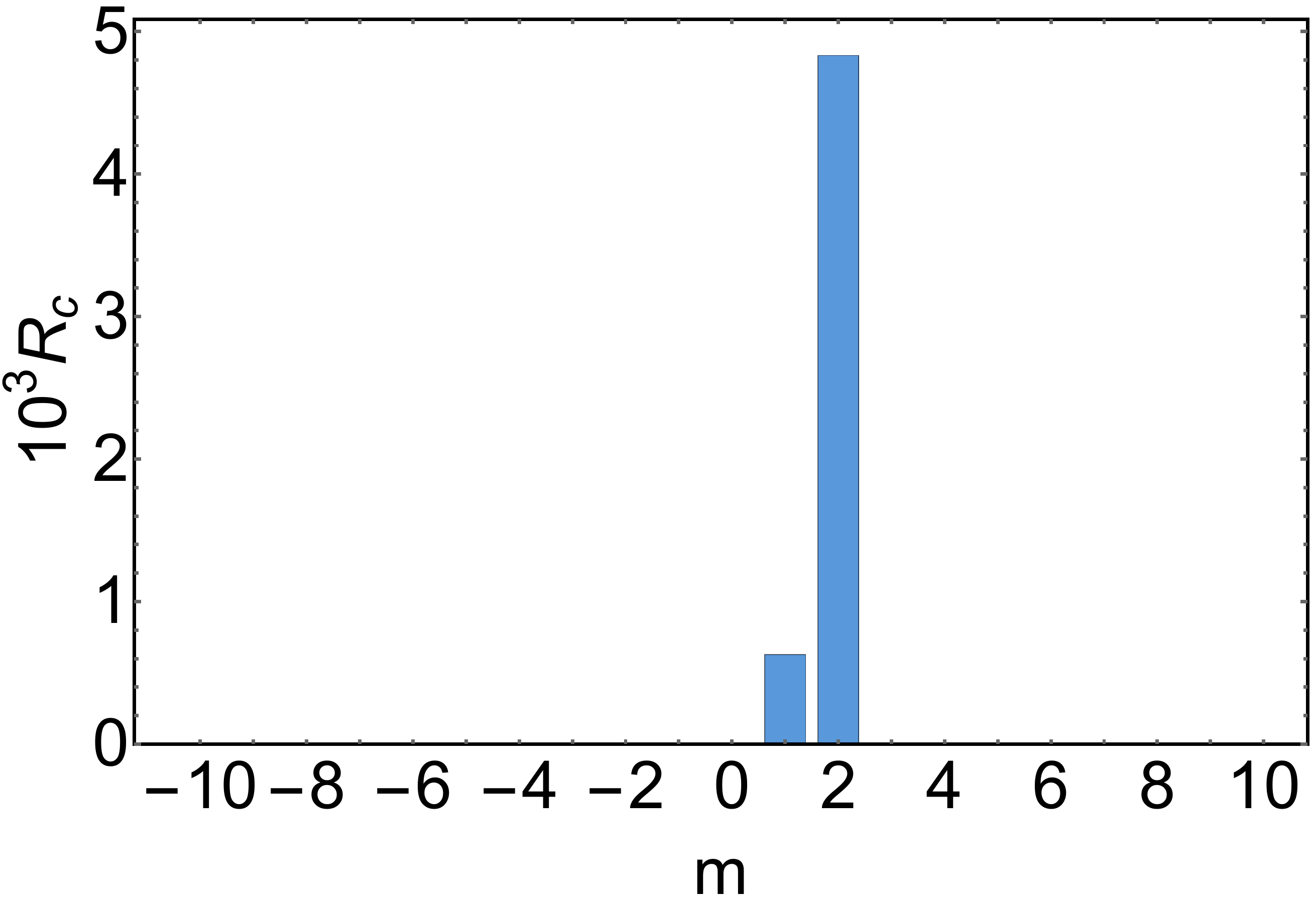}}\,
\raisebox{-0.5\height}{\includegraphics*[width=0.24\linewidth]{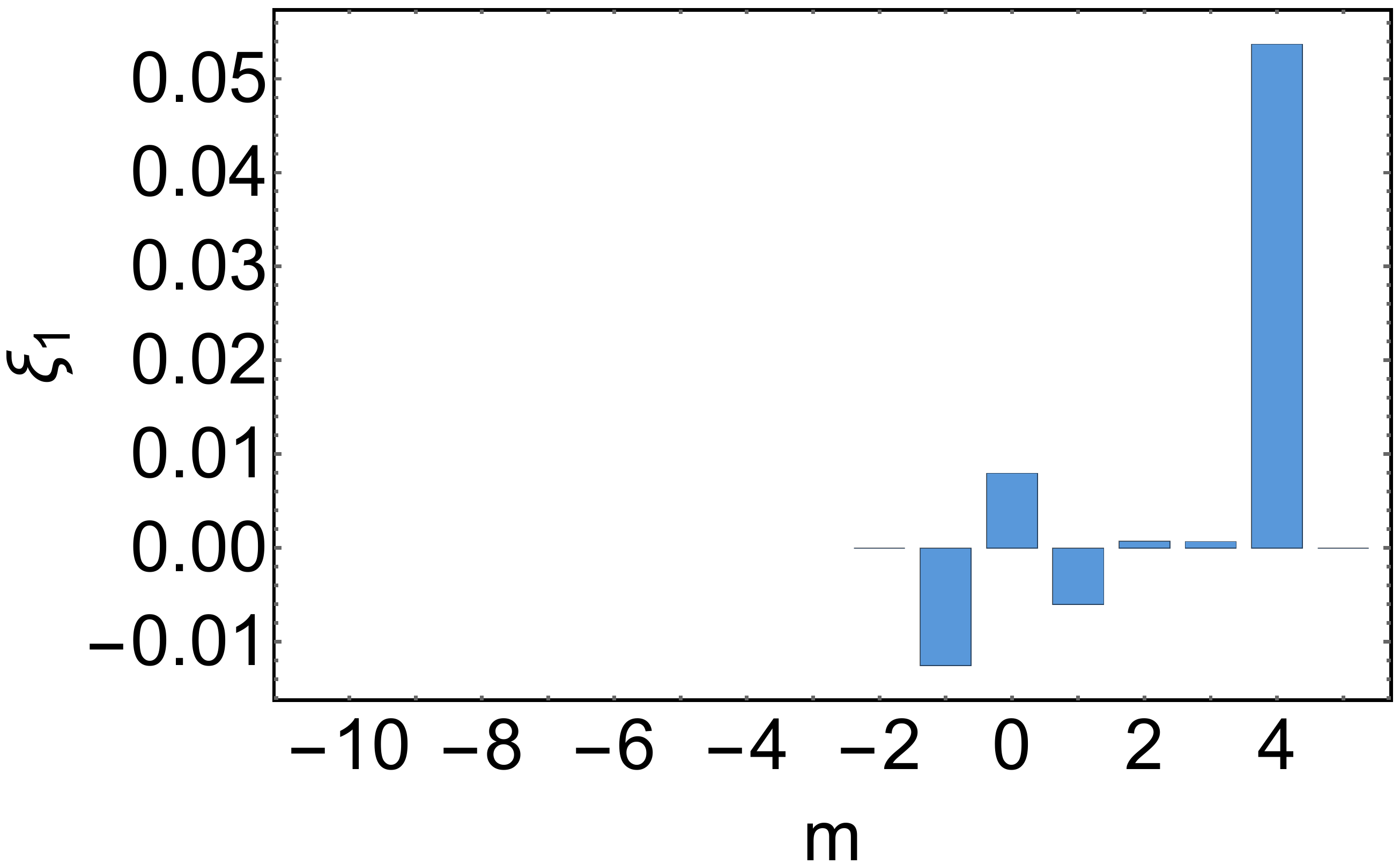}}\,
\raisebox{-0.5\height}{\includegraphics*[width=0.24\linewidth]{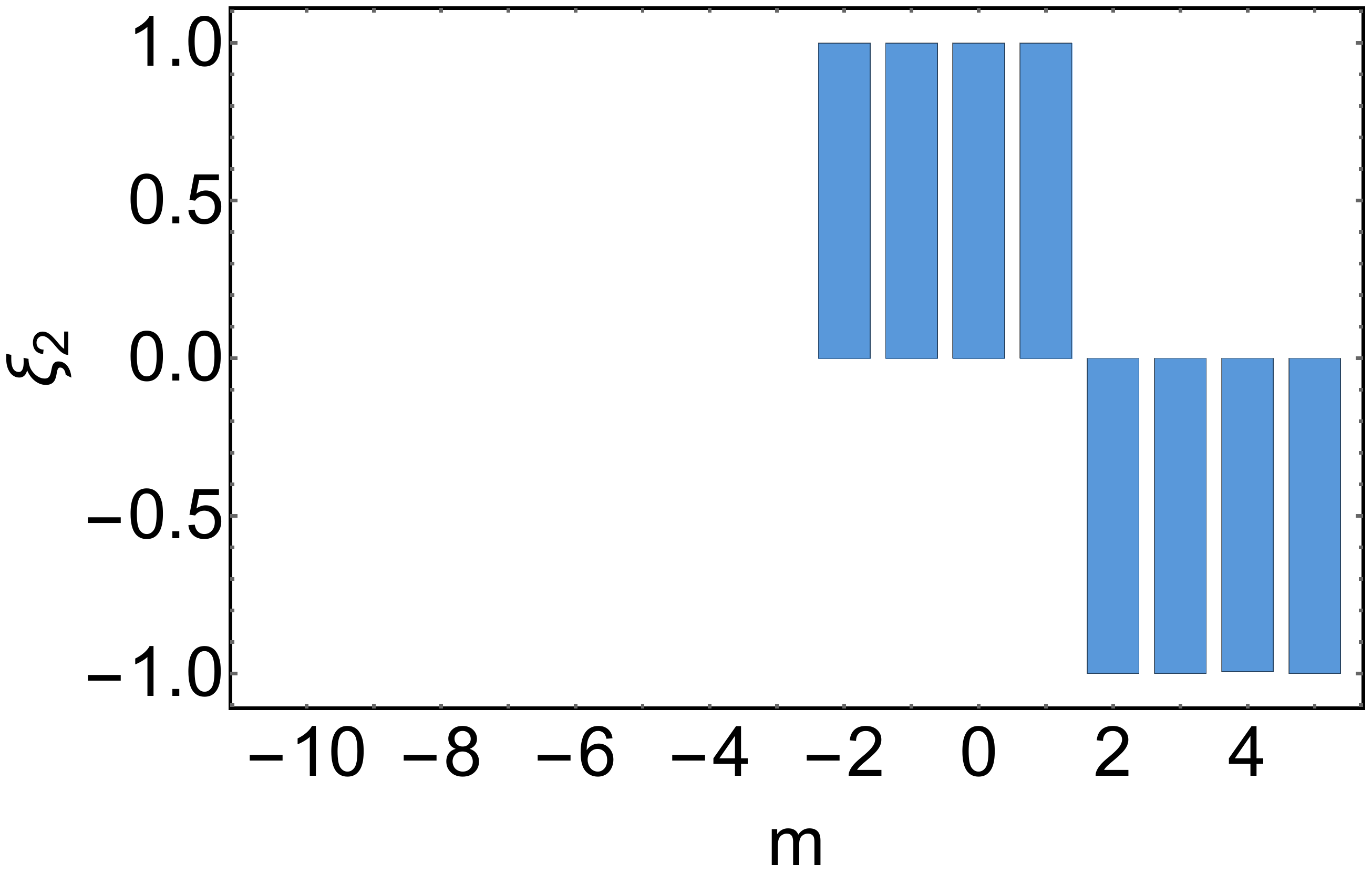}}\,
\raisebox{-0.5\height}{\includegraphics*[width=0.24\linewidth]{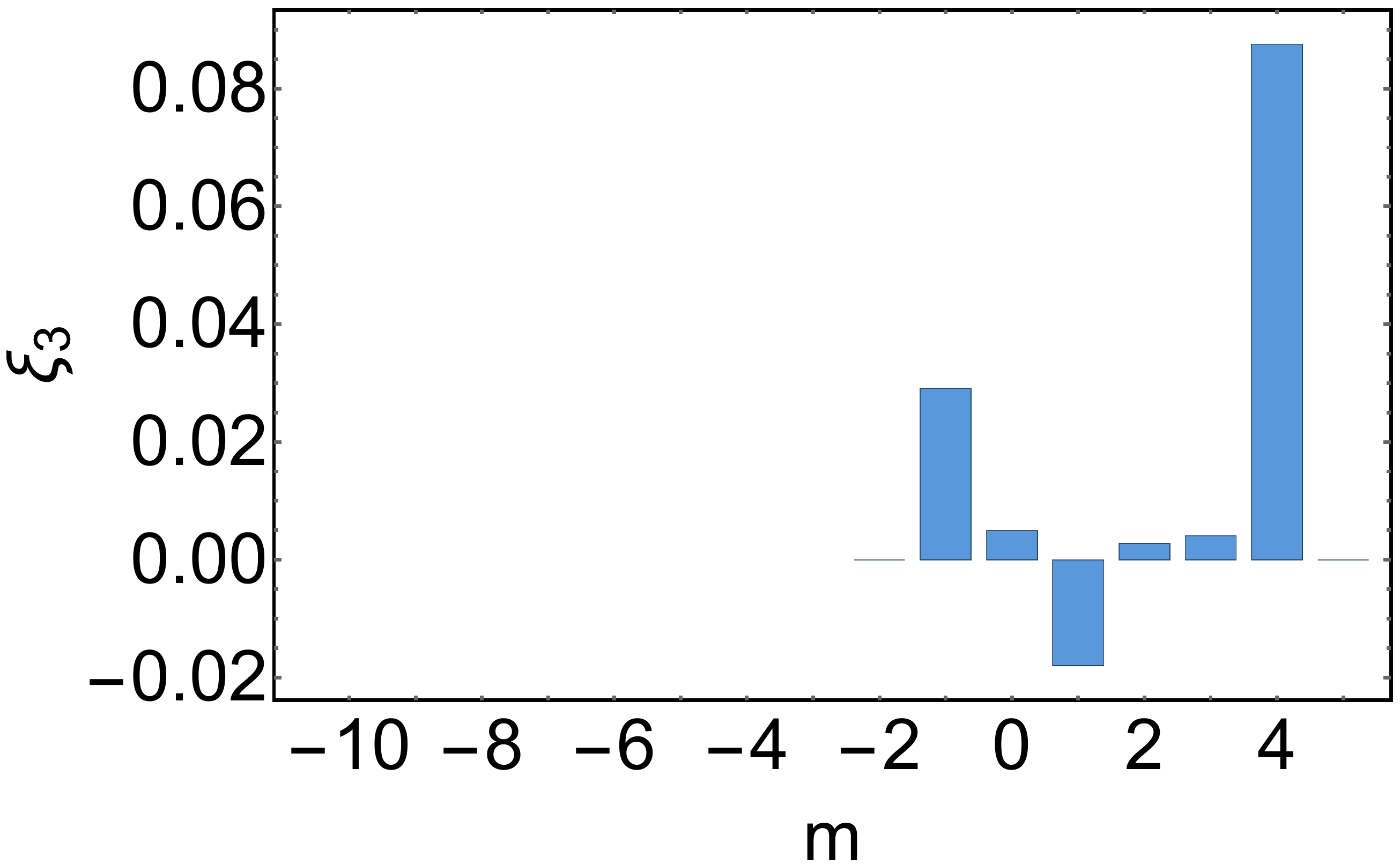}}\\
\raisebox{-0.5\height}{\includegraphics*[width=0.24\linewidth]{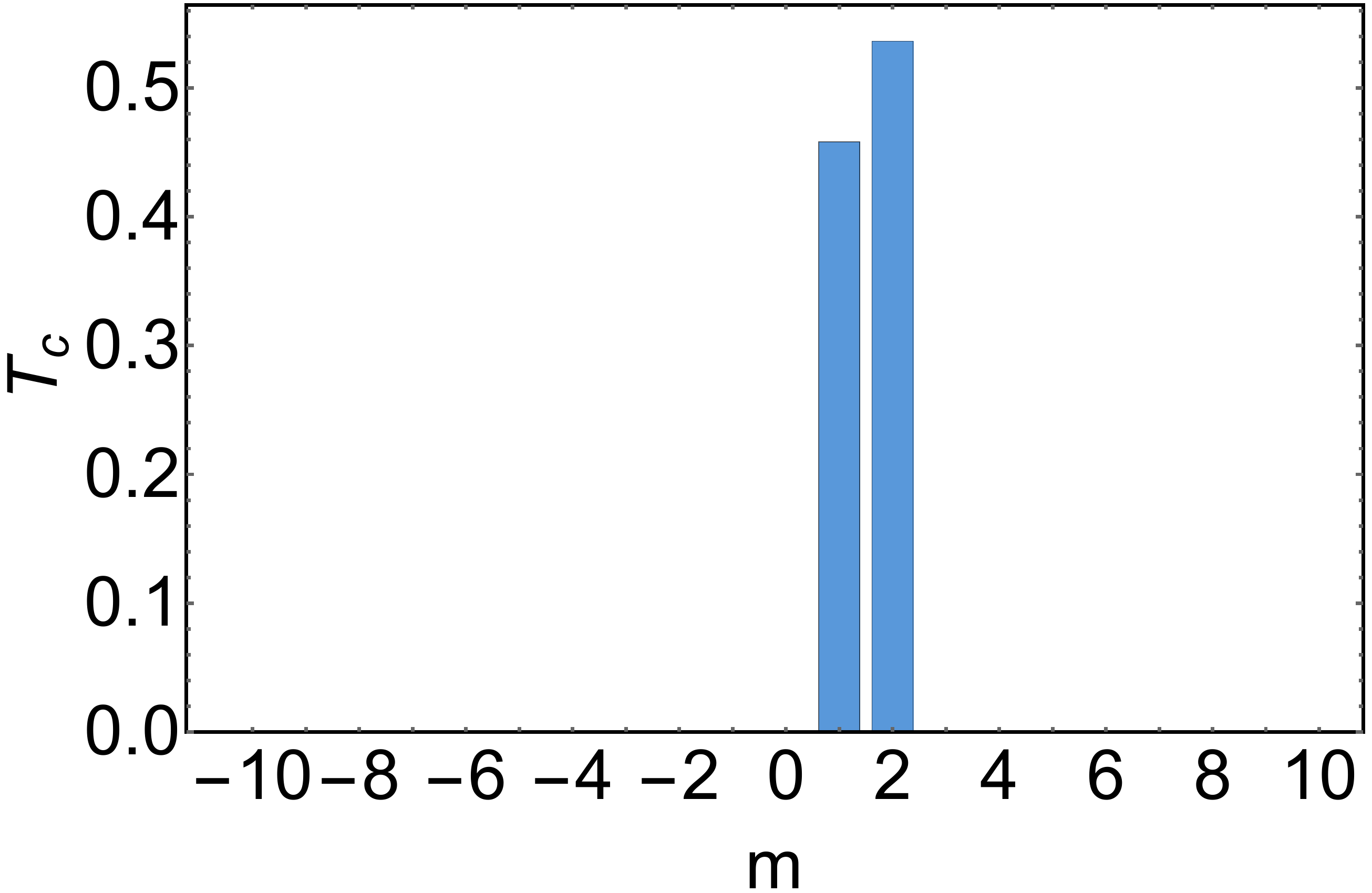}}\,
\raisebox{-0.5\height}{\includegraphics*[width=0.24\linewidth]{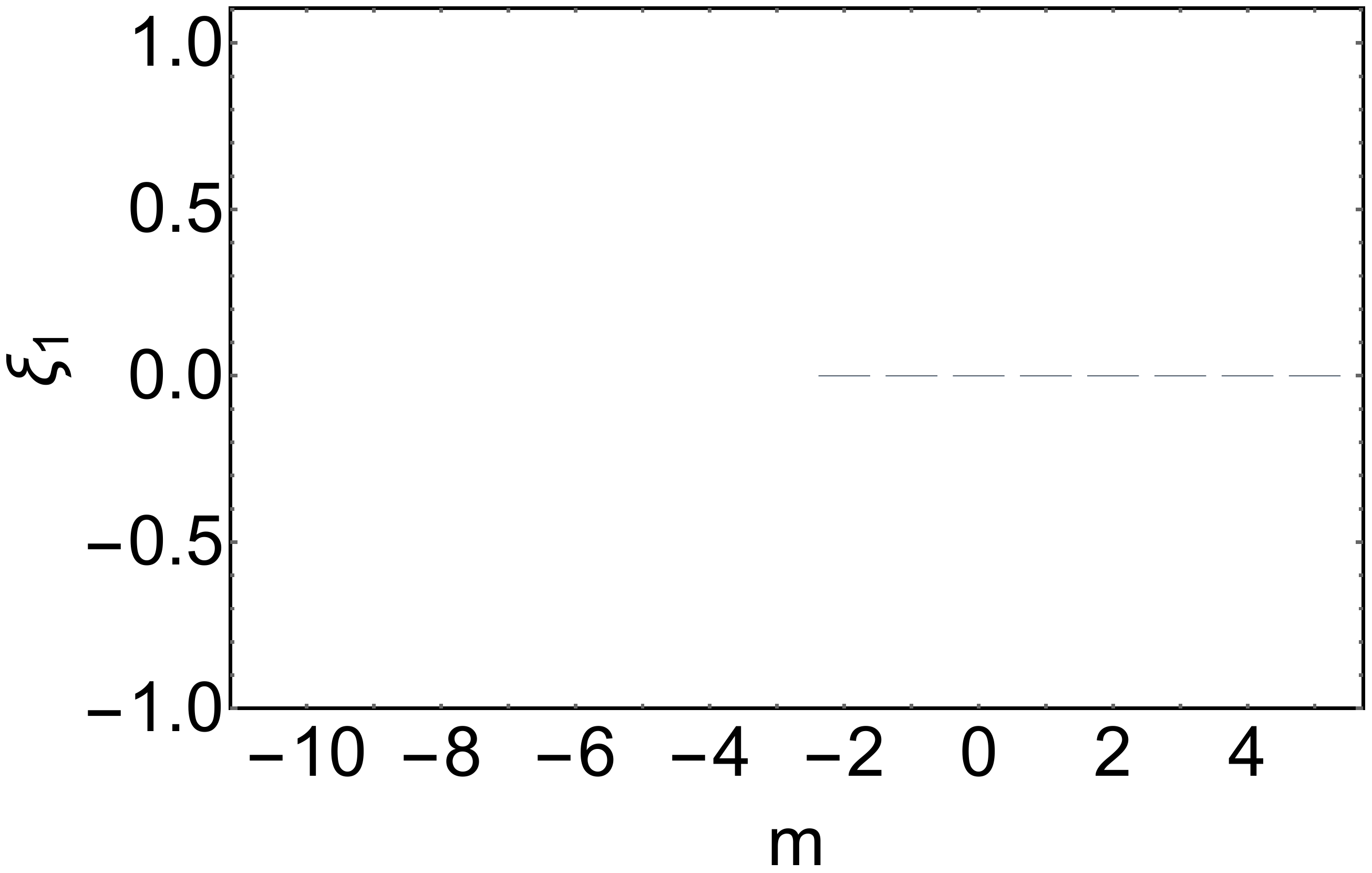}}\,
\raisebox{-0.5\height}{\includegraphics*[width=0.24\linewidth]{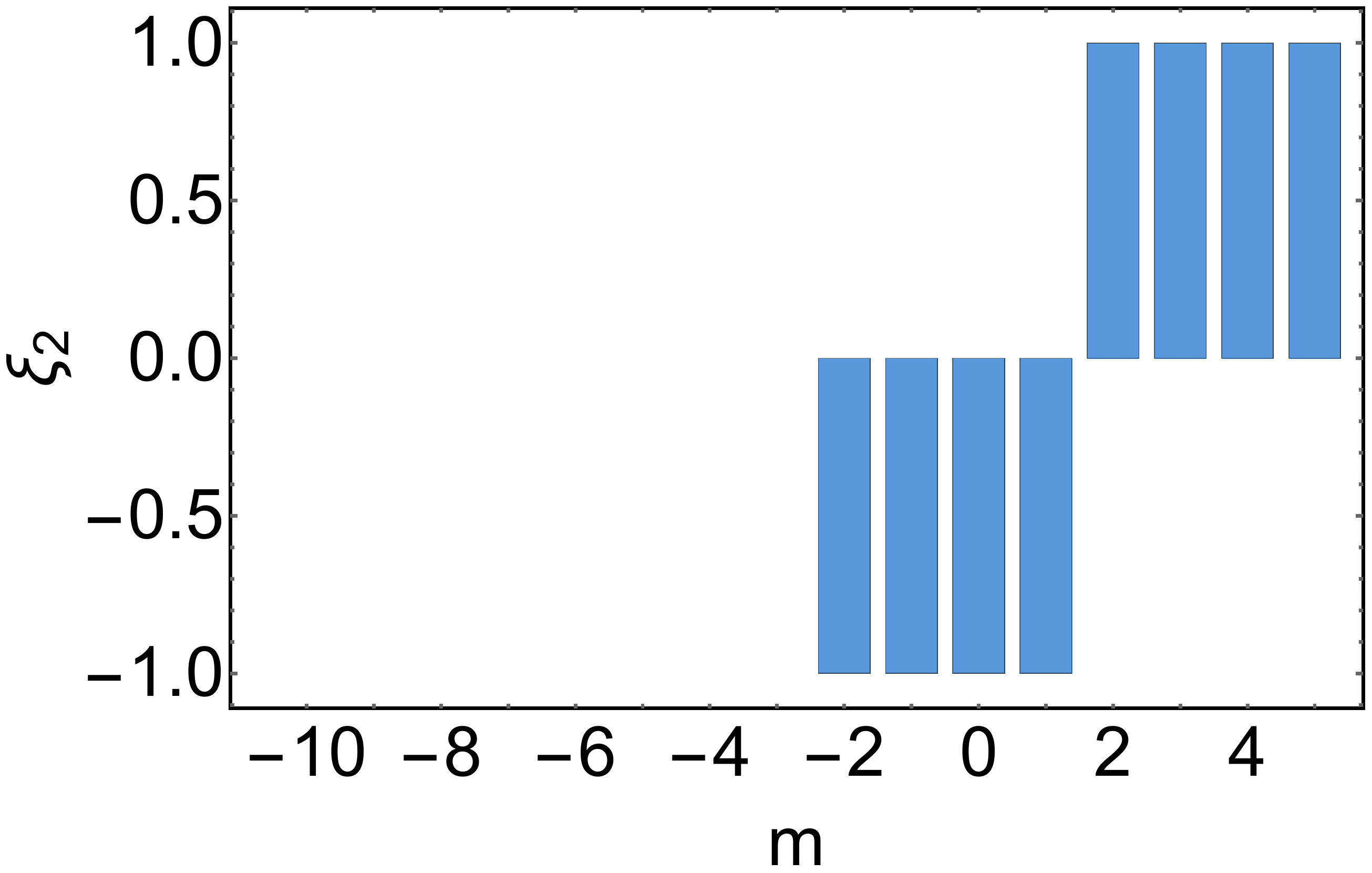}}\,
\raisebox{-0.5\height}{\includegraphics*[width=0.24\linewidth]{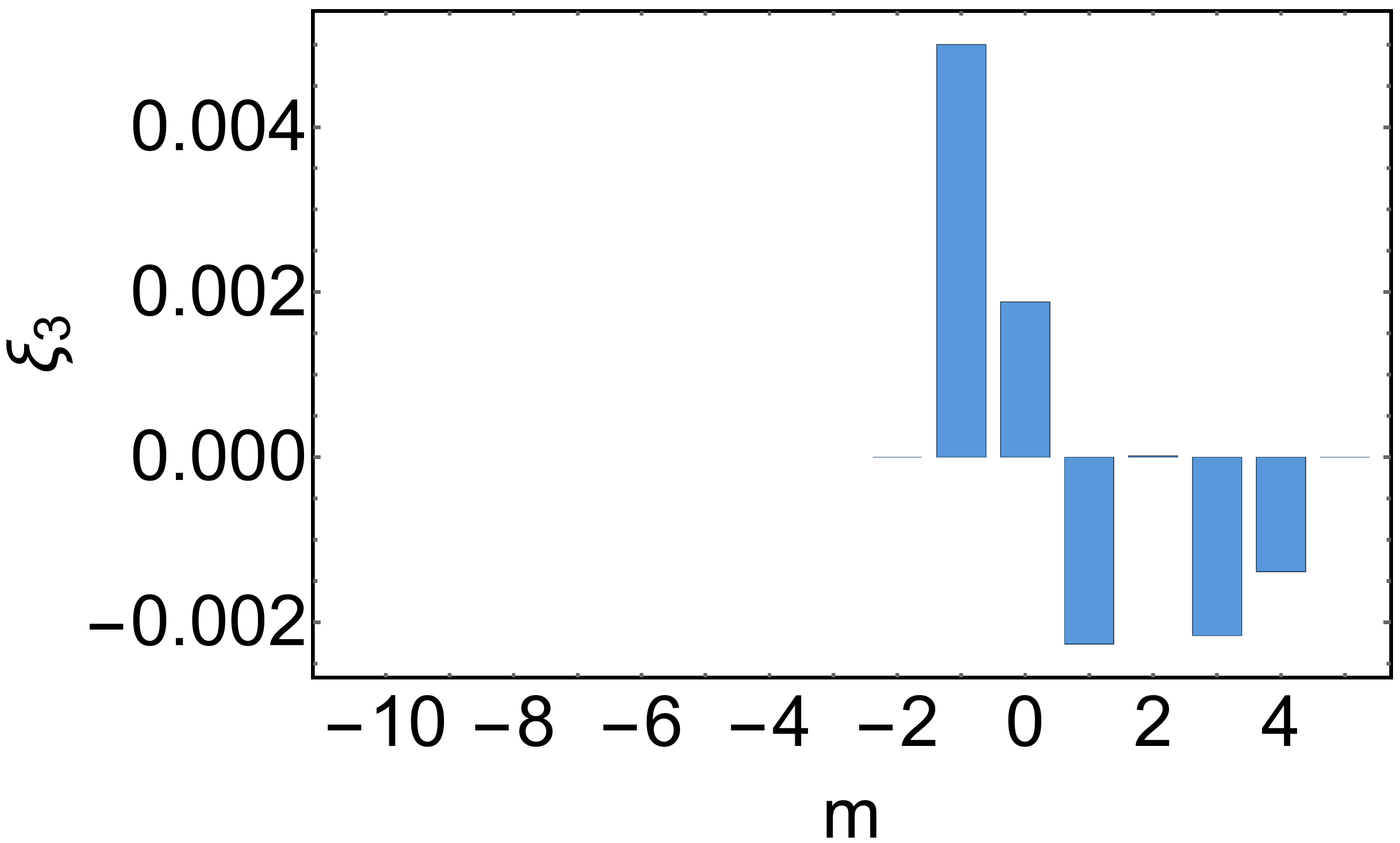}}\\
\raisebox{-0.5\height}{\includegraphics*[width=0.24\linewidth]{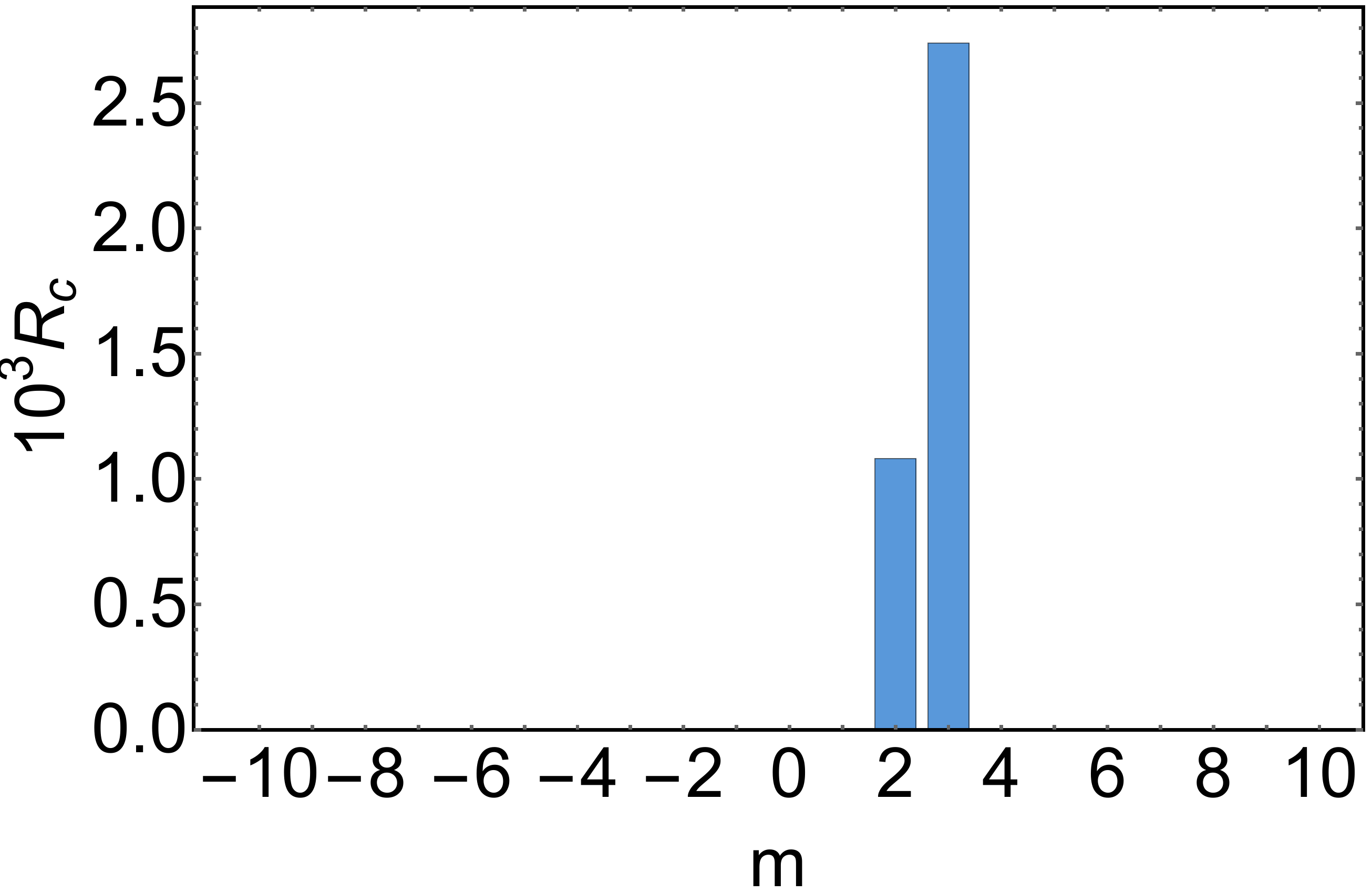}}\,
\raisebox{-0.5\height}{\includegraphics*[width=0.24\linewidth]{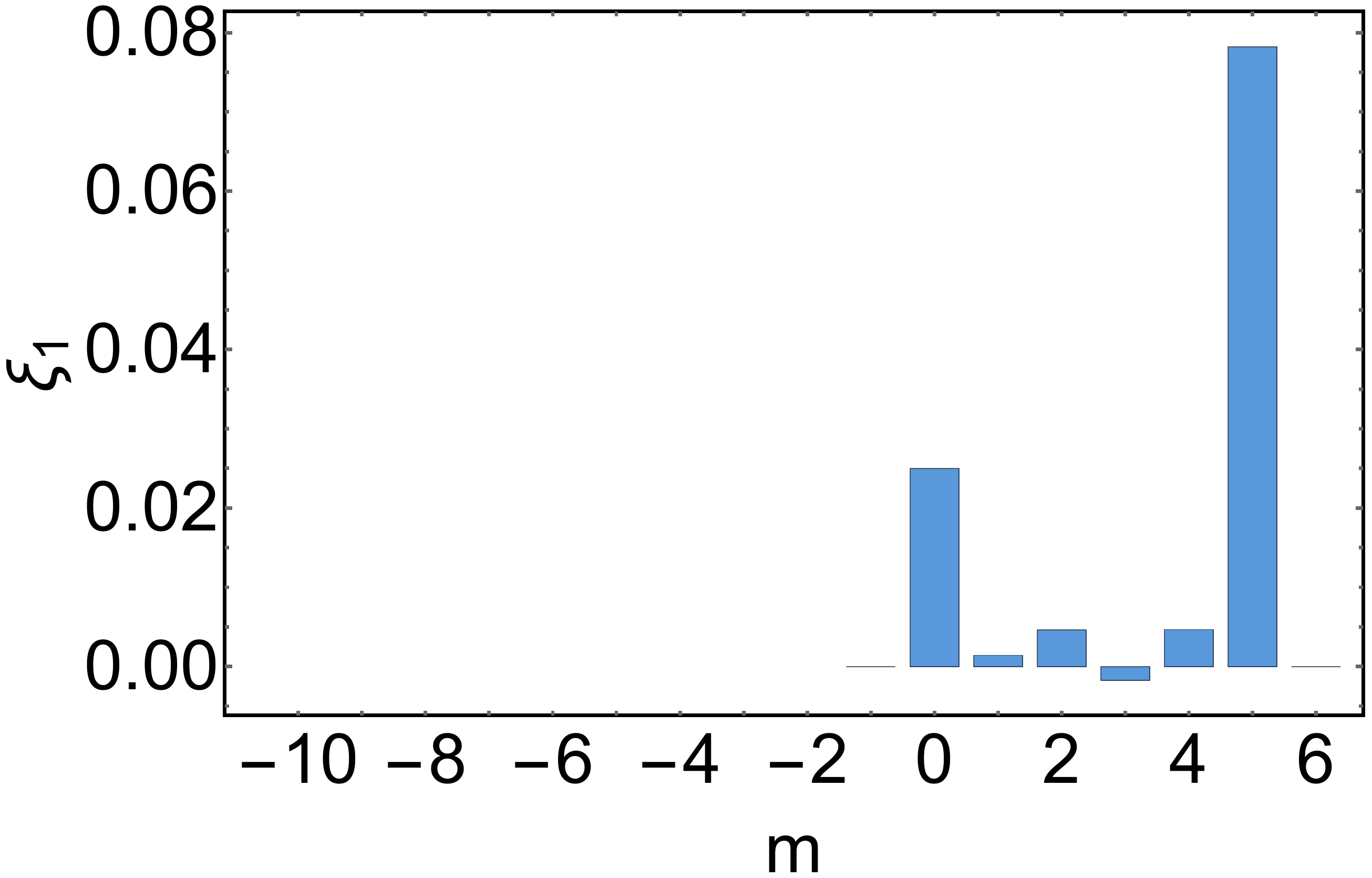}}\,
\raisebox{-0.5\height}{\includegraphics*[width=0.24\linewidth]{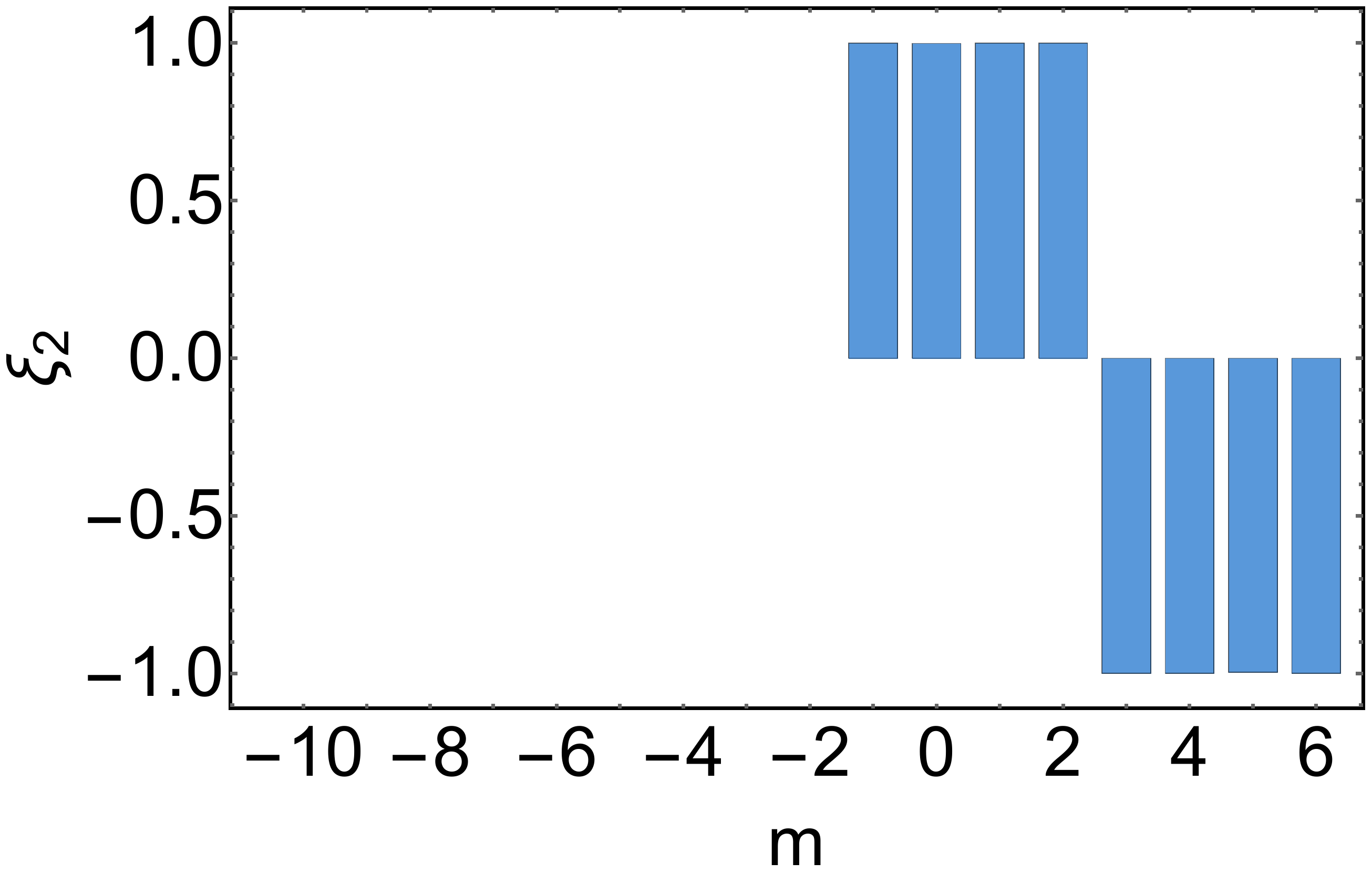}}\,
\raisebox{-0.5\height}{\includegraphics*[width=0.24\linewidth]{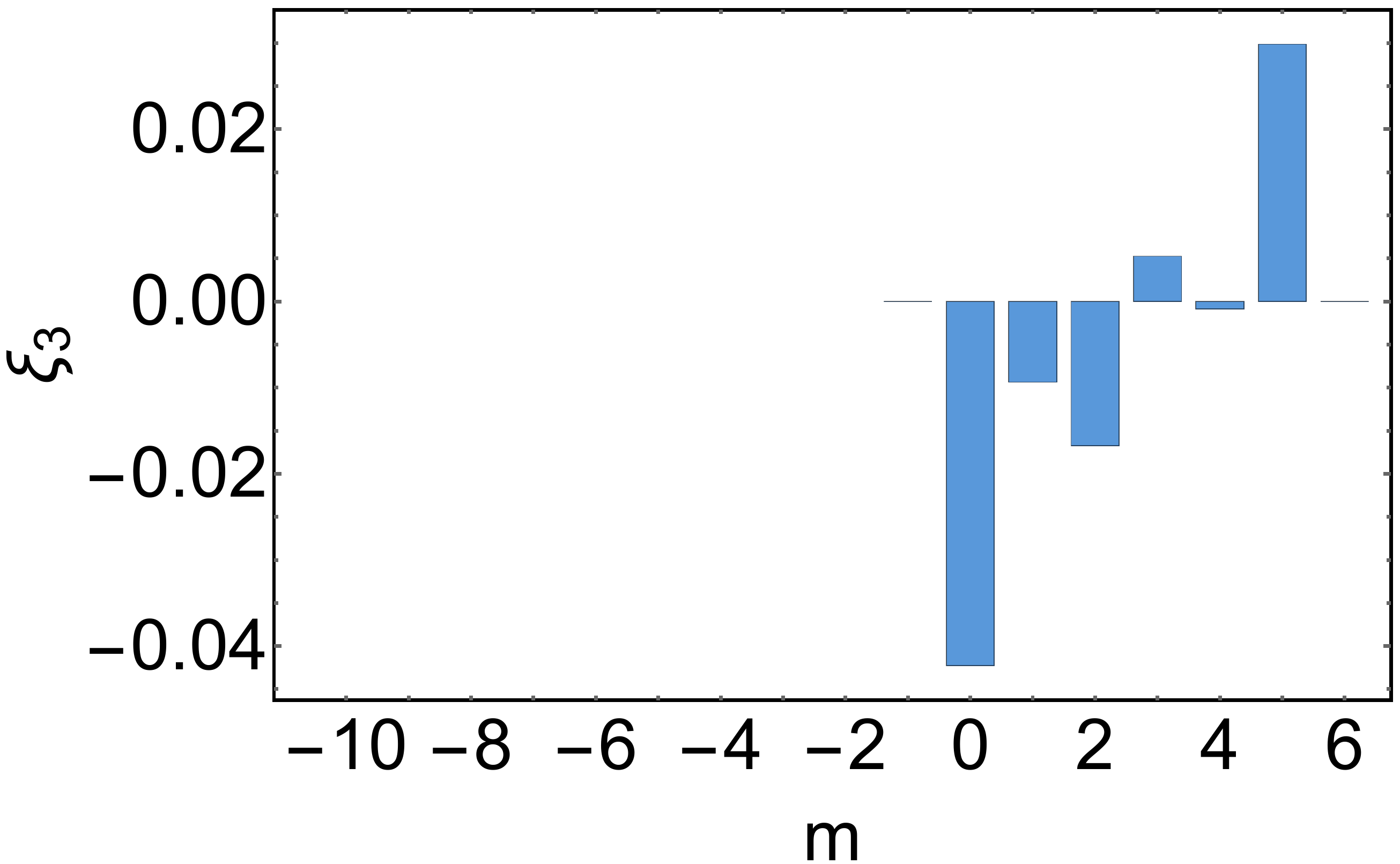}}\\
\raisebox{-0.5\height}{\includegraphics*[width=0.24\linewidth]{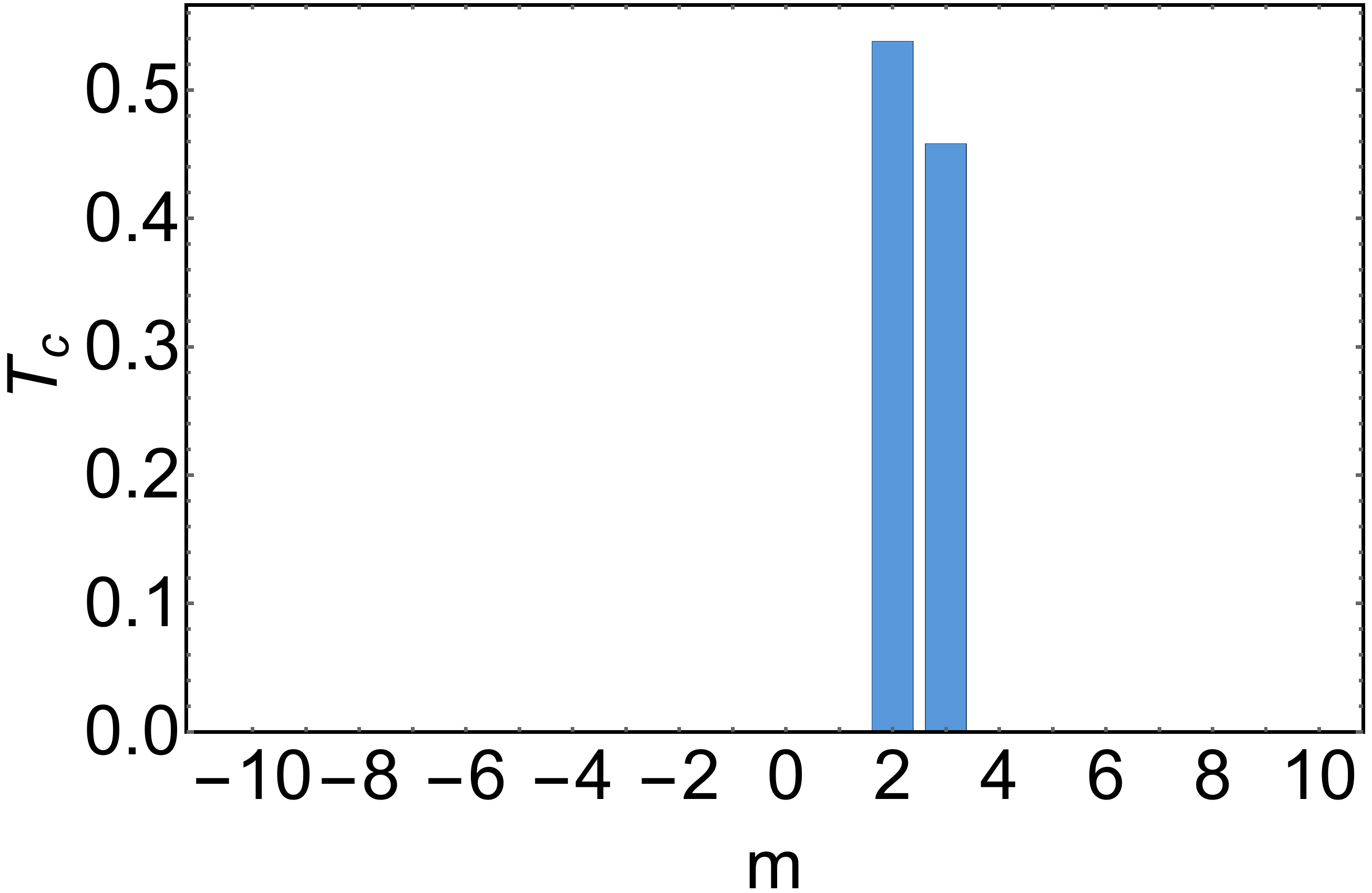}}\,
\raisebox{-0.5\height}{\includegraphics*[width=0.24\linewidth]{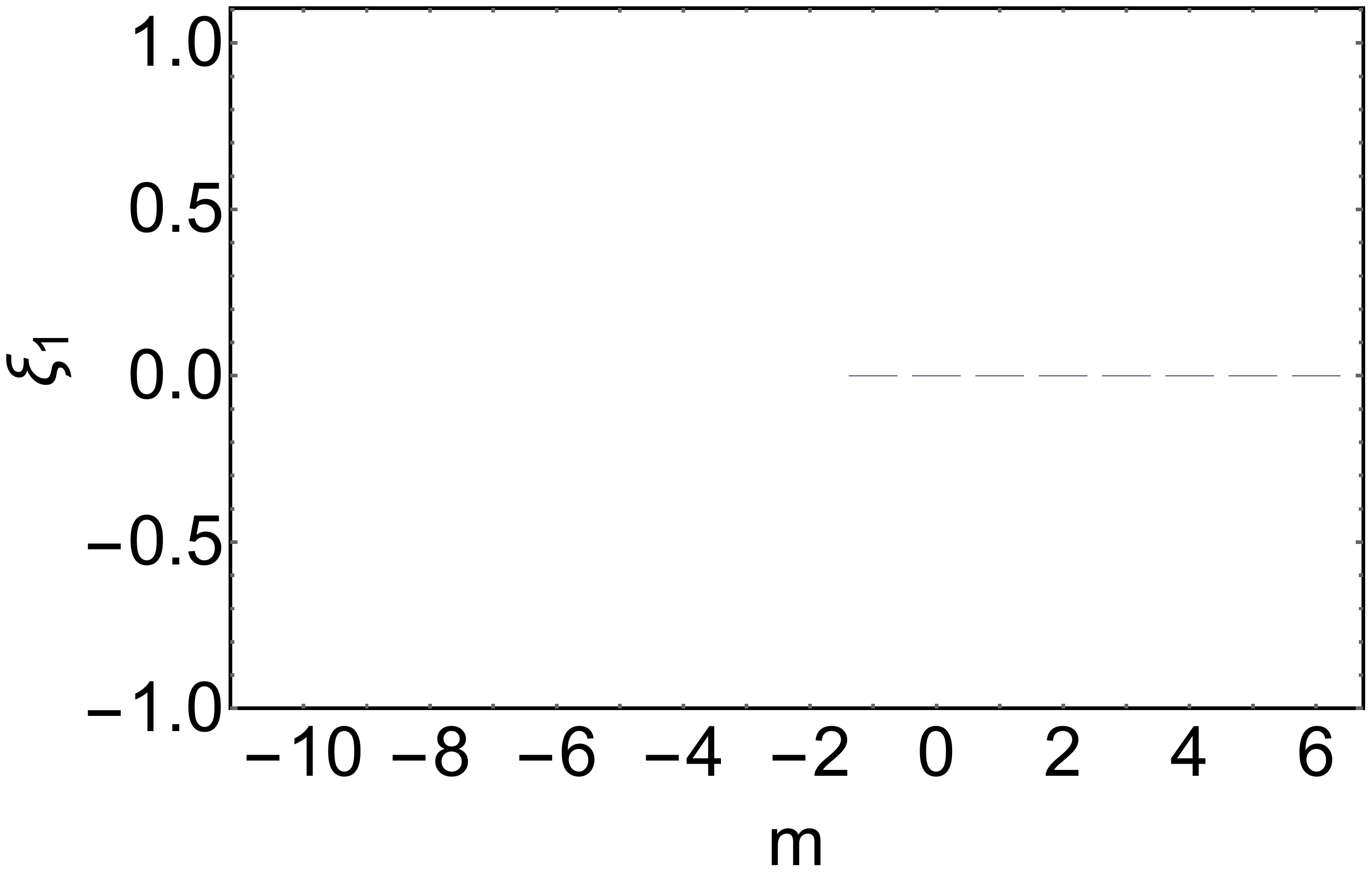}}\,
\raisebox{-0.5\height}{\includegraphics*[width=0.24\linewidth]{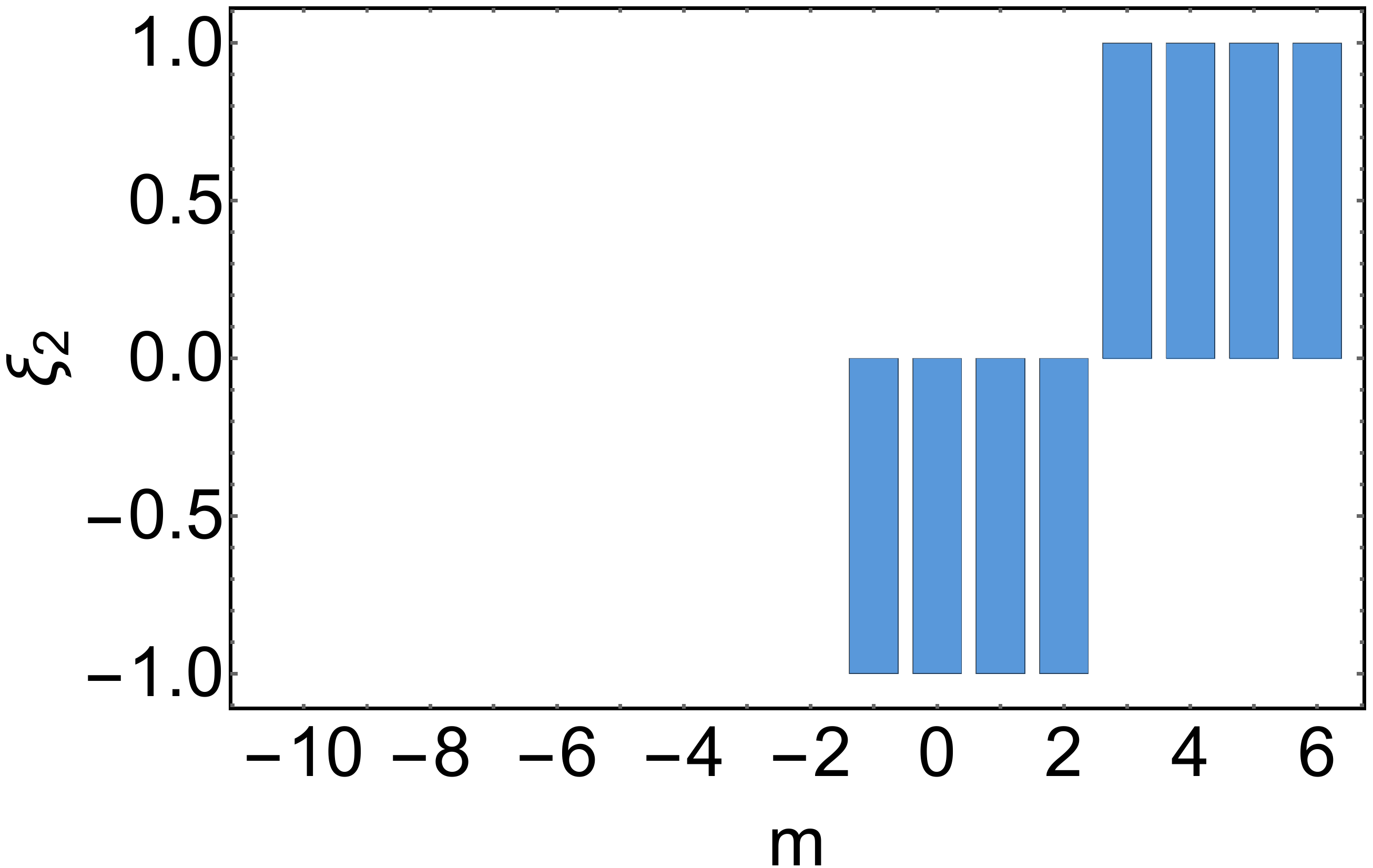}}\,
\raisebox{-0.5\height}{\includegraphics*[width=0.24\linewidth]{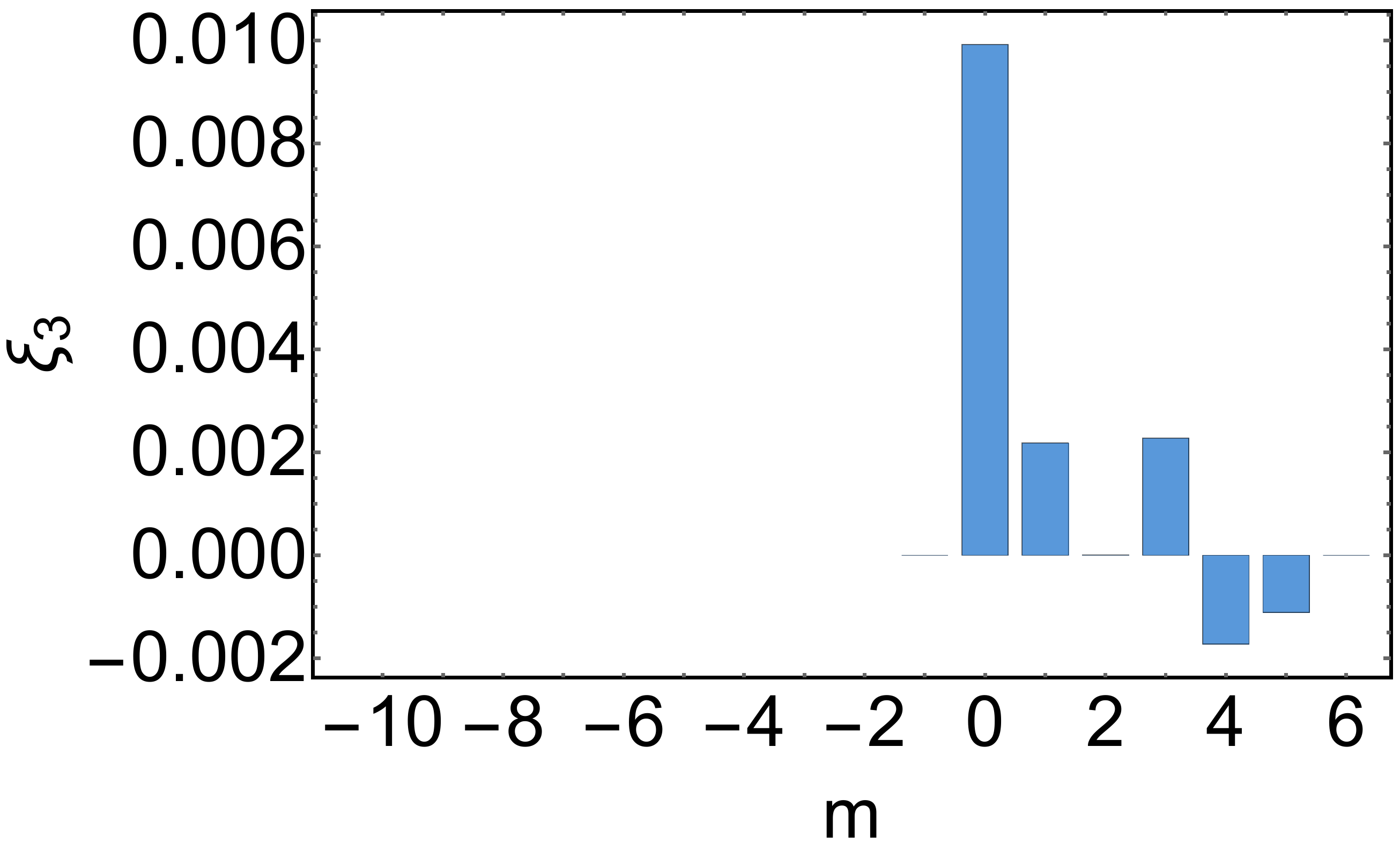}}
\caption{{\footnotesize The same as in Fig. \ref{Scatt_Hels2r_plots} but for the imaginary band gap corresponding to $s_h=\pm1$. The lines $1$-$4$: The case of plane-wave photons scattered in the $(x,z)$ plane is considered. The lines $1$-$2$: The initial photon possesses the helicity $s=1$. The lines $3$-$4$: It has $s=-1$. The lines $5$-$8$: Scattering of the twisted photon with $m=2$ is considered at the energy $k_0=2.7735$ eV belonging to the band gap. The lines $5$-$6$ corresponds to $s=1$, whereas the lines $7$-$8$ are for $s=-1$.  As is seen from the last three plots, there is a considerable part of transmitted twisted photons that acquire the additional projection of the total angular momentum $-s$ and the orbital angular momentum $s$. The projection of the total angular momentum of reflected twisted photons with $qs<0$ is shifted by $1$ in accordance with the selection rule \eqref{sel_rule}.}}
\label{Scatt_Hels1c_plots}
\end{figure}


\newpage
\begin{thebibliography}{999}

%tw telecom%Roadmap16,New18,OAMPM
\bibitem{Roadmap16}
H. Rubinsztein-Dunlop \textit{et al}.,
Roadmap on structured light,
J. Opt. \textbf{19}, 013001 (2017).

\bibitem{New18}
M. Erhard, R. Fickler, M. Krenn, A. Zeilinger,
Twisted photons: new quantum perspectives in high dimensions,
Light Sci. Appl. \textbf{7}, 17146 (2018).


%generated TP  geometric-phase and similar methods %SerboNew,OAMPM,PM,XPlate,CLCTP
\bibitem{OAMPM}
R. Chen, H. Zhou, M. Moretti, X. Wang, J. Li,
Orbital angular momentum waves: Generation, detection and emerging applications,
IEEE Communications Surveys \& Tutorials \textbf{22}, 840 (2020).

%other applications of tw
%applications twisted photons % SerboNew,New18,New19,OAMPM
%tw phot review%PadgOAM25,Roadmap16,SerboNew,New19
\bibitem{PadgOAM25}
M.~J. Padgett,
Orbital angular momentum 25 years on,
Opt. Express \textbf{25}, 11265 (2017).

\bibitem{SerboNew}
B.~A. Knyazev, V.~G. Serbo,
Beams of photons with nonzero projections of orbital angular momenta: New results,
Phys. Usp. \textbf{61}, 449 (2018).

\bibitem{New19}
Y. Shen \textit{et al}.,
Optical vortices 30 years on: OAM manipulation from topological charge to multiple singularities,
Light Sci. Appl. \textbf{8}, 90 (2019).

%detectors of tw phot%BLCBP,RGMMSCFR,Walsh,RGMMSCFR2
\bibitem{BLCBP}
G.~C.~G. Berkhout \textit{et al}.,
Efficient sorting of orbital angular momentum states of light,
Phys. Rev. Lett. \textbf{105}, 153601 (2010).

%detectors of tw phot
\bibitem{RGMMSCFR}
G. Ruffato \textit{et al.},
A compact difractive sorter for high-resolution demultiplexing of orbital angular momentum beams,
Sci. Rep. \textbf{8}, 10248 (2018).

%detectors of tw phot
%\delta l =30
\bibitem{Walsh}
G. F. Walsh \textit{et al}.,
Parallel sorting of orbital and spin angular momenta of light in a record large number of channels,
Opt. Lett. \textbf{43}, 2256 (2018).

%detectors of tw phot
%non-paraxial method
\bibitem{RGMMSCFR2}
G. Ruffato \textit{et al}.,
Non-paraxial design and fabrication of a compact OAM sorter in the telecom infrared,
Opt. Express \textbf{27}, 15750 (2019).

%generation of tw photon%reviews:Barboza2,LiZX21,SerboNew,PadgOAM25,OAMPM,Roadmap16
%q plates
\bibitem{Barboza2}
R. Barboza \textit{et al}.,
Optical vortex induction via light-matter interaction in liquid-crystal media,
Adv. Opt. Photon. \textbf{7}, 635 (2015).

%twisted photons on-chip. review
\bibitem{Fang21}
X. Fang \textit{et al}.,
Nanophotonic manipulation of optical angular momentum for high-dimensional information optics,
Adv. Opt. Photon. \textbf{13}, 772 (2021).

%generated twisted photons LC. review
\bibitem{LiZX21}
Z.-X. Li \textit{et al}.,
Liquid crystal devices for vector vortex beams manipulation and quantum information applications [Invited],
Chin. Opt. Lett. \textbf{19}, 112601 (2021).


%hel media. CLC
%opt prop%Barboza2,deGennProst,YangWu06,BelyakovBook,VetTimShab20
%CLC gen ref. mode functions
\bibitem{deGennProst}
P.~G. de Gennes, J. Prost,
\textsl{The Physics of Liquid Crystals}
(Clarendon Press, Oxford, 1993).

%CLC gen ref. mode functions
\bibitem{YangWu06}
D.-K. Yang, S.-T. Wu,
\textsl{Fundamentals of Liquid Crystal Devices}
(John Wiley \& Sons, Hoboken, 2006).

%optics of CLC. resonances. bound states. gen ref.
\bibitem{BelyakovBook}
V. Belyakov,
\textsl{Diffraction Optics of Complex-Structured Periodic Media}
(Springer, Cham, 2019).

%CLC mode functions. localized modes review
\bibitem{VetTimShab20}
S.~Ya. Vetrov, I.~V. Timofeev, V.~F. Shabanov,
Localized modes in chiral photonic structures,
Phys. Usp. \textbf{63}, 33 (2020).


%q-plates:LiZX21,Barboza2,MarManPap06,KPMS09,Barboza1,Naidoo16,Brasselet18
%q plates. first
\bibitem{MarManPap06}
L. Marrucci, C. Manzo, D. Paparo,
Optical spin-to-orbital angular momentum conversion in inhomogeneous anisotropic media,
Phys. Rev. Lett. \textbf{96}, 163905 (2006).

%q plates. quasiclassical wave propagation
\bibitem{KPMS09}
E. Karimi, B. Piccirillo, L. Marrucci, E. Santamato,
Light propagation in a birefringent plate with topological charge,
Opt. Lett. \textbf{34}, 1225 (2009).

%LC generation of tw. q plates
\bibitem{Barboza1}
R. Barboza \textit{et al}.,
Vortex induction via anisotropy stabilized light-matter interaction,
Phys. Rev. Lett. \textbf{109}, 143901 (2012).

%q plates. generation of entangl. states in OAM
\bibitem{Naidoo16}
D. Naidoo \textit{et al}.,
Controlled generation of higher-order Poincar\'{e} sphere beams from a laser,
Nat. Phot. \textbf{10}, 327 (2016).

%q plates
\bibitem{Brasselet18}
E. Brasselet,
Tunable high-resolution macroscopic self-engineered geometric phase optical elements,
Phys. Rev. Lett. \textbf{121}, 033901 (2018).


%chiral materials:LakhMess05,MacLakh,AskZhaAl14,FarLakh14,SemKhakh
%sculptured thin films. review. book
\bibitem{LakhMess05}
A. Lakhtakia, R. Messier,
Sculptured thin films: Nanoengineered Morphology and Optics
(SPIE, Bellingham, 2005).

%chiral metamaterials. review:
\bibitem{MacLakh}
T.~G. Mackay, A. Lakhtakia,
Negatively refracting chiral metamaterials: a review,
SPIE Reviews \textbf{1}, 018003 (2010).

%opt. prop. of chiral metamaterials
\bibitem{AskZhaAl14}
A.~N. Askarpour, Y. Zhao, A. Al\`{u},
Wave propagation in twisted metamaterials,
Phys. Rev. B \textbf{90}, 054305 (2014).

%Bragg scatt. in chiral sculptured thin films. review. 2 chiral plates
\bibitem{FarLakh14}
M. Faryad, A. Lakhtakia,
The circular Bragg phenomenon,
Adv. Opt. Photon. \textbf{6}, 225 (2014)

%sculptured thin films. metamaterials. review
\bibitem{SemKhakh}
I.~V. Semchenko, S.~A. Khakhomov,
\textsl{Electromagnetic waves in metamaterials and helical structures}
(Belaruskaya navuka, Minsk, 2019)
[in Russian].

%chiral metamaterials. review:HTTFB05,RobBroeBret,HodgWu01,SitBroeBret00,HPSB05,SchmSchuSchu13,OhHess15
\bibitem{OhHess15}
S.~S. Oh, O. Hess,
Chiral metamaterials: enhancement and control of optical activity and circular dichroism,
Nano Convergence \textbf{2}, 24 (2015).

%sculptured thin films. experiments.
\bibitem{SchmSchuSchu13}
D. Schmidt, E. Schubert, M. Schubert,
Generalized ellipsometry characterization of sculptured thin films made by glancing angle deposition,
In \textsl{Ellipsometry at the Nanoscale},
edited by M. Losurdo, K. Hingerl
(Springer, Berlin, 2013),
pp. 341-410.

%sculptured thin films. experiments. defects
\bibitem{HPSB05}
P.~C.~P. Hrudey, A.~C. van Popta, J.~C. Sit, M.~J. Brett,
Photonic device applications of nano-engineered thin film materials,
In \textsl{Nanoengineering: Fabrication, Properties, Optics, and Devices II},
edited by  E.~A. Dobisz, L.~A. Eldada
(SPIE, Bellingham, 2005),
Proc. SPIE \textbf{5931}, p. 593113.

%sculptured thin films. experiments.
\bibitem{SitBroeBret00}
J.~C. Sit, D.~J. Broer, M.~J. Brett,
Liquid crystal alignment and switching in porous chiral thin films,
Adv. Mater. \textbf{12}, 371 (2000).

%sculptured thin films. experiments.
\bibitem{HodgWu01}
I. Hodgkinson, Q.~h. Wu,
Inorganic chiral optical materials,
Adv. Mater. \textbf{13}, 889 (2001).

%sculptured thin films. experiments.
\bibitem{RobBroeBret}
K. Robbie, D.~J. Broer, M.~J. Brett,
Chiral nematic order in liquid crystals imposed by an engineered inorganic nanostructure,
Nature \textbf{399}, 764 (1999).

%sculptured thin films. experiments.
\bibitem{HTTFB05}
P.~C.~P. Hrudey, M. Taschuk, Y.~Y. Tsui, R. Fedosejevs, M.~J. Brett,
Optical properties of porous nanostructured Y${}_2$O${}_3$:Eu thin films,
J. Vac. Sci. Technol. A \textbf{23}, 856 (2005).


%helical dislocations
\bibitem{Weert57}
J. Weertman,
Helical dislocations,
Phys. Rev. \textbf{107}, 1259 (1957).

%gen ref. disl%Weert57,GrilheThes,Friedel64,KRS21
\bibitem{Friedel64}
J. Friedel,
\textsl{Dislocations}
(Pergamon Press, Oxford, 1964).

%helical dislocations
\bibitem{GrilheThes}
J. Grilhe,
Contribution \`{a} l'\'{e}tude des dislocations h\'{e}lico\"{\i}dales,
PhD thesis, Paris, L'Universit\'{e} de Paris, 1965.

%eff. dynamics of dislocations
\bibitem{KRS21}
P.~O. Kazinski, V.~A. Ryakin, A.~A. Sokolov,
Self-interaction of an arbitrary moving dislocation,
arXiv:2109.07331.

%V-Cher. and tr. radiation TF
\bibitem{BKL5}
O.~V. Bogdanov, P.~O. Kazinski, G.~Yu. Lazarenko,
Probability of radiation of twisted photons in the isotropic dispersive medium,
Phys. Rev. A \textbf{100}, 043836 (2019).

%helical medium. paraxial limit. exact solutions
\bibitem{LakhWeigh95}
A. Lakhtakia, W.~S. Weiglhofer,
On light propagation in helicoidal bianisotropic mediums,
Proc. Roy. Soc. London A \textbf{448}, 419 (1995).

%irred tensors
%helical media. irreduc. tensors
\bibitem{PontReyOld02}
S. Ponti, J.~A. Reyes, C. Oldano,
Homogeneous models for bianisotropic crystals,
J. Phys.: Condens. Matter \textbf{14}, 10173 (2002).

%opt prop hel med:LakhWeigh95,FarLakh14,MacLakh,LakhMess05,LaVeMcC00,BitThom05,FurAlex08,Lakht10,TChJhHu17,McCLakh04
%defects in chiral sculptured thin films. states in the gap
\bibitem{LaVeMcC00}
A. Lakhtakia, V.~C. Venugopal, M.~W. McCall,
Spectral holes in Bragg reflection from chiral sculptured thin films: circular polarization filters,
Opt. Commun. \textbf{177}, 57 (2000).

%helical media. additional properties
\bibitem{BitThom05}
I. Bita, E.~L. Thomas,
Structurally chiral photonic crystals with magneto-optic activity: indirect photonic bandgaps, negative refraction, and superprism effects,
J. Opt. Soc. Am. B \textbf{22}, 1199 (2005).

%helical media. pert theory. other approach
\bibitem{FurAlex08}
A.~N. Furs, T.~A. Alexeeva,
Reflection and transmission of weakly inhomogeneous anisotropic and bianisotropic layers calculated by perturbation method,
J. Phys. A: Math. Theor. \textbf{41}, 065203 (2008).

%helical media. additional properties
\bibitem{Lakht10}
A. Lakhtakia,
Reflection of an obliquely incident plane wave by a half space filled by a helicoidal bianisotropic medium,
Phys. Lett. A \textbf{374}, 3887 (2010).

%helical media. additional properties. additional gap.
\bibitem{TChJhHu17}
H.-T. Tung, Y.-K. Chen, P.-L. Jheng, Y.-C. Hung,
Origin and manipulation of band gaps in three-dimensional dielectric helix structures,
Opt. Express \textbf{25}, 17627 (2017).

%QED in medium
%books. medium:BKL5,AbrGorDzyal,GinzbThPhAstr,BKL2,parax,wkb_chol,KazLaz20
\bibitem{AbrGorDzyal}
A.~A. Abrikosov, L.~P. Gorkov, I.~E. Dzyaloshinskii,
\textsl{Methods of Quantum Field Theory in Statistical Physics}
(Prentice Hall Press, New Jersey, 1963).

%trans rad gen ref%Ginzburg,BazylZhev,GinzbThPhAstr
%theory TR and VCh radiation % Ginzburg, VCh, Pbook, x-ray
%anom Doppl eff
\bibitem{GinzbThPhAstr}
V.~L. Ginzburg,
\textsl{Theoretical Physics and Astrophysics}
(Pergamon, London, 1979).

%%tw phots by class currents
\bibitem{BKL2}
O.~V. Bogdanov, P.~O. Kazinski, G.~Yu. Lazarenko,
Probability of radiation of twisted photons by classical currents,
Phys. Rev. A \textbf{97}, 033837 (2018).

%parax
\bibitem{parax}
O.~V. Bogdanov, P.~O. Kazinski, P.~S. Korolev, G.~Yu. Lazarenko,
Radiation of twisted photons from charged particles moving in cholesterics,
J. Mol. Liq. \textbf{326}, 115278 (2021).

%wkb trans radiation
\bibitem{wkb_chol}
O.~V. Bogdanov, P.~O. Kazinski, P.~S. Korolev, G.~Yu. Lazarenko,
Generation of hard twisted photons by charged particles in cholesteric liquid crystals,
Phys. Rev. E \textbf{104}, 024701 (2021).

%quant trans rad
\bibitem{KazLaz20}
P.~O. Kazinski, G.~Yu. Lazarenko,
Transition radiation from a Dirac particle wave packet traversing a mirror,
Phys. Rev. A \textbf{103}, 012216 (2021).

%tw mode functions in medium
\bibitem{BKL6}
O.~V. Bogdanov, P.~O. Kazinski, G.~Yu. Lazarenko,
Generation of twisted photons by undulators filled with dispersive medium,
Eur. Phys. J. Plus \textbf{135}, 901 (2020).

%CLC nonparax bands
%bands at an angle%BerrSchef,RisSchm19
%band structure. theor (numerical). experiment. large angles
\bibitem{BerrSchef}
D.~W. Berreman, T.~J. Scheffer,
Bragg reflection of light from single-domain cholesteric liquid-crystal films,
Phys. Rev. Lett. \textbf{25}, 577 (1970).

%bound states. arb. angles
\bibitem{RisSchm19}
A.~M. Risse, J. Schmidtke,
Angular-dependent spontaneous emission in cholesteric liquid crystal films,
J. Phys. Chem. C \textbf{123}, 2428 (2019).

%3J integrals:Vilenkin3J,JackMax71
\bibitem{Vilenkin3J}
N.~J. Vilenkin,
Special functions and the theory of group representations,
Translations of Mathematical Monographs \textbf{22},
American Mathematical Society, Providence, 1968.

%3J integrals
\bibitem{JackMax71}
A.~D. Jackson, L.~C. Maximon,
Integrals of products of Bessel functions,
SIAM J. Math. Anal. \textbf{3}, 446 (1971).

%paraxial limit. solutions
%helical media. additional properties
\bibitem{McCLakh04}
M.~W. McCall, A. Lakhtakia,
Explicit expressions for spectral remittances of axially excited chiral sculptured thin films,
J. Mod. Opt. \textbf{51}, 111 (2004).

%phys. kinetiks
\bibitem{LLPhysKin}
E.~M. Lifshitz, L.~P. Pitaevskii,
\textsl{Physical Kinetics}
(Butterworth Heinemann, Oxford, 1981).


%other schemes%FarLakh14,RafBrass18,LinT19
%generated twisted photons. CLC plate and mirror
\bibitem{RafBrass18}
M. Rafayelyan, E. Brasselet,
Spin-to-orbital angular momentum mapping of polychromatic light,
Phys. Rev. Lett. \textbf{120}, 213903 (2018).

%generated twisted photons. two CLC plates
\bibitem{LinT19}
T. Lin \textit{et al}.,
Bragg reflective polychromatic vector beam generation from opposite-handed cholesteric liquid crystals,
Opt. Lett. \textbf{44}, 2720 (2019).







\end{thebibliography}
\end{document}